\documentclass{illcdiss}
\makeindex 
\usepackage[utf8]{inputenc}
\usepackage{hyperref}

\usepackage{amsmath}
\usepackage{amssymb}
\usepackage{bbold}
\usepackage{authblk}
\usepackage{bm}
\usepackage{graphicx}
\usepackage{xcolor}
\usepackage{multirow}
\usepackage{caption}  
\usepackage{booktabs}
\usepackage{algorithm}
\usepackage{algpseudocode}

\begin{document}
\pagestyle{plain}
\pagenumbering{roman}

{\pagestyle{empty}
\newcommand{\printtitle}{%
{\Huge\bf Fusing heterogeneous data sets}}  

\begin{titlepage}
\par\vskip 2cm
\begin{center}
\printtitle
\vskip 2cm
{\LARGE\bf Yipeng Song}                           
\vfill
\end{center}
\end{titlepage}

\newpage
\noindent%
{
\vfill
\noindent%
Fusing heterogeneous data sets \\
Yipeng Song \\
PhD thesis \\
ISBN: 978-94-6375-531-3 \\[2ex]     

\noindent%
All rights reserved. No part of this publication may be reproduced in any form
without written permission from the copyright owner.\\[2ex]

\noindent%
Copyright \copyright\ Yipeng Song, 2019 \\[2ex] 

\vskip 2cm
}

\clearpage
\par\vskip 2cm
\begin{center}
\printtitle
\par\vspace {6cm}
{\large \sc Academisch Proefschrift}
\par\vspace {1cm}
{\large ter verkrijging van de graad van doctor\\
aan de Universiteit van Amsterdam\\
op gezag van de Rector Magnificus\\
prof. dr. ir. K.I.J. Maex \\                                 
ten overstaan van een door het College voor Promoties ingestelde\\
\mbox{commissie, in het openbaar te verdedigen in de Aula der Universiteit}\\        
op vrijdag 27 september 2019 om 11.00 uur \\ }        
\par\vspace {1cm} {\large door}
\par \vspace {1cm} 
{\Large Yipeng Song \\}                        
\par\vspace {1cm} 
{\large geboren te Henan} 
\end{center}

\clearpage
\par\vskip 2cm

\noindent%
{\bf Promotiecommisie}\\
\\
\begin{tabular}[t]{@{}lll}
Promotor:      & Prof. dr. A. K. Smilde & Universiteit van Amsterdam \\  
Co-promotor:   & Dr. J. A. Westerhuis   & Universiteit van Amsterdam \\  
\\
Overige leden: & Prof. dr. A.H.C. van Kampen  & Universiteit van Amsterdam \\  
               & Prof. dr. A.H. Zwinderman  & Universiteit van Amsterdam \\  
               & Prof. dr. P.M.A. Sloot  & Universiteit van Amsterdam \\  
               & Prof. dr. M.E. Timmerman & Rijksuniversiteit Groningen \\
               & Prof. dr. ir. M.J.T. Reinders & Technische Universiteit Delft \\
               & Prof. dr. T. Naes  & Nofima
\end{tabular}\\
\\

\vfill

\noindent
Faculteit der Natuurwetenschappen, Wiskunde en Informatica\\[2ex] 

\noindent
The research reported in this thesis was carried out at the Swammerdam Institute
for Life Sciences, Faculty of Science, University of Amsterdam. Yipeng Song gratefully acknowledges the financial support from China Scholarship Council (NO.201504910809).
\vskip 2cm
}

\thispagestyle{plain}
\clearpage
\mbox{}
\vspace{2in}
\begin{center}
{\em To my mother Jin e}
\end{center}

\tableofcontents

\cleardoublepage
\pagestyle{headings}
\pagenumbering{arabic}

\abstract{
In systems biology, it is common to measure biochemical entities at different levels of the same biological system. One of the central problems for the data fusion of such data sets is the heterogeneity of the data. This thesis discusses two types of heterogeneity. The first one is the type of data, such as metabolomics, proteomics and RNAseq data in genomics. These different omics data reflect the properties of the studied biological system from different perspectives. The second one is the type of scale, which indicates the measurements obtained at different scales, such as binary, ordinal, interval and ratio-scaled variables. In this thesis, we developed several statistical methods capable to fuse data sets of these two types of heterogeneity. The advantages of the proposed methods in comparison with other approaches are assessed using comprehensive simulations as well as the analysis of real biological data sets.}

\chapter{Introduction} \label{chapter:1}
In systems biology, it is becoming increasingly common to measure biochemical entities at different levels of the same biological system. Hence, data fusion problems, which focus on analyzing such data sets simultaneously to arrive at a holistic understanding of the studied system, are abundant in the life sciences. With the availability of a multitude of measuring techniques, one of the central problems is the heterogeneity of the data. In this thesis, we mainly discuss two types of heterogeneity. The first one is the type of \emph{data}, such as metabolomics, proteomics and RNAseq data in genomics. These different omics data reflect the properties of the studied biological system from different perspectives. The second one is the type of \emph{scale}, which indicates the measurements obtained at different scales, such as binary, ordinal, interval and ratio-scaled variables. In genomics, an example is the measurements of gene-expression and point mutation status on the same objects. The latter are binary data and gene-expression measurements are quantitative data. Ideally, data fusion methods should consider these two types of heterogeneity of such measurements and this will be the topic of this thesis.

The goal of this thesis is to develop appropriate statistical methods capable to fuse data sets of the two types of heterogeneity. Before going into the details of the developed methods, we begin with a brief introduction of the concept of data fusion in life sciences and the characteristics of the two types of heterogeneity. Another important concept in this thesis is the concave penalty, which is the basis of all the developed methods. Although it is not directly related to the fusion of heterogeneous data sets, it is also introduced in Chapter \ref{chapter:1}. \footnote{This chapter is based on Smilde, A.K., Song, Y., Westerhuis, J.A., Kiers, H.A., Aben, N. and Wessels, L.F., 2019. Heterofusion: Fusing genomics data of different measurement scales. arXiv preprint arXiv:1904.10279.}

\section{Data fusion in life sciences}
With the availability of comprehensive measurements collected in multiple related data sets in the life sciences, the need for a simultaneous analysis of such data to arrive at a global view on the system under study is of increasing importance. There are many ways to perform such a simultaneous analysis and these go also under very different names in different areas of data analysis: data fusion, data integration, global analysis, multi-set or multi-block analysis to name a few. We will use the term \emph{data fusion} in this thesis. Data fusion plays an especially important role in the life sciences, e.g., in genomics it is not uncommon to measure gene-expression (array data or RNAseq data), methylation of DNA and copy number variation. Sometimes, also proteomics and metabolomics measurements are available. All these examples serve to show that having methods in place to integrate these data is not a luxury anymore.

Without trying to build a rigorous taxonomy of data fusion it is worthwhile to distinguish several distinctions in data fusion. The first distinction is between model-based and exploratory data fusion. The former uses background knowledge in the form of models to fuse the data; one example being genome-scale models in biotechnology \cite{zimmermann2017integration}. The latter does not rely on models, since these are not available or poorly known, and thus uses empirical modeling to explore the data. In this thesis, we will focus on exploratory data fusion. The next distinction is between low-, medium-, and high-level data fusion \cite{steinmetz1999methodology}. In low-level data fusion, the data sets are combined at the lowest level, that is, at the level of the (preprocessed) measurements. In medium-level data fusion, each separate data set is first summarized, e.g., by using a dimension reduction method or through variable selection. The reduced data sets are subsequently subjected to the data fusion. In high-level data fusion, each data set is used for prediction or classification of an outcome and the prediction or classification results are then combined, e.g, by using majority voting \cite{doeswijk2011increase}. All these types of data fusion have advantages and disadvantages which are beyond the scope of this thesis. In this thesis, we will focus on low- and medium-level fusion.

The final characteristic of data fusion relevant for this thesis is the heterogeneity of the data sets to be fused. Different types of heterogeneity can be distinguished. The first one is the type of \emph{data}, such as metabolomics, proteomics and RNAseq data in genomics. Clearly, these data relate to different parts of the biological system. The second one is the type of \emph{scale} in which the data are measured present in the fusion problem. In genomics, an example is a data set where gene-expressions are available and mutation data in the form of single nucleotide polymorphisms(SNPs). The latter are binary data and gene-expression measurements are quantitative data. They are clearly measured at a different scale. Ideally, data fusion methods should consider these two levels of heterogeneity in data analysis and this will be the topic of this thesis. In the following section, we will show the characteristics of these two types of heterogeneity and how they affect the data analysis.

\section{Two types of heterogeneity}

\subsection{Heterogeneous measurement scales in multiple data sets}
Multiple data sets measured on the same objects may have different types of measurement scales. The history of measurement scales goes back a long time. A seminal paper drawing attention to this issue appeared in the 40-ties \cite{Stevens1946}. Since then numerous papers, reports and books have appeared \cite{Suppes1962,Krantz1971,Narens1981,Narens1986,Luce1987,Hand1996}. The measuring process assigns numbers to aspects of objects (an \textit{empirical system}), e.g, weights of persons. Hence, measurements can be regarded as a mapping from the empirical system to numbers, and scales are properties of these mappings. In measurement theory, there are two fundamental theorems \cite{Krantz1971}: the representation theorem and the uniqueness theorem. The \textit{representation theorem} asserts the axioms to be imposed on an empirical system to allow for a homomorphism of that system to a set of numerical values. Such a homomorphism into the set of real numbers is called a scale and thus represents the empirical system. A scale possesses \textit{uniqueness} properties: we can measure the weight of persons in kilograms or in grams, but if one person weighs twice as much as another person, this ratio holds true regardless the measurement unit. Hence, weight is a so-called ratio-scaled variable and this ratio is unique. The transformation of measuring in grams instead of kilograms is called a \textit{permissible} transformation since it does not change the ratio of two weights. For a ratio-scaled variable, only similarity transformations are permissible; i.e. $\widetilde{x}=\alpha x; \alpha>0$ where $x$ is the variable on the original scale and $\widetilde{x}$ is the variable on the transformed scale. This is because
\begin{equation*}\label{eRatio}
 \frac{\widetilde{x_i}}{\widetilde{x_j}}=\frac{\alpha x_i}{\alpha x_j}=\frac{x_i}{x_j}.
\end{equation*}
Note that this coincides with the intuition that the unit of measurement is immaterial.

The next level of scale is the interval-scaled measurement. The typical example of such a scale is concentrations of metabolites in metabolomics research and the permissible transformation is affine, i.e. $\widetilde{x}=\alpha x +\beta; \alpha>0$. In that case, the ratio of two intervals is unique because
\begin{equation*}\label{eInterval}
 \frac{\widetilde{x_i}-\widetilde{x_j}}{\widetilde{x_k}-\widetilde{x_l}}=\frac{(\alpha x_i + \beta)-(\alpha x_j + \beta)}{(\alpha x_k + \beta)-(\alpha x_l + \beta)}=\frac{\alpha (x_i-x_j)}{\alpha (x_k-x_l)}=\frac{x_i-x_j}{x_k-x_l}.
\end{equation*}
Stated differently, the zero point and the unit are arbitrary on this scale.

Ordinal-scaled variables represent the next level of measurements. Typical examples are scales of agreement in surveys: strongly disagree, disagree, neutral, agree and strongly agree. There is a rank-order in these answers, but no relationship in terms of ratios or intervals. The permissible transformation of an ordinal-scale is a monotonic increasing transformation since such transformations keep the order of the original scale intact. Nominal-scaled variables are next on the list. These variables are used to encode categories and are sometimes also called categorical. Typical examples are gender, race, brands of cars and the like. The only permissible transformation for a nominal-scaled variable is the one-to-one mapping. A special case of a nominal-scaled variable is the binary (0/1) scale. Binary data can have different meanings; they can be used as categories (e.g. gender) and are then nominal-scale variables. They can also be two points on a higher-level scale, such as absence and presence (e.g. for methylation data).

The above four scales are the most used ones but others exist \cite{Suppes1962,Krantz1971}. Counts, e.g., have a fixed unit and are therefore sometimes called absolute-scaled variables \cite{Narens1986}. Another scale is the one for which the power transformation is permissible; i.e. $\widetilde{x}=\alpha x^\beta; \alpha, \beta>0$ which is called a log-interval scale because a logarithmic transformation of such a scale results in an interval-scale. An example is density \cite{Krantz1971}. Sometimes the scales are lumped in quantitative (i.e. ratio and interval) and qualitative (ordinal and nominal) data.

An interesting aspect of measurement scales is to what extent meaningful statistics can be derived from such scales (see Table 1 in \cite{Stevens1946}). A prototypical example is using a mean of a sample of nominal-scaled variables which is generally regarded as being meaningless. This has also provoked a lot of discussion \cite{Adams1965,Hand1996} and there are nice counter-examples of apparently meaningless statistics that still convey information about the empirical system \cite{Michell1986}. As always, the world is not black or white.

In practice, we also use other taxonomies to classify the types of measurements \cite{agresti2013categorical}. A commonly used one is the Discrete-Continuous variable distinction according to whether or not the possible number of values is countable. Therefore, binary, nominal and ordinal scaled measurements are discrete while ratio and interval scaled measurements are continuous. Another commonly used taxonomy is the Quantitative-Qualitative variable distinction, which depends on whether two different measurements differ in quality or in quantity. Thus nominal scaled measurements are qualitative while ratio and interval scaled measurements are quantitative. And the ordinal scaled measurements have the characteristics of both quantitative and quantitative variables.

\subsection{Heterogeneous information in multiple data sets}
Multiple sets of measurements on the same objects obtained from different platforms may reflect partially complementary information of the studied system. Therefore, these multiple data sets may contain heterogeneous information, the information that is common across all or some of the data sets, and the information which is specific to each data set (often called distinct). The challenge for the data fusion of such data sets is how to separate the common and distinct information existed in multiple data sets. These different sources of information have to be disentangled from every data set to have a holistic understanding of the studied system. Here we focus on using common and distinct components to approximate the common and distinct variation (information) existing in multiple data sets measured on the same objects \cite{smilde2017common}. We will use a simultaneous component analysis (SCA) model with structural sparsity patterns on the loading matrix \cite{gaynanova2017structural} as an example to show the idea.

A classical SCA model tries to discover the common subspace between multiple data sets to represent the common information between these data sets. Suppose the quantitative measurements from $L$ different platforms on the same $I$ objects result into $L$ quantitative data sets, $\left\{\mathbf{X}_l \right\}_{l=1}^{L}$, and the $l^{\text{th}}$ data set $\mathbf{X}_l$($I \times J_l$) has $J_l$ variables. After these data sets are column centered, we can decompose them in the SCA model framework as $\mathbf{X}_l = \mathbf{AB}_l^{\text{T}} + \mathbf{E}_l$, in which $\mathbf{A}$($I\times R$) is the common score matrix; $\mathbf{B}_l$($J_l\times R$) and $\mathbf{E}_l$($I\times J_l$) are the loading matrix and residual term respectively for $\mathbf{X}_l$ and $R$ is the number of components. The common column subspace, which is spanned by the columns of the score matrix $\mathbf{A}$, represents the common information between these $L$ data sets.

The drawback of the SCA model is that only the global common components, which account for the common variation across all the data sets, is modeled. However, the real situation in multiple data sets integration can be far more complex as local common variation across some of the data sets and distinct variation in each data set are expected as well. With the help of the concept of structural sparsity on the loading matrices of a SCA model, we can interpret the common and distinct variation framework as follows. Suppose we construct a SCA model on three column centered quantitative data sets $\left\{\mathbf{X}_l \right\}_{l=1}^{3}$, the common score matrix is $\mathbf{A}$, the corresponding loading matrices are $\left\{\mathbf{B}_l \right\}_{l=1}^{3}$, and $\mathbf{X}_l = \mathbf{A}\mathbf{B}_{l}^{\text{T}} + \mathbf{E}_l$, in which $\mathbf{E}_l$ is the residual term for $l^{\text{th}}$ data set. If the structured sparsity pattern in $\left\{\mathbf{B}_l \right\}_{l=1}^{3}$ is expressed as follows,
\begin{equation*}
\begin{aligned}
   \left(
                 \begin{array}{c}
                   \mathbf{B}_1 \\
                   \mathbf{B}_2 \\
                   \mathbf{B}_3 \\
                 \end{array}
               \right)
               = \left(
                   \begin{array}{cccccccc}
                     \mathbf{b}_{1,1} & \mathbf{b}_{1,2} & \mathbf{b}_{1,3} & \mathbf{0}       & \mathbf{b}_{1,5} & \mathbf{0}       & \mathbf{0}      \\
                     \mathbf{b}_{2,1} & \mathbf{b}_{2,2} & \mathbf{0}       & \mathbf{b}_{2,4} & \mathbf{0}       & \mathbf{b}_{2,6} & \mathbf{0}       \\
                     \mathbf{b}_{3,1} & \mathbf{0}       & \mathbf{b}_{3,3} & \mathbf{b}_{3,4} & \mathbf{0}       & \mathbf{0}       & \mathbf{b}_{3,7} \\
                   \end{array}
                 \right),
\end{aligned}
\end{equation*}
in which $\mathbf{b}_{l,r} \in \mathbf{R}^{J_l}$ indicates the $r^{\text{th}}$ column of the $l^{\text{th}}$ loading matrix $\mathbf{B}_l$, then we have the following relationships,
\begin{equation*}
\begin{aligned}
   \mathbf{X}_1 & = \mathbf{a}_1\mathbf{b}_{1,1}^{\text{T}} &+& \mathbf{a}_2\mathbf{b}_{1,2}^{\text{T}} &+& \mathbf{a}_3\mathbf{b}_{1,3}^{\text{T}} &+& \mathbf{0}                     &+& \mathbf{a}_5\mathbf{b}_{1,5}^{\text{T}} &+& \mathbf{0}                     &+& \mathbf{0}  &+& \mathbf{E}_1                   \\
   \mathbf{X}_2 & = \mathbf{a}_1\mathbf{b}_{2,1}^{\text{T}} &+& \mathbf{a}_2\mathbf{b}_{2,2}^{\text{T}} &+& \mathbf{0}                     &+& \mathbf{a}_4\mathbf{b}_{2,4}^{\text{T}} &+& \mathbf{0}                     &+& \mathbf{a}_6\mathbf{b}_{2,6}^{\text{T}} &+& \mathbf{0}  &+& \mathbf{E}_2                   \\
   \mathbf{X}_3 & = \mathbf{a}_1\mathbf{b}_{3,1}^{\text{T}} &+& \mathbf{0}                     &+& \mathbf{a}_3\mathbf{b}_{3,3}^{\text{T}} &+& \mathbf{a}_4\mathbf{b}_{3,4}^{\text{T}} &+& \mathbf{0}                     &+& \mathbf{0}                     &+& \mathbf{a}_7\mathbf{b}_{3,7}^{\text{T}} &+& \mathbf{E}_3 .
\end{aligned}
\end{equation*}
Here $\mathbf{a}_r$ indicates the $r^{\text{th}}$ column of the common score matrix $\mathbf{A}$. The first component represents the global common variation across three data sets; the $2^{\text{nd}}$, $3^{\text{nd}}$ and $4^{\text{nd}}$ components represent the local common variation across two data sets and the $5^{\text{nd}}$, $6^{\text{nd}}$ and $7^{\text{nd}}$ components represent the distinct variation specific to each single data set. Therefore, the heterogeneity of data (common and distinct information existed in multiple data sets) is disentangled by the common and distinct components. In the above example, we only show a single score and loading vector for any specific variation, but that there can be multiple score and loading vcectors for each (global common, local common or distinct variations).

Except for the above approach, there are also many other methods for distinguishing common and distinct components. However, the above approach has the advantage of estimating the model complexity directly which is problematic in most other methods. Details will be shown in Chapter \ref{chapter:5}.

\section{Using concave penalties to induce sparsity}
Although concave penalties are not directly related to the fusion of heterogeneous data sets, they are the basis of all the developed methods in this thesis. Therefore, it is better to have a general introduction to them at the beginning.

The comprehensive measurements in the current biological research always result in high dimensional data sets, of which the number of variables is much larger than the number of samples. For the analysis of such high dimensional data sets, sparse parameter estimation (many estimated parameters are exactly 0) is always desired since it makes both the data analysis problem feasible and the results easier to be interpreted. Some typical examples of sparse parameter estimation include the sparse regression models \cite{fan2001variable}, the low rank matrix approximation problems \cite{jolliffe2002principal, smilde2017common, gavish2017optimal}, the structure learning problems in graphical models \cite{friedman2008sparse}, and many others \cite{tibshirani2005sparsity, witten2009penalized,huang2012selective}.

We can use a linear regression model as an example to illustrate how to achieve sparse parameter estimation through various penalties. Suppose we have a univariate response variable $y\in \mathbb{R}$ and a multivariate explanatory variable $\mathbf{x} \in \mathbb{R}^{J}$. A standard linear regression model can be expressed as $y = \mathbf{x}^{\text{T}}\bm{\beta} + \epsilon$, in which $\bm{\beta} \in \mathbb{R}^{J}$ is the coefficient vector and $\epsilon \in \mathbb{R}$ is the error term following a normal distribution with mean $0$ and variance $\sigma^2$, $\epsilon \sim N(0,\sigma^2)$. After obtaining $I$ samples of $\{y, \mathbf{x}\}$, we have the data sets $\{y_i, \mathbf{x}_i\}_{i=1}^{I}$, which can be further expressed in their vector and matrix form as $\mathbf{y} \in \mathbb{R}^{I}$ and $\mathbf{X} \in \mathbb{R}^{I\times J}$. The optimization problem associated with the standard linear model is $\min_{\bm{\beta}} ~ \frac{1}{2}||\mathbf{y} - \mathbf{X}\bm{\beta}||^2$, and the analytical form solution is $\hat{\bm{\beta}} = (\mathbf{X}^{\text{T}}\mathbf{X})^{-1}\mathbf{X}^{\text{T}}\mathbf{y}$. Unfortunately, this model is unidentifiable when $J>I$ and ill-conditioned when the explanatory variables are correlated. Also the estimated coefficient vector $\hat{\bm{\beta}}$ is always dense, which makes the interpretation difficult.

\begin{figure}[htbp]
    \centering
    \includegraphics[width=0.5\textwidth]{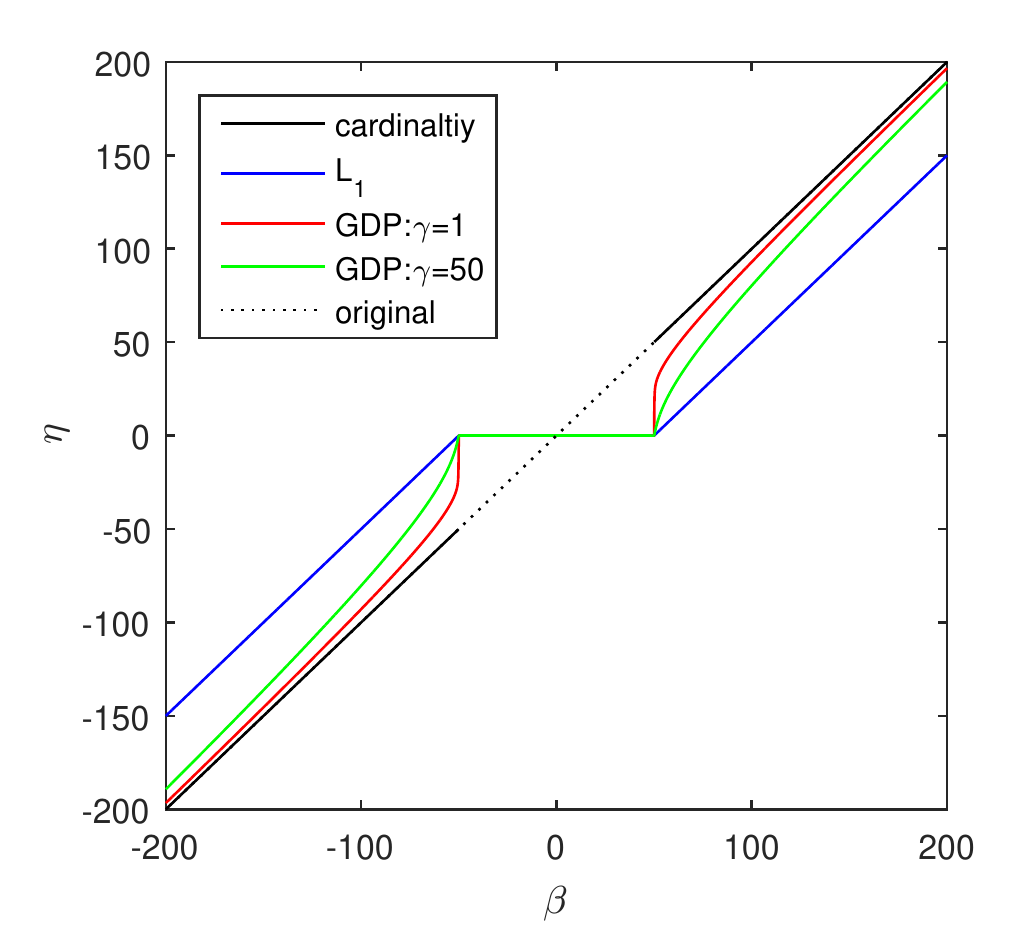}
    \caption{Thresholding properties of the cardinality constraint, $L_{1}$ norm, GDP penalties when the same degree of shrinkage is achieved. Legend cardinality: cardinality constraint, legend $L_{1}$: $L_{1}$ norm penalty, legend original: the original values before thresholding. $\beta$ in $x$ axis indicates the original value of the coefficient while $\eta$ in $y$ axis is the value after thresholding.}
    \label{chapter1_fig:1}
\end{figure}

Cardinality constraint can be imposed on the linear regression model as a hard constraint to induce a sparse parameter estimation of $\bm{\beta}$. If we require only $R$ elements of $\bm{\beta}$ to be nonzero, the associated optimization problem can be expressed as $\min_{\bm{\beta}} ~ \frac{1}{2}||\mathbf{y} - \mathbf{X}\bm{\beta}||^2 ~ \text{subject to} ~ ||\bm{\beta}||_0 = R$, in which $||\bm{\beta}||_0 = R$ is the cardinality constraint, $||~||_0$ indicates the pseudo $L_{0}$ norm and counts the number of nonzero elements. Since the above optimization problem is non-convex and difficult to solve, the cardinality constraint is always replaced by its convex relaxation $L_1$ norm to induce the sparsity, which results in the lasso regression model \cite{tibshirani1996regression}. The optimization problem associated with the lasso regression model is $\min_{\bm{\beta}} ~ \frac{1}{2}||\mathbf{y} - \mathbf{X}\bm{\beta}||^2 + \lambda ||\bm{\beta}||_1$ in which $\lambda$ is the tuning parameter and $||~||_1$ indicates the $L_{1}$ norm. Efficient algorithms exist to solve this convex optimization problem \cite{efron2004least,parikh2014proximal}. However, the $L_{1}$ norm penalty shrinks all the elements of the coefficient vector $\bm{\beta}$ to the same degree. This leads to a biased estimation of the coefficients with large absolute values. This behavior will further make the prediction error or CV error based model selection procedure inconsistent \cite{meinshausen2010stability}. Many non-convex penalties, most of them are concave functions with respect to the absolute value of $\bm{\beta}$, have been proposed to tackle the drawback of the $L_{1}$ norm penalty \cite{fan2001variable,armagan2013generalized}. They can shrink the parameters in a nonlinear manner to achieve both nearly unbiased and sparse parameter estimation. A frequentist version of the generalized double Pareto (GDP) \cite{armagan2013generalized} shrinkage can serve as an example. The optimization problem of a linear regression model with the GDP penalty can be expressed as $\min_{\bm{\beta}} ~ \frac{1}{2}||\mathbf{y} - \mathbf{X}\bm{\beta}||^2 + \lambda \sum_{j}^{J} \log(1+\frac{|\beta_j|}{\gamma})$, in which $\beta_j$ is the $j^{\text{th}}$ element of $\bm{\beta}$, $\gamma$ is a hyper-parameter, and $\log(1+\frac{|\beta_j|}{\gamma})$ is the concave GDP penalty on $\beta_j$. The thresholding properties of the cardinality constraint, the $L_{1}$ norm penalty and the GDP penalty with different values of $\gamma$ are shown in Fig.~\ref{chapter1_fig:1}. The cardinality constraint shrinks all the coefficients, whose absolute values are less than a selected threshold, to 0 while keeping other coefficients unchanged. This thresholding behavior is referred to as hard thresholding \cite{tibshirani1996regression}. The $L_{1}$ norm penalty shrinks all the coefficients to the same degree until some coefficients are 0. And its thresholding behavior is referred to as soft thresholding \cite{tibshirani1996regression}. On the other hand, GDP penalty shrinks the coefficients in a nonlinear manner according to the absolute values of the corresponding coefficients. In this way, coefficients with small absolute values are more likely to be shrunk to 0, while coefficients with large absolute values tend to be shrunk less.

\section{Scope and outline of the thesis}
Principal component analysis (PCA) model is the basis of many commonly used data fusion methods \cite{maage2019performance}. And both PCA and these data fusion methods assume the used data sets are quantitative. Thus, before talking about the data fusion of data sets with heterogeneous measurement scales, we should first introduce the generalizations of PCA for qualitative data sets. We review the extensions of PCA methods for the analysis of multivariate binary data sets in Chapter \ref{chapter:2} and develop a robust logistic PCA method in Chapter \ref{chapter:3}. After that, we are ready for the data fusion of data sets with heterogeneous measurement scales. We generalize the commonly used data fusion method, simultaneous component analysis (SCA), in a probabilistic framework for the data fusion of the multivariate quantitative and binary measurements data sets in Chapter \ref{chapter:4}. Finally, it comes to the data fusion of data sets of the two types of heterogeneity. We develop an exponential family SCA model for the data fusion of multiple data sets of mixed data types, such as quantitative, binary or count, and introduce the nearly unbiased group concave penalty to induce structured sparsity on the loading matrix to separate common (global and local) and distinct variation in such mixed types data sets in Chapter \ref{chapter:5}. Finally, the thesis closes in Chapter \ref{chapter:6} with an outlook into the future of fusing heterogeneous data sets.


\chapter{PCA of binary genomics data} \label{chapter:2}
Genome wide measurements of genetic and epigenetic alterations are generating more and more high dimensional binary data. The special mathematical characteristics of binary data make the direct use of the classical PCA model to explore low dimensional structures less obvious. Although there are several PCA alternatives for binary data in the psychometric, data analysis and machine learning literature, they are not well known to the bioinformatics community. In this chapter we introduce the motivation and rationale of some parametric and nonparametric versions of PCA specifically geared for binary data. Using both realistic simulations of binary data as well as mutation, CNA and methylation data of the Genomic Determinants of Sensitivity in Cancer 1000 (GDSC1000) the methods are explored for their performance with respect to finding the correct number of components, overfit, finding back the correct low dimensional structure, variable importance etc. The results show that if a low dimensional structure exists in the data that most of the methods can find it. When assuming a probabilistic generating process is underlying the data, we recommend to use the parametric logistic PCA model (using the projection based approach), while when such an assumption is not valid and the data is considered as given, the nonparametric Gifi model is recommended.
\interfootnotelinepenalty=10000
\footnote{This chapter is based on Song, Y., Westerhuis, J.A., Aben, N., Michaut, M., Wessels, L.F. and Smilde, A.K., 2017. Principal component analysis of binary genomics data. Briefings in bioinformatics, 20(1), pp.317-329.}

\section{Background}
Binary measurements only have two possible outcomes, such as presence and absence, or true and false, which are usually labeled as ``1'' and ``0''. In many research problems, objects are characterized by multiple binary features, each depicting a different aspect of the object. In biological research, several examples of binary data sets can be found. Genome wide measurements of genetic and epigenetic alterations are generating more and more high dimensional binary data \cite{mclendon2008comprehensive, iorio2016landscape}. One example is the high throughput measurements of point mutation. Here, a feature is labeled as ``1'' when it is classified as mutated in a sample, ``0'' when it is not. Another often observed binary data set is the binarized version of copy number aberrations (CNA), which are gains and losses of segments in chromosomal regions. Segments are labeled as ``1'' when aberration is presents in a sample, otherwise ``0'' \cite{wu2014detecting}. DNA methylation data can also be discretized as binary features, where ``1'' indicates a high level of methylation and ``0'' means a low level \cite{iorio2016landscape}.

Compared to commonly used quantitative data, binary data has some special mathematical characteristics, which should be taken into account during the data analysis. In binary measurements, ``0'' and ``1'' are abstract representations of two exclusive categories rather than quantitative values 0 and 1. These two categories can also be encoded to any other two different labels, like ``-1'' and ``1'' or ``-'' and ``+'', without changing the meaning. Because ``1'' and ``0'' are only an abstract representation of two categories, they cannot be taken interpreted as quantitative data. Furthermore, the measurement error of binary data is discrete in nature. Binary measurement error occurs when the wrong label is assigned to an object, such as when a mutated gene is mis-classified as wild type. Therefore, the by default used Gaussian error assumption for continuous data in many statistical models is inappropriate for binary data analysis. Another aspect of binary data is that there can be an order in the two categories. For example, presence is often considered more important than absence. Finally, binary data can be generated from a discrete measurement process, but also a continuous measurement process \cite{young1980quantifying}.

PCA is one of the most popular methods in dimension reduction with numerous applications in biology, chemistry and many other disciplines \cite{jolliffe2002principal}. PCA can map data points, which are in a high dimensional space, to a low dimensional space with minimum loss of variation. The derived low dimensional features, which provide a parsimonious representation of the original high dimensional data, can be used in data visualization or for further statistical analysis.

Classical linear PCA methods are appropriate for quantitative data. The direct use of linear PCA on binary data does not take into account the distinct mathematical characteristics of binary data. In this chapter, we are going to introduce, compare and evaluate some of the PCA alternatives for binary data. First, the theory of the different approaches is introduced together with their model properties and how the different models are assessed. Then we will introduce three binary genomics data sets on which the models will be applied. Besides the real data, realistic simulations of binary data are used to uncover some of the properties of the different models.

\section{Theory}
There exist two separate directions in extending PCA for binary data; parametric and nonparametric. Parametric approaches are represented by logistic PCA methods, originating from the machine learning literature. In these methods, PCA is extended to binary data from a probabilistic perspective in a similar way as linear regression is extended to logistic linear regression \cite{collins2001generalization, schein2003generalized, landgraf2015generalized}. Nonparametric methods, originating from the psychometric and data analysis communities, include optimal scaling \cite{de2009gifi}, multiple correspondence analysis \cite{mori2016nonlinear} and many others \cite{kiers1989three}. In this direction, PCA is extended to binary data from a geometric perspective without probabilistic assumptions. The details for the motivation and rationale of above approaches for binary data will be explained later in this section. We will start by introducing classical PCA.

\subsection{Classical PCA}
Classical PCA can be expressed as a projection based approach (finding the low dimensional space that best represents a cloud of high dimensional points) following Pearson \cite{pearson1901lines}. The measurements of $J$ quantitative variables on $I$ objects result into a matrix $\mathbf{X}$($I \times J$) with $I$ rows and $J$ columns. The column vector form of the $i^{\text{th}}$ row of $\mathbf{X}$ is $\mathbf{x}_i \in \mathbb{R}^J$, which is taken as a point in $J$ dimensional space. Suppose that a low dimensional space is spanned by the columns of an orthogonal loading matrix $\mathbf{B}$($J \times R)$, $R \ll \text{min}(I,J)$. The orthogonal projection of $\mathbf{x}_i$ on this low dimensional space is $\mathbf{B}\mathbf{B}^{\text{T}}\mathbf{x}_i$. We find $\mathbf{B}$ by minimizing the Euclidean distance between the centered high dimensional points $\mathbf{x}_i, i=1\cdots I$, and their low dimensional projections:
\begin{equation}\label{chapter2_eq:1}
\begin{aligned}
& \underset{\bm{\mu},\mathbf{B}}{\text{min}}
& &\sum_{i}^{I}((\mathbf{x_i} - \bm{\mu}) - \mathbf{BB}^{\text{T}}(\mathbf{x_i} - \bm{\mu}))^{2}\\
& \text{subject to}
& &\mathbf{B}^{\text{T}}\mathbf{B} = \mathbf{I},
\end{aligned}
\end{equation}
in which the column offset term $\bm{\mu}$($J \times 1$) is included to center the samples, and $\mathbf{I}$ is the identity matrix. The exact position of the centered $i^{\text{th}}$ data point $\mathbf{x}_i$ in this low dimensional space is represented by its $R$ dimensional score vector $\hat{\mathbf{a_i}}$, $\hat{\mathbf{a_{i}}} = \hat{\mathbf{B}}^{\text{T}}( \mathbf{x_i} - \hat{\bm{\mu}})$, where $\hat{\mathbf{B}}$ and $\hat{\bm{\mu}}$ are the estimated values of equation \ref{chapter2_eq:1}. In matrix form, we have $\hat{\mathbf{A}}=(\mathbf{X} - \mathbf{1}\hat{\bm{\mu}}^{\text{T}})\hat{\mathbf{B}}$, $\hat{\mathbf{a_i}}$ is the $i^{\text{th}}$ row of $\hat{\mathbf{A}}$; $\mathbf{1}$ is an $I$ dimensional vector of ones; the estimated offset $\hat{\bm{\mu}}$ contains the column means of $\mathbf{X}$ and $\mathbf{X} - \mathbf{1}\hat{\bm{\mu}}^{\text{T}}$ is the column centered $\mathbf{X}$.

Another approach to explain PCA is the reconstruction based approach \cite{zou2006sparse}. A high dimensional data point $\mathbf{x_i} \in \mathbb{R}^J$ is approximated by a linear function of the latent low dimensional score $\mathbf{a_i} \in \mathbb{R}^R$ with orthogonal coefficients $\mathbf{B}$, $\mathbf{x_i} \approx \bm{\mu} + \mathbf{Ba_i}$, $\mathbf{B}^{\text{T}}\mathbf{B} = \mathbf{I}$, $\bm{\mu}$ is the offset term. Now, $\bm{\mu}$, $\mathbf{A}$ and $\mathbf{B}$ can be found simultaneously by minimizing the Euclidean distance between centered $\mathbf{x_i}, i\cdots I$, and their low dimensional linear approximations $\bm{\mu} + \mathbf{Ba_i}, i \cdots I$:
\begin{equation}\label{chapter2_eq:2}
\begin{aligned}
&\underset{\bm{\mu},\mathbf{A}, \mathbf{B}}{\text{min}}
&&\sum_{i}^{I}{\mathbf{( x_i-\bm{\mu} -Ba_i)^{2}}}\\
&\text{subject to}
&&\mathbf{B}^{\text{T}}\mathbf{B} = \mathbf{I}.
\end{aligned}
\end{equation}

It is well known that the above two approaches for classical PCA are equivalent and the global optimal solution can be obtained from the $R$ truncated singular value decomposition (SVD) of the column centered $\mathbf{X}$ \cite{ten1993least}. The solution $\hat{\bm{\mu}}$ contains the column means of $\mathbf{X}$; $\hat{\mathbf{A}}$ is the product of the first $R$ left singular vectors and the diagonal matrix of first $R$ singular values; $\hat{\mathbf{B}}$ contains the first $R$ right singular vectors.

Above, the classical PCA was derived from a geometrical perspective. Bishop et al.\cite{tipping1999probabilistic} have derived another explanation for PCA from a probabilistic perspective, called probabilistic PCA. A high dimensional point $\mathbf{x_i}$ can be regarded as a noisy observation of the true data point $\bm{\theta_i} \in \mathbb{R}^J$ , which lies in a low dimensional space. The model can be expressed as $\mathbf{x_i} = \bm{\theta_i} + \bm{\epsilon_i}$ and $\bm{\theta_i} = \bm{\mu} + \mathbf{Ba_i}$, $\bm{\mu}$ is the offset term as before; $\mathbf{B}$ contains the coefficients; $\mathbf{a_{i}}$ represents the low dimensional score vector. The noise term $\bm{\epsilon_{i}}$ is assumed to follow a multivariate normal distribution with mean $\mathbf{0}$ and constant variance $\sigma^2$, $\bm{\epsilon_i} \sim N(0,\sigma^2\mathbf{I})$. Thus the conditional distribution of $\mathbf{x_{i}}$ is a normal distribution with mean $\bm{\theta_i}$ and constant variance, $\mathbf{x_i}|\bm{\mu},\mathbf{A},\mathbf{B} \sim N(\bm{\mu} + \mathbf{Ba_i},\sigma^2\mathbf{I})$. $\bm{\mu}$, $\mathbf{A}$ and $\mathbf{B}$ can be obtained by maximum likelihood estimation.
\begin{equation}\label{chapter2_eq:3}
\begin{aligned}
&\underset{\bm{\mu},\mathbf{A},\mathbf{B}}{\text{max}}
&& \sum_i^I{\log(p(\mathbf{x_i}|\bm{\mu},\mathbf{a_i},\mathbf{B}))}\\
&
&&=\sum_i^I{\log(N(\mathbf{x_i}|\bm{\mu} + \mathbf{Ba_i},\sigma^2\mathbf{I}))}\\
&\text{subject to}
&&\mathbf{B}^{\text{T}}\mathbf{B} = \mathbf{I}.
\end{aligned}
\end{equation}

The above maximum likelihood estimation is equivalent to the least squares minimization in classical PCA from the perspective of frequentist statistics \cite{collins2001generalization}. One important implication is that all the elements in the observed matrix $\mathbf{X}$ are conditionally independent of each other given the offset $\bm{\mu}$, the score matrix $\mathbf{A}$ and the loading matrix $\mathbf{B}$, which is the key point for the further extension to binary data.

\subsection{Logistic PCA}
The probabilistic interpretation of PCA under multivariate normal distribution for the observed data provides a framework for the further generalization to other data types \cite{tipping1999probabilistic}. As the Gaussian assumption is only appropriate for continuous quantitative data, it is necessary to replace the Gaussian assumption by the Bernoulli distribution for binary observations in a similar way as from linear regression to logistic linear regression \cite{collins2001generalization, landgraf2015generalized, de2006principal}. The $ij^{\text{th}}$ element in observed matrix $\mathbf{X}$, $x_{ij}$, is a realization of the Bernoulli distribution with parameter $p_{ij}$, which is the $ij^{\text{th}}$ element in the probability matrix $\mathbf{\Pi}$. Specifically, the probability that $x_{ij}$ equals ``1" is $p_{ij}$. Similar to probabilistic PCA, all the elements in the observed matrix $\mathbf{X}$ are conditionally independent of each other given the parameter matrix $\mathbf{\Pi}$($I\times J$). The log likelihood for observation $\mathbf{X}$ given the probability matrix $\mathbf{\Pi}$ is as follows:
\begin{equation}\label{chapter2_eq:4}
l(\mathbf{\Pi}) = \sum_{i}^{I}\sum_{j}^{J}x_{ij}\log(p_{ij})+(1-x_{ij})\log(1-p_{ij}).
\end{equation}
The log-odds of $p_{ij}$ is $\theta_{ij}$, where $\theta_{ij} = \log(\frac{p_{ij}}{1-p_{ij}})$, which is the natural parameter of the Bernoulli distribution expressed in the exponential family form. Thus $p_{ij} = \phi(\theta_{ij}) = (1+ e^{-\theta_{ij}})^{-1}$ and $\phi()$ is called the logistic function. The log likelihood for observation $\mathbf{X}$ given log-odds $\mathbf{ \Theta }$ is represented as:
\begin{equation}\label{chapter2_eq:5}
l(\mathbf{\Theta}) = \sum_{i}^{I}\sum_{j}^{J}x_{ij}\log(\phi(\theta_{ij}))+(1-x_{ij})\log(1-\phi(\theta_{ij})).
\end{equation}
A low dimensional structure can be assumed to exist in the log-odds $\mathbf{\Theta}$($I\times J$) as $\mathbf{\Theta} = \mathbf{A}\mathbf{B}^{\text{T}} + \mathbf{1}\bm{\mu}^{\text{T}}$. Here $\mathbf{A}$ is the object score matrix for the log-odds $\mathbf{ \Theta}$; $\mathbf{B}$ is the loading matrix; $\bm{\mu}$ is the offset.

There are mainly two approaches to fit the model (equation \ref{chapter2_eq:5}), logistic PCA \cite{de2006principal} and projection based logistic PCA (logistic PPCA) \cite{landgraf2015generalized}. The main difference between these two approaches is whether $\mathbf{A}$ and $\mathbf{B}$ are estimated  simultaneously or sequentially. In the logistic PCA model, the score matrix $\mathbf{A}$ and loading matrix $\mathbf{B}$ are estimated simultaneously by alternating minimization \cite{collins2001generalization, udell2016generalized} or by a majorization-minimization (MM) algorithm \cite{de2006principal}.

On the other hand, logistic PPCA only estimates $\mathbf{B}$ directly. After $\mathbf{B}$ is estimated, $\mathbf{A}$ is obtained by a projection based approach in the same manner as classical PCA in equation \ref{chapter2_eq:1} \cite{landgraf2015generalized}. Score matrix $\mathbf{A}$ is the low dimensional representation of the log-odds $\mathbf{\widetilde{\Theta}}$ of the saturated model in the subspace spanned by $\mathbf{B}$. Details of the log-odds $\mathbf{\widetilde{\Theta}}$ from the saturated model will be shown latter. In matrix form, $\mathbf{A} = (\mathbf{\widetilde{\Theta }}-\mathbf{1}\bm{\mu}^{\text{T}})\mathbf{B}$, $\bm{\mu}$ is the offset term. Then the log-odds $\mathbf{\Theta}$ in equation \ref{chapter2_eq:5} can be represented as $\mathbf{\Theta} = (\mathbf{\widetilde{\Theta }}-\mathbf{1}\bm{\mu}^{\text{T}})\mathbf{B}\mathbf{B}^{\text{T}} + \mathbf{1}\bm{\mu}^{\text{T}}$. The estimation of parameters $\hat{\bm{\mu}}$ and $\hat{\mathbf{B}}$, can be obtained by maximizing the conditional log likelihood $l(\mathbf{\Theta})$ in equation \ref{chapter2_eq:5}. Then, the solution for the score matrix is $\hat{\mathbf{A}} = (\mathbf{\widetilde{\Theta }}-\mathbf{1}\hat{\bm{\mu}}^{\text{T}})\hat{\mathbf{B}}$.

Compared to logistic PCA, logistic PPCA has fewer parameters to estimate, and thus is less prone to overfitting. In addition, the estimation of the scores of new samples in logistic PCA involves an optimization problem while for logistic PPCA, it is a simple projection of the new data on $\hat{\mathbf{B}}$.

In a saturated model, there is a separate parameter for every individual observation. The model is over-parameterized and has perfect fit to the observed data. For quantitative data, the parameters of the saturated model are simple the observed data. For example, the parameters of the saturated PCA model on observed data matrix $\mathbf{X}$ are the matrix $\mathbf{X}$ itself. For binary data, the parameter (probability of success) of a saturated model for the observation ``1" is 1; for the observation ``0" is 0. Thus, the $ij^{\text{th}}$ element in $\mathbf{\widetilde{\Theta}}$ from the saturated logistic PCA model is $\widetilde{\theta}_{ij} = \log(\frac{x_{ij}}{1-x_{ij}})$. It is negative infinity when $x_{ij} = 0$; positive infinity when $x_{ij} = 1$. In order to project $\mathbf{\widetilde{\Theta}}$ onto the low dimensional space spanned by $\mathbf{B}$, one needs a finite $\mathbf{\widetilde{\Theta}}$. In logistic PPCA, positive and negative infinities in $\mathbf{\widetilde{\Theta}}$ are approximated by large numbers $m$ and $-m$. When $m$ is too large, the elements in the estimated probability matrix $\hat{\mathbf{\Pi}}$ for generating the binary observation $\mathbf{X}$ are close to 1 or 0; when $m$ is close to 0, the elements in $\hat{\mathbf{\Pi}}$ are close to 0.5. In the original paper of logistic PPCA, CV is used to select $m$ \cite{landgraf2015generalized}. In this paper we select $m = 2.94$ which corresponds to a probability of success 0.95. This can be interpreted as using probabilities 0.95 and 0.05 to approximate the probabilities 1 and 0 in the saturated model.

\subsection{Theory of nonlinear PCA with optimal scaling}
Another generalization of PCA to binary data is nonlinear PCA with optimal scaling (the Gifi method). This method was primarily developed for categorical data, of which binary data is a special case \cite{gifi_B_90, de2009gifi}. The basic idea is to quantify the binary variables to quantitative values by minimizing some loss functions. The quantified variables are then used in a linear PCA model. The $j^{\text{th}}$ column of $\mathbf{X}$, $\mathbf{ X_{*j}}$, is encoded into an $ I \times 2 $ indicator matrix $\mathbf{G_j}$. $\mathbf{G_j}$ has two columns, ``1'' and ``0''. If the $i^{\text{th}}$ object belongs to column ``1'', the corresponding element of $\mathbf{ G_j }$  is 1, otherwise it is 0. $\mathbf{A}$ is the $I \times R$ object score matrix, which is the representation of $\mathbf{X}$ in a low dimensional Euclidean space; $\mathbf{ Q_j }$ is a $2 \times R$ quantification matrix, which quantifies this $j^{\text{th}}$ binary variable to a quantitative value. For binary data, the rank of the quantification matrix $\mathbf{ Q_j }$ is constrained to 1.~This is the PCA solution in the Gifi method.~$\mathbf{ Q_j }$ can be expressed as $\mathbf{ Q_j} = \mathbf{z_j}\mathbf{w_j}^{\text{T}}$, where $\mathbf{z_j}$ is a two dimensional column vector with binary quantifications and $\mathbf{w_j}$ is the vector of weights for $R$ principal components. The loss function is expressed as:
\begin{equation}
\underset{\mathbf{A,z_j,w_j}}{\text{min}} \sum_{j=1}^{J} {\mathbf{ ( A-G_j z_j w_j )}^2},
\end{equation}
in which the score matrix $\mathbf{A}$ is forced to be centered and orthogonal, $\mathbf{1}^{\text{T}}\mathbf{A}=0$, $\mathbf{A}^{\text{T}}\mathbf{A}=\mathbf{I}$, to avoid trivial solutions. The loss function is optimized by alternating least squares algorithms. For binary data, nonlinear PCA with optimal scaling is equivalent to multiple correspondence analysis and to PCA on standardized variables \cite{kiers1989three}.

\section{Model properties}
\subsection{Offset}
Including the column offset term $\bm{\mu}$ in component models also implies that the column mean of score matrix is $\mathbf{0}$, i.e. $\mathbf{1}^{\text{T}}\mathbf{A}=\mathbf{0}$. Otherwise, the model is unidentifiable. In PCA and the Gifi method, the estimated $\hat{\bm{\mu}}$ equals the column mean of $\mathbf{X}$. Therefore, including $\bm{\mu}$ in the model has the same effect as column centering of $\mathbf{X}$. In logistic PPCA and logistic PCA, the $j^{\text{th}}$ element of $\bm{\mu}$, $\mu_{j}$, can be interpreted as the log-odds of the marginal probability of the $j^{\text{th}}$ variable. When only the offset $\bm{\mu}$ is included in the model, $\mathbf{\Theta} = \mathbf{1}\bm{\mu}^{\text{T}}$, the $j^{\text{th}}$ element of the solution $\hat{\bm{\mu}}$, $\hat{\mu_{j}}$, is the log-odds of the empirical marginal probability of the $j^{\text{th}}$ variable (the proportion of ``1" in the $j^{\text{th}}$ column). When more components are included, $\mathbf{\Theta} = \mathbf{1}\bm{\mu}^{\text{T}} + \mathbf{A}\mathbf{B}^{\text{T}}$, the solution $\hat{\bm{\mu}}$ is not unique. If an identical offset is required for comparing component models with a different number of components, one can fix the offset term to the log-odds of the empirical marginal probability during the maximum likelihood estimation.

\subsection{Orthogonality}
Similar to PCA, the orthogonality constraint $\mathbf{B}^{\text{T}}\mathbf{B} = \mathbf{I}$ in logistic PPCA and logistic PCA actually is inactive. If $\mathbf{B}$ is not orthogonal, it can be made orthogonal by subjecting $\mathbf{A}\mathbf{B}^{\text{T}}$ to an SVD algorithm. $\mathbf{B}$ equals the right hand singular vectors and $\mathbf{A}$ equals the product of the left hand singular vectors and the diagonal matrix of singular values. This extra step will not change the objective value. Table \ref{chapter2_table:01} gives the orthogonality properties of the scores and loadings of the four methods discussed above.
\begin{table}[htbp]
\centering
\caption{Orthogonality properties of the scores and loadings of the four methods. O: the columns of this matrix are orthonormal vectors, $\mathbf{B}^{\text{T}}\mathbf{B} = \mathbf{I}$; D: the columns of this matrix are orthogonal vectors, $\mathbf{B}^{\text{T}}\mathbf{B} = \mathbf{D}$, $\mathbf{D}$ is a $R \times R$ diagonal matrix.}
\label{chapter2_table:01}
\begin{tabular}{|c|c|c|c|c|}
  \hline
                                & PCA & Gifi & logistic PCA & logistic PPCA \\
    score matrix $\mathbf{A}$   & D & O & D & D \\
    loading matrix $\mathbf{B}$ & O & D & O & O \\
    \hline
\end{tabular}
\end{table}

\subsection{Nestedness}
Linear PCA models are nested in the number of components, which means the first $R$ principal components in the $R+1$ components model are exactly the same as the $R$ components model. For the Gifi method, this property only holds for the binary data case but not in general. For logistic PPCA and logistic PCA, this property does not hold.

\section{Model assessment}

\subsection{Error metric}
To make a fair comparison between linear PCA, the Gifi method, logistic PPCA and logistic PCA, the training error is defined as the average misclassification rate in using the derived low dimensional structure to fit the training set $\mathbf{X}$. Each of the four methods provides an estimation of the offset term, score matrix and loading matrix, $\bm{\hat{\mu}}$, $\mathbf{\hat{A}}$ and $\mathbf{\hat{B}}$. For linear PCA and the Gifi method, we take $\mathbf{1}\bm{\hat{\mu}}^{\text{T}} + \mathbf{\hat{A}}\mathbf{\hat{B}}^{\text{T}}$ as an approximation of the binary matrix $\mathbf{X}$; for logistic PCA and logistic PPCA, $\phi(\mathbf{1}\bm{\hat{\mu}}^{\text{T}} + \mathbf{\hat{A}}\mathbf{\hat{B}}^{\text{T}})$ is used as an approximation for the probability matrix $\mathbf{\Pi}$, of which the observed matrix $\mathbf{X}$ was generated. Since both approximations are continuous, we need to select a threshold to discretize them to binary fitting.

In the discretization process, two misclassification errors exist. ``0" can be misclassified as ``1", which we call $err0$ and ``1" can be misclassified as ``0", which we call $err1$. $N_{err0}$ is the number of $err0$ in this process, and $N_{err1}$ is the number of $err1$; $N_0$ is the number of ``0" in the observed binary matrix $\mathbf{X}$, and $N_1$ is the number of ``1" in $\mathbf{X}$. A commonly used total error rate is given by $(N_{err0} + N_{err1})/(N_0 + N_1)$, which gives equal weights to these two errors. However, this can lead to undesirable results for imbalanced binary data, i.e. when the proportions of ``1" and ``0" are extreme. Usually, imbalanced binary data sets are common in real applications, where sometimes the proportion of ``1" in the observed matrix $\mathbf{X}$ can be less than 5\%. In such a case, $err0$ is more likely to occur than $err1$, and hence it seems inappropriate to give them equal weights. In imbalanced cases, a balanced error rate $0.5\times(N_{err0}/N_0 + N_{err1}/N_1)$ is more appropriate \cite{wei2013role}. To decide whether the predicted quantitative value represents a ``0" or a ``1", a threshold value has to be selected. This threshold value can be selected by minimizing the balanced error rate in a training set after which it can be applied to a test set in order to prevent biased (too optimistic) results.

\subsection{Cross validation}
The training error is an overly optimistic estimator of the generalization error, which can be intuitively understood as the average misclassification rate in predicting an independent test set. Thus, we use cross validation (CV) to approximate the generalization error. In this chapter, we use the CV algorithm named EM-Wold \cite{wold1978cross,Bro2008}. In this approach, validation sets of elements of the matrix $\mathbf{X}$ are selected in a diagonal style rather than a row wise style. The left out part is considered as missing. In this way the prediction of the left out part is independent of the left out part itself. It is possible to use this approach as all the component models in this paper can handle missing data. A 7-fold CV procedure was used for all calculations in this paper. In each of these folds, a component model is developed taking the missing data into account. The model is then used to make a prediction of the missing elements. This is repeated until all elements of $\mathbf{X}$ have been predicted in this way. The threshold of converting the continuous predictions to binary predictions in CV was the same as the one used in computing the training error.

\section{Data Sets}
\subsection{Real data sets}
The data we used is from the Genomic Determinants of Sensitivity in Cancer 1000 (GDSC1000) \cite{iorio2016landscape}. To facilitate the interpretability of the results, only three cancer types are included in the data analysis: BRCA (breast invasive carcinoma, 48 human cell lines), LUAD (lung adenocarcinoma, 62 human cell lines) and SKCM (skin cutaneous melanoma, 50 human cell lines). Each cell line is a sample in the data analysis. For these samples, three different binary data sets are available: mutation, copy number aberration (CNA) and methylation data. For the mutation data, there are 198 mutation variables. Each variable is a likely cancer driver or suppressor gene. A gene is labeled as ``1" when it is classified as mutated in a sample and as ``0" when classified as wild type. The mutation data is very imbalanced (supplemental Fig.~S2.1 a): roughly $2\%$ of the data matrix is labeled as ``1". The CNA data has 410 observed CNA variables. Each variable is a copy number region in a chromosome. It is labeled as ``1" for a specific sample when it is identified as aberrated and it is labeled as ``0" otherwise. The CNA data set is also imbalanced (supplemental Fig.~S2.1 b): roughly $7\%$ of the data matrix is labeled as ``1". For the methylation data, there are 38 methylation variables. Each variable is a CpG island located in gene promotor region. In each variable, ``1" indicates a high level of methylation and ``0" indicates a low level. The methylation data set is relatively balanced compared to other data sets (supplemental Fig.~S2.1 c): roughly $27\%$ of the data matrix is labeled as ``1".

\subsection{Simulated binary data sets}
Binary data matrices with an underlying low dimensional structure can be simulated either from a latent variable model or as the noise corrupted version of a structured binary data set. In the first case the data generating process is considered to provide a quantitative data set while there is a binary read out. In the second case the data generating process is considered to provide a binary data set. We use both of these two approaches to study the properties of different binary PCA methods.

\subsubsection{Simulated binary data based on the logistic PCA model}
Data sets with different degrees of imbalance and with low dimensional structures were simulated according to the logistic PCA model. The offset term $\bm{\mu}$ is used to control the degree of imbalance and the log-odds $\mathbf{\Theta}$ is defined to have a low dimensional structure. The observed binary matrix $\mathbf{X}$ is generated from the corresponding Bernoulli distributions.

Each element in the $J \times R$ loading matrix $\mathbf{B}$ is sampled from the standard normal distribution. The Gram-Schmidt algorithm is used to force $\mathbf{B}^{\text{T}}\mathbf{B}=\mathbf{I}_R$. $R$ is set to 3. The simulated $I \times R$ score matrix $\mathbf{A}$ has three group structures in the samples. $I$ samples are divided into three groups of equal size. The three group means are set manually to force sufficient difference between the groups. The first two group means are set to $\mathbf{a_1^*} = [2,-1,3]^{\text{T}}$ and $\mathbf{a_2^*} = [-1,3,-2]^{\text{T}}$. The third group mean is $\mathbf{a_3^*} = [0,0,0]^{\text{T}} - \mathbf{a_1^*} - \mathbf{a_2^*}$. The scores in first group are sampled from the multivariate normal distribution $N(\mathbf{a_1^*}, \mathbf{I}_R)$, the scores in second group from $N(\mathbf{a_2^*},\mathbf{I}_R)$ and the scores in the third group from $N(\mathbf{a_3^*}, \mathbf{I}_R)$. In this way, scores between groups are sufficiently different and scores within the same group are similar.

When the elements in $\mathbf{A}\mathbf{B}^{\text{T}}$ are close to $0$, the corresponding probabilities are close to 0.5. In this case, the binary observations are almost a random guess. When their absolute values are large, the corresponding probabilities are close to 1 or 0, the binary observations are almost deterministic. The scale of $\mathbf{A}\mathbf{B}^{\text{T}}$ should be in a reasonable interval, not too large and not too small. A constant $C$ is multiplied to $\mathbf{A}\mathbf{B}^{\text{T}}$ to control the scale for generating proper probabilities. In addition, the offset term $\bm{\mu}$ is included to control the degree of imbalance in the simulated binary data set. After $\mathbf{\Theta} = C\mathbf{A}\mathbf{B}^{\text{T}} + \mathbf{1}\bm{\mu}^{\text{T}}$ is simulated as above, it is transformed to the probability matrix $\mathbf{\Pi}$ by the logistic function $\phi()$ and $x_{ij}$ in $\mathbf{X}$ is a realization of Bernoulli distribution with parameter $p_{ij}$, which is the $ij^{\text{th}}$ element of probability matrix $\mathbf{\Pi}$.

\subsubsection{Simulated binary data based on noise corruption of pre-structured binary data}
Another approach of simulating binary data is by the noise corruption of a pre-structured data set. Compared to the latent variable model, this approach provides an intuitive understanding of the low dimensional structure in the observed binary data. Pre-structured binary data set $\mathbf{X}_{\text{true}}$ has structured and unstructured parts. The goal of the current simulation is to find the variables that belong to the structured part, while the goal of the previous simulation is to see how well the whole data set can be approximated. The structured part is simulated as follows. $R$ different $I$ dimensional binary vectors are simulated first, each element is sampled from the Bernoulli distribution with probability $p$, which is the degree of imbalance in the binary data simulation. Each of these $R$ binary vectors is replicated 10 times to form the structured part. In the unstructured part of $\mathbf{X}_{\text{true}}$, all the elements are randomly sampled from Bernoulli distribution with probability $p$. The observed binary data $\mathbf{X}$ is a noise corrupted version of the pre-structured binary data set $\mathbf{X}_{\text{true}}$. If the noise level is set to 0.1, all the elements in the binary data $\mathbf{X}_{\text{true}}$ have a probability of 0.1 to be bit-flipped. The observed binary matrix $\mathbf{X}$, has $R$ groups of 10 highly correlated variables and the other variables are not correlated. The $R$ groups are taken as the low dimensional structure. The above simulation process is illustrated in the supplemental Fig.~S2.4.\\

\section{Results}
All the computations are done in R \cite{RProject}. The linear PCA model is fitted using the SVD method after centering the data \cite{stacklies2007pcamethods}. The Gifi method is fitted using the alternating least squares approach by Homals package \cite{de2009gifi}. The logistic PCA and logistic PPCA models are fitted using an MM algorithm with offset term \cite{de2006principal, landgraf2015generalized}. The default stopping criterion is used for all the approaches.

\subsection{Balanced simulation}
\begin{figure}[htbp]
    \centering
    \includegraphics[width=0.9\textwidth]{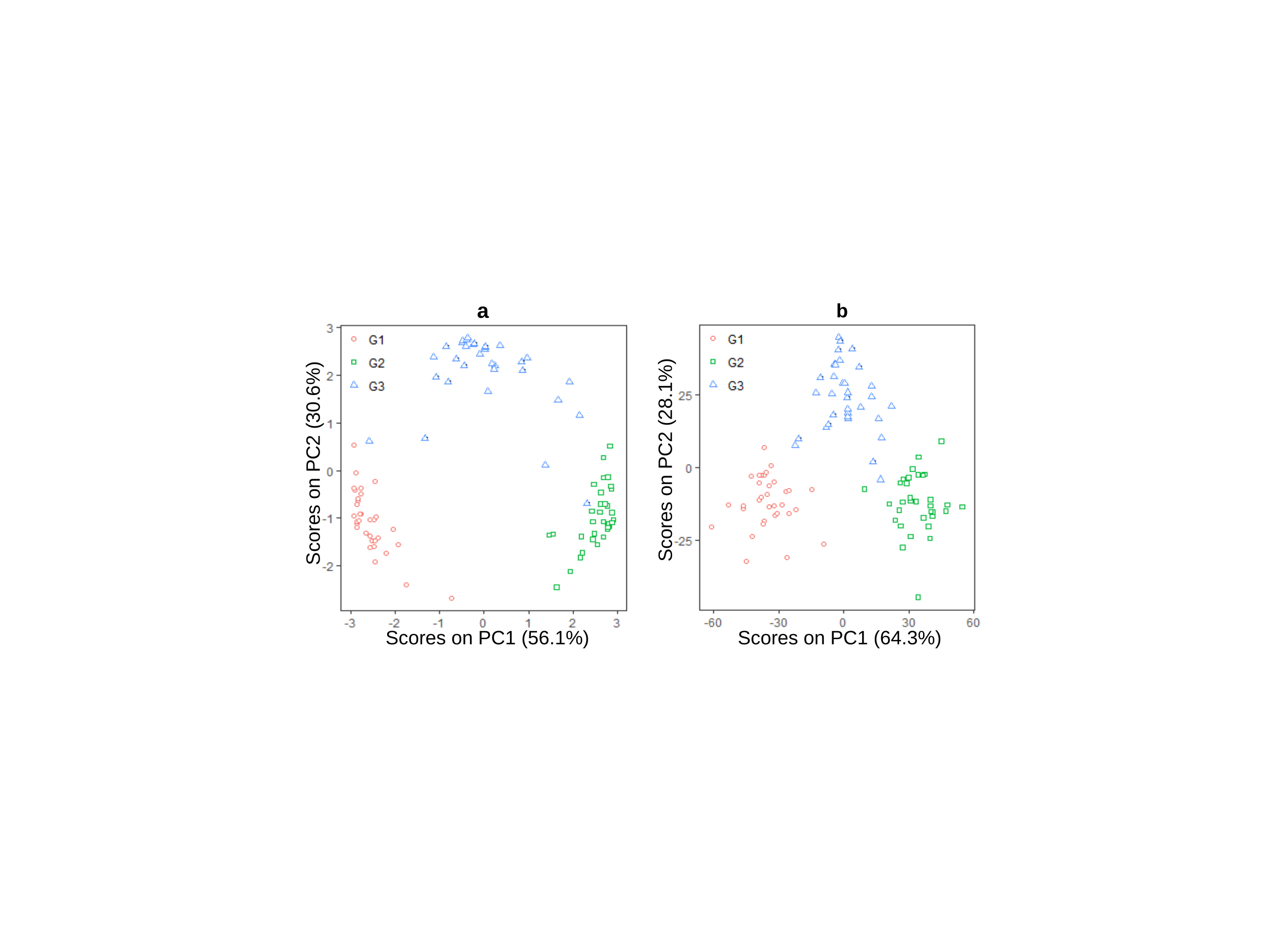}
    \caption{Score plot of the first two principal components (PCs) derived from linear PCA on the probability matrix $\mathbf{\Pi}$ (left) and  log-odds matrix $\mathbf{\Theta}$ (right) used in the binary data simulation. G1, G2 and G3 are three simulated groups in the samples.}
    \label{chapter2_fig:1}
\end{figure}

The goal of this simulation is to evaluate the abilities of the four approaches in finding back the embedded low dimensional structures in the sample space and variable space. The simulation process is based on the logistic PCA model. The offset term $\bm{\mu}$ is set to $\mathbf{0}$ to simulate balanced binary data. The parameters are set to $I = 99, J = 50, R = 3, C = 10$. The simulated balanced binary data are shown in supplemental Fig.~S2.2. First a classical PCA on the simulated probability matrix $\mathbf{\Pi}$ and the log-odds $\mathbf{\Theta}$ was performed. Fig.~\ref{chapter2_fig:1} shows the score plots of these two PCA analyses. The difference between the score plots of linear PCA on $\mathbf{\Pi}$ (Fig.~\ref{chapter2_fig:1} a) and on log-odds $\mathbf{\Theta}$ (Fig.~\ref{chapter2_fig:1} b) is obvious. The scores of the linear PCA model on $\mathbf{\Pi}$ lie in the margin of the figure, while for $\mathbf{\Theta}$, they lie more in the center of the figure. This difference is related to the nonlinear logistic function $\phi()$, which transforms $\mathbf{\Theta}$ to $\mathbf{\Pi}$. Furthermore, PCA on the log-odds matrix describes more variation in the first two PCs.

\begin{figure}[htbp]
    \centering
    \includegraphics[width=0.9\textwidth]{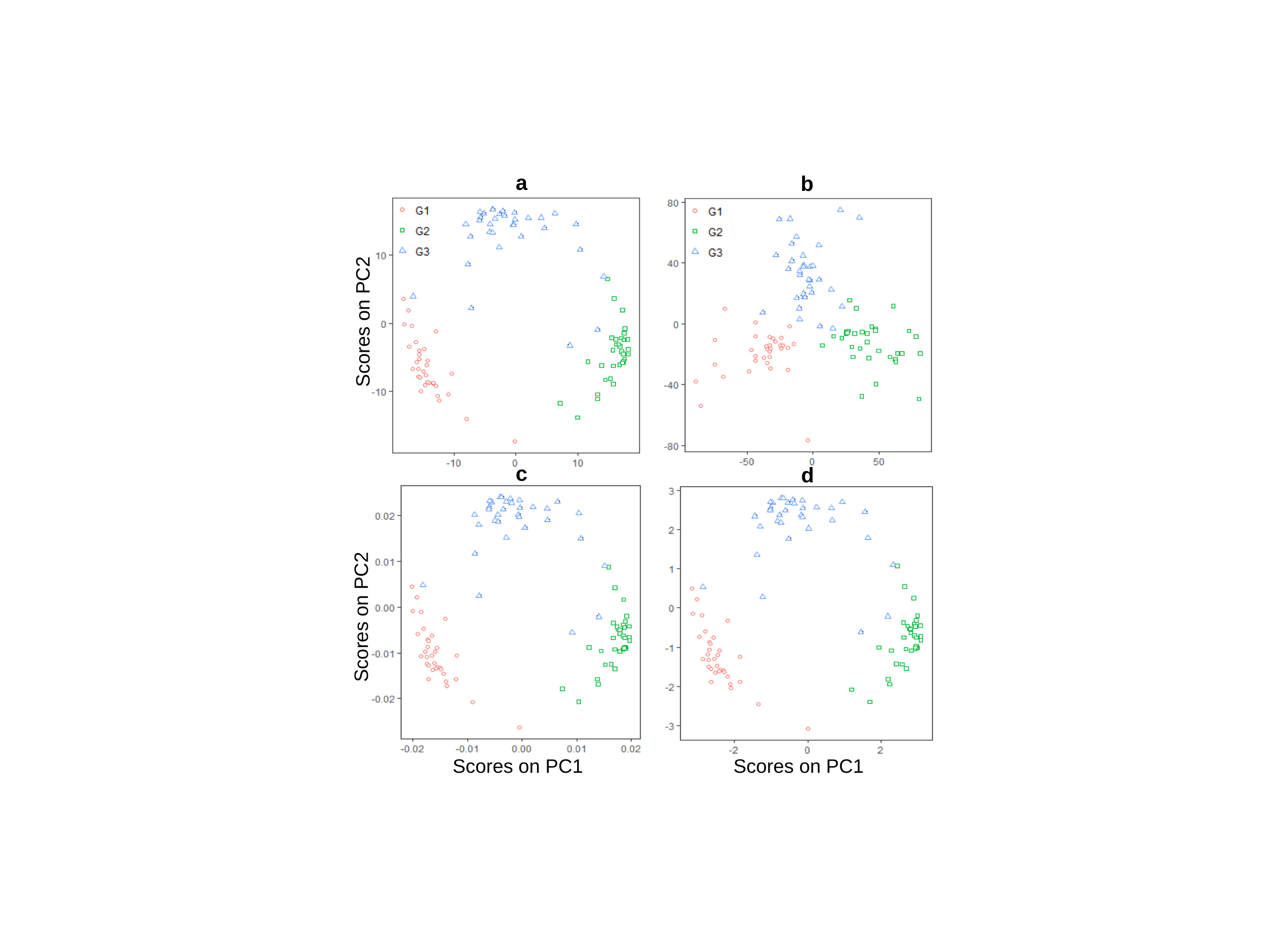}
    \caption{Score plot of first two PCs produced by the four different approaches. \textbf{a}: logistic PPCA; \textbf{b}: logistic PCA; \textbf{c}: the Gifi method; \textbf{d}: PCA. G1, G2 and G3 are three simulated groups in the samples.}
    \label{chapter2_fig:2}
\end{figure}

Logistic PCA, logistic PCA, Gifi and linear PCA are used to model the binary matrix $\mathbf{X}$. Two principal components are used. Offset terms are included in the model. The score plots produced by these different approaches are shown in Fig.~\ref{chapter2_fig:2}. The similarity between Fig.~\ref{chapter2_fig:1} and Fig.~\ref{chapter2_fig:2} indicates that the logistic PCA model approximates the underlying log-odds $\mathbf{\Theta}$ from the binary observation $\mathbf{X}$, while the other approaches approximate the probability matrix $\mathbf{\Pi}$.

Another observation is that the score plots derived from logistic PPCA (Fig.~\ref{chapter2_fig:2} a), Gifi (Fig.~\ref{chapter2_fig:2} c) and linear PCA (Fig.~\ref{chapter2_fig:2} d) are very similar except for some scale differences. The similarity between the Gifi and linear PCA for balanced binary data set is understandable. For binary data, the Gifi method is equivalent to PCA on standardized binary variables. Since the proportion of ``1'' and ``0'' of each binary variable are similar in a balanced simulated data set, the column mean and standard deviation of each binary variable are close to 0.5. Thus the standardization of each binary variable will change 0 and 1 binary data to -1 and 1 data. Therefore, except for the difference in scale, Gifi and linear PCA are almost the same for balanced binary data. For logistic PPCA, the score matrix $\mathbf{A}$ is a low dimensional representation of the log-odds $\mathbf{ \widetilde{\Theta } }$ from the saturated model, $\mathbf{A} = (\mathbf{\widetilde{\Theta }}-\mathbf{1}\bm{\mu}^{\text{T}})\mathbf{B}$, and the $\mathbf{ \widetilde{\Theta } }$ is estimated by $2m(\mathbf{X}-1)$. This is equivalent to changing 0 and 1 to $-m$ and $m$. Thus, the true difference between linear PCA and logistic PPCA is how to find the low dimension spanned by loading matrix $\mathbf{B}$. Logistic PPCA finds it by minimizing the logistic loss and linear PCA finds it by minimizing the least squares loss.

\begin{figure}[htbp]
    \centering
    \includegraphics[width=0.9\textwidth]{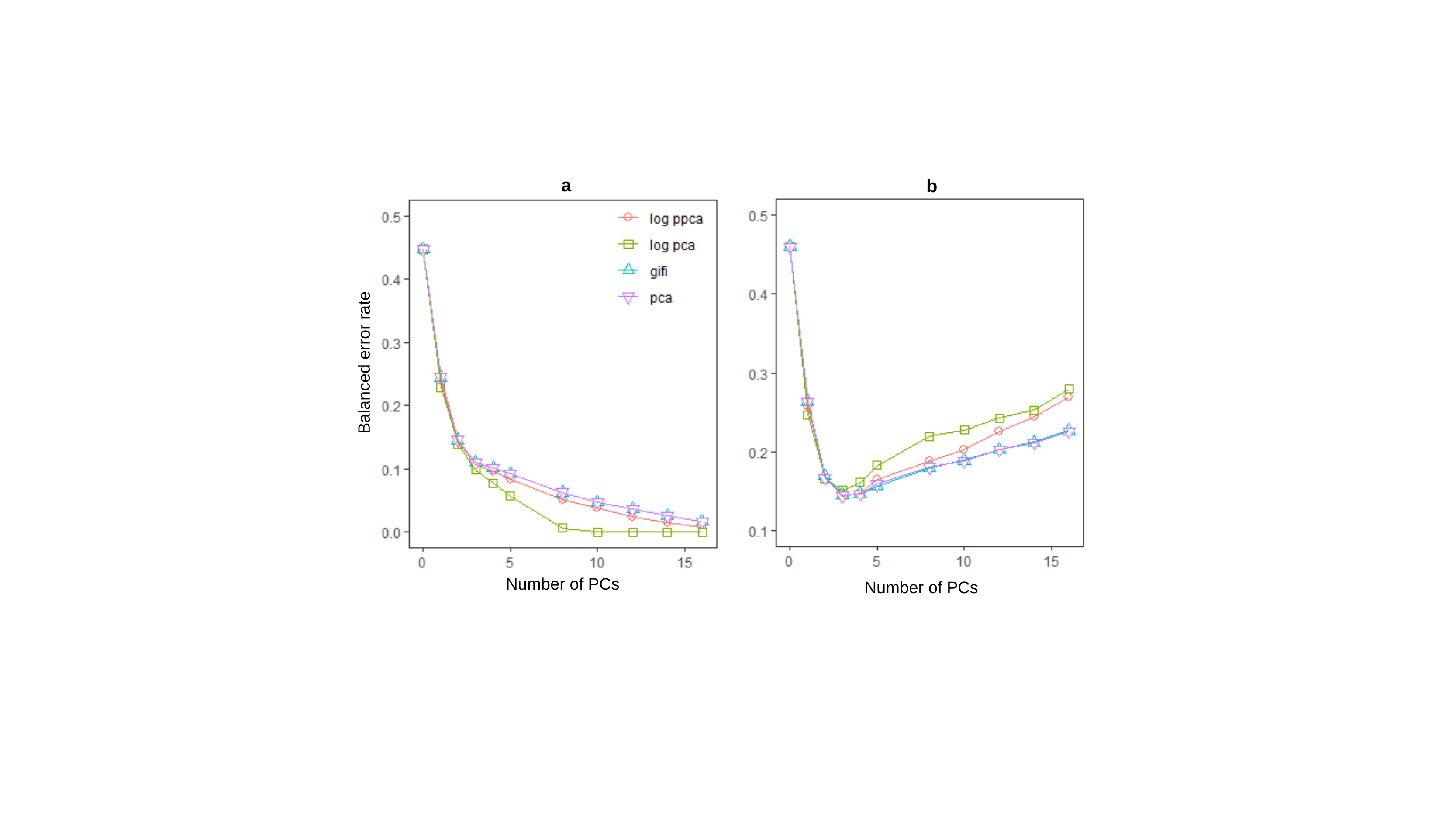}
    \caption{The balanced training error (\textbf{a}) and CV error (\textbf{b}) for the balanced simulated data set produced by four different approaches with different number of components. \textbf{a}: training error; \textbf{b}: CV error. log ppca: logistic PPCA; log pca: logistic PCA; gifi: the Gifi method; pca: linear PCA.}
    \label{chapter2_fig:3}
\end{figure}

The training error and CV error for different models are shown in Fig.~\ref{chapter2_fig:3}. We add the zero component model in which only the offset term $\bm{\mu}$ is included, as the baseline for evaluating the different methods with different numbers of components. The estimated offset $\hat{\bm{\mu}}$ in the zero component model is the column mean of $\mathbf{X}$ for PCA and the Gifi method, while it is the logit transform of the column mean of $\mathbf{X}$ for logistic PPCA and logistic PCA. All approaches successfully find the three components truely underlying the data. It can also be observed that logistic PCA is more eager to overfit the data. It shows a lower balanced error rate, but a higher CV error rate for more than three components compared to the other methods.

\subsection{Imbalanced simulation}
\begin{figure}[htbp]
    \centering
    \includegraphics[width=0.9\textwidth]{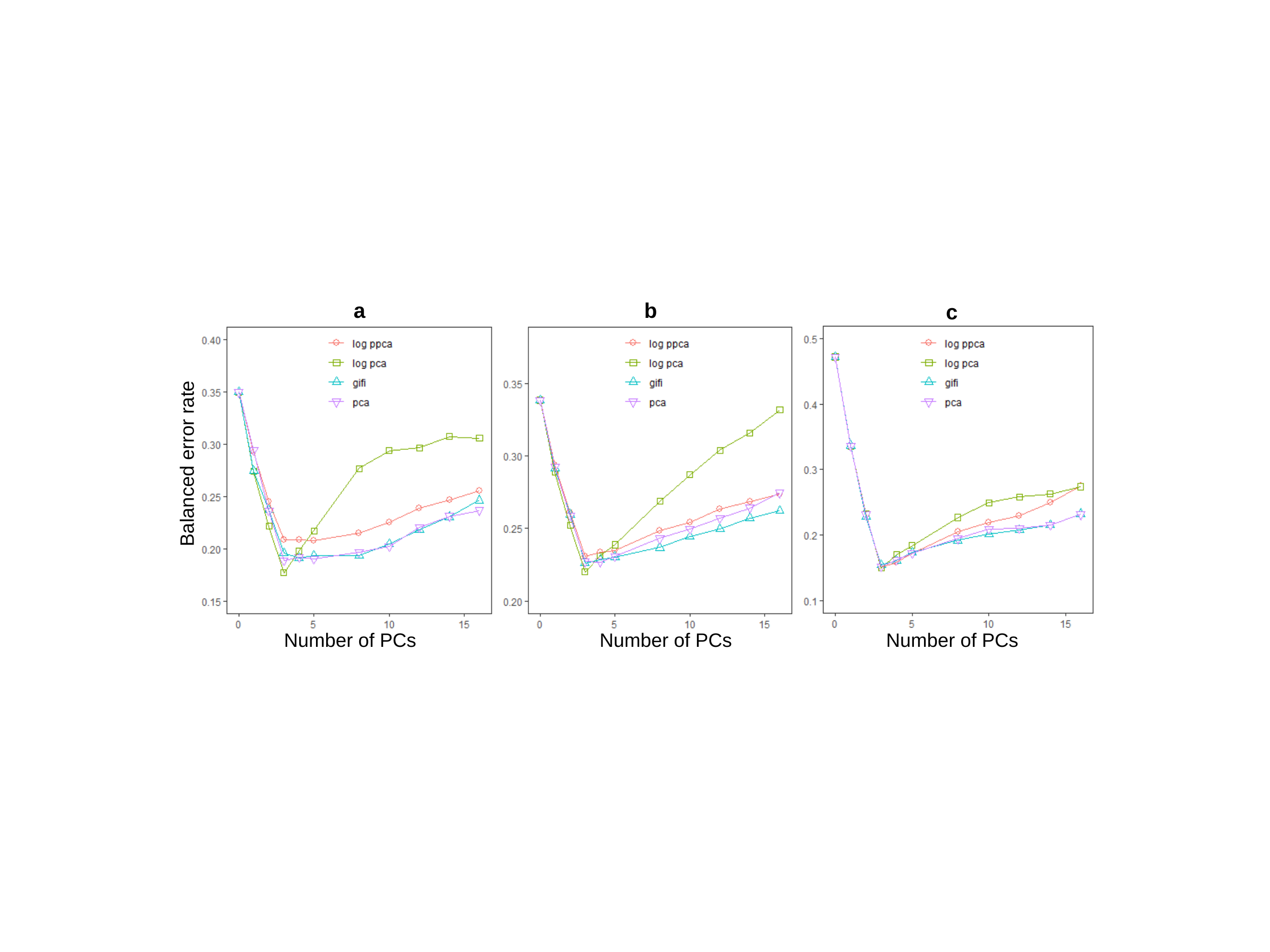}
    \caption{CV error plot for simulated imbalanced data sets with different degrees of imbalance. \textbf{a}: similar as mutation data; \textbf{b}: similar as CNA data; \textbf{c}: similar as methylation data.}
    \label{chapter2_fig:4}
\end{figure}

The goal of the imbalanced simulation is to evaluate the effect of imbalanced simulated data on the ability of the four approaches in finding back the underlying low dimensional structures in variable space. Since the offset $\bm{\mu}$ in logistic PCA model can be interpreted as the log-odds of marginal probabilities, we can use the log-odds of the empirical marginal probabilities from the real data sets with different degrees of imbalance as the offset in the simulation. The simulation process is based on the logistic PCA model. The offset term $\bm{\mu}$ is set to log-odds of column means of real data to simulate imbalanced binary data. The parameters $I, J$ are set to the size of corresponding real data. The constant $C$ is selected as 20, $R$ is set to 3. The simulated data is shown in supplemental Fig.~S2.3. We evaluate the effect of imbalanced binary data set on the different models' abilities of finding back the simulated low dimensional structure. The CV error plots of different models are shown in Fig.~\ref{chapter2_fig:4}. All the approaches are successful in finding back three significant PCs.

\subsection{Feature selection}
\begin{figure}[htbp]
    \centering
    \includegraphics[width=0.9\textwidth]{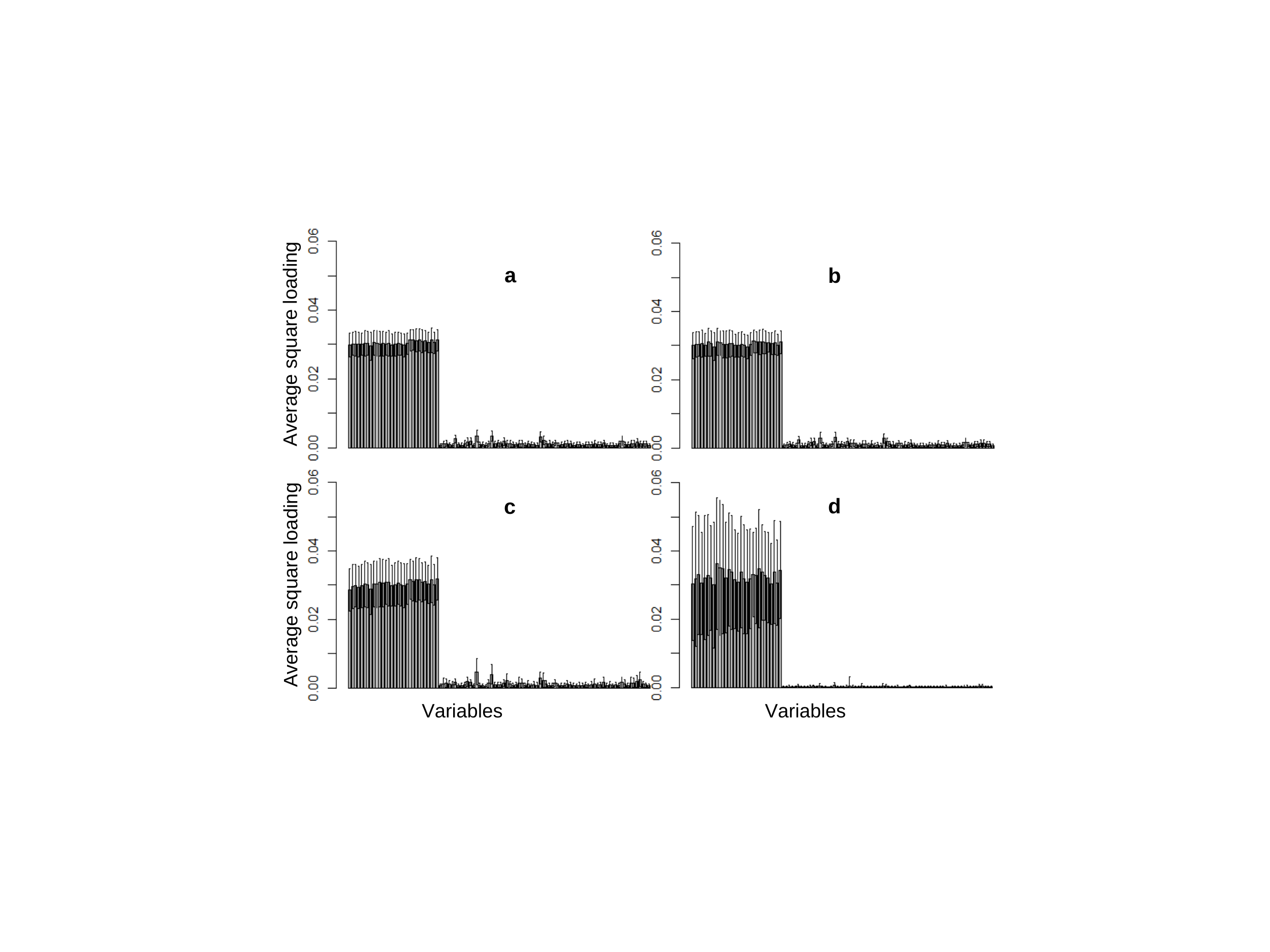}
    \caption{Barplot with one standard deviation error bar of the mean square loadings of linear PCA model (\textbf{a}), the Gifi method (\textbf{b}), logistic PPCA (\textbf{c}) and logistic PCA (\textbf{d}).}
    \label{chapter2_fig:5}
\end{figure}

For the assessment of feature importance, the binary data is simulated by noise corruption of a pre-structured binary data set. $I$ is 198; $J$ is 100; $R$ is set to 3. The degree of imbalance is set to 0.2, and the noise level is 0.1. The simulated data is shown in supplemental Fig.~S2.4. There are noisy corrupted structures in the first 30 variables. For feature selection purposes we estimate the importance of each feature in the model. This is performed as follows, $\frac{1}{3}(b_{j1}^2+b_{j2}^2+ b_{j3}^2)$, where $b_{j2}$ is the loading in the $2^{\text{nd}}$ PCs for $j^{\text{th}}$ variable, is taken as the importance measure. The process is repeated 15 times, the mean and standard deviation of the average squared loading for the 100 variables are shown in Fig.~\ref{chapter2_fig:5}. It can be observed that highly correlated binary variables have large loadings. The variance of the loadings derived from logistic PCA is much higher than other approaches. This indicates that the logistic PCA model cannot make stable estimation of loadings.

\subsection{Real data}
\begin{figure}[htbp]
    \centering
    \includegraphics[width=0.9\textwidth]{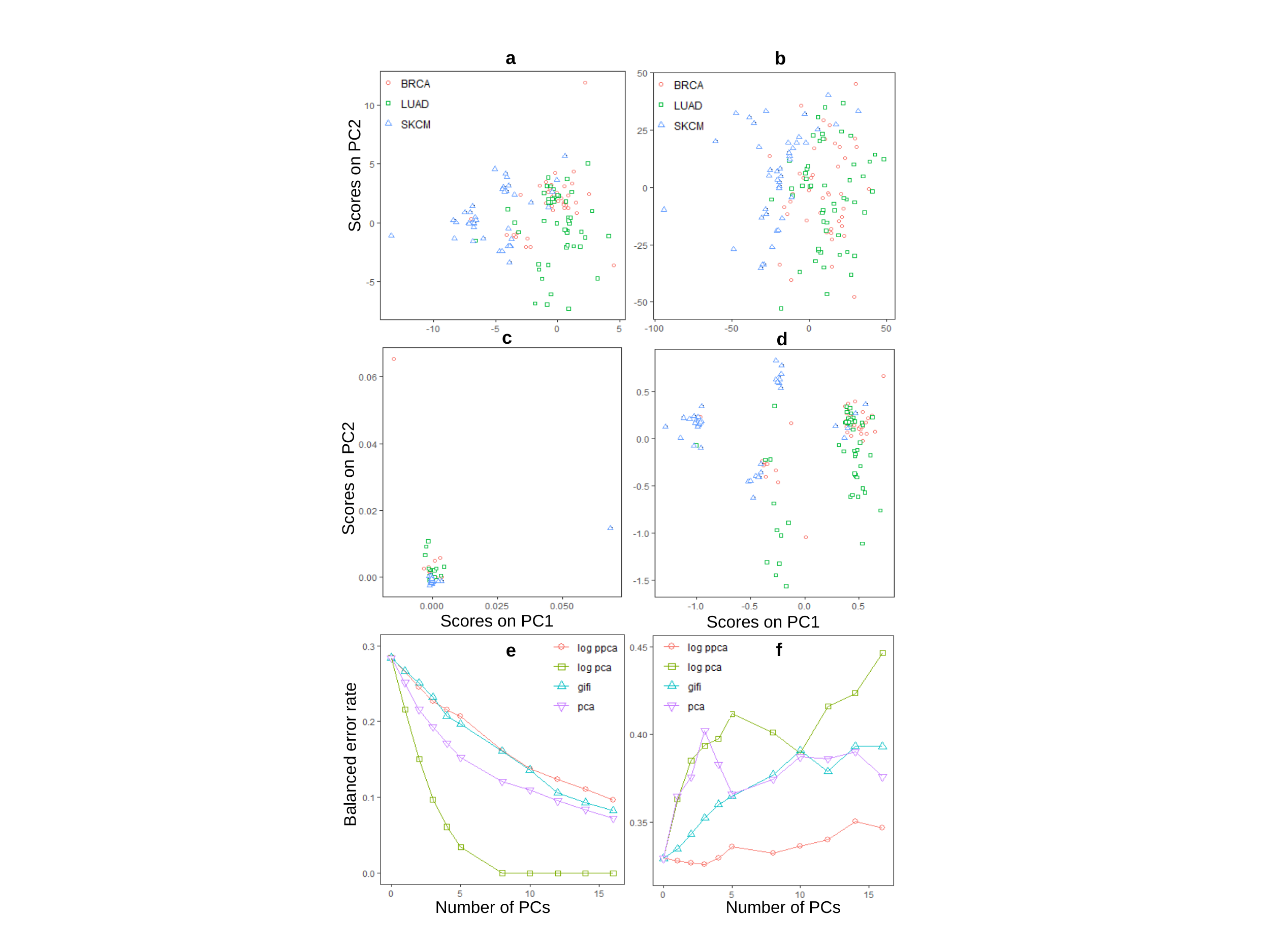}
    \caption{Score plot of the first two PCs, training and CV error plot of the four different approaches for the mutation data. \textbf{a}: score plot of logistic PPCA ; \textbf{b}: score plot of logistic PCA; \textbf{c}: score plot of the Gifi method; \textbf{d}: score plot of PCA; \textbf{e}: training error plot; \textbf{f}: CV error plot. BRCA: breast invasive carcinoma; LUAD: lung adenocarcinoma; SKCM: skin cutaneous melanoma. The legend of the training error and CV error plot is the same as Fig.~\ref{chapter2_fig:3}.}
    \label{chapter2_fig:6}
\end{figure}

The binary mutation, CNA and methylation data sets are analysed using the four different approaches. The score plots and error plots from different approaches on the real mutation data set are shown in Fig.~\ref{chapter2_fig:6}. The CV results of PCA, Gifi and logistic PCA in Fig.~\ref{chapter2_fig:6} f do not support the assumption that a low dimensional structure exists in the mutation data. For the CV result of logistic PPCA (Fig.~\ref{chapter2_fig:6} f), the minimum CV error was achieved using three components. However, this minimum was only slightly lower than the zero component model. Although the CV result of logistic PPCA is ambiguous, we can observe four clusters in the score plot.

\begin{figure}[htbp]
    \centering
    \includegraphics[width=0.9\textwidth]{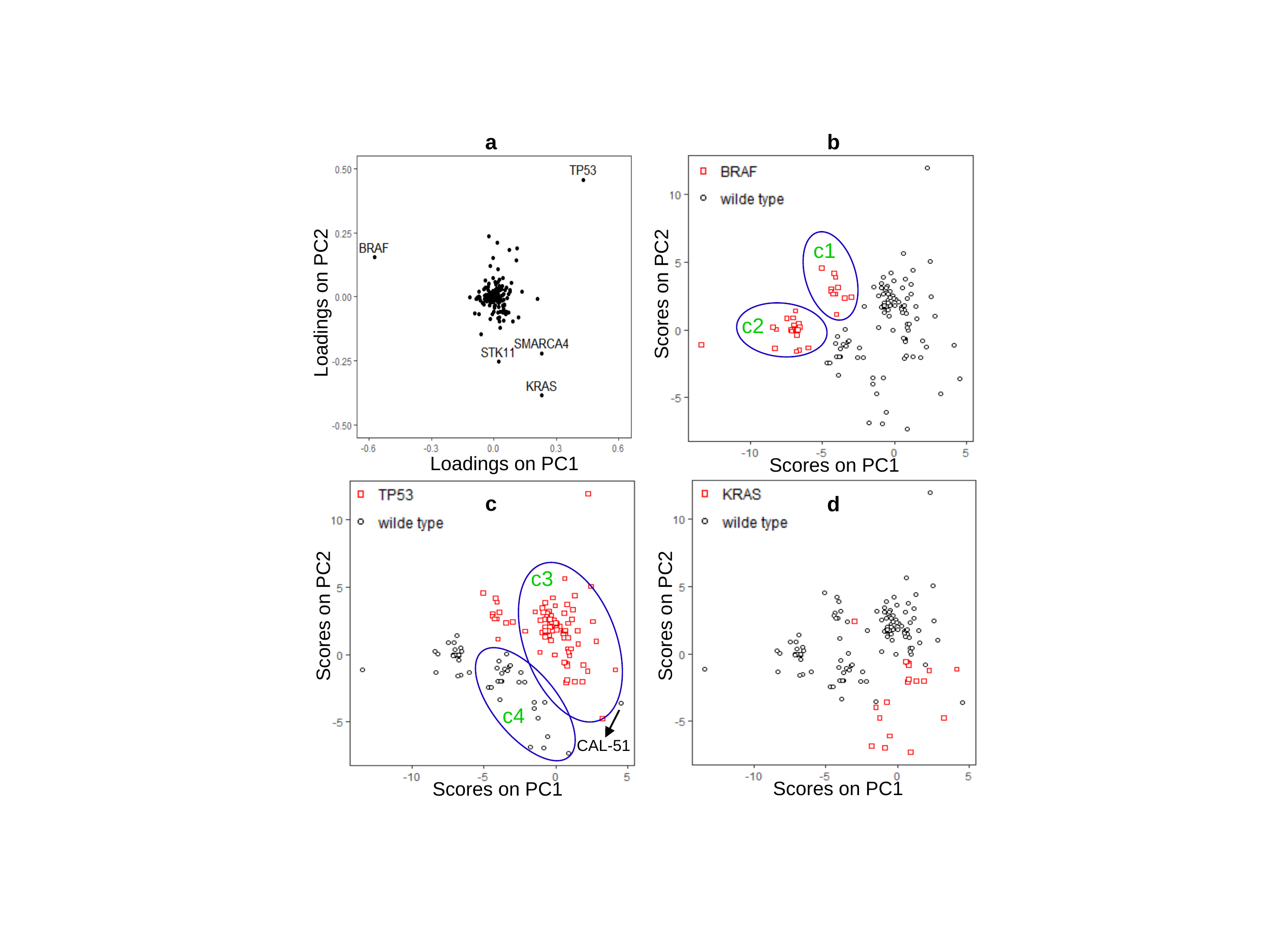}
    \caption{Loading plot (\textbf{a}) and score plots of the first two PCs derived from logistic PPCA model on mutation data. The score plots (\textbf{b, c, d}) are labeled according to the mutation patterns. \textbf{b}: BRAF mutation labeled score plot; \textbf{c}: TP53 mutation labeled score plot; \textbf{d}: KRAS mutation labeled score plot. Red square: mutated; black dot: wild type. c1, c2, c3 and c4 are the plausible four clusters in the samples on mutation data.}
    \label{chapter2_fig:7}
\end{figure}

To explore the clusters in more detail, Fig.~\ref{chapter2_fig:7} shows the loading plot and score plots with different mutation status of the logistic PPCA model. With the corresponding loading values (Fig.~\ref{chapter2_fig:7} a) we determined that these clusters were largely defined by TP53, BRAF and KRAS mutation status. Interestingly, these genes also have the highest mutational load, suggesting that variables with a higher aberration frequency contain more information. Cluster 1 (c1 in Fig.~\ref{chapter2_fig:7} b) is BRAF-mutant and TP53-mutant type; while cluster 2 (c2 in Fig.~\ref{chapter2_fig:7} b) is BRAF-mutant and TP53-wild type. Cluster 3 (c3 in Fig.~\ref{chapter2_fig:7} c) mostly consists of BRAF-wild and TP53-mutant cell lines, a configuration that often occurs in all three analyzed cancer types. Cluster 4 (c4 in Fig.~\ref{chapter2_fig:7} c) contains BRAF-wild and TP53-wild type cell lines, which again is a configuration that occurs across cancer types. Finally, we observed sub-clusters of LUAD cell lines towards the bottom of cluster 3 and 4, which consist of KRAS-mutant cell lines (Fig.~\ref{chapter2_fig:7} d). As BRAF and KRAS mutations both activate the MAPK pathway in a similar fashion, double mutants are redundant and hence rarely observed. Our results are in line with this mutual exclusivity pattern: with the exception of a single BRAF/KRAS double mutant, we find BRAF mutations only in cluster 1 and 2 and KRAS mutations only in cluster 3 and 4. One notable exception of the above characterization of the clusters is CAL-51 (labeled in Fig.~\ref{chapter2_fig:7} c). Given its TP53 wild-type status, CAL-51 would be expected in cluster 4, but it actually resides in the bottom-left of cluster 3. This shift left is likely due to mutations in both SMARCA4 and PIK3CA, which have the third and fourth most negative loading values on PC1.

\begin{figure}[htbp]
    \centering
    \includegraphics[width=0.9\textwidth]{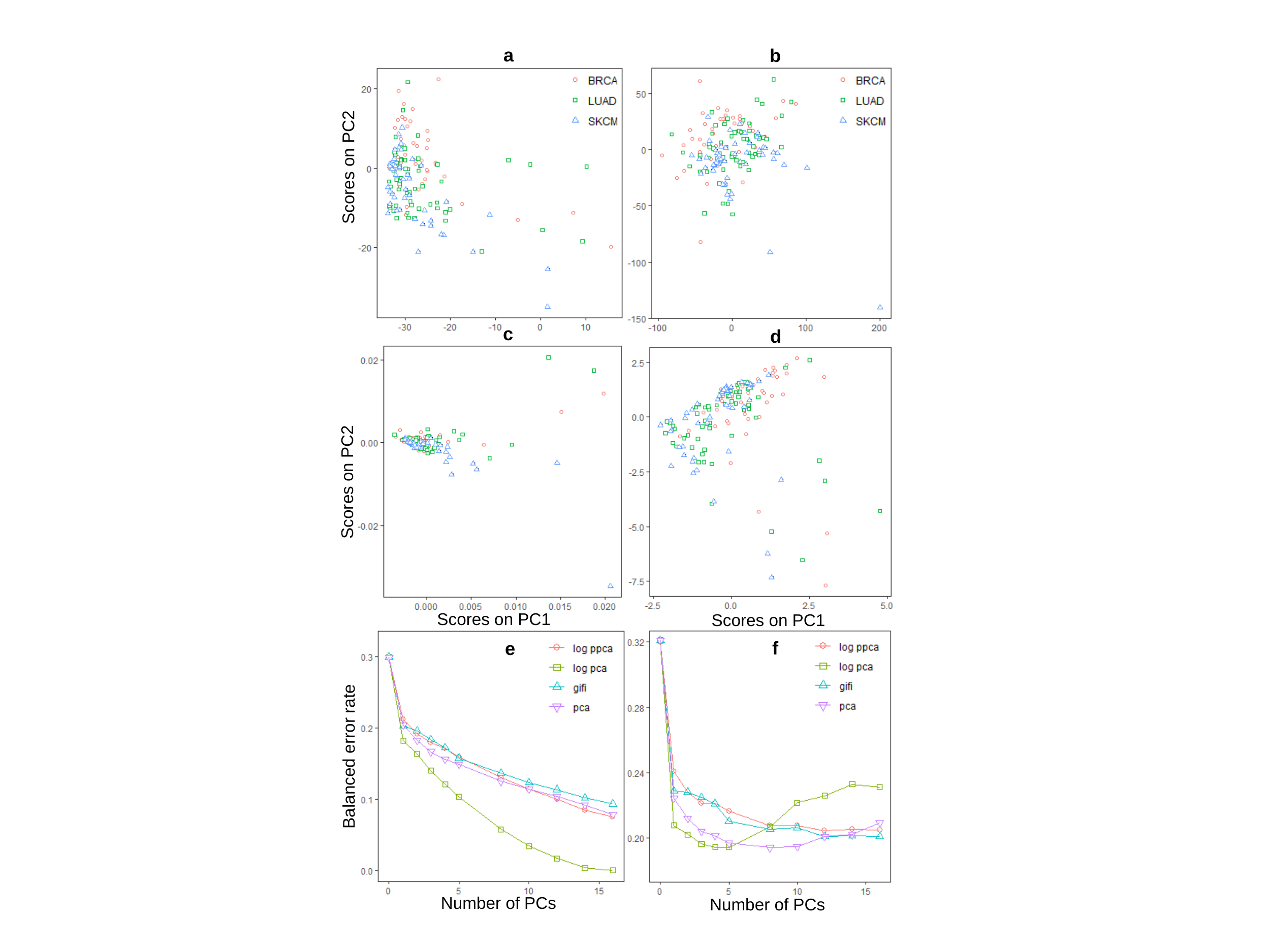}
    \caption{Score plot of the first two PCs, training and CV error plot of the four different approaches for the CNA data. \textbf{a}: score plot of logistic PPCA; \textbf{b}: score plot of logistic PCA; \textbf{c}: score plot of the Gifi method; \textbf{d}: score plot of PCA; \textbf{e}: training error plot; \textbf{f}: CV error plot. The legends for the score plot and training error plot are the same as Fig.~\ref{chapter2_fig:4}.}
    \label{chapter2_fig:8}
\end{figure}

The score plots and error plots from different approaches on CNA data are shown in Fig.~\ref{chapter2_fig:8}. There is some evidence from the CV results from all the models in Fig.~\ref{chapter2_fig:8} f for a five dimensional structure in the data. However, in the score plots of Fig.~\ref{chapter2_fig:8}, the samples with different cancer types are not well separated and there is no clear evidence of natural clusters. Therefore, we do not zoom in further on this data type.

\begin{figure}[htbp]
    \centering
    \includegraphics[width=0.9\textwidth]{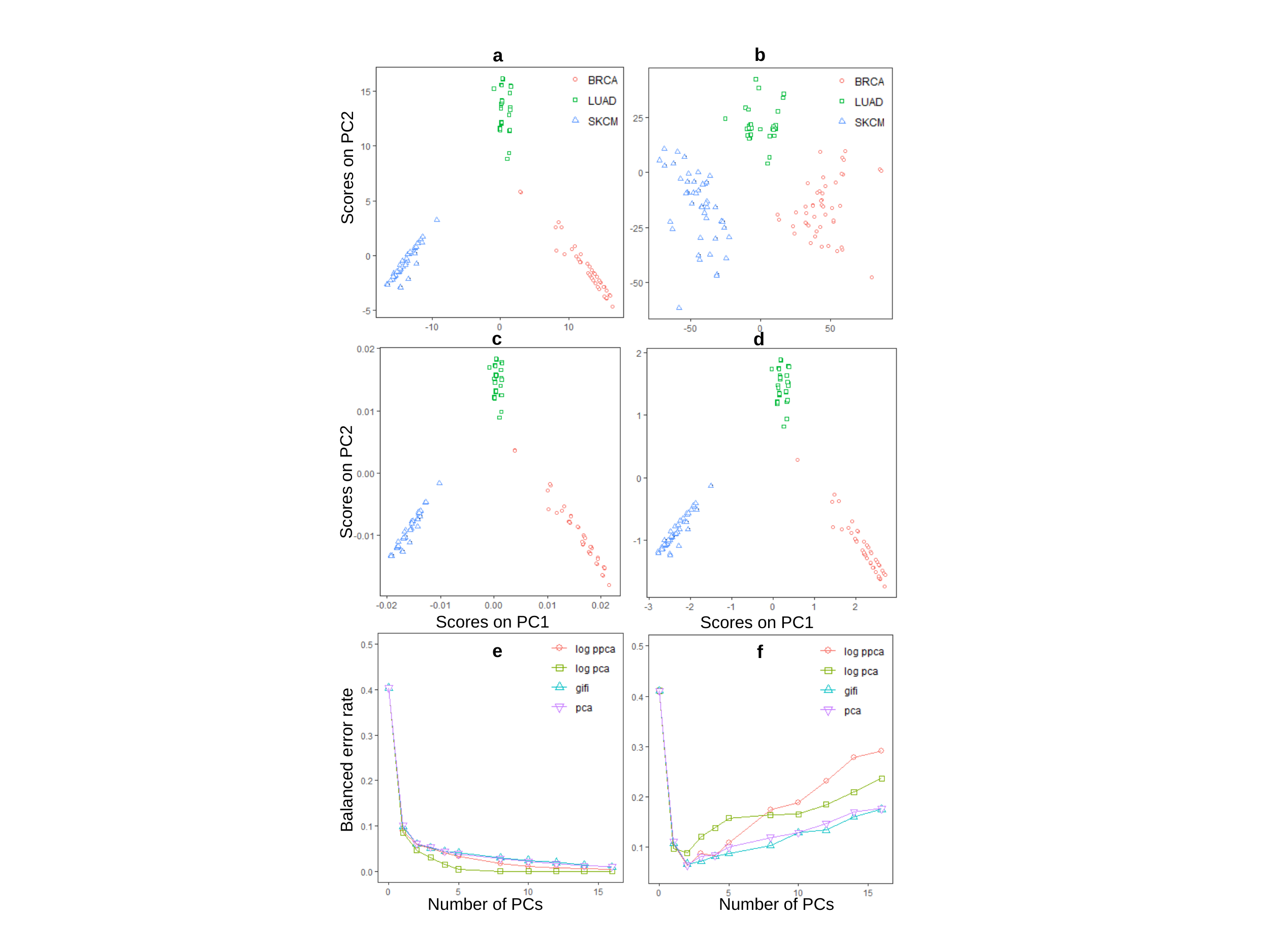}
    \caption{Score plot of the first two PCs, training and CV error plot of the four different approaches for the methylation data. \textbf{a}: score plot of logistic PPCA ; \textbf{b}: score plot of logistic PCA; \textbf{c}: score plot of the Gifi method; \textbf{d}: score plot of PCA; \textbf{e}: training error plot; \textbf{f}: CV error plot. The legends for the score plot and training error plot are the same as Fig.~\ref{chapter2_fig:4}.}
    \label{chapter2_fig:9}
\end{figure}

The score plots and error plots from different approaches on methylation data are shown in Fig.~\ref{chapter2_fig:9}. The three cancer types are well separated in all the score plots. The similar and specific structure in the score plots of logistic PPCA, the Gifi method and linear PCA may be related to the unique structure of the methylation data (supplemental Fig.~S2.1 c). Different cancer types have very different methylation patterns, represented by unique sets of features. In addition, there is some evidence from the CV results from all the models in Fig.~\ref{chapter2_fig:9} f for a two dimensional structure in the data.

\begin{figure}[htbp]
    \centering
    \includegraphics[width=0.5\textwidth]{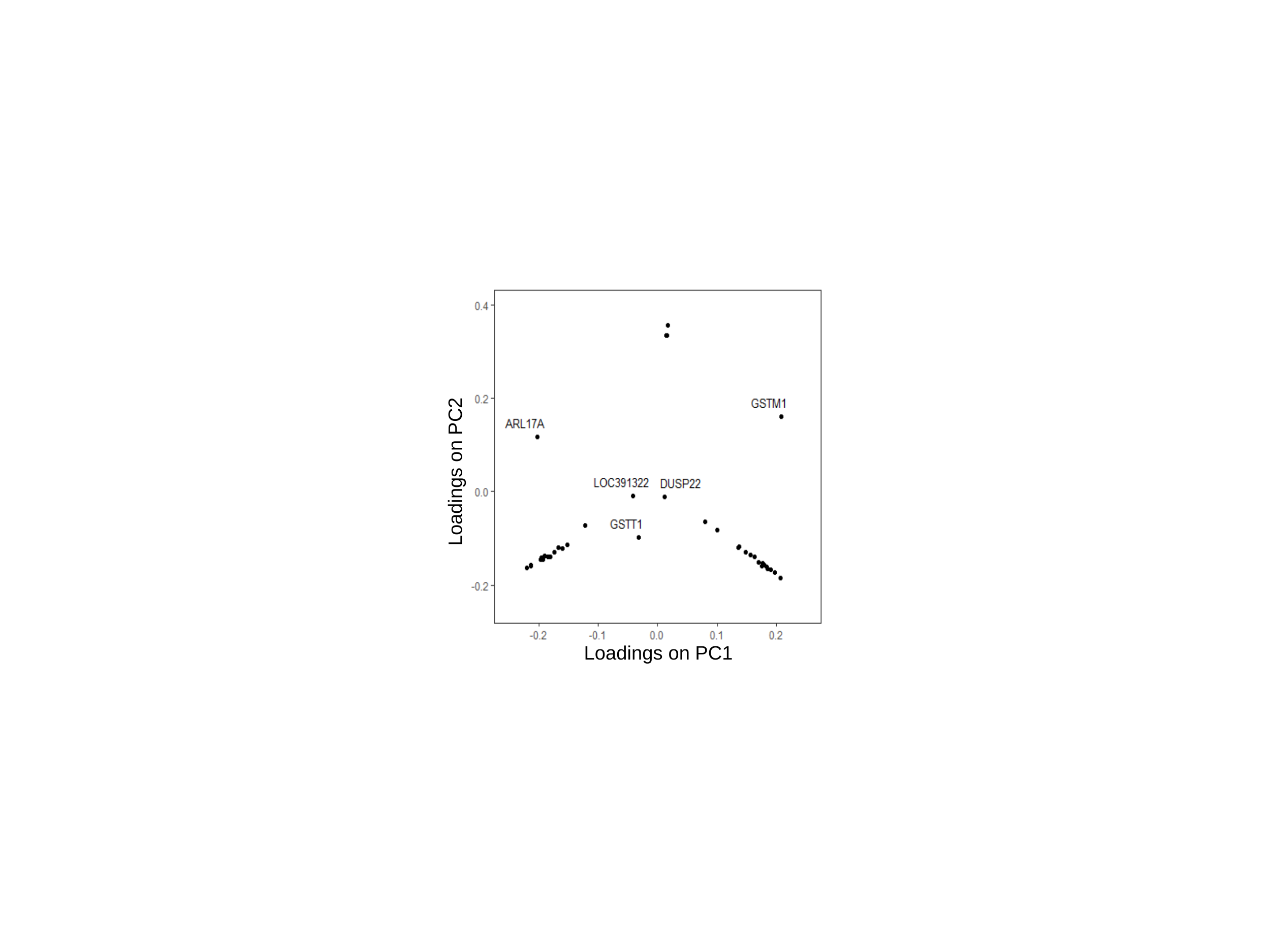}
    \caption{Loading plot of logistic PPCA model on methylation data. The gene names corresponding to the methylation variables, which are interpreted in the paper, are labeled on the plot.}
    \label{chapter2_fig:10}
\end{figure}

We use the score plot derived from logistic PPCA model on methylation data as an example to interpret the result. The first two principal components from the logistic PPCA applied to the methylation data show three clusters, which perfectly represent the three cancer types (Fig.~\ref{chapter2_fig:9} a). The corresponding loading values also roughly fall into three cancer type specific clusters (Fig.~\ref{chapter2_fig:10}), as most variables are exclusively non-zero in a single cancer type. Notable exceptions are GSTM1 and ARL17A, which are non-zero in two cancer types and hence each reside between two clusters, and variables GSTT1 and DUSP22, which are non-zero in all three cancer types and hence reside towards the center of the plot.

\section{Discussion}
In this chapter, four methods were discussed that all aim to explore binary data using low dimensional scores and loadings. It was shown that each of the methods has different goals and therefore produces slightly different results. Linear PCA (without the standardization processing step) treats the binary data as quantitative data and tries to use low rank score and loading matrices to fit the quantitative data. For binary data, the quantification process in the Gifi method is simply a standardization of the binary variables. After quantification, the Gifi method tries to use low rank score and loading matrices to fit the quantified binary variables. Both logistic PPCA and logistic PCA assume that the binary data follows a Bernoulli distribution, and try to find an optimal estimation of the log-odds matrix, which lies in the low dimensional space. Logistic PCA tries to estimate the low dimensional log-odds matrix directly; while logistic PPCA estimates this matrix by the projection of the log-odds matrix from the saturated model on a approximated low dimensional space.

For all the four approaches it is not the degree of imbalance which is the problem. As shown in our simulation results, the low dimensional structure of binary variables, which can be simulated from a latent variable model or by noise corruption of a pre-structured binary data set, is the key issue for the results of data analysis, rather than the degree of imbalance of the data set. When there is a low dimensional structure in our simulation process, all the approaches can successfully find the correct number of components with different degrees of imbalance.

In both the analysis of the simulated data and of the real data, the performance of the linear PCA method, in the criteria of training error and CV error, is similar to other specially designed algorithms for binary data. In addition, since the global optimum of the linear PCA model can always be achieved, the solution is very stable. However, the linear PCA model on binary data obviously contradicts the mathematical characteristics of binary data and the assumptions of the linear PCA model itself. In addition, the fitted values, elements in the product of score and loading matrix, can only be regarded as an approximation to quantitative 0 and 1, and are thus difficult to interpret.

The results of linear PCA and the Gifi method are very similar, especially when the degree of imbalance in each variable is approximately equal. Furthermore, there are signs of overfitting in the analysis of the CNA data by the Gifi model. However, compared to linear PCA, the interpretability of the Gifi method is better. The mathematical characteristics of binary data are taken into account from the geometrical perspective and the solutions can be interpreted as an approximation of the optimally quantified binary variables.

On the other hand, logistic PPCA and logistic PCA methods take into account the mathematical characteristics of binary data from the probabilistic perspective. Fitted values, elements in the product of the derived score and loading matrices, can be interpreted as the log-odds for generating the binary data, and the log-odds can again be transformed to probability. The problem for logistic PCA is that it is not robust with respect to the score and loading estimation, although it is able to select the correct number of components \cite{de2006principal}. Since both score matrix $\mathbf{A}$ and loading matrix $\mathbf{B}$ are free parameters to fit in the optimization, the estimation of $\mathbf{A}\mathbf{B}^{\text{T}}$ will not hesitate to move to infinity to minimize the loss function. This represents itself in such a way that logistic PCA is prone to overfit (as can be seen from the CV results) and the large variation in the loading estimation. The non-robustness problem is mitigated in the logistic PPCA model. Since only the loading matrix $\mathbf{B}$ is freely estimated in logistic PPCA to find the optimal model, while the score matrix $\mathbf{A}$ is fixed given the loadings, the logistic PPCA model is less prone to overfitting. Thus, the estimation of the loadings of binary variables is more stable compared to logistic PCA. Furthermore, since the fitted values are the linear projection of the log-odds matrix of the saturated model, its scale is constrained by the scale of the approximate log-odds matrix of the saturated model, which can be specified in advance.

When assuming a probabilistic generating process is underlying the binary data we recommend to use the parametric logistic PPCA model. When such an assumption is not valid and the data is considered as given, the nonparametric Gifi model is recommended.

\section*{Acknowledgements}
Y.S. gratefully acknowledges the financial support from China Scholarship Council (NO.201504910809). The research leading to these results has received funding from the European Research Council under the European Union's Seventh Framework Programme (FP7/2007-2013) / ERC synergy grant agreement $\text{n}^{0}$ 319661 COMBATCANCER.

\clearpage
\section{Supplementary information}
\begin{figure}[htbp]
    \centering
    \includegraphics[width=0.9\textwidth]{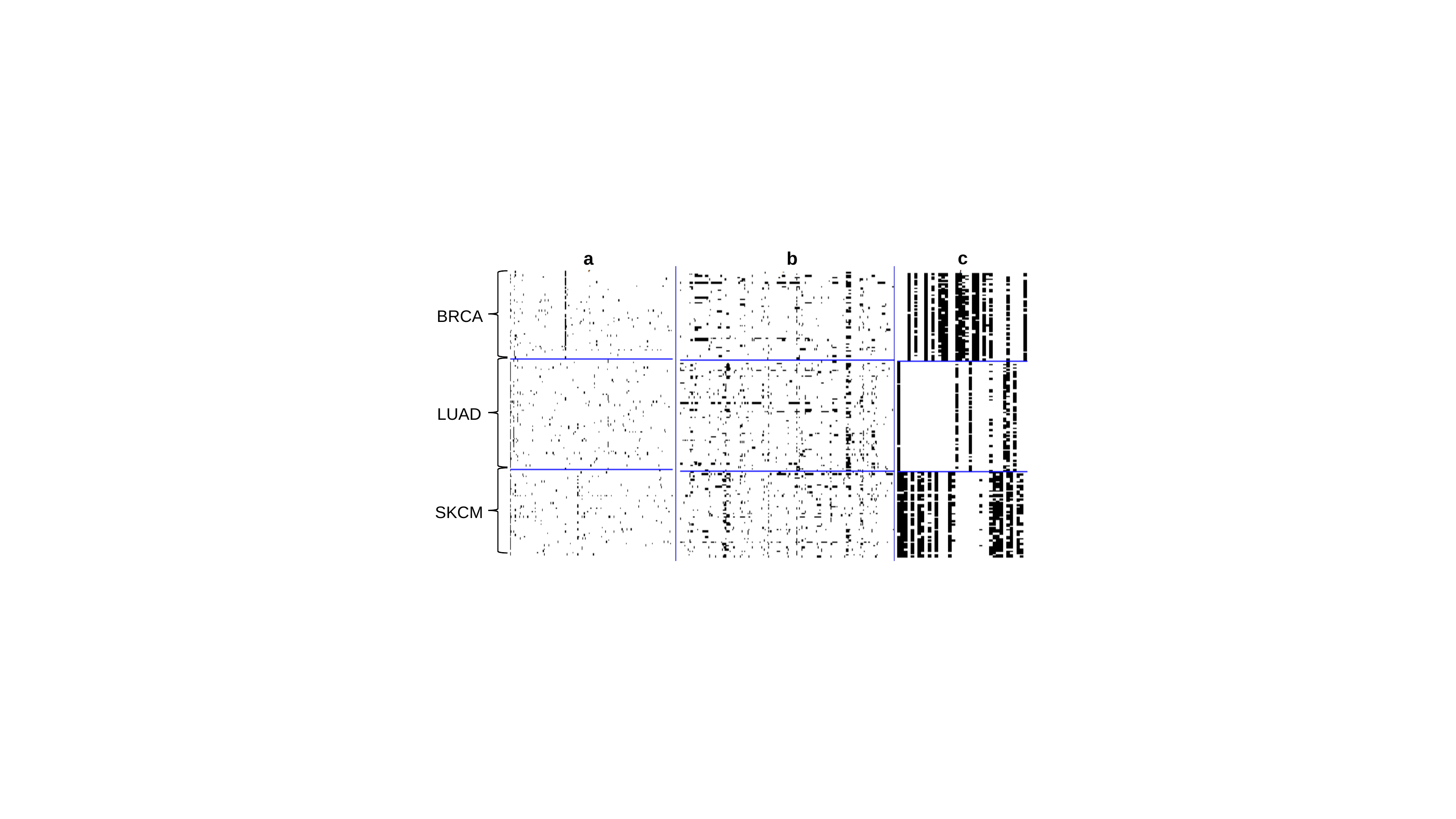}
    \label{chapter2_fig:S1}
    \caption*{Figure S2.1: Heatmap of the real data sets. White color: ``0''; black color: ``1''. \textbf{a}: mutation data; \textbf{b}: CNA data; \textbf{c}: methylation data. BRCA: breast invasive carcinoma; LUAD: lung adenocarcinoma; SKCM: skin cutaneous melanoma.}
\end{figure}

\begin{figure}[htbp]
    \centering
    \includegraphics[width=0.5\textwidth]{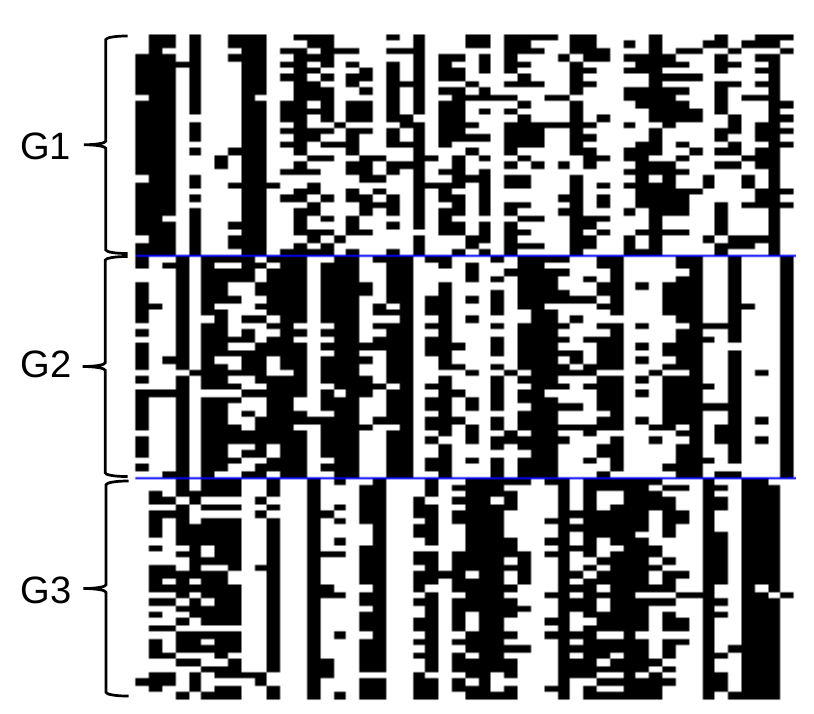}
    \caption*{Figure S2.2: Heatmap of the simulated balanced data set with low dimensional structure. White color: ``0''; black color: ``1''. G1, G2 and G3 are three groups in the samples during simulation.}
    \label{chapter2_fig:S2}
\end{figure}

\begin{figure}[htbp]
    \centering
    \includegraphics[width=0.9\textwidth]{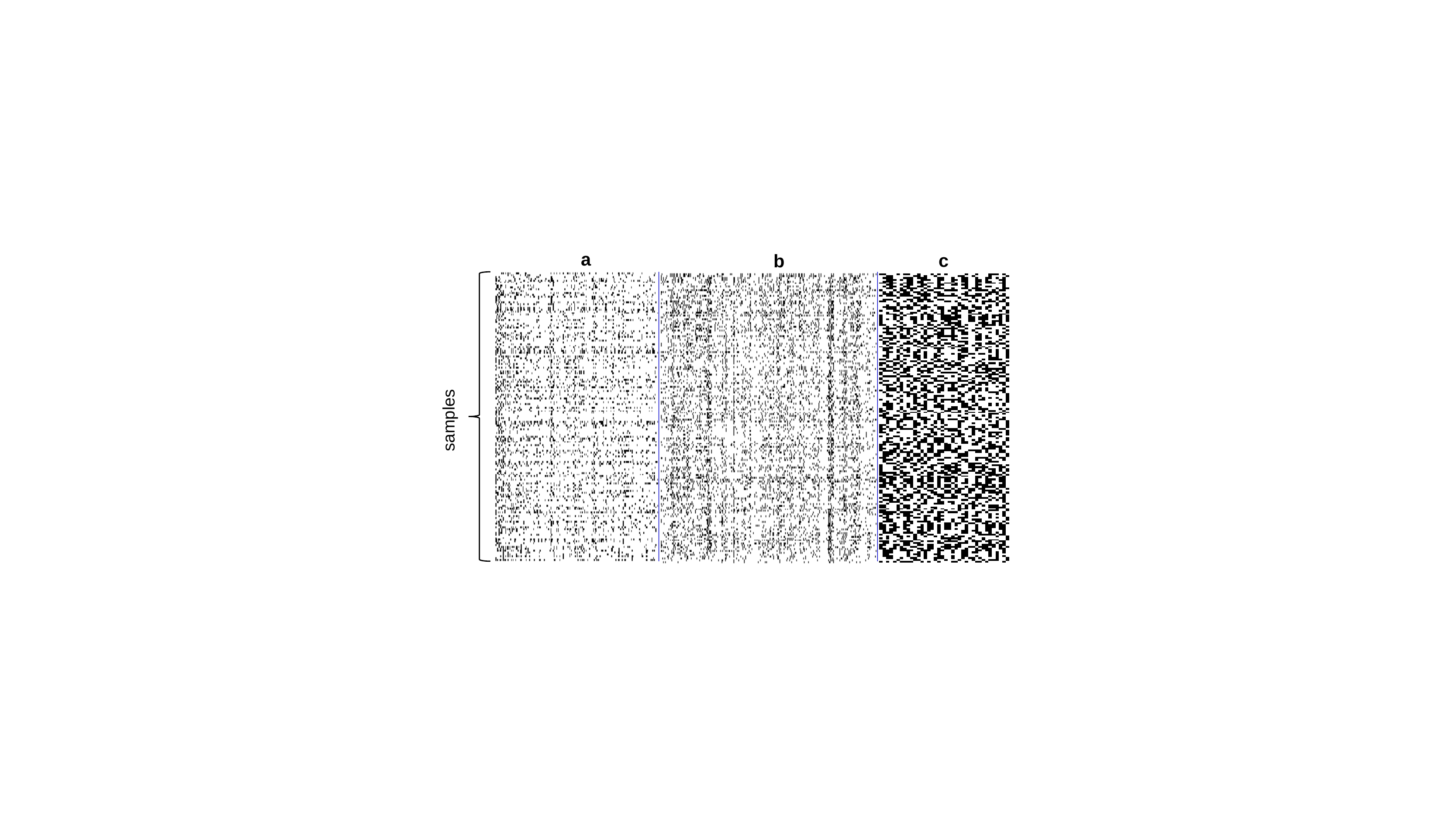}
    \caption*{Figure S2.3: Heatmap of simulated imbalanced binary data with different degrees of imbalance. White color: ``0''; black color: ``1''. \textbf{a}: similar as mutation data; \textbf{b}: similar as CNA data; \textbf{c}: similar as methylation data. No group structures are in the samples.}
    \label{chapter2_fig:S3}
\end{figure}

\begin{figure}[htbp]
    \centering
    \includegraphics[width=0.9\textwidth]{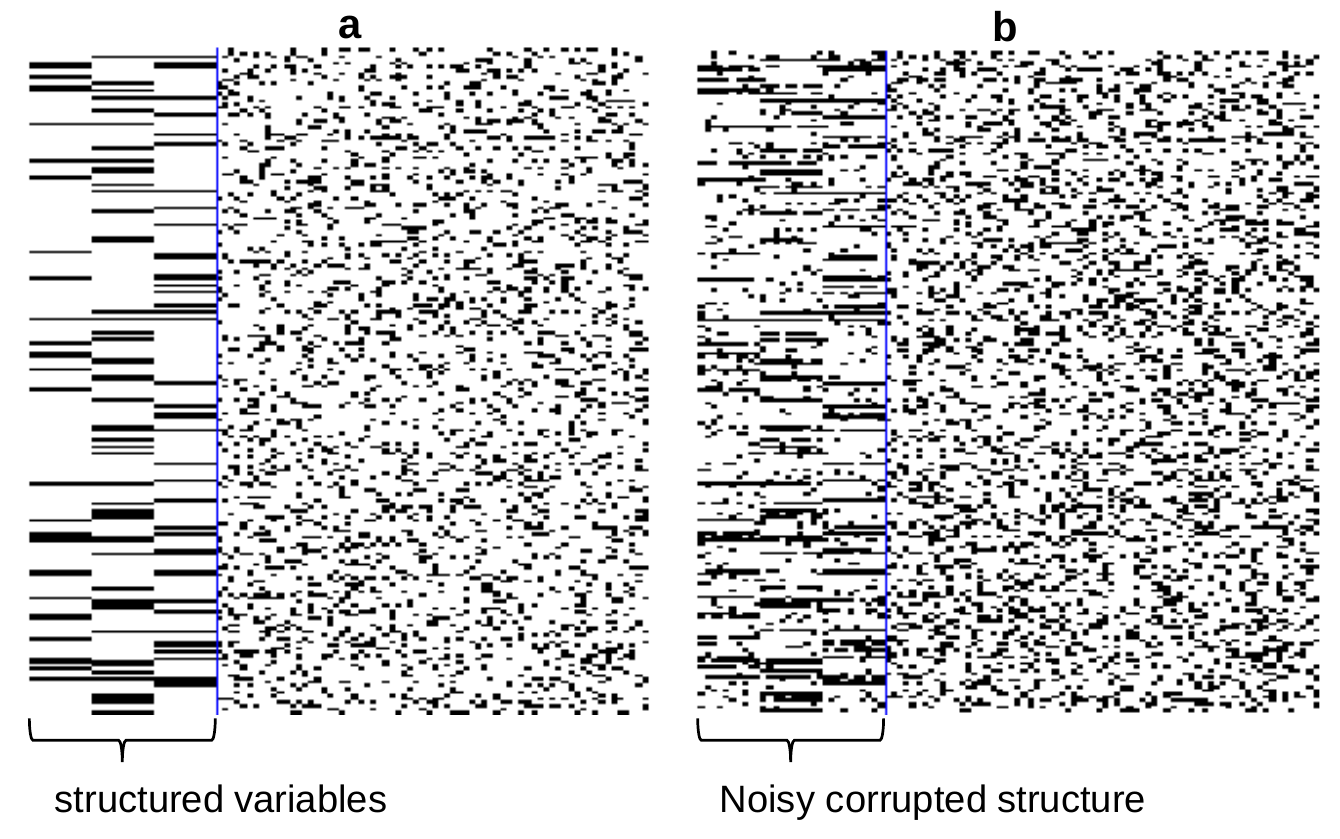}
    \caption*{Figure S2.4: Heatmap of the pre-structured binary data (\textbf{a}) and noise corrupted binary data (\textbf{b}). White color: ``0''; black color: ``1''.}
    \label{chapter2_fig:S4}
\end{figure}

\chapter{Robust logistic PCA model} \label{chapter:3}
Although the projection based logistic PCA model works well in our previous data analysis, the idea of projection in the context of multivariate binary data is not straightforward. Furthermore, the results of the model are related to the specification of the parameter $m$, which is the approximation of the infinity in the saturated model. In addition, this approach is difficult to be generalized to the data fusion of multiple data sets with different data types. Therefore, in this chapter we focus on developing a robust version of logistic PCA model without the involvement of projection.

The non-robustness of the standard logistic PCA model, manifested as some estimated parameters diverge towards infinity, is because of the used exact low rank constraint, which is specified as the multiplication of low rank score and loading matrices. Therefore, we propose to fit a logistic PCA model through non-convex singular value thresholding to alleviate the non-robustness issue. An efficient MM algorithm is implemented to fit the model and a missing value based CV procedure is introduced for the model selection. In addition, we re-express the logistic PCA model based on the latent variable interpretation of the generalized linear model on binary data. The multivariate binary data set is assumed to be the sign observation of an unobserved quantitative data set, on which a low rank structure is assumed to exist. The latent variable interpretation of the logistic PCA model not only makes the assumption of low rank structure easier to understand, but also provides us a way to define signal-to-noise ratio in multivariate binary data simulation. Our experiments on realistic simulations of imbalanced binary data and low signal-to-noise ratio show that the CV error based model selection procedure is successful in selecting the proposed model. And the selected model demonstrates superior performance in recovering the underlying low rank structure compared to models with convex nuclear norm penalty and exact low rank constraint. Finally, a binary CNA data set is used to illustrate the proposed methodology in practice.
\footnote{This chapter is based on Song, Y., Westerhuis, J.A. and Smilde, A.K., 2019. Logistic principal component analysis via non-convex singular value thresholding. arXiv preprint arXiv:1902.09486.}

\section{Background}
Logistic PCA \cite{schein2003generalized,de2006principal} is motivated from the probabilistic interpretation of the classical PCA model with Gaussian distributed error. The extension of the classical PCA model to the logistic PCA model is similar to the extension of linear regression to logistic linear regression. In the classical PCA model, the low rank constraint is imposed on the conditional mean of the observed quantitative data set, while in the logistic PCA model, the low rank constraint is imposed on the logit transform of the conditional mean of the observed binary data. Therefore, the logistic PCA model can also be re-expressed in a similar way as the latent variable interpretation of the generalized linear models (GLMs) on binary data \cite{agresti2013categorical}. In logistic PCA, the observed binary data set can be assumed as the sign observation of an unobserved quantitative data set, on which low rank structure is assumed to exist. This intuitive latent variable interpretation not only facilitates the understanding of the low rank structure in the logistic PCA model, but also provides a way to define the signal-to-noise ratio (SNR) in the simulation of multivariate binary data.

However, the standard logistic PCA model with the exact low rank constraint, which is expressed as the multiplication of two low rank matrices, is prone to overfitting, leading to divergence of some estimated parameters towards infinity \cite{de2006principal, groenen2016multinomial}. The same overfitting problem also happens for the logistic linear regression model. If two classes of the outcome are linearly separable with respect to an explanatory variable, the corresponding coefficient of this variable tends to go to infinity \cite{agresti2013categorical}. A common trick is adding a ridge regression (quadratic) type penalty on the coefficient vector to alleviate the overfitting issue. If we apply the same trick on the logistic PCA model, the quadratic penalty on the loading matrix is equivalent to a quadratic penalty on the singular values of a matrix, which is the multiplication of the score and loading matrices. Details will be shown later. Therefore, it is possible to derive a robust logistic PCA model via the regularization of the singular values. \cite{davenport20141} proposed to use a nuclear norm penalty in the low rank matrix approximation framework for the binary matrix completion problem. The proposed method is similar to the logistic PCA model except that the column offset term is not included and the exact low rank constraint is replaced by its convex relaxation, the nuclear norm penalty. The nuclear norm penalty, which is equivalent to applying a lasso penalty on the singular values of a matrix, induces low rank estimation and constrains the scale of non-zeros singular values simultaneously. However, a lasso type penalty shrinks all parameters to the same degree, leading to biased parameter estimation. This behavior will further make the CV error or the prediction error based model selection procedure inconsistent \cite{meinshausen2010stability}. On the other hand, non-convex penalties, many of which are concave functions, are capable to simultaneously achieve nearly unbiased parameter estimation and sparsity \cite{fan2001variable,armagan2013generalized}. Recent research \cite{shabalin2013reconstruction, josse2016adaptive} has also shown the superiority of non-convex singular value thresholding (applying non-convex penalties on the singular values of a matrix) in recovering the true signal in a low rank approximation framework under Gaussian noise. In this chapter, we propose to fit the logistic PCA model via non-convex singular value thresholding as a way to alleviate the overfitting problem and to induce low rank estimation simultaneously. A MM algorithm is implemented to fit the proposed model and an option for missing values is included. In the developed algorithm, the updating of all the parameters has an analytical form solution, and the loss function is guaranteed to decrease in each iteration. After that, a missing value based CV procedure is introduced for the model selection.

Based on the latent variable interpretation of the logistic PCA model, realistic multivariate binary data sets (low SNR, imbalanced binary data) are simulated to evaluate the performance of the proposed model and the corresponding model selection procedure. It turns out that the CV error based model selection procedure is successful in the selection of the proposed model, and the selected model has superior performance in recovering the underlying low rank structure compared to the model with convex nuclear norm penalty and exact low rank constraint. Furthermore, the performance of the logistic PCA model as a function of the SNR in multivariate binary data simulation is fully characterized. Finally, a binary CNA data set is used to illustrate the proposed methodology in practise.

\section{Latent variable interpretation of models on binary data}
\subsection{Latent variable interpretation of the GLMs on binary data}
A univariate binary response variable $y$ is assumed to follow a Bernoulli distribution with parameter $\pi$, $y \sim \text{Bernoulli}(\pi)$. $\mathbf{x}$ is a multivariate explanatory variable and $\mathbf{x} \in \mathbb{R}^J$. For the GLMs on binary data, we assume that the nonlinear transformation of the conditional mean of $y$ is a linear function of $\mathbf{x}$, $h(\text{E}(y|\mathbf{x})) = \mathbf{x}^{\text{T}} \bm{\beta}$, in which $h()$ is the link function, $\text{E}(y|\mathbf{x})$ is the conditional mean, and $\bm{\beta}$ is a $J$ dimensional coefficient vector. If the inverse function of $h()$ is $\phi()$, we have $\text{E}(y|x) = \phi(x^{\text{T}}\bm{\beta})$. If the logit link is used, $\phi(\theta) = (1+\exp(-\theta))^{-1}$, which is the logistic linear regression model, and $x^{\text{T}}\bm{\beta}$ can be interpreted as the log-odds, which is the natural parameter of Bernoulli distribution expressed in exponential family distribution form. If the probit link is used, $\phi(\theta) = \Phi(\theta)$, in which $\Phi(\theta)$ is the cumulative density function (CDF) of the standard normal distribution, which is the probit linear regression model.

The fact that the inverse link function $\phi()$ can be interpreted as the CDF of a specific probability distribution, motivates the latent variable interpretation of the logistic or probit linear regression \cite{agresti2013categorical}. $y$ can be assumed as the sign observation of a quantitative latent variable $y^{\ast}$, which has a linear relationship with the explanatory variable $\mathbf{x}$. Taking the probit linear regression as an example, the latent variable interpretation can be expressed as,
\begin{equation*}
\begin{split}
        y^{\ast} &= \mathbf{x}^{\text{T}}\beta + \epsilon\\
 \epsilon &\sim \text{N}(0,1) \\
        y &= \mathbb{1}{(y^{\ast}>0)},
\end{split}
\end{equation*}
in which $y^{\ast}$ is the latent variable, $\epsilon$ is the error term, and $\mathbb{1}()$ is the indicator function. The probability for the observation $y = 1$ is $\text{Pr}(y=1|\mathbf{x}^{\text{T}}\bm{\beta}) = \text{Pr}(y^{\ast} \geq 0)= \Phi(\mathbf{x}^{\text{T}}\bm{\beta})$. A similar latent variable interpretation can be applied to the logistic linear regression model by assuming that the error term $\epsilon$ follows the standard logistic distribution. The probability density function of the logistic distribution can be expressed as,
\begin{equation*}
  p(\epsilon) = \frac{\exp(- \frac{\epsilon-\mu}{\sigma})}{\sigma(1+\exp(-\frac{\epsilon-\mu}{\sigma}))^2},
\end{equation*}
in which $\mu$ and $\sigma$ are the location and scale parameters. In the standard logistic distribution, $\mu=0$, $\sigma=1$. The inverse-logit function $\phi()$ is the CDF of the standard logistic distribution. The assumption of $\mu=0$ for the $\epsilon$ is reasonable since we want to use the linear function $\mathbf{x}^{\text{T}}\bm{\beta}$ to capture the conditional mean of $y^{\ast}$. The assumption of $\sigma=1$ for the $\epsilon$ seems restrictive, however scaling the estimated $\hat{\bm{\beta}}$ by a positive constant as $\hat{\bm{\beta}}/\sigma$ will not change the conclusion of the model. Since the assumption of logistic distributed noise is not very straightforward, the latent variable interpretation of the logistic linear regression model is not widely used.

The above latent variable interpretation of the GLMs on binary data is naturally connected to the generating process of binary data \cite{young1980quantifying}. Binary data can be discrete in nature, for example when females and males are classified as ``1'' and ``0''. Another possibility is that there is a continuous process underlying the binary observation. For example in a toxicology study, the binary outcome of a subject being dead or alive relates to the dosage of a toxin used and the subject's tolerance level. The tolerance varies for different subjects, and the status (dead or alive) of a specific subject depends on whether its tolerance is higher than the used dosage or not. Thus, a continuous tolerance level is underlying the binary outcome \cite{agresti2013categorical}. If we assume our binary data set is generated from a continuous process, it is natural to use the latent variable interpretation of the probit link; or if it is assumed from a discrete process, we can use the logit link, and interpret it from the probabilistic perspective rather than the latent variable perspective. However, usually, the difference between the results derived from the GLMs using logit or probit link is negligible \cite{agresti2013categorical}.

\subsection{Latent variable interpretation of the logistic PCA model}
The measurement of $J$ binary variables on $I$ samples results in a binary matrix $\mathbf{X}$($I\times J$), whose $ij^{\text{th}}$ element $x_{ij}$ equals ``1'' or ``0''. The logistic PCA model on $\mathbf{X}$ can be interpreted as follows. Conditional on the low rank structure assumption, which is used to capture the correlations observed in $\mathbf{X}$, elements in $\mathbf{X}$ are independent realizations of the Bernoulli distributions, whose parameters are the corresponding elements of a probability matrix $\mathbf{\Pi}$($I \times J$), $\text{E}(\mathbf{X}|\mathbf{\Pi}) = \mathbf{\Pi}$. Assuming the natural parameter matrix, which is the logit transform of the probability matrix $\mathbf{\Pi}$, is $\mathbf{\Theta}$($I \times J$), we have $h(\mathbf{\Pi}) = \mathbf{\Theta}$ and $\mathbf{\Pi} = \phi(\mathbf{\Theta})$, in which $h()$ and $\phi()$ are the element-wise logit and inverse logit functions. The low rank structure is imposed on $\mathbf{\Theta}$ in the same way as in a classical PCA model, $\mathbf{\Theta} = \mathbf{1}\bm{\mu}^{\text{T}} + \mathbf{A}\mathbf{B}^{\text{T}}$, in which $\bm{\mu}$($J\times 1$) is the $J$ dimensional column offset term and can be interpreted as the logit transform of the marginal probabilities of the binary variables. $\mathbf{A}$ ($I \times R$) and $\mathbf{B}$($J \times R$) are the corresponding low rank score and loading matrices, and $R$, $R \ll \text{min}(I,J)$, is the low rank. Therefore, for the logistic PCA model, we have $\text{E}(\mathbf{X}|\mathbf{\Theta}) = \phi(\mathbf{\Theta}) = \phi(\mathbf{1}\bm{\mu}^{\text{T}} + \mathbf{A}\mathbf{B}^{\text{T}})$. On the other hand, in a classical PCA model, we have $\text{E}(\mathbf{X}|\mathbf{\Theta}) = \mathbf{\Theta} = \mathbf{1}\bm{\mu}^{\text{T}} + \mathbf{A}\mathbf{B}^{\text{T}}$, which is equivalent to using the identity link function. Furthermore, unlike in the classical PCA model, the column offset $\bm{\mu}$ has to be included into the logistic PCA model to do the model based column centering. The reason is that the commonly used column centering processing step is not allowed to be applied on the binary data set as the column centered binary data is not binary anymore.

The logistic PCA model can be re-expressed in the same way as the latent variable interpretation of the GLMs on binary data. Our binary observation $\mathbf{X}$ is assumed to be the sign observation of an underlying quantitative data set $\mathbf{X}^{\ast}$($I\times J$), and for the $ij^{\text{th}}$ element, we have $x_{ij} = 1$ if  $x^{\ast}_{ij} \geq 0$ and $x_{ij} = 0$ \textit{vice versa}. The low rank structure is imposed on the latent data set $\mathbf{X}^{\ast}$ as $\mathbf{X}^{\ast} = \mathbf{\Theta} + \mathbf{E}$, in which $\mathbf{E}$($I\times J$) is the error term, and its elements follow a standard logistic distribution. The latent variable interpretation of the logistic PCA model can be expressed as,
\begin{equation*}
\begin{split}
                  \mathbf{X}^{\ast} &= \mathbf{\Theta} + \mathbf{E} \\
                  \epsilon_{ij} & \sim \text{Logistic}(0,1), i = 1 \cdots I, j = 1 \cdots J \\
                  x_{ij} & = \mathbb{1}{(x^{\ast}_{ij}>0)}, i = 1 \cdots I, j = 1 \cdots J.
\end{split}
\end{equation*}

Similar to the latent variable interpretation of the logistic linear model, the assumption of $\epsilon_{ij} \sim \text{Logistic}(0,1)$ is not restrictive, since scaling the estimated $\hat{\mathbf{\Theta}}$ by a positive constant $\sigma$ will not change the conclusions from the model. When the standard normal distributed error is used in the above derivation, we get the probit PCA model. The latent variable interpretation of the logistic or probit PCA not only facilitates our understanding of the low rank structure underlying a multivariate binary data, but also provides a way to define the SNR in multivariate binary data simulation.

\section{Logistic PCA via singular value thresholding}
\subsection{The standard logistic PCA model}
Assume the column centered $\mathbf{\Theta}$ is $\mathbf{Z}$, $\mathbf{Z} = \mathbf{\Theta} - \mathbf{1}\bm{\mu}^{\text{T}} = \mathbf{AB}^{\text{T}}$. In the standard logistic PCA model, the exact low rank constraint is imposed on $\mathbf{Z}$ as the multiplication of two rank $R$ matrices $\mathbf{A}$ and $\mathbf{B}$. The negative log likelihood of fitting the observed $\mathbf{X}$ conditional on the low rank structure assumption on $\mathbf{\Theta}$ is used as the loss function. We also introduce a weight matrix $\mathbf{W}$($I \times J$) to tackle the potential missing values in $\mathbf{X}$. The $ij^{\text{th}}$ element of $\mathbf{W}$, $w_{ij}$, equals 0 when the corresponding element in $\mathbf{X}$ is missing; while it is 1 \textit{vice versa}. The optimization problem of the standard logistic PCA model can be expressed as,
\begin{equation}\label{chapter3_eq:1}
\begin{aligned}
\min_{\bm{\mu}, \mathbf{Z}} \quad & -\log(p(\mathbf{X}|\mathbf{\Theta},\mathbf{W}))\\
               &= -\log(\prod_{i}^{I}\prod_{j}^{J} (p(x_{ij}|\theta_{ij}))^{w_{ij}})\\
               &= -\sum_{i}^{I}\sum_{j}^{J} w_{ij} \left[x_{ij}\log(\phi(\theta_{ij})) + (1-x_{ij})\log(1-\phi(\theta_{ij}))\right] \\
           \text{subject to} \quad   \mathbf{\Theta} &= \mathbf{1}\bm{\mu}^{\text{T}} + \mathbf{Z}\\
                               \text{rank}(\mathbf{Z}) &= R \\
                               \mathbf{1}^{\text{T}}\mathbf{Z} &= \mathbf{0},
\end{aligned}
\end{equation}
in which the constraint $\mathbf{1}^{\text{T}}\mathbf{Z} = \mathbf{0}$ is imposed to make $\bm{\mu}$ identifiable. Unfortunately, the classical logistic PCA model tends to overfit the observed binary data. In order to decrease the loss function in equation \ref{chapter3_eq:1}, $\theta_{ij}$ tends to approach positive infinity when $x_{ij}=1$, and negative infinity when $x_{ij} = 0$. This overfitting problem will be explored in more detail below. In logistic linear regression, this overfitting problem can be solved by adding a quadratic penalty on the coefficient vector to regularize the estimated parameters. A similar idea can be applied to the logistic PCA model by taking it as a regression type problem. The columns of the score matrix $\mathbf{A}$ are taken as the unobserved explanatory variables, while the loading matrix $\mathbf{B}$ are the coefficients. If we decompose $\mathbf{Z}$ into a $R$ truncated SVD as $\mathbf{Z}=\mathbf{UDV}^{\text{T}}$, then $\mathbf{A}=\mathbf{U}$ and $\mathbf{B}=\mathbf{VD}^{\text{T}}$. It is easy to show that the quadratic penalty $||\mathbf{B}||_F^2 = \sum_{r} \sigma_{r}^2$, in which $\sigma_{r}$ is the $r^{\text{th}}$ singular value of $\mathbf{Z}$. Therefore, it is possible to derive a robust logistic PCA model by thresholding the singular values of $\mathbf{Z}$.

\subsection{Logistic PCA via non-convex singular value thresholding}
\begin{figure}[htbp]
    \centering
    \includegraphics[width=0.9\textwidth]{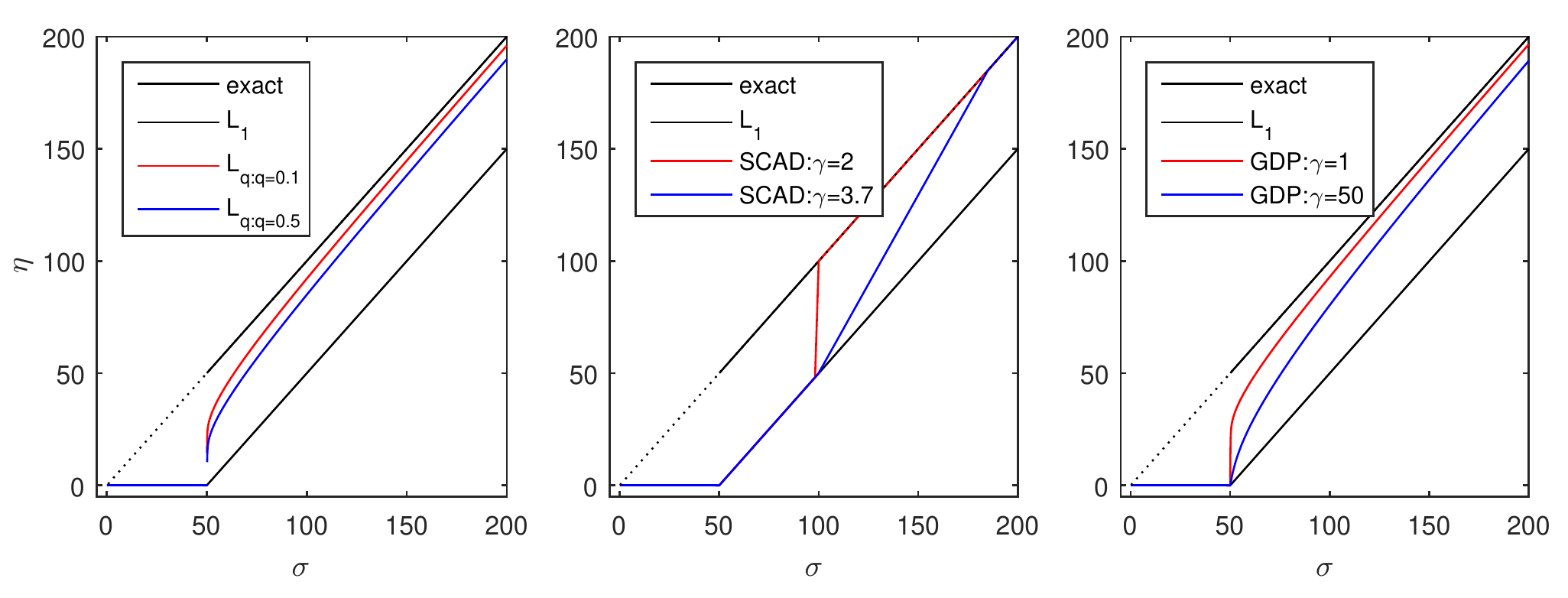}
    \caption{Thresholding properties of the exact low rank constraint, $L_{q}$, SCAD and GDP penalties when the same degree of shrinkage is achieved. exact: exact low rank constraint, $L_{1}$: nuclear norm penalty. $\sigma$ indicates the original singular value while $\eta$ is the value after thresholding. Note that in contrast to SCAD and GDP, the $L_{q: 0 < q < 1}$ penalty has a small discontinuity region, thus continuous thresholding can not be achieved.}
    \label{chapter3_fig:1}
\end{figure}

The most commonly used penalty function in thresholding singular values is the nuclear norm penalty, and it has been used in solving many low rank approximation problems \cite{candes2009exact,mazumder2010spectral,davenport20141,groenen2016multinomial}. If the SVD decomposition of matrix $\mathbf{Z}$ is $\mathbf{Z}=\mathbf{UDV}^{\text{T}}$, the nuclear norm penalty can be expressed as $\sum_{r} \sigma_r$, in which $\sigma_r$ is the $r^{\text{th}}$ singular value. The nuclear norm penalty is the convex relaxation of the exact low rank constraint and can be regarded as applying a lasso penalty on the singular values of a matrix. Therefore, the nuclear norm penalty has the same problem as the lasso penalty, it shrinks all singular values to the same degree. This leads to a biased estimation of the large singular values. This behavior will further make the prediction error or CV error based model selection procedure inconsistent \cite{meinshausen2010stability}. As an alternative, non-convex penalties can shrink the parameters in a nonlinear manner to achieve both nearly unbiased and sparse parameter estimation \cite{fan2001variable,armagan2013generalized}. Therefore, we propose to replace the exact low rank constraint in the logistic PCA model by a concave penalty on the singular values of $\mathbf{Z}$ to achieve a low rank estimation and to alleviate the overfitting issue. We include the frequentist version of the generalized double Pareto (GDP) \cite{armagan2013generalized} shrinkage, the smoothly clipped absolute deviation (SCAD) penalty \cite{fan2001variable} and the $L_{q:0<q \leq 1}$ penalty \cite{fu1998penalized} as examples of concave penalties in our implementation. The concave penalty on the singular values of $\mathbf{Z}$ can be expressed as $g(\mathbf{Z}) = \sum_{r} g(\sigma_r)$, in which $g()$ is a concave function in Table \ref{chapter3_tab:1}, $\sigma_{r}$ is the $r^{\text{th}}$ singular value of $\mathbf{Z}$. The thresholding properties of the exact low rank constraint, the nuclear norm penalty, and various concave penalties with different values of hyper-parameter are shown in Fig.~\ref{chapter3_fig:1}. Since the nuclear norm penalty is a linear function of the singular values, it is both convex and concave with respect to the singular values. Also, it is a special case of the $L_q$ penalty when setting $q=1$. Thus, the algorithm developed in this chapter also applies to the model with a nuclear norm penalty. The penalized negative log likelihood for fitting the observed binary data $\mathbf{X}$ of the logistic PCA with a concave penalty can be shown as,
\begin{equation}\label{chapter3_eq:2}
\begin{aligned}
\min_{\bm{\mu}, \mathbf{Z}} \quad & -\log(p(\mathbf{X}|\mathbf{\Theta},\mathbf{W})) + \lambda g(\mathbf{Z}) \\
           \text{subject to} \quad   \mathbf{\Theta} &= \mathbf{1}\bm{\mu}^{\text{T}} + \mathbf{Z}\\
                               \mathbf{1}^{\text{T}}\mathbf{Z} &= \mathbf{0},
\end{aligned}
\end{equation}
in which $\log(p(\mathbf{X}|\mathbf{\Theta},\mathbf{W}))$ and $g(\mathbf{Z})$ are as described above, and $\lambda$ is the tuning parameter.

\begin{table}[htbp]
\centering
\caption{Some commonly used concave penalty functions and their supergradients. $\sigma$ is taken as the singular value and $q$, $\lambda$ and $\gamma$ are tuning parameters. The supergradient is the counter-concept of the subgradient of a convex function in concave analysis, and it is mainly used in the developed algorithms.}
\label{chapter3_tab:1}
\begin{tabular}{lll}
  \toprule
Penalty & Formula & Supergradient \\
  \midrule
 Nuclear norm & $ \lambda \sigma $ & $\lambda$ \\

$L_{q}$ & $ \lambda \sigma^q $ & $\left\{ \begin{array}{ll} +\infty &\textrm{$\sigma=0$}\\
                                 \lambda q \sigma^{q-1} &\textrm{$\sigma>0$}\\ \end{array} \right.$ \\

SCAD & $\left\{ \begin{array}{ll} \lambda \sigma &\textrm{$\sigma \leq \lambda$}\\
 \frac{-\sigma^2+2\gamma \lambda \sigma - \lambda^2}{2(\gamma-1)} &\textrm{$\lambda < \sigma \leq \gamma \lambda$}\\
 \frac{\lambda^2(\gamma+1)}{2} &\textrm{$\sigma > \gamma \lambda$}\\ \end{array} \right.$ &
                          $\left\{ \begin{array}{ll} \lambda &\textrm{$\sigma \leq \lambda$}\\
 \frac{\gamma \lambda - \sigma}{\gamma-1} &\textrm{$\lambda < \sigma \leq \gamma \lambda$}\\
 0 &\textrm{$\sigma > \gamma \lambda$}\\ \end{array} \right.$ \\

GDP & $ \lambda \log(1+\frac{\sigma}{\gamma}) $ & $\frac{\lambda}{\gamma + \sigma}$ \\
  \bottomrule
\end{tabular}
\end{table}

\section{Algorithm}
Based on the MM principle \cite{de1994block,hunter2004tutorial}, a MM algorithm is derived to fit the logistic PCA model via non-convex singular value thresholding. The derived algorithm is guaranteed to decrease the objective function in equation \ref{chapter3_eq:2} during each iteration and the analytical form for updates of all the parameters in each iteration and are presented below. Although the following derivation focuses on using the logit link function, the option for probit link is included in our implementation.

\subsection{The majorization of the negative log-likelihood}
The negative log-likelihood $f(\mathbf{\Theta}) = -\log(p(\mathbf{X}|\mathbf{\Theta},\mathbf{W}))$ can be majorized to a quadratic function of $\mathbf{\Theta}$ by exploiting the upper-bound of the second order gradient of $f(\mathbf{\Theta})$. Suppose $f_{ij}(\theta_{ij}) = -\left[x_{ij}\log(\phi(\theta_{ij})) + (1-x_{ij})\log(1-\phi(\theta_{ij}))\right]$, in which $x_{ij}$ and $\theta_{ij}$ are the $ij^{\text{th}}$ elements of $\mathbf{X}$ and $\mathbf{\Theta}$, $f(\mathbf{\Theta})$ can be expressed as $f(\mathbf{\Theta}) = \sum_{i}^{I}\sum_{j}^{J} w_{ij} f_{ij}(\theta_{ij})$. When the logit link is used, the following results can be easily derived out, $\nabla f_{ij}(\theta_{ij}) = \phi(\theta_{ij}) - x_{ij}$, $\nabla^2 f_{ij}(\theta_{ij}) = \phi(\theta_{ij})(1-\phi(\theta_{ij}))$. Assume that $\nabla^2 f_{ij}(\theta_{ij})$ is upper bounded by a constant $L$. Since $\nabla^2 f_{ij}(\theta_{ij}) \leq 0.25$ when the logit link is used \cite{de2006principal} we can set $L=0.25$. Take $f(\theta)$ as the general representation of $f_{ij}(\theta_{ij})$, according to the Taylor's theorem and the assumption that $\nabla^2 f(\theta) \leq L$ for $\theta \in \text{domain}f$, we have the following inequality,
\begin{equation}\label{chapter3_eq:3}
\begin{aligned}
f(\theta) &= f(\theta^k) + <\nabla f(\theta^k), \theta-\theta^k> + \frac{1}{2}(\theta-\theta^k)^{\text{T}} \nabla^{2}f(\theta^k + t(\theta-\theta^k))(\theta-\theta^k) \\
          &\leq f(\theta^k) + <\nabla f(\theta^k), \theta-\theta^k> + \frac{L}{2}(\theta-\theta^k)^2 \\
          &= \frac{L}{2}(\theta-\theta^k + \frac{1}{L}\nabla f(\theta^k))^2 + c,\\
\end{aligned}
\end{equation}
where $\theta^k$ is the $k^{\text{th}}$ approximation of $\theta$, $t\in[0,1]$ is an unknown constant, $c$ is an unknown constant doesn't depend on $\mathbf{\Theta}$. Therefore, we have the following inequality about $f_{ij}(\theta_{ij})$,
$f_{ij}(\theta_{ij}) \leq \frac{L}{2}(\theta_{ij} - \theta_{ij}^k + \frac{1}{L}\nabla f_{ij}(\theta_{ij}^k ))^2 + c$. Assume $\nabla f(\mathbf{\Theta}^k)$ is the matrix forms of $\nabla f_{ij}(\theta_{ij}^k )$, $\nabla f(\mathbf{\Theta}^k)) = \phi(\mathbf{\Theta}^{k}) - \mathbf{X}$. The inequality of $f(\mathbf{\Theta})$ can be derived out as $f(\mathbf{\Theta})\leq \frac{L}{2} \sum_{i}^{I}\sum_{j}^{J} w_{ij}[(\theta_{ij} - \theta_{ij}^k + \frac{1}{L}\nabla f_{ij}(\theta_{ij}^k ))^2] + c = \frac{L}{2} ||\mathbf{W} \odot (\mathbf{\Theta} - \mathbf{\Theta}^k + \frac{1}{L} \nabla f(\mathbf{\Theta}^k))||_F^2 + c$, in which $\odot$ indicates element-wise matrix multiplication. Following \cite{kiers1997weighted}, we further majorize the weighted least-squares upper bound into a quadratic function of $\mathbf{\Theta}$ as
\begin{equation}\label{chapter3_eq:4}
\begin{aligned}
              &\frac{L}{2} ||\mathbf{W} \odot(\mathbf{\Theta} - \mathbf{\Theta}^k + \frac{1}{L} \nabla f(\mathbf{\Theta}^k))||_F^2 \\
             &\leq  \frac{L}{2} ||\mathbf{\Theta}-\mathbf{H}^k||_F^2 + c \\
              \mathbf{H}^k &= \mathbf{\Theta}^k - \frac{1}{L} (\mathbf{W}\odot \nabla f(\mathbf{\Theta}^k)).
\end{aligned}
\end{equation}

\subsection{The majorization of the non-convex penalty}
Suppose $\sigma_r$ is the $r^{\text{th}}$ singular value of $\mathbf{Z}$, and $g()$ is a concave function. From the definition of concavity \cite{boyd2004convex}, we have $g(\sigma_r) \leq g(\sigma_r^k) + \omega_r^k(\sigma_r - \sigma_r^k) = \omega_r^k \sigma_r + c$, in which $\sigma_r^k$ is the $r^{\text{th}}$ singular value of the $k^{\text{th}}$ approximation $\mathbf{Z}^k$ and $c$ is an unknown constant. Also, $\omega_r^k \in \partial g(\sigma_r^k)$ and $\partial g(\sigma_r^k)$ is the set of supergradients of function $g()$ at $\sigma_r^k$. For all the concave penalties used in Table \ref{chapter3_tab:1}, their supergradient is unique, thus $\omega_r^k = \partial g(\sigma_r^k)$. Therefore, $g(\mathbf{Z})= \sum_{r}g(\sigma_r(\mathbf{Z}))$ can be majorized as follows
\begin{equation}\label{chapter3_eq:5}
\begin{aligned}
g(\mathbf{Z}) &= \sum_{r}g(\sigma_r)\\
              &\leq \sum_{r}\omega_{r}^k \sigma_r + c\\
       \omega_r^k &= \partial g(\sigma_r^k).
\end{aligned}
\end{equation}

\subsection{Block coordinate descent}
Summarizing the above two majorization steps, we have the following majorized problem during the $k^{\text{th}}$ iteration.
\begin{equation}\label{chapter3_eq:6}
\begin{aligned}
\min_{\bm{\mu},\mathbf{Z}} \quad & \frac{L}{2}||\mathbf{\Theta}-\mathbf{H}^{k}||_F^2 + \lambda \sum_{r} \omega_r^k\sigma_{r}\\
           \text{subject to} \quad   \mathbf{\Theta} &= \mathbf{1}\bm{\mu}^{\text{T}} + \mathbf{Z}\\
                               \mathbf{1}^{\text{T}}\mathbf{Z} &= \mathbf{0} \\
                                \mathbf{H}^k &= \mathbf{\Theta}^k - \frac{1}{L} (\mathbf{W}\odot \nabla f(\mathbf{\Theta}^k)) \\
                               \omega_r^k &= \partial g(\sigma_r^k).
\end{aligned}
\end{equation}
This majorized problem during the $k^{\text{th}}$ iteration can be solved by the block coordinate descent algorithm.

\subsubsection*{Updating $\bm{\mu}$}
When fixing $\mathbf{Z}$ in equation \ref{chapter3_eq:6}, the analytical form solution of $\bm{\mu}$ is the column mean of $\mathbf{H}^k$, $\bm{\mu} = \frac{1}{I} (\mathbf{H}^k)^{\text{T}} \mathbf{1}$.

\subsubsection*{Updating $\mathbf{Z}$}
After deflating the offset set term $\bm{\mu}$ in equation \ref{chapter3_eq:6}, the optimization problem of $\mathbf{Z}$ becomes $\min_{\mathbf{Z}} \frac{L}{2}||\mathbf{Z}-\mathbf{J}\mathbf{H}^k||_F^2 + \lambda \sum_{r} \omega_r^k\sigma_{r}$, in which $\mathbf{J}$ is the column centering operator $\mathbf{J} = \mathbf{I} - \frac{1}{I}\mathbf{11}^{\text{T}}$. This optimization problem is equivalent to finding the proximal operator of the weighted sum of singular values, for which the analytical form global solution exists \cite{lu2015generalized}. If the SVD decomposition of $\mathbf{J}\mathbf{H}^k$ is $\mathbf{J}\mathbf{H}^k = \mathbf{UDV}^{\text{T}}$, the analytical form solution of $\mathbf{Z}$ is $\mathbf{Z} = \mathbf{U}\mathbf{D}_{z} \mathbf{V}^{\text{T}}$, in which $\mathbf{D}_{z} = \text{Diag}\{\text{max}(0, d_{r}-\frac{\lambda \omega_r^k}{L}) \}$, and $d_{r}$ is $r^{\text{th}}$ element of the diagonal of $\mathbf{D}$.

\subsubsection*{Initialization}
The initialization $\mathbf{Z}^0$ and $\mu^0$ can be set according to the user imputed values, or by using the following random initialization strategy. All the elements in $\mathbf{Z}^0$ can be sampled from the standard uniform distribution and $\mu^0$ can be set to $\mathbf{0}$. In the following algorithm, $f^k$ indicates the objective value in equation \ref{chapter3_eq:2} during the $k^{\text{th}}$ iteration, the relative change of the objective value is used as the stopping criteria. $\epsilon_f$ indicates the tolerance for the relative change of the objective value. Pseudocode of the algorithm described above is shown in Algorithm \ref{alg:logisticPCA}.

\begin{algorithm}[htb]
  \caption{An MM algorithm to fit the logistic PCA model via non-convex singular value thresholding.}
  \label{alg:logisticPCA}
  \begin{algorithmic}[1]
    \Require
      $\mathbf{X}$, $\lambda$, $\gamma$;
    \Ensure
      $\bm{\mu}$, $\mathbf{A}$, $\mathbf{B}$;
    \State $k = 0$;
    \State Compute $\mathbf{W}$ for missing values;
    \State Initialize $\mu^0$, $\mathbf{Z}^0$;

    \While{$(f^{k-1}-f^{k})/f^{k-1}>\epsilon_f$}
        \State $\mathbf{\Theta}^k = \mathbf{1}(\bm{\mu}^k)^{\text{T}} + \mathbf{Z}^k$;
        \State $\nabla f(\mathbf{\Theta}^{k})= \phi(\mathbf{\Theta}^k)-\mathbf{X}$;
        \State $\mathbf{H}^k = \mathbf{\Theta}^k - \frac{1}{L} (\mathbf{W}\odot \nabla f(\mathbf{\Theta}^k))$;
        \State $\omega_r^k = \partial g(\sigma_r^k)$;
        \State $\bm{\mu}^{k+1} = \frac{1}{I}(\mathbf{H}^{k})^{\text{T}} \mathbf{1}$;
        \State $\mathbf{JH}=\mathbf{J}\mathbf{H}^{k}$;
        \State $\mathbf{UDV}^{\text{T}} = \mathbf{JH}$;
        \State $\mathbf{D}_{z} = \text{Diag}\{\text{max}(0, d_{r}-\frac{\lambda \omega_r^k}{L}) \}$;
        \State $\mathbf{Z}^{k+1} = \mathbf{U}\mathbf{D}_{z}\mathbf{V}^{\text{T}}$;
        \State $\mathbf{\Theta}^{k+1} = \bm{\mu}^{k+1} + \mathbf{Z}^{k+1}$;
        \State $k=k+1$;
    \EndWhile
    \State $\mathbf{A} = \mathbf{U}$;
    \State $\mathbf{B} = \mathbf{V}\mathbf{D}_{z}$;
  \end{algorithmic}
\end{algorithm}

\section{Real data set and simulation process}
\subsection{Real data set}
The CNA data sets in Chapter \ref{chapter:1} is used as an example of real data sets to show the results. The characterization of the CNA data set is shown in supplemental Fig.~S1 b.

\subsection{Simulation process}
Multivariate binary data $\mathbf{X}$ is simulated according to the logistic PCA model in a similar way as Chapter \ref{chapter:2} except that SNR is defined and there are no group structures in the sample space. Based on the latent variable interpretation of the logistic PCA model, we can define the SNR as $\text{SNR}=\frac{||\mathbf{Z}||_F^2}{||\mathbf{E}||_F^2}$, in which $\mathbf{E}$ is the error term, and its elements are sampled from the standard logistic distribution. The column offset term $\bm{\mu}$ can be set in the same way as in Chapter \ref{chapter:1}. However the rank $R$ matrix $\mathbf{Z}=\mathbf{AB}^{\text{T}}$ is simulated in a slightly different way. We first express $\mathbf{Z}$ in a SVD type as $\mathbf{Z}= \mathbf{UDV}^{\text{T}}$, in which $\mathbf{U}^{\text{T}}\mathbf{U} = \mathbf{I}_{R}$, $\mathbf{V}^{\text{T}}\mathbf{V} = \mathbf{I}_{R}$ and the diagonal of $\mathbf{D}$ contains the singular values. Elements in $\mathbf{U}$ and $\mathbf{V}$ are first sampled from $N(0,1)$. After that, the column mean of $\mathbf{U}$ is deflated to have $\mathbf{1}^{\text{T}}\mathbf{U}=\mathbf{0}$, and the SVD is used to force $\mathbf{U}$ being orthogonal. Also, $\mathbf{V}$ is forced to orthogonality by the Gram-Schmidt algorithm. Then, the diagonal matrix $\mathbf{D}_{pre}$, whose $R$ diagonal elements are the sorted absolute values of the samples from $N(1,0.5)$, is simulated. We express $\mathbf{D}$ as $\mathbf{D} = c\mathbf{D}_{pre}$, in which $c$ is a constant used to adjust the SNR in the simulation of the multivariate binary data. Then $\mathbf{\Theta} = \mathbf{1}\bm{\mu}^{\text{T}} + \mathbf{Z}$ according to the logistic PCA model and $\mathbf{X}^{\ast} = \mathbf{\Theta} + \mathbf{E}$ according to the latent variable interpretation. The probability matrix $\mathbf{\Pi}$ is generated as $\mathbf{\Pi} = \phi(\mathbf{\Theta})$, in which $\phi()$ by the inverse logit link function. Finally, the multivariate binary data set $\mathbf{X}$ is generated from Bernoulli distributions with the corresponding parameters in $\mathbf{\Pi}$.

\section{Model assessment and model selection}
In this chapter we focus on evaluating the model's performance in estimating the simulated parameters $\mathbf{\Theta}$, $\mathbf{\Pi}$, $\bm{\mu}$ and $\mathbf{Z}$. Also, the CV error is defined based on the log likelihood of fitting binary data rather than misclassifying the binary data.

\subsection{Model assessment}
After a logistic PCA model is constructed on the simulated binary data, we have the estimated parameters $\hat{\bm{\mu}}$, $\hat{\mathbf{A}}$ and $\hat{\mathbf{B}}$, and $\hat{\mathbf{\Theta}} = \mathbf{1}\hat{\bm{\mu}}^{\text{T}} + \hat{\mathbf{AB}^{\text{T}}}$ and $\hat{\mathbf{\Pi}} = \phi(\hat{\mathbf{\Theta}})$ can also be computed.
The model's ability in recovering the true $\mathbf{\Theta}$ can be evaluated by the relative mean squares error (RMSE), which is defined as $\text{RMSE}(\mathbf{\Theta}) = \frac{||\mathbf{\Theta}-\hat{\mathbf{\Theta}}||_F^2}{||\mathbf{\Theta}||_F^2}$, where $\mathbf{\Theta}$ is the true parameter. The RMSEs in estimating $\bm{\mu}$ and $\mathbf{Z}$ are defined in the same way. In some cases the mean Hellinger distance (MHD) to quantify the similarity between the true probability matrix $\mathbf{\Pi}$ and the estimated $\hat{\mathbf{\Pi}}$ is used. Hellinger distance \cite{le2012asymptotics} is a symmetric measure to quantify the similarity between two probability distributions. Assuming the parameter of a Bernoulli distribution is $\pi$ and its estimation is $\hat{\pi}$, the Hellinger distance is defined as $\text{HD}(\pi,\hat{\pi}) = \frac{1}{\sqrt{2}}\sqrt{(\sqrt{\pi}-\sqrt{\hat{\pi}})^2 + (\sqrt{1-\pi}-\sqrt{1-\hat{\pi}})^2}$. The mean Hellinger distance between the probability matrix $\mathbf{\Pi}$ and its estimate $\hat{\mathbf{\Pi}}$ is defined as $\text{MHD}(\mathbf{\Pi}) =  \frac{1}{I\times J} \sum_{i,j}^{I,J}\text{HD}(\pi_{ij},\hat{\pi}_{ij})$.

\subsection{Model selection}
For the model selection on real data, a missing value based cross validation (CV) procedure is proposed. The CV error is computed as follows. First, elements in $\mathbf{X}$ are split into the training and test sets as follows: $10\%$ ``1''s and ``0''s of $\mathbf{X}$ are randomly selected as the test set $\mathbf{X}^{\text{test}}$, which are set to missing values, and the resulting data set is taken as $\mathbf{X}^{\text{train}}$. After getting an estimation of $\hat{\mathbf{\Theta}}$ from a logistic PCA on the $\mathbf{X}^{\text{train}}$, we can index the elements, which are corresponding to the test set $\mathbf{X}^{\text{test}}$, as $\hat{\mathbf{\Theta}}^{\text{test}}$. Then the CV error is defined as the negative log-likelihood of using $\hat{\mathbf{\Theta}}^{\text{test}}$ to fit $\mathbf{X}^{\text{test}}$.

There are two tuning parameters, $\gamma$ and $\lambda$, during the model selection of the logistic PCA model with a concave penalty. However, the performance of the model is rather insensitive to the selection of $\gamma$ for some concave penalties, which will be shown below. After fixing the value of $\gamma$, we can use a grid search to select a proper value of $\lambda$ based on the minimum CV error. First, a sequence of $\lambda$ values can be selected from a proper searching range, after which logistic PCA models will be fitted with the selected $\lambda$ values on the training set $\mathbf{X}^{\text{train}}$. A warm start strategy, using the results of a previous model as the initialization of the next model, is used to accelerate the model selection process. The model with the minimum CV error is selected and then it is re-fitted on the full data set $\mathbf{X}$. Because the proposed model is non-convex and its result is sensitive to the used initialization, it is recommended to use the results derived from the selected model as the initialization of the model to fit the full data sets.

\section{Results}
\subsection{Standard logistic PCA model tends to overfit the data}
\begin{figure}[htbp]
    \centering
    \includegraphics[width=0.9\textwidth ]{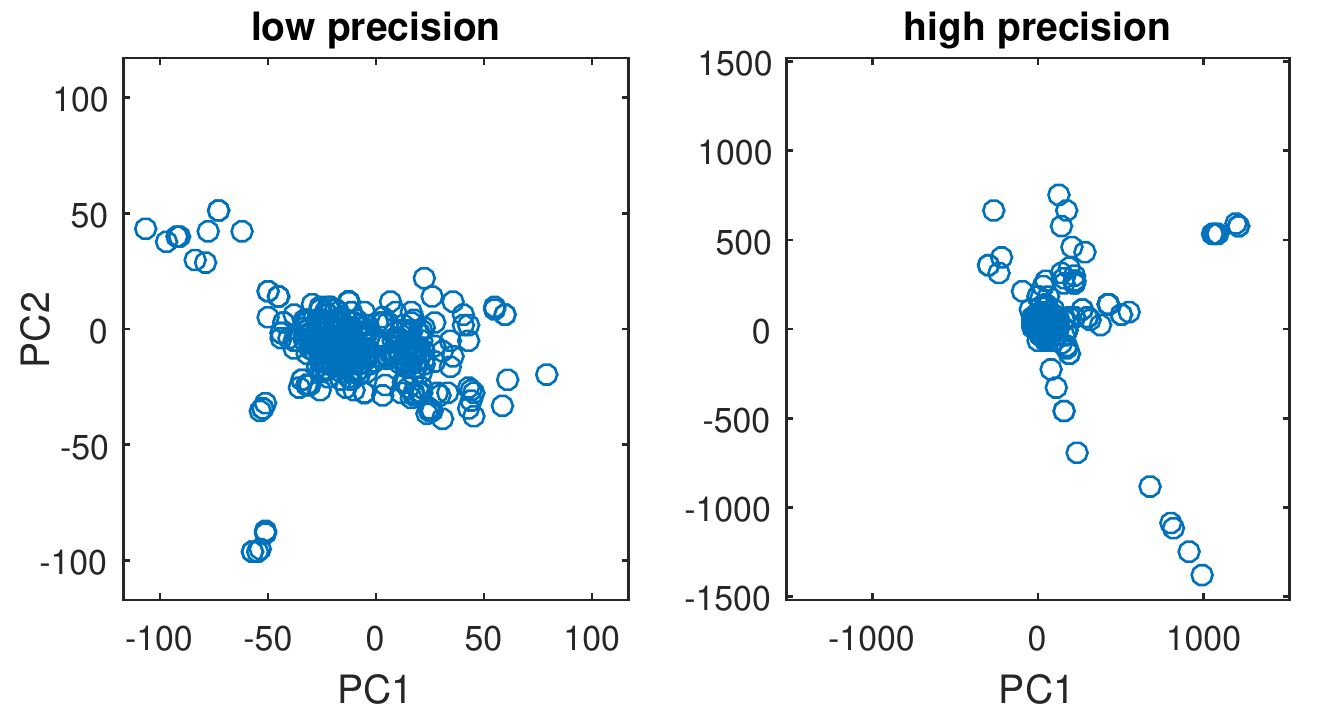}
    \caption{The loading plots of the first two components derived from the low precision (left) and high precision (right) standard logistic PCA models.}
    \label{chapter3_fig:2}
\end{figure}

In this first section we will use the CNA data as an example. The algorithm from \cite{de2006principal} is implemented to fit the standard logistic PCA model. Constraints, $\mathbf{A}^{\text{T}}\mathbf{A} = \mathbf{I}$ and $\mathbf{1}^{\text{T}}\mathbf{A}$, are imposed. Two different standard logistic PCA models  are constructed of the CNA data, both with three components. The first model is obtained with low precision (stopping criteria was set to $\epsilon_f=10^{-4}$) while for the other model a high precision was used ($\epsilon_f=10^{-8}$). The initialization was the same for these two models. The low precision model converged already after 220 iterations, while the high precision model did not convergence even after 50000 iterations. The difference between the final objective values of these two models is not large, $8.12e+03$ and $7.58e+03$ respectively. However, as shown in Fig.~\ref{chapter3_fig:2}, the scale of the loading plots derived from these two models is very different. When a high precision stopping criteria is used, some of the elements from the estimated loading matrix from the standard logistic PCA model tend to become very large.

\subsection{Model selection of the logistic PCA model with a concave GDP penalty}
\begin{figure}[htbp]
    \centering
    \includegraphics[width=0.9\textwidth ]{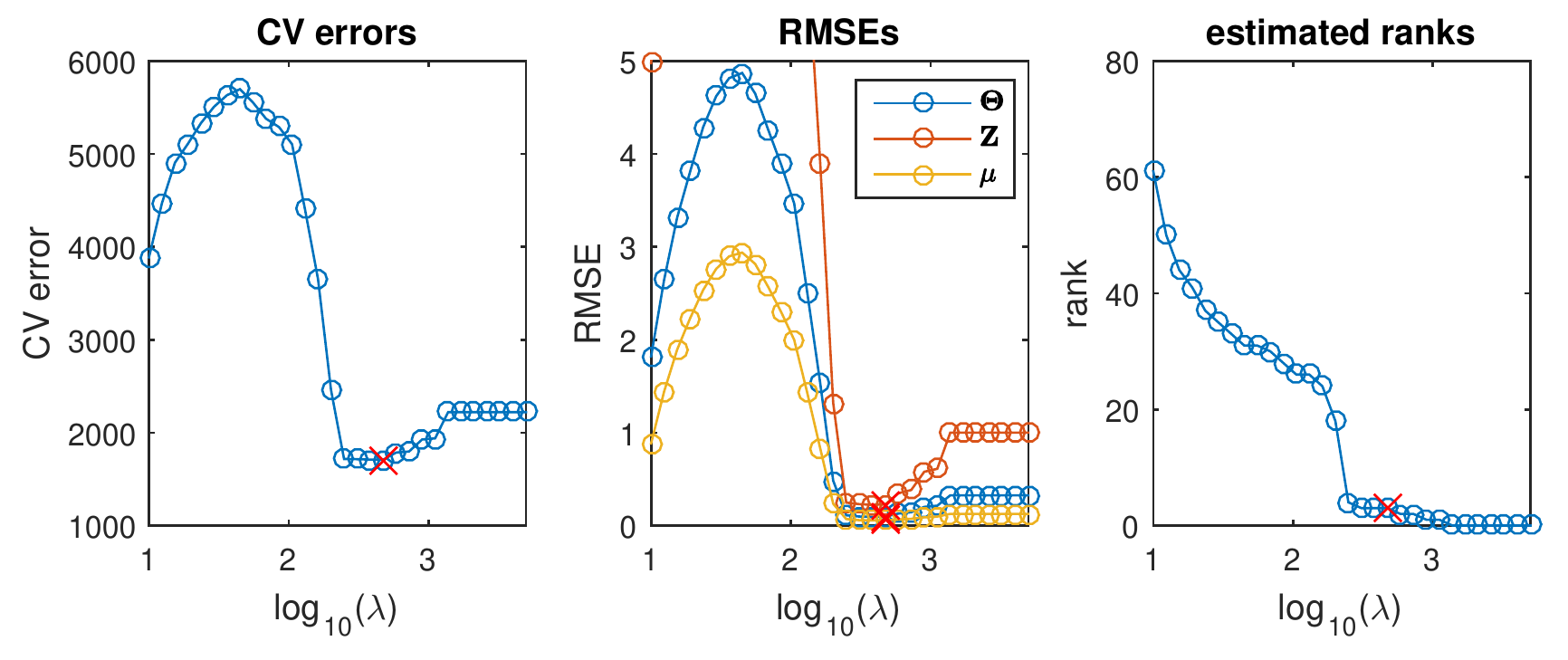}
    \caption{Model selection and performance of the logistic PCA model with a GDP penalty. The CV error, RMSE of estimating $\mathbf{\Theta}$, $\mathbf{Z}$ and $\bm{\mu}$ and the estimated rank as a function of $\lambda$. The increased CV error and RMSEs for small $\lambda$ are the result of non-converged models after 500 iterations. The red cross marker indicates the $\lambda$ value where minimum CV error is achieved.}
	\label{chapter3_fig:3}
\end{figure}

Here we use logistic PCA model with a concave GDP penalty as an example to show the CV error based model selection procedure. The data set is simulated as follows. The offset term $\bm{\mu}$ is set to the logit transform of the empirical marginal probabilities of the CNA data to simulate an imbalanced binary data set. Other parameters used in the simulation are $I=160$, $J=410$, $\text{SNR}=1$ and $R=5$. First we will show the model selection procedure of $\lambda$ while the hyper-parameter $\gamma$ is fixed to $\gamma=1$. After splitting the simulated binary data set $\mathbf{X}$ into the training set $\mathbf{X}^{\text{train}}$ and the test set $\mathbf{X}^{\text{test}}$, 30 $\lambda$ values are selected from the searching range $[10,5000]$ with equal distance in log-space. For each $\lambda$ value, a logistic PCA with a GDP penalty ($\epsilon_f = 10^{-6}$, maximum iteration is 500) is constructed on $\mathbf{X}^{\text{train}}$ and for each model we evaluate its performance in estimating the simulated parameters. As shown in the model selection results (Fig.~\ref{chapter3_fig:3}), the selected model with minimum CV error can also achieve approximately optimal RMSEs in estimating the simulated $\mathbf{\Theta}$, $\mathbf{Z}$ and $\bm{\mu}$ . However, the rank of the estimated $\mathbf{Z}$ from the selected model is 3, which is different from the simulated rank $R=5$. The reason will be discussed later. The selected model is re-fitted on the full simulated data $\mathbf{X}$, and the RMSEs of estimating $\mathbf{\Theta}$, $\mathbf{Z}$ and $\bm{\mu}$ are 0.0797, 0.2064 and 0.0421 respectively.

\begin{figure}[htbp]
    \centering
    \includegraphics[width=0.9\textwidth ]{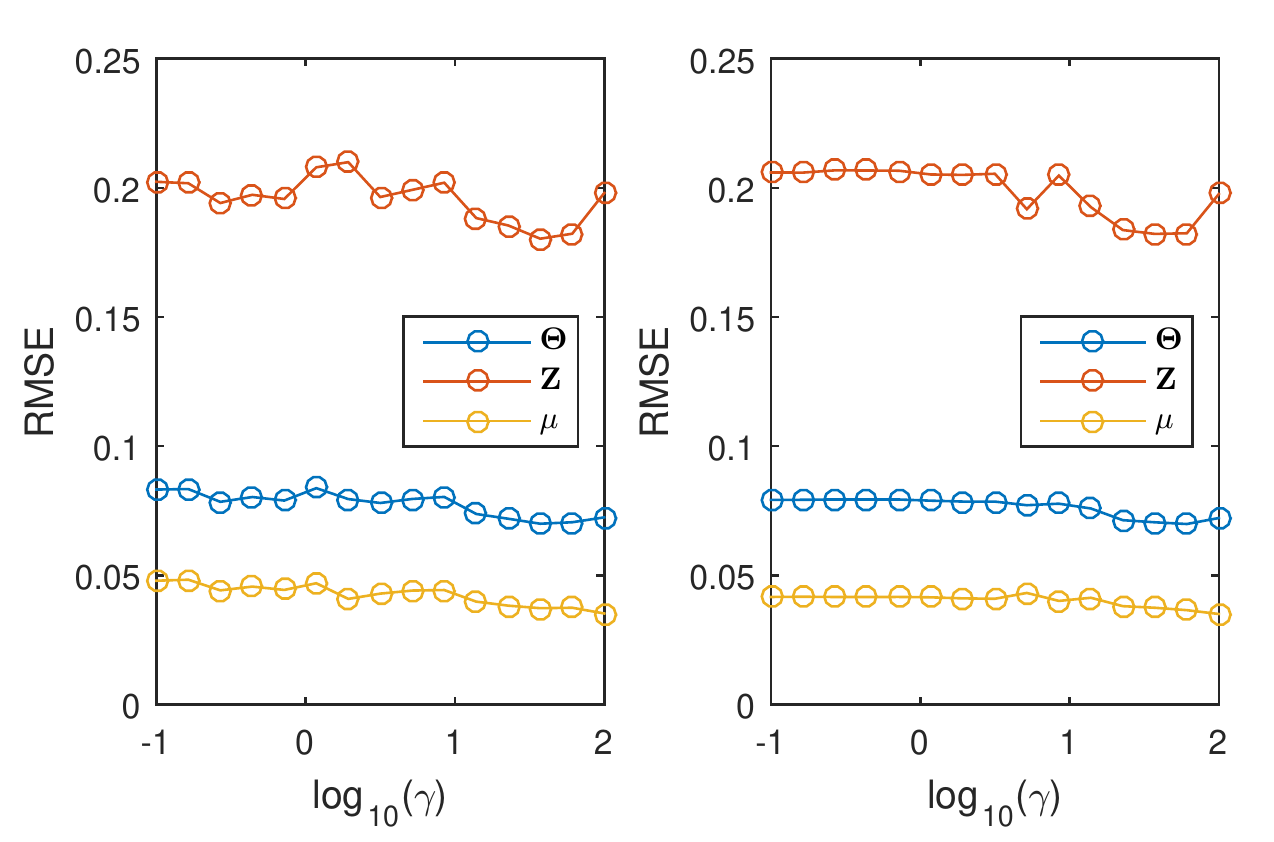}
    \caption{The RMSE of estimating $\mathbf{\Theta}$, $\mathbf{Z}$ and $\bm{\mu}$ as a function of hyper-parameter $\gamma$ of GDP penalty. Results on the left side are obtained when the optimal model is selected based on minimum CV error while on the right hand side model selection was based on minimum  $\text{RMSE}(\mathbf{\Theta})$.}
    \label{chapter3_fig:4}
\end{figure}

Next, we will show the model selection process of both $\gamma$ and $\lambda$. The simulated data $\mathbf{X}$ is split into the $\mathbf{X}^{\text{train}}$ and the $\mathbf{X}^{\text{test}}$ in the same way as described above. 30 $\gamma$ values are selected from the range $[10^{-1},10^{2}]$ equidistant in log-space. For each $\gamma$, 30 values of $\lambda$ are selected from a proper searching range, which is determined by an automatic procedure. For each value of $\gamma$, the model selection of $\lambda$ is done on the $\mathbf{X}^{\text{train}}$ in the same way as described above, after which the selected model is re-fitted on the full data $\mathbf{X}$. Therefore, for each value of $\gamma$, we have a selected model, which is optimal with respect to the CV error. As shown in Fig.~\ref{chapter3_fig:4}(left), the difference between the RMSEs derived from these selected models is very small. This can be caused by two reasons: the model is insensitive to the selection of $\gamma$ or the CV error based model selection procedure is not successful in selecting $\gamma$. To clarify the correct reason, we also fit $30 \times 30$ models on the full data $\mathbf{X}$ in the same way as the above experiment. For each value of $\gamma$, the model with minimum $\text{RMSE}(\mathbf{\Theta})$ is selected. As shown in Fig.~\ref{chapter3_fig:4}(right), the value of $\gamma$ does not have a large effect on the RMSEs of the selected models, which are optimal with respect to the $\text{RMSE}(\mathbf{\Theta})$. Therefore, it can be concluded that the performance of the model is insensitive to the model selection of $\gamma$. Therefore, the strategy can be to set a default value for $\gamma$ and focus on the selection of $\lambda$.

\subsection{Model selection of the logistic PCA model with other concave penalties}
\begin{figure}[htbp]
    \centering
    \includegraphics[width=0.9\textwidth ]{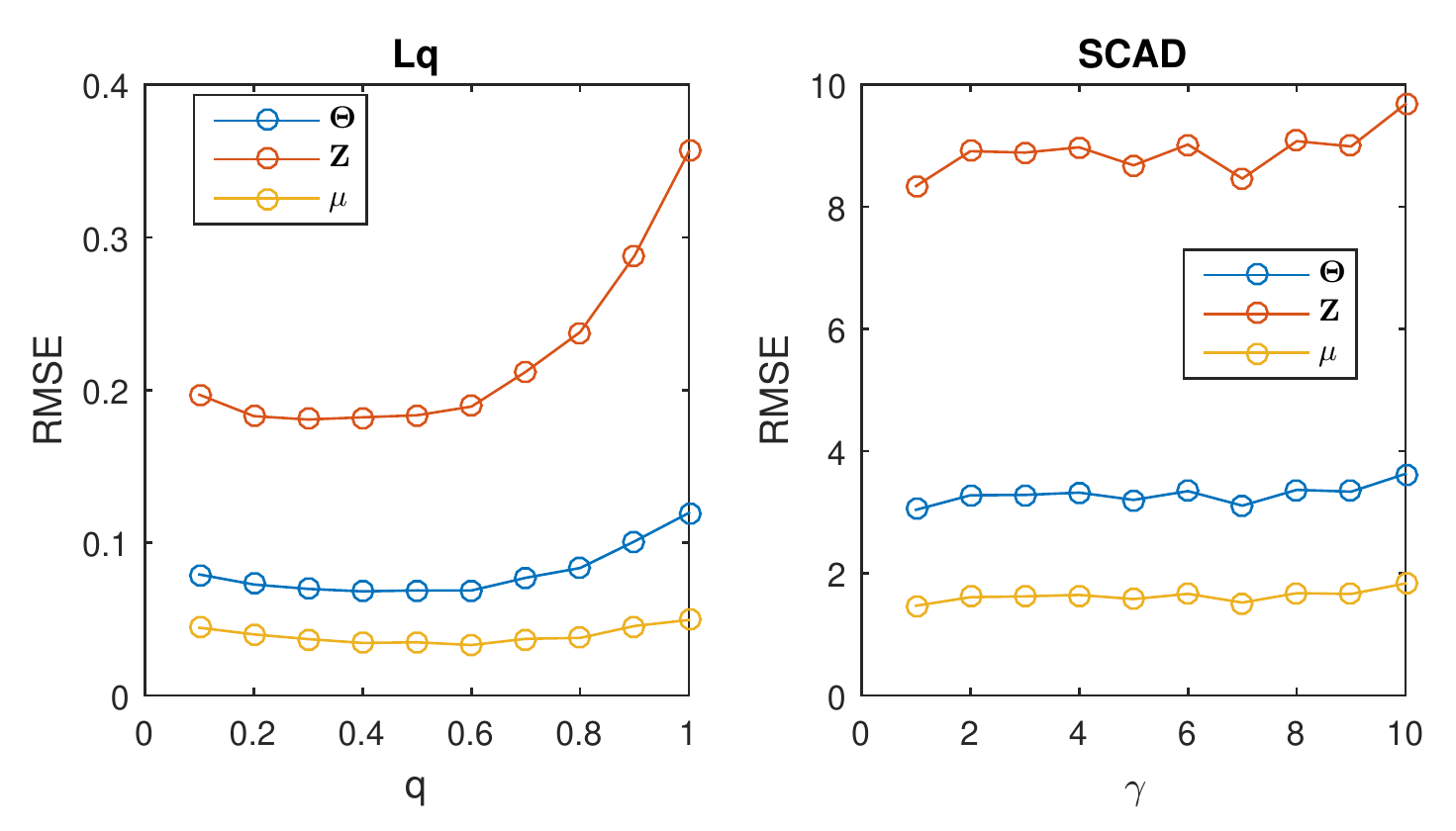}
    \caption{The RMSE of estimating $\mathbf{\Theta}$, $\mathbf{Z}$ and $\bm{\mu}$ as a function of hyper-parameter $q$ in $L_{q}$ penalty (left) and $\gamma$ in SCAD penalty (right). The corresponding logistic PCA models are optimal with respect to CV error.}
    \label{chapter3_fig:5}
\end{figure}

For the logistic PCA models with $L_{q}$ and SCAD penalty, the model is selected and re-fitted on the full data sets in the same way as above. Fig.~\ref{chapter3_fig:5} shows how the value of hyper-parameter $q$ or $\gamma$ effects the RMSEs in estimating $\mathbf{\Theta}$, $\mathbf{Z}$ and $\bm{\mu}$ from the logistic PCA models, which are optimal with respect to CV error, with $L_{q}$ penalty (left) and SCAD penalty (right). The model with $L_{q}$ penalty can achieve similar performance as the model with GDP penalty when proper value of $q$ is selected, while the model with SCAD penalty tends to have very large RMSEs in estimating $\mathbf{\Theta}$, $\mathbf{Z}$ and $\bm{\mu}$ for all the values of hyper-parameter $\gamma$.

\subsection{The performance of the logistic PCA model using different penalties}\label{chapter3_section_3_7}
In this section, we compare the performance of the logistic PCA models with the exact low rank constraint, the nuclear norm penalty, SCAD ($\gamma=3.7$), $L_{q}$ ($q=0.5$) and the GDP ($\gamma=1$) penalty. Random initialization is used and the maximum number of iterations is set to 10000 for all the models. Furthermore, all models are fitted using both $\epsilon_f = 10^{-6}$ and $\epsilon_f = 10^{-8}$ to test the model's robustness to the stopping criteria. For the standard logistic PCA model using the exact low rank constraint, 5 components are used. For the models with GDP ($\gamma=1$), nuclear norm, SCAD ($\gamma=3.7$) and $L_{q}$ ($q=0.5$) penalties, the models are selected (the model selection results are shown in Fig.~\ref{chapter3_fig:3} and the supplemental Fig.~S3.1) and re-fitted on full data set in the same way as was described above. In addition, according to the latent variable interpretation of the logistic PCA model, the unobserved quantitative data set $\mathbf{X}^{\ast} = \mathbf{\Theta} + \mathbf{E}$ is available in our simulation. We constructed a 5 components PCA model (with offset term) on this latent data $\mathbf{X}^{\ast}$, and this model is called the full information model. The results of above experiment are shown in Table \ref{chapter3_tab:2}. Since the logistic PCA model with nuclear norm penalty is a convex problem, the global solution can be achieved. The results from this model are taken as the baseline to compare other approaches. The drawback of the model with the nuclear norm penalty is that the proposed CV error based model selection procedure tends to select a too complex model to compensate for the biased estimation caused by the nuclear norm penalty (supplemental Fig.~S3.2, Table \ref{chapter3_tab:2}). Compared to the model with nuclear norm penalty, the logistic PCA model with exact low rank constraint and SCAD penalty tends to overfit the data, thus have bad performance in estimating the simulated parameters $\mathbf{\Theta}$, $\mathbf{Z}$ and $\bm{\mu}$. Also these models are not robust to the stopping criteria. Compared to the model with nuclear norm penalty, the logistic PCA models with a GDP penalty or a $L_{q}$ penalty perform well in estimating the simulated parameters, and their results are even close to the full information model.
\begin{table}[htbp]
\centering
\caption{Comparison of the logistic PCA models with the exact low rank constraint, the nuclear norm penalty, the SCAD penalty, the $L_{q}$ penalty, the GDP penalty, and the full information model. The RMSEs of estimating $\mathbf{\Theta}$, $\mathbf{Z}$ and $\bm{\mu}$, as well as the mean Hellinger distance (MHD) of estimating the simulated probability matrix $\mathbf{\Pi}$ and the rank estimation of $\hat{\mathbf{Z}}$ are shown in the table.}
\label{chapter3_tab:2}
\begin{tabular}{lllllll}
  \toprule
penalty & $\epsilon_f$ & $\text{RMSE}(\mathbf{\Theta})$ & $\text{RMSE}(\mathbf{Z})$ & $\text{RMSE}(\bm{\mu})$ & MHD & rank \\
  \midrule
  \multirow{2}{0.5em}{exact} & $10^{-6}$  &3.8017    &7.8491    &2.6023    &0.0726    &5      \\
                             & $10^{-8}$  &8.4955    &17.0129   &5.9715    &0.0733    &5      \\
  \hline
  \multirow{2}{0.5em}{nuclear norm} & $10^{-6}$  &0.1407    &0.3788    &0.0701    &0.0670   &27      \\
                                    & $10^{-8}$  &0.1405    &0.3783    &0.0700    &0.0670   &27     \\
  \hline
  \multirow{2}{0.5em}{GDP} & $10^{-6}$   &0.0797    &0.2064    &0.0421    &0.0515    &3    \\
                           & $10^{-8}$   &0.0786    &0.2063    &0.0408    &0.0514    &3    \\
  \hline
  \multirow{2}{0.5em}{SCAD} & $10^{-6}$  &3.1495    &8.5635    &1.5452    &0.2026   &88      \\
                            & $10^{-8}$  &5.6032    &14.1106    &3.0821    &0.2109   &88    \\
  \hline
  \multirow{2}{0.5em}{$L_{q}$} & $10^{-6}$   &0.0672    &0.1836    &0.0327    &0.0483  &4    \\
                           & $10^{-8}$   &0.0671    &0.1834    &0.0327    &0.0484    &4    \\
  \hline
  full &   &0.0120    &0.0465    &0.0017    &0.0258    &5     \\

  \bottomrule
\end{tabular}
\end{table}

\begin{figure}[htbp]
    \centering
    \includegraphics[width = 0.9\textwidth]{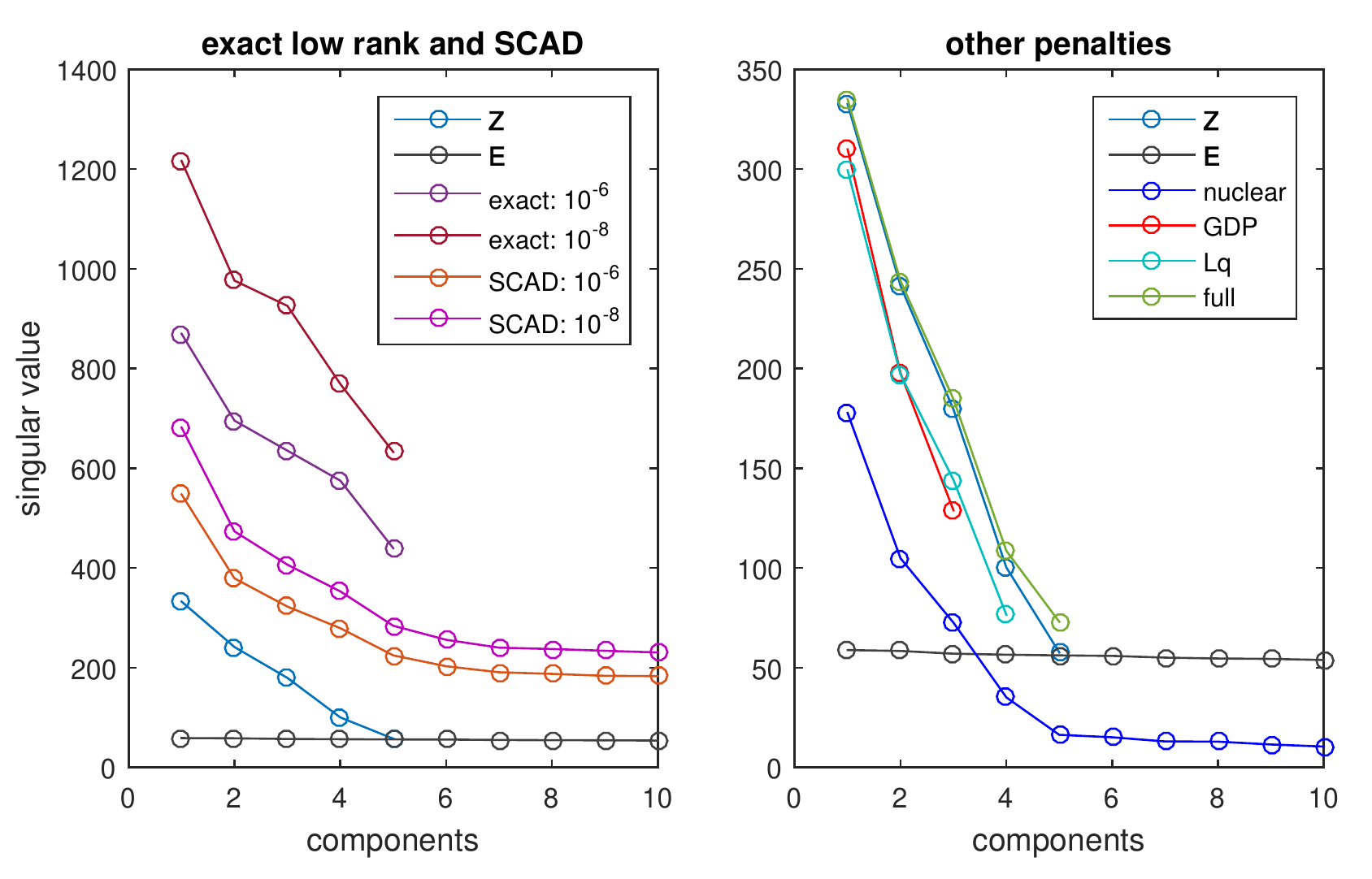}
    \caption{Left: the singular values of the simulated $\mathbf{Z}$ and $\mathbf{E}$, and the singular values of the estimated $\hat{\mathbf{Z}}$ from logistic PCA models ($\epsilon_f = 10^{-6}$ and $\epsilon_f = 10^{-8}$) with exact low rank constraint and SCAD penalty. Right: the singular values of the simulated $\mathbf{Z}$ and $\mathbf{E}$, and the singular values of the estimated $\hat{\mathbf{Z}}$ from the logistic PCA models ($\epsilon_f = 10^{-6}$) with a nuclear norm, GDP and $L_{q}$ penalty, and from the full information model. Only the first 10 components are shown to increase the resolution of the plot.}
    \label{chapter3_fig:6}
\end{figure}

The difference in the performance of the logistic PCA models with different penalties (Table \ref{chapter3_tab:2}) are mainly related to how these penalties shrink the singular values. Therefore, we also compared the singular values of the simulated $\mathbf{Z}$ and their estimations from the logistic PCA models with different penalties, and its estimation from the full information model. The results are shown in Fig.~\ref{chapter3_fig:6}. The simulated low rank is 5, however the last component is overwhelmed by the noise. Furthermore, the $4^{\text{th}}$ component is less than 2 times noise level and therefore cannot be expected to be distinguished from the noise. From Fig.~\ref{chapter3_fig:6}(left) it becomes clear that the logistic PCA models with exact low rank constraint and SCAD penalty clearly overestimate the singular values of $\mathbf{Z}$. And when the more strict stopping criterion is used, the overestimation problem becomes even worse. Fig.~\ref{chapter3_fig:6}(right) shows that the logistic PCA model with nuclear norm penalty underestimated the singular values of $\mathbf{Z}$, and includes too many small singular values into the model. The logistic PCA model with GDP penalty and $L_{q}$ penalty have very accurate estimation of the first three and four singular values of $\mathbf{Z}$. These results are in line with their performance measures in Table \ref{chapter3_tab:2} and their thresholding properties in Fig.~\ref{chapter3_fig:1}. The bad performance of the model with exact low rank constraint is mainly because the non-zero singular values are not regularized at all (Fig.~\ref{chapter3_fig:1}). Similarly, some of the non-zero singular values (the values larger than $\gamma \lambda$) are also not regularized at all for the SCAD penalty (Fig.~\ref{chapter3_fig:1}). This property can be more problematic for the logistic PCA model with SCAD penalty because the selected model depends on the model with the smallest $\lambda$ value due to the used warm start strategy during the model selection process. The low performance of the model with nuclear norm penalty is because this penalty will over shrink the larger singular values, and the model selected based on CV error is too complex (Fig.~\ref{chapter3_fig:1}). On the contrary, both the models with $L_{q}$ and GDP penalties have nice thresholding properties and the corresponding logistic models have superior performance. However, unlike SCAD and GDP, the $L_{q}$($0<q<1$) penalty has a small discontinuity region, continuous thresholding can not be achieved, which could results in instable prediction \cite{fan2001variable}. Therefore, even though the model with $L_{q}$ penalty achieves slight better performance, we still recommend to use the GDP penalty for the non-convex singular thresholding in the logistic PCA model.

\subsection{Performance of the logistic PCA model as a function of SNR in the binary simulation.}
\begin{figure}[htbp]
    \centering
    \includegraphics[width = 0.9\textwidth]{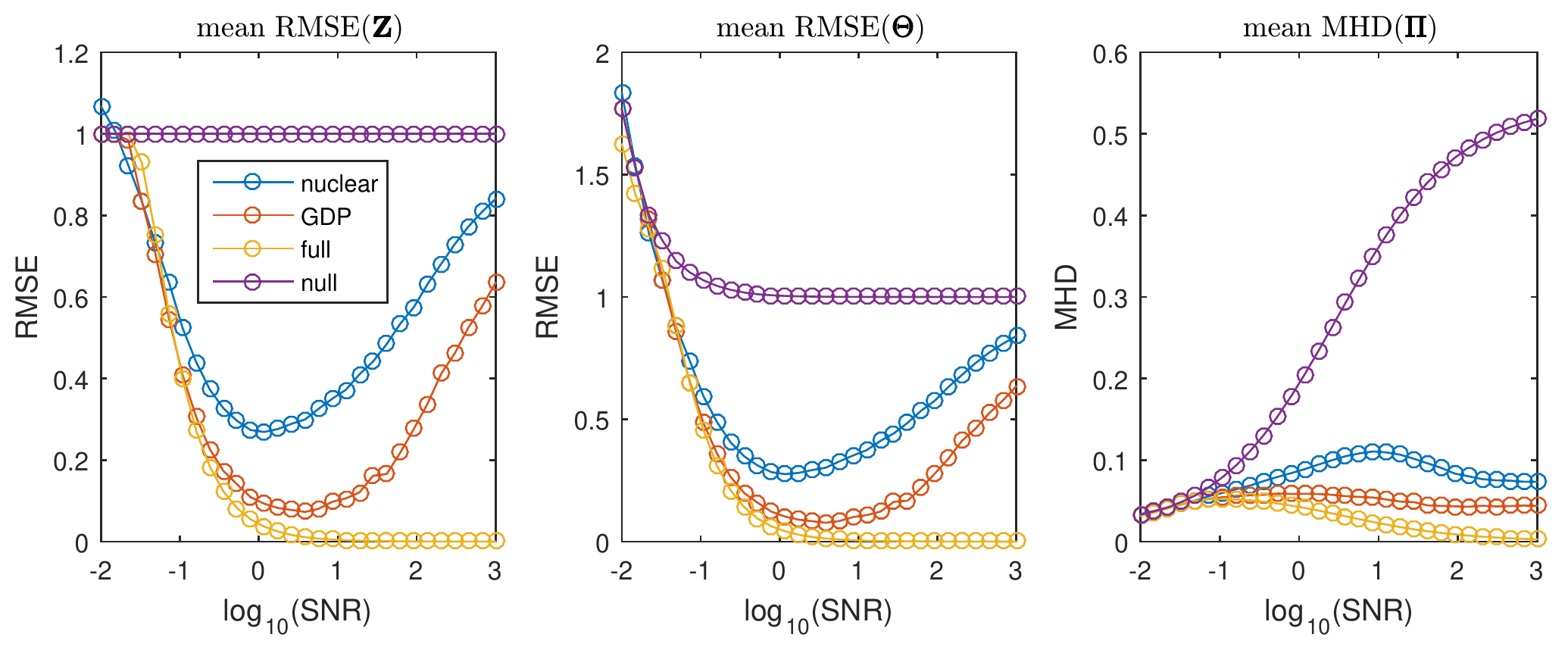}
    \caption{RMSE of $\mathbf{Z}$ (left) and $\mathbf{\Theta}$ (middle) and the MHD of $\mathbf{\Pi}$ as a function of increasing SNR values for simulated balanced binary data.}
    \label{chapter3_fig:7}
\end{figure}

In the analysis of simulated quantitative data sets using the PCA model, an increase in SNR makes the estimation of the true underlying low rank structure easier. Unfortunately, this is not true in the estimation of the true underlying logistic PCA model for simulated binary data. To illustrate this, the following experiment was performed using logistic PCA model with GDP penalty as an example. 30 SNR values are selected from the interval $[10^{-2}, 10^{3}]$ equidistant in log-space. The simulated offset term $\bm{\mu}$ is set to $\mathbf{0}$ to simulate balanced binary data sets, the number of samples, variables, and the low rank are the same as the experiment described above. For the binary data simulations with different SNRs, only the constant $c$, which is used to adjust the SNR, changes with the SNR. All other parameters are kept the same. For each simulated $\mathbf{X}$ with a specific SNR, logistic PCA models with GDP ($\gamma=1$) penalty and with nuclear norm penalty are selected and re-fitted. In addition, PCA models with different numbers of components are fitted on the latent quantitative data set $\mathbf{X}^{\ast}$, and the model with the minimum $\text{RMSE}(\mathbf{\Theta})$ is selected. In addition, the null model, i.e. the logistic PCA model with only the column offset term, is used to provide a baseline for comparison of the different approaches. The above experiments are repeated 10 times, and their results are shown in Fig.~\ref{chapter3_fig:7}. Results obtained from a similar experiment but performed on imbalanced data simulation are shown in supplemental Fig.~S3.2. There, the simulated $\bm{\mu}$ is set according to the marginal probabilities of the CNA data set. Overall, the logistic PCA models with different penalties can always achieve better performance than the null model, and the model with a GDP penalty demonstrates superior performance with respect to all the used metrics compared to the model with convex nuclear norm penalty.

Fig.~\ref{chapter3_fig:7} (left and center) shows that with increasing SNR, the estimation of the quantitative full model improves as expected. However, for the parameters estimated from the binary data this is not the case. First the estimation of the simulated parameters $\mathbf{\Theta}$ and $\mathbf{Z}$ improves, but when the SNR increases even further, the estimation deteriorates again leading to a bowl shaped pattern. This pattern has been observed before in binary matrix completion using nuclear norm penalty \cite{davenport20141}. In order to understand this effect, considering the S-shaped logistic curve (supplemental Fig.~S3.3), the plot of the function $\text{E}(x|\theta) = \phi(\theta) = (1+\exp(-\theta))^{-1}$, in which $x$ and $\theta$ are a typical element of $\mathbf{X}$ and $\mathbf{\Theta}$ respectively. This curve almost becomes flat when $\theta$ becomes very large. There is no resolution anymore in these flat regimes. A large deviation in $\theta$ has almost no effect on the logistic response. When the SNR becomes extremely large, the scale of the simulated parameter $\theta$ is very extreme, then even if we have a good estimation of the probability $\hat{\pi} = \phi(\hat{\theta})$, the scale of estimated $\hat{\theta}$ can be far away from the simulated $\theta$. That is why we observed that even though the model is able to recovered the simulated $\mathbf{\Pi}$ based on the logistic PCA model almost exactly (Fig.~\ref{chapter3_fig:7} right), the estimation of $\mathbf{\Theta}$ and $\mathbf{Z}$ are not accurate (Fig.~\ref{chapter3_fig:7} left and center). We refer to \cite{davenport20141} for a detailed interpretation of this phenomenon.

\subsection{Real data analysis}
\begin{figure}[htbp]
    \centering
    \includegraphics[width = 0.9\textwidth]{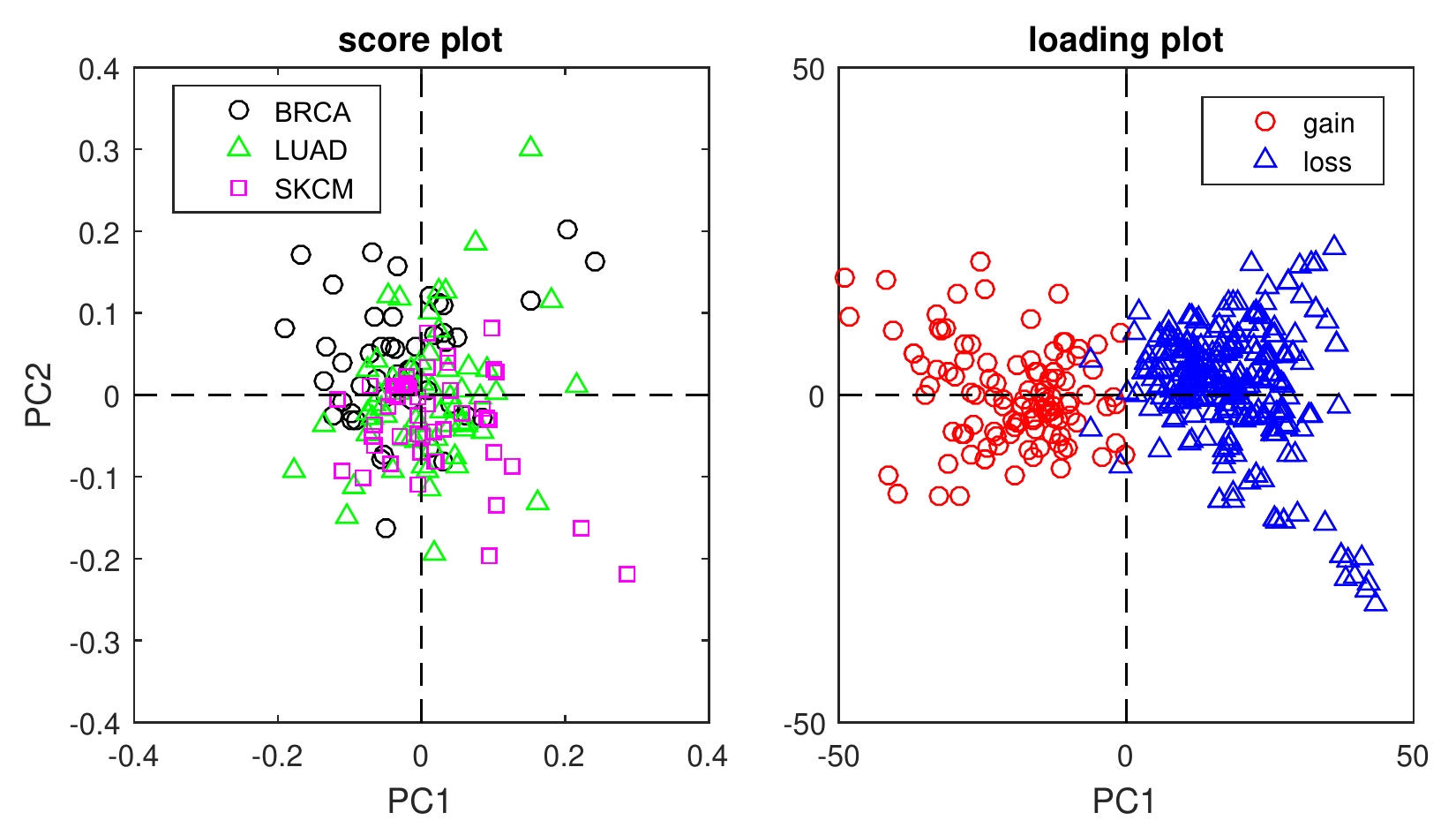}
    \caption{The score and loading plots of the first 2 components of the logistic PCA model on the CNA data. The legend, BRCA, LUAD and SKCM, indicates the corresponding three cancer types. The legend, gain, loss, indicates the gain or loss of a segment in the chromosome region corresponding to the CNA measurement.
    }
    \label{chapter3_fig:8}
\end{figure}

We demonstrate the proposed logistic PCA model with a GDP ($\gamma=1$) penalty and the corresponding model selection procedure on the CNA data set. The model selection is done in the same way as was described above, and the result is shown in supplemental Fig.~S3.4. After that, the selected 4 components model is re-fitted on the full data set. The score and loading plot of the first 2 components are shown in Fig.~\ref{chapter3_fig:8}. As was explained before in \cite{song2018generalized}, the CNA data set is not discriminative for the three cancer types (illustrated in the score plot of Fig.~\ref{chapter3_fig:8} left). The structure in the loading plot (Fig.~\ref{chapter3_fig:8} right) mainly explains the technical characteristics of the data. Fig.~\ref{chapter3_fig:8} (right) shows that the gains and losses of the segments in the chromosomal regions corresponding to the CNA measurements are almost perfectly separated from each other in the first component. Therefore, the variation explained of the first component is mainly because of the difference of gains and losses in CNA measurements.

\section{Discussion}
To study the properties of the logistic PCA model with different penalties, we need to have the ability to simulate the multivariate binary data set with an underlying low rank structure, and the simulated structure should have a proper SNR so that the model can find it back. The latent variable interpretation of the logistic PCA model not only makes the assumption of low rank structure easier to understand, but also provides us a way to define SNR in multivariate binary data simulation.

The standard logistic PCA model using the exact low rank constraint has an overfitting problem. The overfitting issue manifests itself in a way that some of the elements in the estimated loading matrix $\hat{\mathbf{B}}$ (the orthogonality constraint is imposed on $\mathbf{A}$) have the tendency to approach infinity, and the non-zero singular values of the $\hat{\mathbf{Z}} = \hat{\mathbf{A}}\hat{\mathbf{B}}^{\text{T}}$ are not upper-bounded when strict stopping criteria are used. This overfitting issue can be alleviated by regularizing the singular values of $\mathbf{Z}$. Both convex nuclear norm penalties and some of the concave penalties can induce low rank estimation and simultaneously constrain the scale of the non-zero singular values. Therefore, logistic PCA models with these penalties do not suffer from the overfitting problem.

However, the logistic PCA model with a GDP or a $L_{q}$ penalty has several advantages compared to the model with the nuclear norm penalty. Since the nuclear norm penalty applies the same degree of shrinkage on all the singular values, the large singular values are shrunken too much. Therefore, the implemented CV error based model selection procedure tends to select a very complex model with too many components to compensate for the biased estimations. On the contrary, both the GDP penalty and the $L_{q}$ penalty achieve nearly unbiased estimation. Thus the CV error based model selection is successful in selecting the logistic PCA model with the a GDP penalty. Furthermore, the selected logistic PCA model with a GDP or a $L_{q}$ penalty has shown superior performance in recovering the simulated low rank structure compared to the model with the nuclear norm penalty, and the exact low rank constraint.

One exception of the used concave penalties is the SCAD penalty, which leads to a logistic PCA model with poor performance. As stated in the Section \ref{chapter3_section_3_7}, the poor performance of the model with the SCAD penalty is mainly because of that some of the large singular values are not regularized at all. And this drawback is exaggerated by the wart start strategy used during the model selection process. The poor performance of the models with the exact low rank constraint and the SCAD penalty reminds us the importance of regularizing all the singular values simultaneously in inducing the low rank structure for a logistic PCA model.

\section*{Acknowledgements}
Y.S. gratefully acknowledges the financial support from China Scholarship Council (NO.201504910809).

\clearpage
\section{Supplementary information}
\begin{figure}[htbp]
    \centering
    \includegraphics[width=0.9\textwidth ]{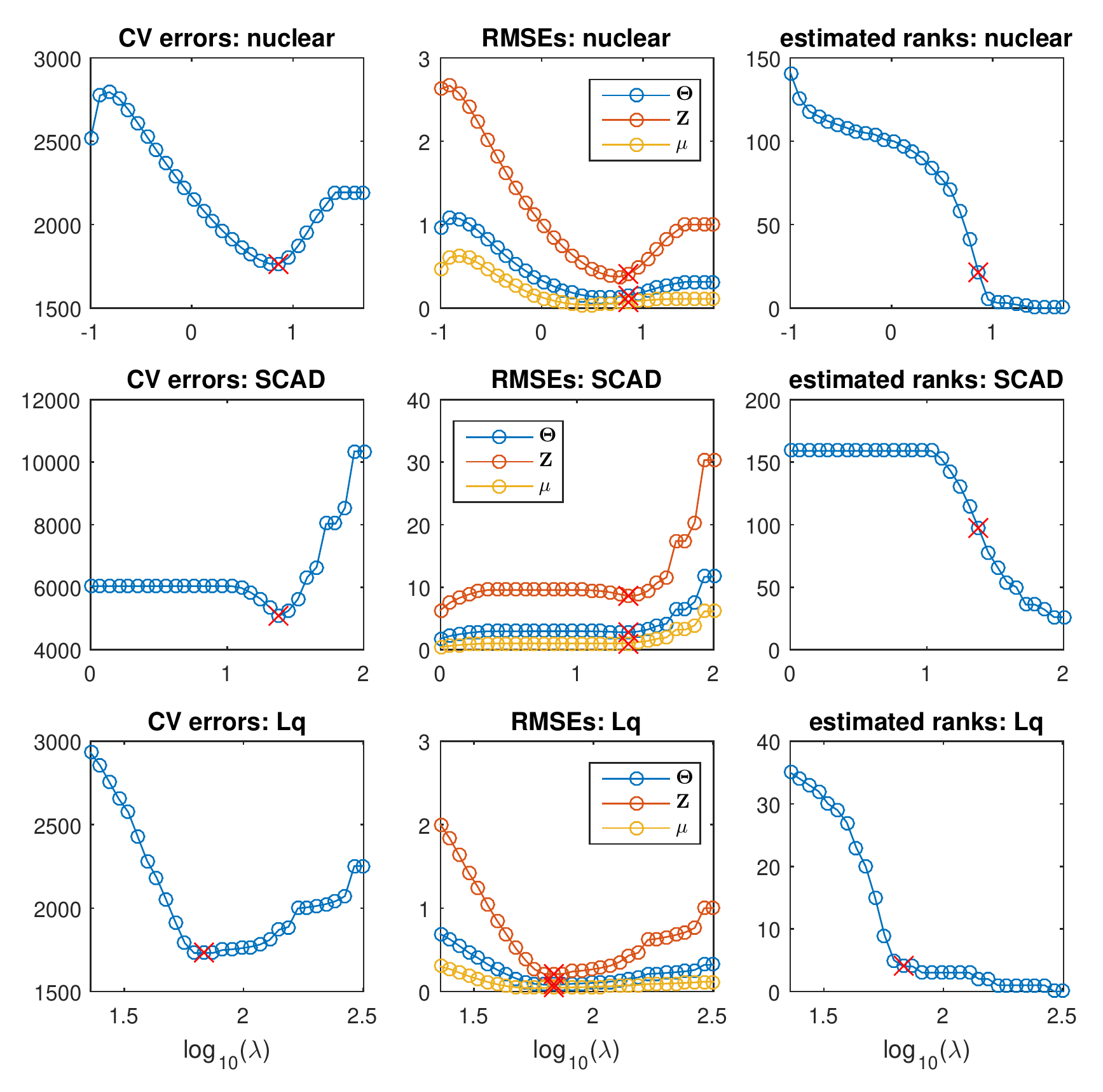}
    \caption*{Figure S3.1: Model selection of the logistic PCA model with the nuclear norm penalty, the SCAD penalty and $L_{q}$ penalty. The CV error, RMSE of estimating $\mathbf{\Theta}$, $\mathbf{Z}$ and $\bm{\mu}$ and the estimated rank as a function of $\lambda$. The red cross marker indicates the $\lambda$ value where minimum CV error is achieved.}
    \label{chapter3_fig:S1}
\end{figure}

\begin{figure}[htbp]
    \centering
    \includegraphics[width = 0.9\textwidth]{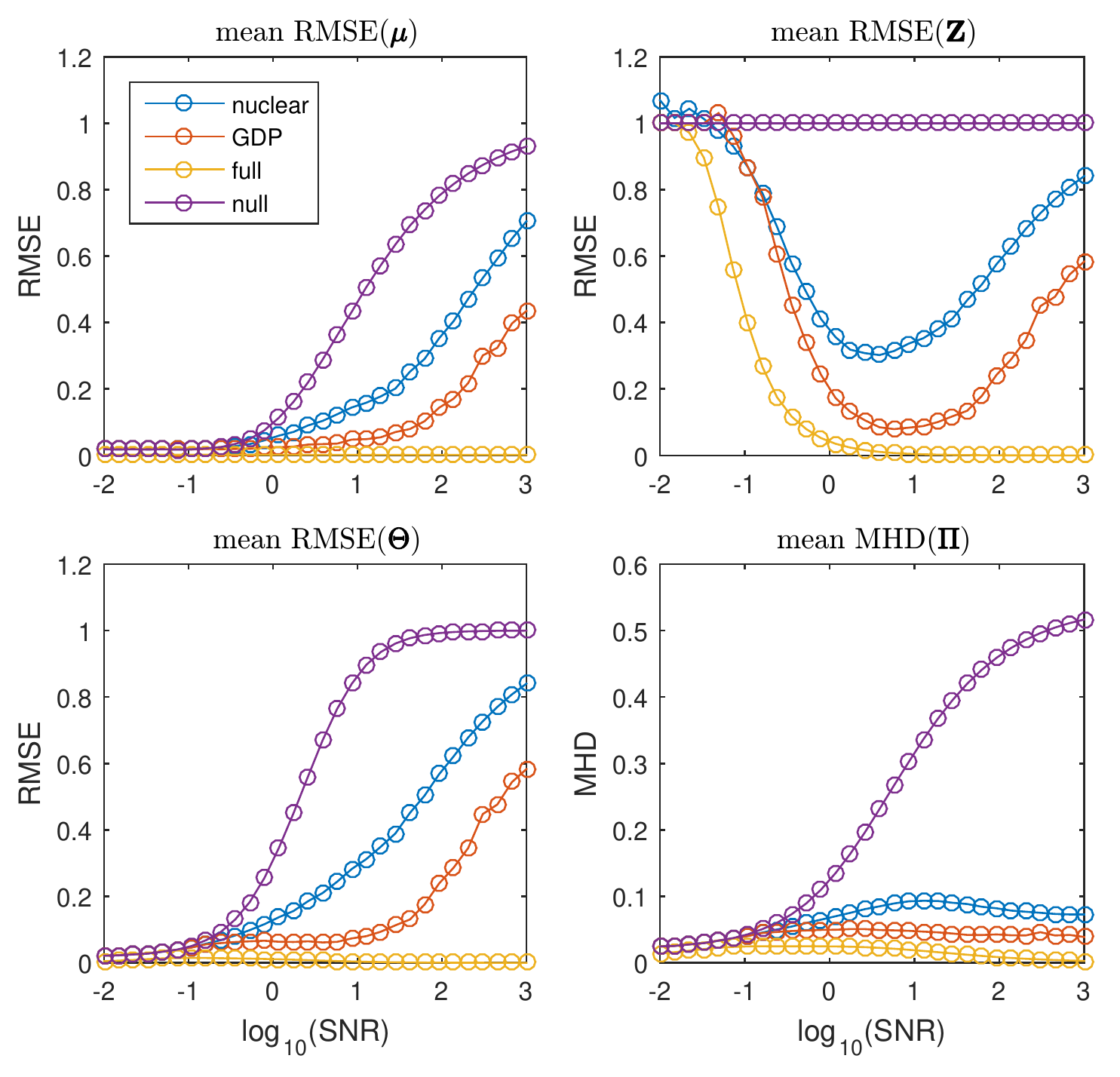}
    \caption*{Figure S3.2: How the SNR in imbalanced binary data simulation affects the performance of the logistic PCA models with different penalties, and the full information model.}
    \label{chapter3_fig:S2}
\end{figure}

\begin{figure}[htbp]
    \centering
    \includegraphics[width = 0.5\textwidth]{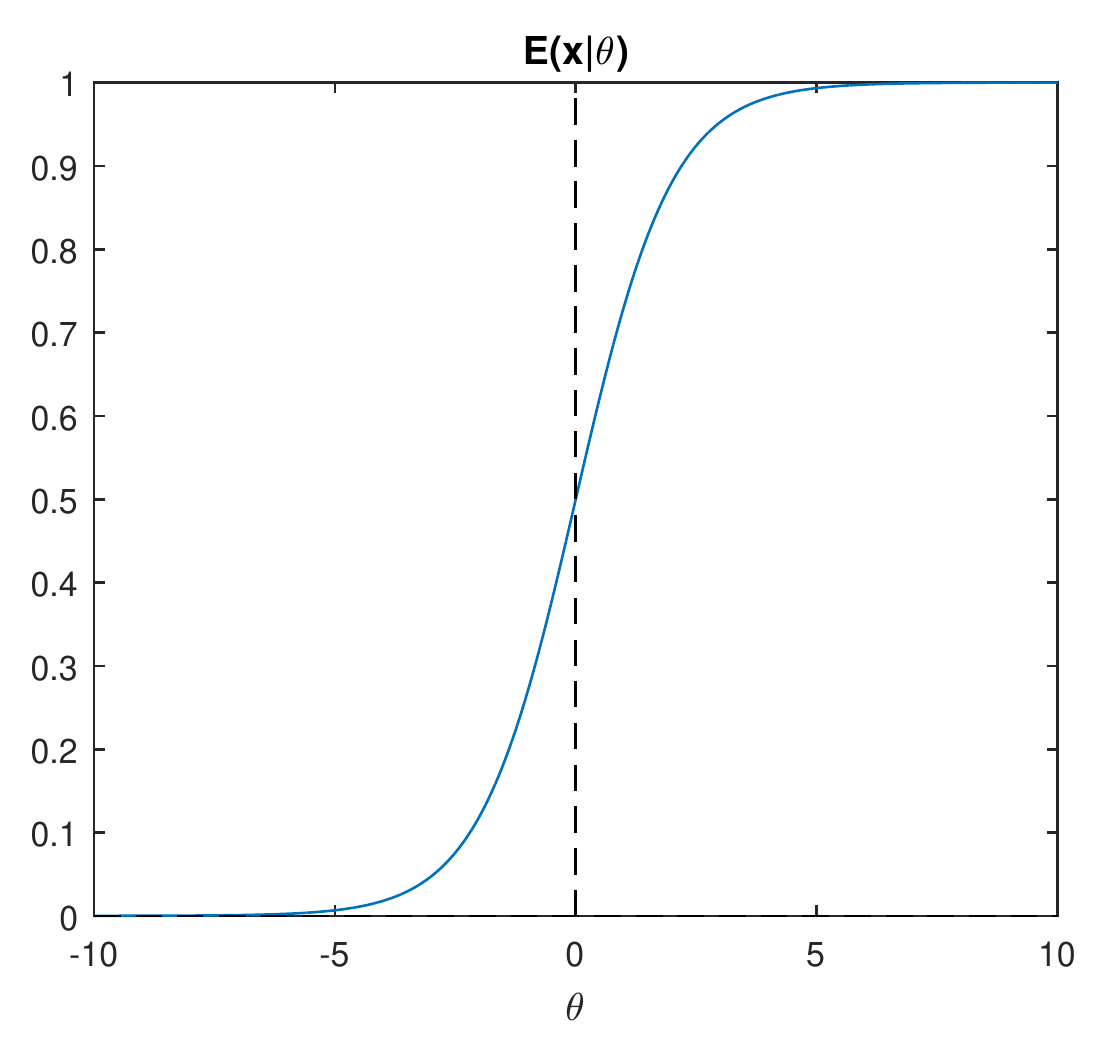}
    \caption*{Figure S3.3: The relationship of $\text{E}(x|\theta) = \phi(\theta)$, in which $x$ and $\theta$ are a typical element of $\mathbf{X}$ and $\mathbf{\Theta}$ respectively.}
    \label{chapter3_fig:S3}
\end{figure}

\begin{figure}[htbp]
    \centering
    \includegraphics[width = 0.9\textwidth]{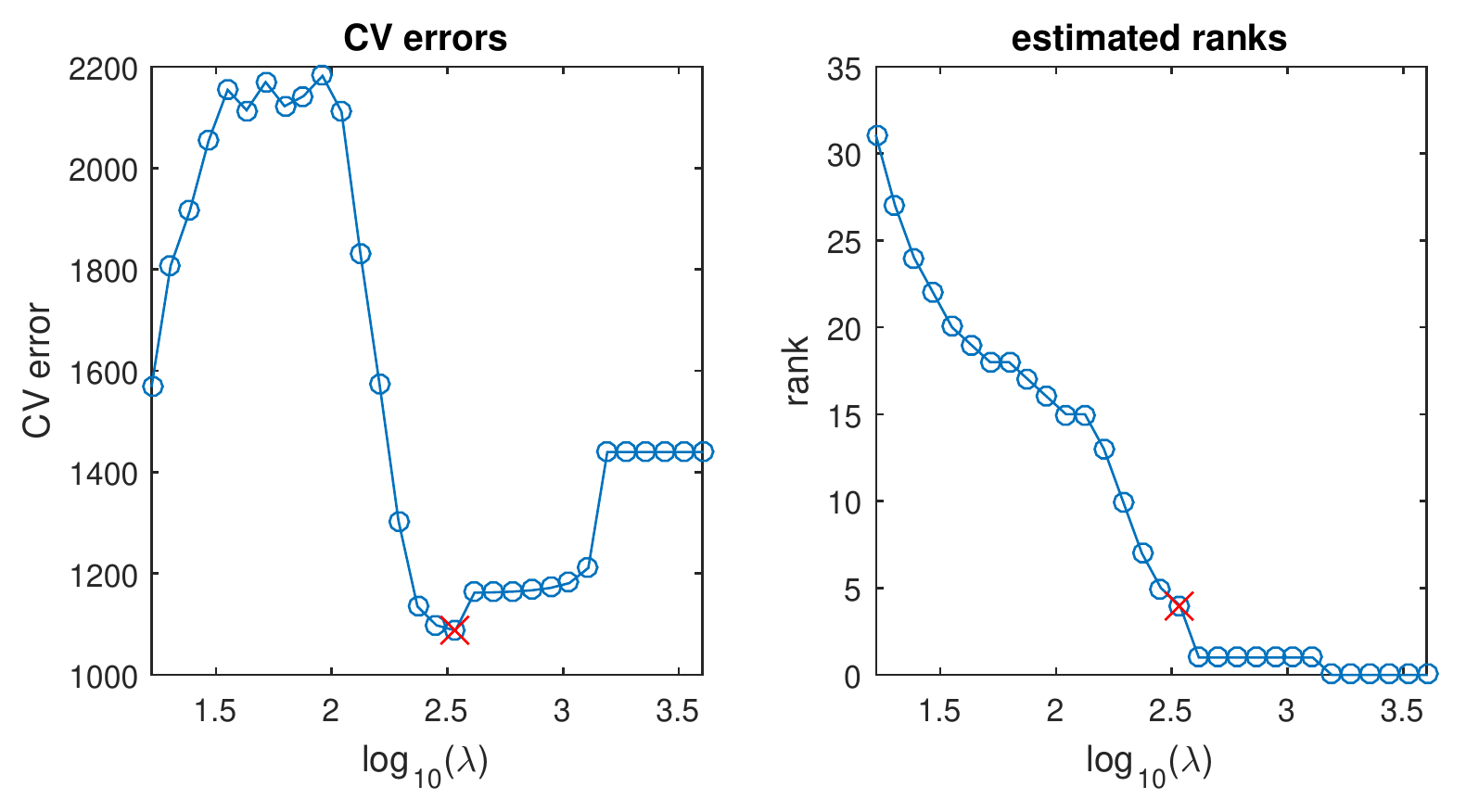}
    \caption*{Figure S3.4: How $\lambda$ effects the CV error (left) and the rank estimation (right) in the model selection process of the logistic PCA model with a GDP penalty on the CNA data set.}
    \label{chapter3_fig:S4}
\end{figure}

\chapter{Fusing binary and quantitative data sets} \label{chapter:4}
In the current era of systems biological research there is a need for the integrative analysis of binary and quantitative genomics data sets measured on the same objects. One standard tool of exploring the underlying dependence structure present in multiple quantitative data sets is simultaneous component analysis (SCA) model. However, it does not have any provisions when a part of the data are binary. To this end, we propose the generalized SCA (GSCA) model, which takes into account the distinct mathematical properties of binary and quantitative measurements in the maximum likelihood framework. Like in the SCA model, a common low dimensional subspace is assumed to represent the shared information between these two distinct types of measurements. However, the GSCA model can easily be overfitted when a rank larger than one is used, leading to some of the estimated parameters to become very large. To achieve a low rank solution and combat overfitting, we propose to use non-convex singular value thresholding. An efficient majorization algorithm is developed to fit this model with different concave penalties. Realistic simulations (low signal-to-noise ratio and highly imbalanced binary data) are used to evaluate the performance of the proposed model in recovering the underlying structure. Also, a missing value based cross validation procedure is implemented for model selection. We illustrate the usefulness of the GSCA model for exploratory data analysis of quantitative gene expression and binary copy number aberration (CNA) measurements obtained from the GDSC1000 data sets.
\footnote{This chapter is based on Song, Y., Westerhuis, J.A., Aben, N., Wessels, L.F., Groenen, P.J. and Smilde, A.K., 2018. Generalized Simultaneous Component Analysis of Binary and Quantitative data. arXiv preprint arXiv:1807.04982.}

\section{Background} \label{section:4.1}
In biological research it becomes increasingly common to have measurements of different aspects of information on the same objects to study complex biological systems. The resulting coupled data sets should be analyzed simultaneously to explore the dependency between variables in different data sets and to reach a global understanding of the underlying biological system. The SCA model is one of the standard methods for the integrative analysis of such coupled data sets in different areas, from psychology to chemistry and biology \cite{van2009structured}. SCA discovers the common low dimensional column subspace of the coupled quantitative data sets, and this subspace represents the shared information between them.

Next to the quantitative measurements (such as gene expression data), it is common in biological research to have additional binary measurements, in which distinct categories differ in quality rather than in quantity (such as mutation data). Typical examples include the measurements of point mutations, the binary version of CNA and DNA methylation measurements \cite{iorio2016landscape}. Compared to quantitative measurement, a binary measurement only has two mutually exclusive outcomes, such as presence vs absence (or true vs false), which are usually labeled as ``1'' and ``0''. However, ``1'' and ``0'' indicate abstract representations of two categories rather than quantitative values 1 and 0. As such, the special mathematical properties of binary data should be taken into account in the data analysis. In most biological data sets, the number of ``0''s is significantly larger than the number of ``1''s for most binary variables making the data imbalanced. Therefore, an additional requirement of the data analysis method is that it should be able to handle imbalanced data.

There is a need for statistical methods appropriate for doing an integrative analysis of coupled binary and quantitative data sets in biology research. The standard SCA models \cite{van2009structured, van2009integrating} that use column centering processing steps and least-squares loss criteria are not appropriate for binary data sets. Recently, iClusterPlus \cite{mo2013pattern} was proposed as a factor analysis framework to model discrete and quantitative data sets simultaneously by exploiting the properties of exponential family distributions. In this framework, the special properties of binary, categorical, and count variables are taken into account in a similar way as in generalized linear models. The common low dimensional latent variables and data set specific coefficients are used to fit the discrete and quantitative data sets. For the binary data set, the Bernoulli distribution is assumed and the canonical logit link function is used. The sum of the log likelihood is then used as the objective function. Furthermore, the approach allows the use of a lasso type penalty for feature selection. The Monte Carlo Newton–Raphson algorithm for this general framework, however, involves a very slow Markov Chain Monte Carlo simulation process. Both the high complexity of the model and the algorithmic inefficiency limit its use for large data sets and exploring its properties through simulations.

In this chapter, we generalize the SCA model to binary and quantitative data from a probabilistic perspective similar as in Collins \cite{collins2002generalization} and Mo \cite{mo2013pattern}. However, the generalized SCA model can easily lead to overfitting by using a rank restriction larger than $1$, leading to some of the parameters to become very large. Therefore, a penalty on the singular values of the matrix is used to simultaneously induce the low rank structure in a soft manner and to control the scale of estimated parameters. A natural choice is the convex nuclear norm penalty, which is widely used in low rank approximation problems \cite{koltchinskii2011nuclear, groenen2016multinomial, wu2015fast}. However, the nuclear norm penalty shrinks all singular values to the same degree, leading to biased estimates of the important latent factors. Hence, we would like to reduce the shrinkage for the most important latent factors while increasing the shrinkage for unimportant latent factors. This nonlinear shrinkage strategy has shown its superiority in recent work of low rank matrix approximation problems under the presence of Gaussian noise \cite{gavish2017optimal, josse2016adaptive}. Therefore, we will explore the nonlinear shrinkage of the latent factors through concave penalties in our GSCA model. The fitting of the resulting GSCA model is a penalized maximum likelihood estimation problem. We derive a MM \cite{de1994block,hunter2004tutorial} algorithm to solve it. Simple closed form updates for all the parameters are derived for each step in the algorithm. A missing value based cross validation procedure is also implemented to do model selection. Our algorithm is easy to implement and is guaranteed to decrease the loss function monotonically in each iteration.

\section{The GSCA model} \label{section:4.2}
Before the GSCA model is introduced, consider the standard SCA model. The quantitative measurements on the same $I$ objects from two different platforms result into two data sets $\mathbf{X}_1$($I\times J_1$) and $\mathbf{X}_2$($I\times J_2$), in which $J_1$ and $J_2$ are the number of variables. Assume both $\mathbf{X}_1$ and $\mathbf{X}_2$ are column centered. The standard SCA model can be expressed as
\begin{equation}\label{chapter4_eq:1}
\begin{aligned}
\mathbf{X}_1 &= \mathbf{AB}_1^{\text{T}} + \mathbf{E}_1\\
\mathbf{X}_2 &= \mathbf{AB}_2^{\text{T}} + \mathbf{E}_2,\\
\end{aligned}
\end{equation}
where $\mathbf{A}$($I\times R$) denotes the common component scores (or latent variables), which span the common column subspace of $\mathbf{X}_1$ and $\mathbf{X}_2$, $\mathbf{B}_1$($J_1\times R$) and $\mathbf{B}_2$($J_2\times R$) are the data set specific loading matrices for $\mathbf{X}_1$ and $\mathbf{X}_2$ respectively, $\mathbf{E}_1$($I\times J_1$) and $\mathbf{E}_2$($I\times J_2$) are residuals, $R$, $R \ll \{I,J_1,J_2\}$, is an unknown low rank. Orthogonality is imposed on $\mathbf{A}$ as $\mathbf{A}^{\text{T}}\mathbf{A}=\mathbf{I}_{R}$, where $\mathbf{I}_{R}$ indicates the $R\times R$ identity matrix, to have a unique solution. $\mathbf{A}$, $\mathbf{B}_1$ and $\mathbf{B}_2$ are estimated by minimizing the sum of the squared residuals $\mathbf{E}_1$ and $\mathbf{E}_2$.

\subsection{The GSCA model of binary and quantitative data sets}
Following the probabilistic interpretation of the PCA model \cite{tipping1999probabilistic}, the high dimensional quantitative data set $\mathbf{X}_2$ can be assumed to be a noisy observation from a deterministic low dimensional structure $\mathbf{\Theta}_2$($I\times J_2$) with independent and identically distributed measurement noise, $\mathbf{X}_2 = \mathbf{\Theta}_2 + \mathbf{E}_2$. Elements in $\mathbf{E}_2$($I\times J_2$) follow a normal distribution with mean 0 and variance $\sigma^2$, $\epsilon_{2ij} \sim \text{N}(0,\sigma^2$). In the same way, following the interpretation of the exponential family PCA on binary data \cite{collins2002generalization}, we assume there is a deterministic low dimensional structure $\mathbf{\Theta}_1$($I\times J_1$) underlying the high dimensional binary observation $\mathbf{X}_1$. Elements in $\mathbf{X}_1$ follow the Bernoulli distribution with parameters $\phi(\mathbf{\Theta}_1)$, $x_{1ij} \sim \text{Ber}(\phi(\theta_{1ij}))$. Here $\phi()$ is the element wise inverse link function in the generalized linear model for binary data; $x_{1ij}$ and $\theta_{1ij}$ are the $ij^{\text{th}}$ element of $\mathbf{X}_1$ and $\mathbf{\Theta}_1$ respectively. If the logit link is used, $\phi(\theta) = (1+\exp(-\theta))^{-1}$, while if the probit link is used, $\phi(\theta) = \Phi(\theta)$, where $\Phi$ is the cumulative density function of the standard normal distribution. Although in this chapter, we only use the logit link in deriving the algorithm and in setting up the simulations, the option for the probit link is included in our implementation. The two link functions are similar, but their interpretations can be quite different \cite{agresti2013categorical}.

In the same way as in the standard SCA model, $\mathbf{\Theta}_1$ and $\mathbf{\Theta}_2$ are assumed to lie in the same low dimensional subspace, which represents the shared information between the coupled matrices $\mathbf{X}_1$ and $\mathbf{X}_2$. The commonly used column centering is not appropriate for the binary data set as the centered binary data will not be ``1'' and ``0'' anymore. Every column will still have only 2 values but these values are different for different columns. Therefore, we include column offset terms $\bm{\mu}_1$($J_1 \times 1$) and $\bm{\mu}_2$($J_2 \times 1$) for a model based centering. The above ideas are modeled as
\begin{equation}\label{chapter4_eq:2}
\begin{aligned}
\mathbf{\Theta}_1 &= \mathbf{1}\bm{\mu}_1^{\text{T}} + \mathbf{AB}_1^{\text{T}}\\
\mathbf{\Theta}_2 &= \mathbf{1}\bm{\mu}_2^{\text{T}} + \mathbf{AB}_2^{\text{T}},\\
\end{aligned}
\end{equation}
where, $\mathbf{1}$($I\times 1$) is a $I$ dimensional vector of ones; the parameters $\mathbf{A}$, $\mathbf{B}_1$ and $\mathbf{B}_2$ have the same meaning as in the standard SCA model. Constraints $\mathbf{A}^{\text{T}}\mathbf{A}=\mathbf{I}_R$ and $\mathbf{1}^{\text{T}}\mathbf{A} = \mathbf{0}$ are imposed to have a unique solution.

For the generalization to quantitative and binary coupled data, we follow the maximum likelihood estimation framework. The negative log likelihood for fitting coupled binary $\mathbf{X}_1$ and quantitative $\mathbf{X}_2$ is used as the objective function. In order to implement a missing value based cross validation procedure \cite{bro2008cross}, we introduce two weight matrices $\mathbf{W}_1$($I\times J_1$) and $\mathbf{W}_2$($I\times J_2$) to handle the missing elements. The $ij^{\text{th}}$ element of $\mathbf{W}_1$, $w_{1ij}$ equals 0 if the $ij^{\text{th}}$ element in $\mathbf{X}_1$ is missing, while it equals 1 \textit{vice versa}. The same rules apply to $\mathbf{W}_2$ and $\mathbf{X}_2$. The loss functions $f_1(\mathbf{\Theta}_1)$ for fitting $\mathbf{X}_1$ and $f_2(\mathbf{\Theta}_2,\sigma^2)$ for fitting $\mathbf{X}_2$ are defined as follows:
\begin{equation}\label{chapter4_eq:3}
\begin{aligned}
f_1(\mathbf{\Theta}_1) &= -\sum_{i}^{I}\sum_{j}^{J_1} w_{1ij} \left[x_{1ij}\log(\phi(\theta_{1ij})) + (1-x_{1ij})\log(1-\phi(\theta_{1ij}))\right] \\
f_2(\mathbf{\Theta}_2,\sigma^2)  &= \frac{1}{2\sigma^2}
              ||\mathbf{W}_2 \odot (\mathbf{X}_2-\mathbf{\Theta}_2)||_F^2 +     \frac{1}{2} ||\mathbf{W}_2||_0 \log(2\pi \sigma^2),\\
\end{aligned}
\end{equation}
where $\odot$ indicates element-wise multiplication; $||\quad||_F$ is the Frobenius norm of a matrix; $||\quad||_0$ is the pseudo $L_0$ norm of a matrix, which equals the number of nonzero elements.

The shared information between $\mathbf{X}_1$ and $\mathbf{X}_2$ is assumed to be fully represented by the low dimensional subspace spanned by the common component score matrix $\mathbf{A}$. Thus, $\mathbf{X}_1$ and $\mathbf{X}_2$ are conditionally independent given that the low dimensional structures $\mathbf{\Theta}_1$ and $\mathbf{\Theta}_2$ lie in the same low dimensional subspace. Therefore, the joint loss function is the direct sum of the negative log likelihood functions for fitting $\mathbf{X}_1$ and $\mathbf{X}_2$.
\begin{equation}\label{chapter4_eq:4}
\begin{aligned}
f(\mathbf{\Theta}_1,\mathbf{\Theta}_2,\sigma^2) &= -\log(p(\mathbf{X}_1,\mathbf{X}_2|\mathbf{\Theta}_1,\mathbf{\Theta}_2,\sigma^2))\\
& = -\log(p(\mathbf{X}_1|\mathbf{\Theta}_1) p(\mathbf{X}_2|\mathbf{\Theta}_2,\sigma^2))\\
              &= -\log(p(\mathbf{X}_1|\mathbf{\Theta}_1)-\log(p(\mathbf{X}_2|\mathbf{\Theta}_2,\sigma^2))\\
              &= f_1(\mathbf{\Theta}_1) + f_2(\mathbf{\Theta}_2,\sigma^2).\\
\end{aligned}
\end{equation}

\subsection{Concave penalties as surrogates for low rank constraint}
To arrive at meaningful solutions for the GSCA model, it is necessary to introduce penalties on the estimated parameters. If we take $\mathbf{\Theta} = [\mathbf{\Theta}_1 ~ \mathbf{\Theta}_2]$, $\bm{\mu} = [\bm{\mu}_1^{\text{T}} \bm{\mu}_2^{\text{T}}]^{\text{T}}$, and $\mathbf{B} = [\mathbf{B}_1^{\text{T}} ~ \mathbf{B}_2^{\text{T}}]^{\text{T}}$, equation \ref{chapter4_eq:2} in the GSCA model can be expressed as $\mathbf{\Theta} = \mathbf{1}\bm{\mu}^{\text{T}} + \mathbf{AB}^{\text{T}}$. In the above interpretation of the GSCA model, the low rank constraint on the column centered $\mathbf{\Theta}$ is expressed as the multiplication of two rank $R$ matrices $\mathbf{A}$, $\mathbf{B}$, $\mathbf{Z} = \mathbf{\Theta} - \mathbf{1}\bm{\mu}^{\text{T}} = \mathbf{AB}^{\text{T}}$. However, using an exact low rank constraint in the GSCA model has some issues. First, the maximum likelihood estimation of this model easily leads to overfitting. Given the constraint that $\mathbf{A}^{\text{T}}\mathbf{A}=\mathbf{I}$, overfitting represents itself in a way that some elements in $\mathbf{B}_1$ tend to diverge to plus or minus infinity. In addition, the exact low rank $R$ in the GSCA model is commonly unknown and its selection is not straightforward.

In this chapter, we take a penalty based approach to control the scale of estimated parameters and to induce a low rank structure simultaneously. The low rank constraint on $\mathbf{Z}$ is obtained by a penalty function $g(\mathbf{Z})$, which shrinks the singular values of $\mathbf{Z}$ to achieve a low rank structure. The most widely used convex surrogate of a low rank constraint is the nuclear norm penalty, which is simply the sum of singular values, $g(\mathbf{Z}) = \sum_{r} \sigma_r(\mathbf{Z})$ \cite{koltchinskii2011nuclear}, where $\sigma_r(\mathbf{Z})$ represents the $r^{\text{th}}$ singular value of $\mathbf{Z}$. The nuclear norm penalty was also used in a related work \cite{wu2015fast}. Although the convex nuclear norm penalty is easy to optimize, the same amount of shrinkage is applied to all the singular values, leading to biased estimates of the large singular values. Recent work \cite{gavish2017optimal, lu2015generalized} already showed the superiority of concave surrogates of a low rank constraint under Gaussian noise compared to the nuclear norm penalty. We take $g(\mathbf{Z}) = \sum_{r} g(\sigma_r(\mathbf{Z}))$ as the concave surrogate of a low rank constraint on $\mathbf{Z}$, where $g(\sigma_r)$ is a concave penalty function of $\sigma_r$. After replacing the low rank constraint in equation \ref{chapter4_eq:4} by $g(\mathbf{Z})$, the model becomes,
\begin{equation}\label{chapter4_eq:5}
\begin{aligned}
    \min_{\bm{\mu},\mathbf{Z},\sigma^2} \quad & f_1(\mathbf{\Theta}_1) + f_2(\mathbf{\Theta}_2,\sigma^2) + \lambda g(\mathbf{Z}) \\
    \text{subject to} \quad \mathbf{\Theta} &= \mathbf{1}\bm{\mu}^{\text{T}} + \mathbf{Z} \\
     \mathbf{\Theta} &= [\mathbf{\Theta}_1 ~ \mathbf{\Theta}_2] \\
     \mathbf{1}^{\text{T}}\mathbf{Z} &= \mathbf{0}.
\end{aligned}
\end{equation}

The most commonly used non-convex surrogates of a low rank constraint are concave functions, including $L_{q:0 < q < 1}$ (bridge penalty) \cite{fu1998penalized,liu2007support}, smoothly clipped absolute deviation (SCAD) \cite{fan2001variable}, a frequentist version of the generalized double Pareto (GDP) shrinkage \cite{armagan2013generalized} and others \cite{lu2015generalized}. We include the first three concave penalties in the algorithm. Their formulas and their thresholding properties are shown in Table \ref{chapter3_tab:1} and Fig.~\ref{chapter3_fig:1}.

\section{Algorithm}  \label{section:4.3}
Based on the Majorization-Minimization (MM) principle \cite{de1994block,hunter2004tutorial}, an MM algorithm is derived to fit the GSCA model with concave penalties. The derived algorithm is guaranteed to decrease the objective function in equation \ref{chapter4_eq:5} during each iteration and the analytical form for updates of all the parameters in each iteration exist.

\subsection{The majorization of the penalized negative lilkelihood}
When fixing $\sigma^2$, we can majorize $f(\mathbf{\Theta}) = f_1(\mathbf{\Theta}_1) + f_2(\mathbf{\Theta}_2)$ to a quadratic function of the parameter $\mathbf{\Theta}$. In addition, the concave penalty function $g(\mathbf{Z})$ can be majorized to a linear function of the singular values of $\mathbf{Z}$ by exploiting the concavity. The resulting majorized problem can be analytically solved by weighted singular value thresholding \cite{lu2015generalized}. The derivation process is the same as the algorithm in Chapter \ref{chapter:3}, therefore we will only show the result here.
\begin{equation}\label{chapter4_eq:6}
\begin{aligned}
f(\mathbf{\Theta}) &\leq \frac{L_k}{2}||\mathbf{\Theta}-\mathbf{H}^k||_F^2 + c\\
g(\mathbf{Z}) &\leq \sum_{r}\omega_{r}^k \sigma_r + c\\
              \mathbf{H}^k &= \mathbf{\Theta}^k - \frac{1}{L_k} (\mathbf{W}\odot \nabla f(\mathbf{\Theta}^k))\\
\omega_r^k &= \partial g(\sigma_r^k),
\end{aligned}
\end{equation}
in which $\nabla f(\mathbf{\Theta}^k) = [\nabla f_1(\mathbf{\Theta}_1^k) ~ \nabla f_2(\mathbf{\Theta}_2^k)]$, $\nabla f_1(\mathbf{\Theta}_1^k) =  \phi(\mathbf{\Theta}_1^k - \mathbf{X}_1)$ and $\nabla f_2(\mathbf{\Theta}_2^k) = \frac{1}{\sigma^2} (\mathbf{\Theta}_2^k - \mathbf{X}_2)$; $L_k$ is the upper-bound of the second order gradient of $f(\mathbf{\Theta})$ during the $k^{\text{th}}$ iteration, and can always set to $L_k = \text{max}(0.25, 1/(\sigma^2)^k)$; $(\sigma^2)^k$ is the approximation of parameter $\sigma^2$ during the $k^{\text{th}}$ iteration; $\sigma_r^k$ is the $r^{\text{th}}$ singular value of $\mathbf{Z}^{k}$, which is an approximation of $\mathbf{Z}$ during the $k^{\text{th}}$ iteration; $\mathbf{\Theta}^{k}$ is the approximation of $\mathbf{\Theta}$ during the $k^{\text{th}}$ iteration; $c$ is a constant doesn't depend on any unknown parameters. Summarizing these two majorization steps, we have the following majorized problem during the $k^{\text{th}}$ iteration.
\begin{equation}\label{chapter4_eq:7}
\begin{aligned}
\min_{\bm{\mu},\mathbf{Z}} \quad & \frac{L_k}{2}||\mathbf{\Theta}-\mathbf{H}^{k}||_F^2 + \lambda \sum_{r} \omega_r^k\sigma_{r}\\
           \text{subject to} \quad   \mathbf{\Theta} &= \mathbf{1}\bm{\mu}^{\text{T}} + \mathbf{Z}\\
                               \mathbf{1}^{\text{T}}\mathbf{Z} &= \mathbf{0} \\
                               \mathbf{H}^k &= \mathbf{\Theta}^k - \frac{1}{L_k} (\mathbf{W}\odot \nabla f(\mathbf{\Theta}^k)) \\
                               \omega_r^k &= \partial g(\sigma_r^k).
\end{aligned}
\end{equation}

\subsection{Block coordinate descent}
We optimize $\bm{\mu}$, $\mathbf{Z}$ and $\sigma^2$ alternatingly while fixing the other parameters. However, updating $\bm{\mu}$ and $\mathbf{Z}$ depend on solving the majorized problem in equation \ref{chapter4_eq:7} rather than solving the original problem in equation \ref{chapter4_eq:5}. Because of the MM principle, this step will also monotonically decrease the original loss function in equation \ref{chapter4_eq:5}.

\subsubsection*{Updating $\bm{\mu}$}
The analytical solution of $\bm{\mu}$ in equation \ref{chapter4_eq:7} is simply the column mean of $\mathbf{H}^k$, $\bm{\mu} = \frac{1}{I} (\mathbf{H}^k)^{\text{T}} \mathbf{1}$.

\subsubsection*{Updating $\mathbf{Z}$}
After deflating the offset term $\bm{\mu}$, the loss function in equation \ref{chapter4_eq:7} becomes $\frac{L_k}{2} ||\mathbf{Z} - \mathbf{J} \mathbf{H}^k||_F^2 + \lambda \sum_{r}\omega_{r}^k \sigma_r$, in which $\mathbf{J} = \mathbf{I} - \frac{1}{I} \mathbf{1} \mathbf{1}^{\text{T}}$ is the column centering matrix. The solution of the resulting problem is equivalent to the proximal operator of the weighted sum of singular values, which has an analytical form solution \cite{lu2015generalized}. Suppose $\mathbf{USV}^{\text{T}} = \mathbf{J} \mathbf{H}^k$ is the SVD decomposition of $\mathbf{J} \mathbf{H}^k$, the analytical form solution of $\mathbf{Z}$ is $\mathbf{Z} = \mathbf{US}_{z}\mathbf{V}^{\text{T}}$, in which $\mathbf{S}_{z} = \text{Diag}\{(s_{r}-\lambda \omega_r /L_k)_{+}\}$ and $s_{r}$ is the $r^{\text{th}}$ diagonal element in $\mathbf{S}$.

\subsubsection*{Updating $\sigma^2$}
By setting the gradient of $f(\mathbf{\Theta},\sigma^2)$ in equation \ref{chapter4_eq:5} with respect to $\sigma^2$ to be 0, we have the following analytical solution of $\sigma^2$,  $\sigma^2= \frac{1}{||\mathbf{W}_2||_0} ||\mathbf{W}_2 \odot (\mathbf{X}_2 - \mathbf{\Theta}_2)||_F^2$. When no low rank estimation of $\mathbf{Z}$ can be achieved, the constructed model is close to a saturated model and the estimated $\hat{\sigma^2}$ is close to 0. In that case, when $\hat{\sigma^2}<0.05$, the algorithm stops and gives a warning that a low rank estimation has not been achieved.

\subsubsection*{Initialization and stopping criteria}
Random initialization is used. All the elements in $\mathbf{Z}^0$ are sampled from the standard uniform distribution, $\bm{\mu}^0$ is set to 0 and $(\sigma^2)^0$ is set to 1. The relative change of the objective value is used as the stopping criteria. Pseudocode of the algorithm described above is shown in Algorithm \ref{alg:GSCA}. $\epsilon_f$ is the tolerance of relative change of the loss function.

\begin{algorithm}[htb]
  \caption{A MM algorithm for fitting the GSCA model with concave penalties.}
  \label{alg:GSCA}
  \begin{algorithmic}[1]
    \Require
      $\mathbf{X}_1$, $\mathbf{X}_2$, penalty, $\lambda$, $\gamma$;
    \Ensure
      $\hat{\bm{\mu}}$, $\hat{\mathbf{Z}}$, $\hat{\sigma^2}$;
    \State Compute $\mathbf{W}_1$, $\mathbf{W}_2$ for missing values in $\mathbf{X}_1$ and $\mathbf{X}_2$, and $\mathbf{W} = [\mathbf{W}_1 ~ \mathbf{W}_2]$;
    \State Initialize $\bm{\mu}^0$, $\mathbf{Z}^0$, $(\sigma^2)^0$;
    \State $k = 0$;
    \While{$(f^{k-1}-f^{k})/f^{k-1}>\epsilon_f$}
        \State $\nabla f_1(\mathbf{\Theta}_1^k) = \phi(\mathbf{\Theta}_1^k) - \mathbf{X}_1$; $\nabla f_2(\mathbf{\Theta}_2^k) = \frac{1}{(\sigma^2)^k} (\mathbf{\Theta}_2^k - \mathbf{X}_2)$;
        \State $\nabla f(\mathbf{\Theta}^k) = [\nabla f_1(\mathbf{\Theta}_1^k) ~ \nabla f_2(\mathbf{\Theta}_2^k)]$;
        \State $L_k=\text{max}(0.25,1/(\sigma^2)^k)$;
        \State $\mathbf{H}^{k} = \mathbf{\Theta}^{k}- \frac{1}{L_{k}} (\mathbf{W} \odot \nabla f(\mathbf{\Theta}^{k}))$;
        \State $\omega_r^k = \partial g(\sigma_r^k)$;
        \State $\bm{\mu}^{k+1} = \frac{1}{I} (\mathbf{H}^{k})^{\text{T}} \mathbf{1}$;
        \State $\mathbf{USV}^{\text{T}} = \mathbf{J}\mathbf{H}^{k}$;
        \State $\mathbf{S}_{z} = \text{Diag}\{ (s_{r} - \lambda \omega_r^k /L_{k})_{+}\}$;
        \State $\mathbf{Z}^{k+1} = \mathbf{US}_{z}\mathbf{V}^{\text{T}}$;
        \State $\mathbf{\Theta}^{k+1} = \mathbf{1}(\bm{\mu}^{k+1})^{\text{T}} + \mathbf{Z}^{k+1}$;
        \State $[\mathbf{\Theta}_1^{k+1} ~ \mathbf{\Theta}_2^{k+1}] = \mathbf{\Theta}^{k+1}$;
        \State $(\sigma^2)^{k+1} = \frac{1}{||\mathbf{W}_2||_0} ||\mathbf{W}_2 \odot (\mathbf{X}_2 - \mathbf{\Theta}_2^{k+1})||_F^2$
        \State $k=k+1$;
    \EndWhile
  \end{algorithmic}
\end{algorithm}

\section{Simulation}  \label{section:4.4}
To see how well the GSCA model is able to reconstruct data generated according to the model, we do a simulation study with similar characteristics as a typical empirical data set. We first simulate the imbalanced binary $\mathbf{X}_1$ and quantitative $\mathbf{X}_2$ following the GSCA model with logit link and low signal-to-noise ratio (SNR). After that, we evaluate the GSCA model with respect to 1) the quality of the reconstructed low rank structure from the model, and 2) the reconstruction of true number of dimensions.

\subsection{Data generating process}
The SNR for generating binary data is defined according to the latent variable interpretation of the logistic PCA model. Elements in $\mathbf{X}_1$ are independent and indirect binary observations of the corresponding elements in an underlying quantitative matrix $\mathbf{X}_1^{\ast}$, $x_{1ij} = 1$ if $x_{1ij}^{\ast}>0$ and $x_{1ij} = 0$ otherwise. $\mathbf{X}_1^{\ast}$ can be expressed as $\mathbf{X}_1^{\ast} = \mathbf{\Theta}_1 + \mathbf{E}_1$, in which $\mathbf{\Theta}_1 = \mathbf{1}\bm{\mu}_1^{\text{T}} + \mathbf{AB}_1^{\text{T}}$, and elements in $\mathbf{E}_1$ follow the standard logistic distribution, $\epsilon_{1ij} \sim \text{Logistic}(0,1)$. The SNR for generating binary data $\mathbf{X}_1$ is defined as $\text{SNR}_1 = ||\mathbf{AB}_1^{\text{T}}||_{F}^2/||\mathbf{E}_1||_{F}^2$. Assume the quantitative $\mathbf{X}_2$ is simulated as $\mathbf{X}_2 = \mathbf{\Theta}_2 + \mathbf{E}_2$, in which $\mathbf{\Theta}_2 = \mathbf{1}\bm{\mu}_2^{\text{T}} + \mathbf{AB}_2^{\text{T}}$ and elements in $\mathbf{E}_2$ follow a normal distribution with 0 mean and $\sigma^2$ variance, $\epsilon_{2ij} \sim N(0,\sigma^2)$. The SNR for generating quantitative $\mathbf{X}_2$ is defined as $\text{SNR}_2 = ||\mathbf{AB}_2^{\text{T}}||_{F}^2/||\mathbf{E}_2||_{F}^2$.

After the definition of the SNR, we simulate the coupled binary $\mathbf{X}_1$ and quantitative $\mathbf{X}_2$ as follows. $\bm{\mu}_1$ represents the logit transform of the marginal probabilities of binary variables and $\bm{\mu}_2$ represents the mean of the marginal distributions of quantitative variables. They will be simulated according to the characteristics of a real biological data set. The score matrix $\mathbf{A}$ and loading matrices $\mathbf{B}_1$, $\mathbf{B}_2$ are simulated as follows. First, we express $\mathbf{A}\mathbf{B}_1^{\text{T}}$ and $\mathbf{A}\mathbf{B}_2^{\text{T}}$ in a SVD type as $\mathbf{A}\mathbf{B}_1^{\text{T}} = \mathbf{U}\mathbf{D}_1\mathbf{V}_1^{\text{T}}$ and $\mathbf{A}\mathbf{B}_2^{\text{T}} = \mathbf{U}\mathbf{D}_2\mathbf{V}_1^{\text{T}}$, in which $\mathbf{U}^{\text{T}}\mathbf{U} = \mathbf{I}_R$, $\mathbf{D}_1$ and $\mathbf{D}_2$ are diagonal matrices, $\mathbf{V}_1^{\text{T}}\mathbf{V}_1 = \mathbf{I}_R$ and $\mathbf{V}_2^{\text{T}}\mathbf{V}_2 = \mathbf{I}_R$. All the elements in $\mathbf{U}$, $\mathbf{V}_1$ and $\mathbf{V}_2$ are independently sampled from the standard normal distribution. Then, $\mathbf{U}$, $\mathbf{V}_1$ and $\mathbf{V}_2$ are orthogonalized by the QR algorithm. The diagonal matrix $\mathbf{D}$($R\times R$) is simulated as follows. $R$ elements are sampled from standard normal distribution, their absolute values are sorted in decreasing order. To satisfy the pre-specified $\text{SNR}_1$ and $\text{SNR}_2$, $\mathbf{D}$ is scaled by positive scalars $c_1$ and $c_2$ as $\mathbf{D}_1 = c_1\mathbf{D}$ and $\mathbf{D}_2 = c_2\mathbf{D}$. Then, binary elements in $\mathbf{X}_1$ are sampled from the Bernoulli distribution with corresponding parameter $\phi(\theta_{1ij})$, in which $\phi()$ is inverse logit function and $\mathbf{\Theta}_1 = \mathbf{1}\bm{\mu}_1^{\text{T}} + \mathbf{AB}_1^{\text{T}}$. Quantitative data set $\mathbf{X}_2$ is generated as $\mathbf{X}_2 = \mathbf{\Theta}_2 + \mathbf{E}_2$, in which $\mathbf{\Theta}_2 = \mathbf{1}\bm{\mu}_2^{\text{T}} + \mathbf{AB}_2^{\text{T}}$ and elements in $\mathbf{E}_2$ are sampled from $N(0,\sigma^2)$. Take $\mathbf{Z} = \mathbf{A}\mathbf{B}^{\text{T}}$, $\mathbf{B} = [\mathbf{B}_1 ~ \mathbf{B}_2]$. In order to make $\mathbf{1}^{\text{T}}\mathbf{Z} = \mathbf{0}$, we further deflate the column offset of $\mathbf{Z}$ to the simulated $\bm{\mu}$, $\bm{\mu} = [\bm{\mu}_1^{\text{T}} ~ \bm{\mu}_2^{\text{T}}]^{\text{T}}$. This step will not change the value of $\mathbf{\Theta}_1$ and $\mathbf{\Theta}_2$, thus does not affect the simulation of $\mathbf{X}_1$ and $\mathbf{X}_2$.

\subsection{Evaluation metric and model selection}
As for simulated data sets, the true parameters $\mathbf{\Theta} = [\mathbf{\Theta}_1~\mathbf{\Theta}_2]$, $\bm{\mu} = [\bm{\mu}_1^{\text{T}} \bm{\mu}_1^{\text{T}}]^{\text{T}}$ and $\mathbf{Z} = \mathbf{A}\mathbf{B}^{\text{T}}$ are available. Therefore, the generalization error of the constructed model can be evaluated by comparing the true parameters and their model estimates. Thus, the evaluation metric is defined as the relative mean squared error (RMSE) of the model parameters. The RMSE of estimating $\mathbf{\Theta}$ is defined as $\text{RMSE}(\mathbf{\Theta}) = ||\mathbf{\Theta}-\hat{\mathbf{\Theta}}||_F^2/||\mathbf{\Theta}||_F^2$, where $\mathbf{\Theta}$ represents the true parameter and $\hat{\mathbf{\Theta}}$ its GSCA model estimate. The RMSE of $\bm{\mu}$ and $\mathbf{Z}$, are expressed as $\text{RMSE}(\bm{\mu})$ and $\text{RMSE}(\mathbf{Z})$ and they are defined in the same way as for $\mathbf{\Theta}$.

For real data sets, missing value based cross validation (CV) is used to estimate the generalization error of the constructed model. Also in order to have an estimation of the uncertainty of the CV error, we will use a K-fold CV procedure. To make the prediction of the left out fold elements independent to the constructed model based on the reminding folds, the data is partitioned into K folds of elements which are selected in a diagonal style rather than row wise from $\mathbf{X}_1$ and $\mathbf{X}_2$ respectively, similar to the leave out patterns described by Wold \cite{wold1978cross, bro2008cross}. The test set elements of each fold in $\mathbf{X}_1$ and $\mathbf{X}_2$ are taken as missing values, and the remaining data are used to construct a GSCA model. After estimation of $\hat{\mathbf{\Theta}}$ and $\hat{\sigma^2}$ are obtained from the constructed GSCA model, the negative log likelihood of using $\hat{\mathbf{\Theta}}$, $\hat{\sigma^2}$ to predict the missing elements (left out fold) is recorded. This negative log likelihood is scaled by the number of missing elements. This process is repeated K times until all the K folds have been left out once. The mean of the K scaled negative log likelihoods is taken as the CV error.

When we define $\mathbf{X}=[\mathbf{X}_1 ~ \mathbf{X}_2]$ and $J=J_1+J_2$, the penalty term $\lambda g(\mathbf{Z})$ is not invariant to the number of non-missing elements in $\mathbf{X}$, as the joint loss function (equation \ref{chapter4_eq:4}) is the sum of the log likelihoods for fitting all the non-missing elements in the data $\mathbf{X}$. Therefore, we effectively follow a similar approach as Fan \cite{fan2001variable} by adjusting the penalty strength parameter $\lambda$ for the relative number observations. By setting one fold of elements to be missing during the CV process, $\lambda||\mathbf{X}||_0/(I\times J)$ rather than $\lambda$ is used as the amount of penalty. During the K-fold CV process, a warm start strategy, using the results of previous constructed model as the initialization of next model, is applied. In this way, the K-fold CV can be greatly accelerated.

In the model selection process, the tuning parameter $\lambda$ and hyper-parameters ($q$ in $L_{q}$ and $\gamma$ in SCAD and GDP) can be selected by a grid search. However, previous work of using these penalty functions in supervised learning context \cite{fu1998penalized,fan2001variable,armagan2013generalized} and our experiments have shown that the results are not very sensitive to the selection of these hyper-parameters, and thus a default value can be set. On the other hand, the selection of tuning parameter $\lambda$ does have a significant effect on the results, and should be optimized by the grid search.

\subsection{Experiments}
In the loading or score plots of the following experiments and real data analysis, we will multiply the estimated score matrix $\hat{\mathbf{A}}$ by $\sqrt{I}$, and the estimated loading matrix $\hat{\mathbf{B}}$ by $1/\sqrt{I}$ to make the loading and score plots have the same scale.

\subsubsection{Overfitting of the GSCA model with a fixed rank and no penalty}
\begin{figure}[htbp]
    \centering
    \includegraphics[width=0.9\textwidth]{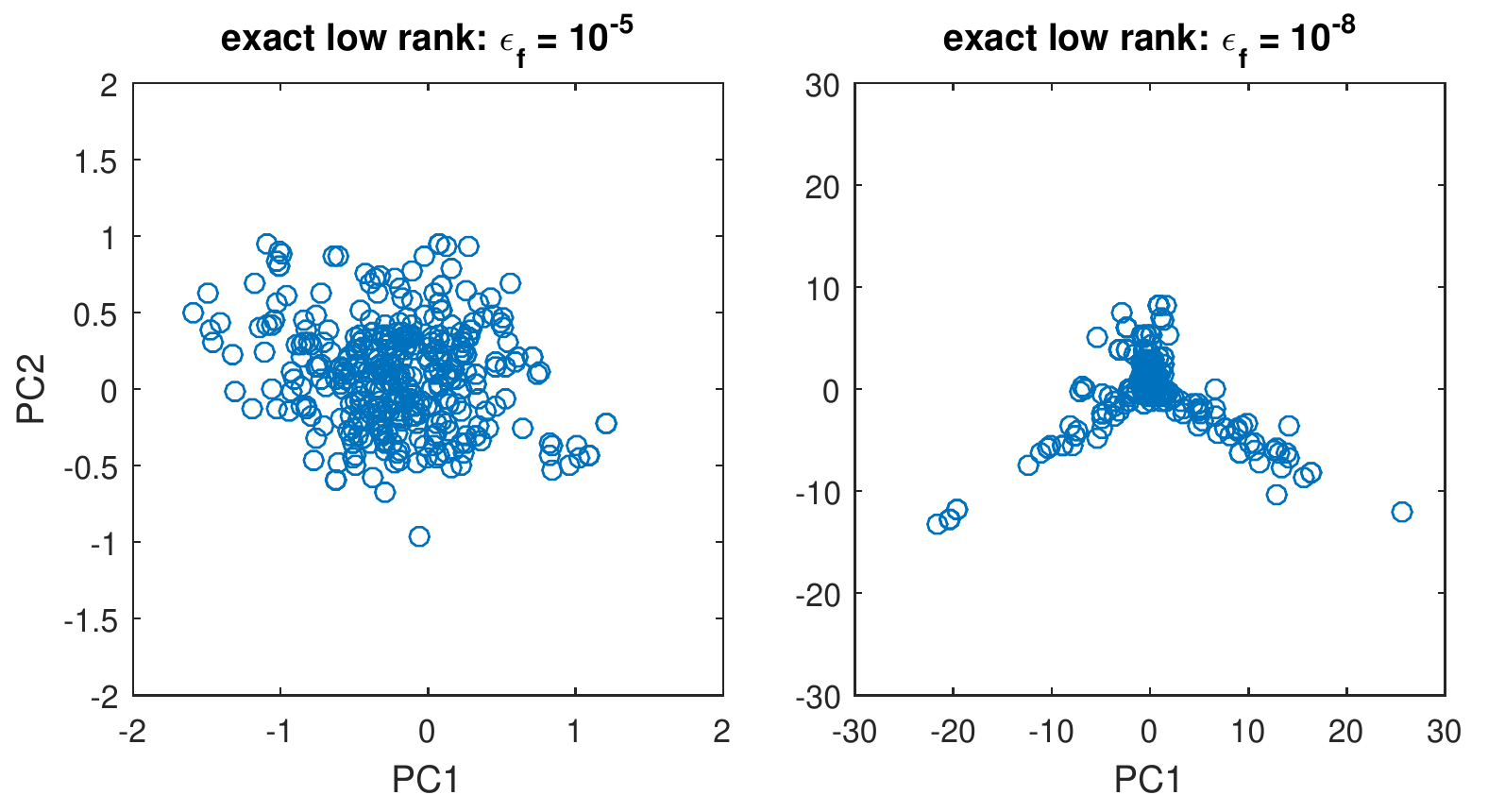}
    \caption{Loading plots of estimated $\hat{\mathbf{B}}_1$ from the GSCA models with exact low rank constraint using two different stopping criteria $\epsilon_f =10^{-5}$ and $\epsilon_f=10^{-8}$. Note that the scales of the coordinates for $\epsilon_f = 10^{-8}$ (right) is over ten times larger than those for $\epsilon_f = 10^{-5}$ (left).}
    \label{chapter4_fig:1}
\end{figure}

The real data sets from the Section \ref{section:4.5} are used to show how the GSCA model with a fixed rank and no penalty will overfit the data. The algorithm (details are in the supplemental material) used to fit the GSCA model (with an exact low rank constraint and orthogonality constraint $\mathbf{A}^{\text{T}}\mathbf{A} = I\mathbf{I}$) is a modification of the developed algorithm in Section \ref{section:4.3}. GSCA models with three components are fitted using stopping criteria $\epsilon_f = 10^{-5}$ and $\epsilon_f=10^{-8}$. Exactly the same initialization is used for these two models. As shown in Fig.~\ref{chapter4_fig:1}, different stopping criteria can greatly affect the estimated $\hat{\mathbf{B}}_1$ from the GSCA models. Furthermore, the number of iterations to reach convergence increases from 141 to 23991.

Related phenomena have been observed in logistic linear regression model and logistic PCA model \cite{de2006principal, song2017principal} where some estimated parameters tend to diverge towards infinity. The overfitting issue of the GSCA model with exact low rank constraint can be interpreted in the same way by taking the columns of score matrix $\mathbf{A}$ as the latent variables and the loading matrix $\mathbf{B}_1$ as the coefficients to fit the binary $\mathbf{X}_1$. This result suggests that if an exact low rank constraint is preferred in the GSCA model, an extra scale penalty should be added on $\mathbf{B}_1$ to avoid overfitting.

\subsubsection{Comparing the generalization errors of the GSCA models with nuclear norm and concave penalties}
To evaluate the performance of the GSCA model in recovering the underlying structure, we set up the realistic simulation (strongly imbalanced binary data and low SNR) as follows. The simulated $\mathbf{X}_1$ and $\mathbf{X}_2$ have the same size as the real data sets in the Section \ref{section:4.5}, $I=160$, $J_1=410$, $J_2 = 1000$. The logit transform of the empirical marginal probabilities of the CNA data set in the Section \ref{section:4.5} is set as $\bm{\mu}_1$. Elements in $\bm{\mu}_2$ are sampled from the standard normal distribution. The simulated low rank is set to $R=10$; $\sigma^2$ is set to 1; $\text{SNR}_1$ and $\text{SNR}_2$ are set to 1. After the simulation of $\mathbf{X}_1$, there are two columns only have ``0'' elements, which are removed as they provide no information (no variation).

\begin{figure}[htbp]
    \centering
    \includegraphics[width=0.9\textwidth]{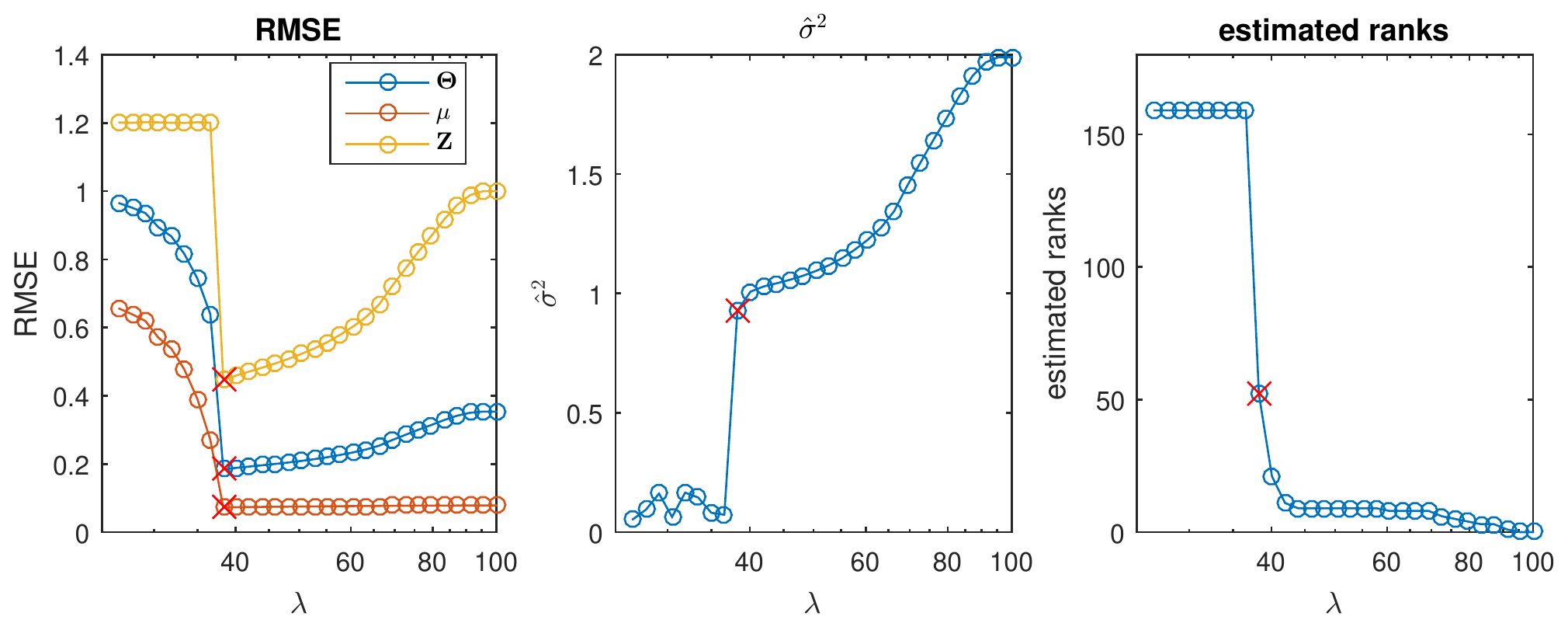}
    \caption{RMSEs in estimating $\mathbf{\Theta}$, $\bm{\mu}$, $\mathbf{Z}$ (left), the estimated $\hat{\sigma^2}$ (center) and the estimated $\text{rank}(\hat{\mathbf{Z}})$ (right) from the GSCA model with nuclear norm penalty as a function of the tuning parameter $\lambda$. Red cross marker indicates the model with minimum $\text{RMSE}(\mathbf{\Theta})$.}
    \label{chapter4_fig:2}
\end{figure}

As the GSCA model with the nuclear norm penalty is a convex problem, a global optimum can be obtained. The nuclear norm penalty is therefore used as the baseline in the comparison with other penalties. An interval from $\lambda_0$, which is large enough to achieve an estimated rank of at most rank 1, to $\lambda_{t}$, which is small enough to achieve an estimated rank of 159, is selected based on low precision models ($\epsilon_f=10^{-2}$). 30 log-spaced $\lambda$s are selected equally from the interval $[\lambda_{t},\lambda_{0}]$. The convergence criterion is set as $\epsilon_f = 10^{-8}$. The results are shown in Fig.~\ref{chapter4_fig:2}. With decreasing $\lambda$, the estimated rank of $\hat{\mathbf{Z}}$ increased from 0 to 159, and the estimated $\hat{\sigma^2}$ decreased from 2 to close to 0. The minimum $\text{RMSE}(\mathbf{\Theta})$ of 0.184 (the corresponding $\text{RMSE}(\mathbf{\Theta}_1)=0.229$, $\text{RMSE}(\mathbf{\Theta}_2)=0.054$, $\text{RMSE}(\bm{\mu}) = 0.072$ and $\text{RMSE}(\mathbf{Z})=0.446$) can be achieved at $\lambda=38.3$, which corresponds to $\text{rank}(\hat{\mathbf{Z}})=52$ and $\hat{\sigma^2}=0.9271$. There are sharp transitions in all the three subplots near the point $\lambda=40$. The reason is that when the penalty is not large enough, the estimated rank becomes 159, and the constructed GSCA model is almost a saturated model. Thus the model has high generalization error and the estimated $\hat{\sigma^2}$ also becomes close to 0. Given that we only have indirect binary observation $\mathbf{X}_1$ and highly noisy observation $\mathbf{X}_2$ of the underlying structure $\mathbf{\Theta}$, the performance of the GSCA model with nuclear norm penalty is reasonable. However, results can be greatly improved by using concave penalties.

\begin{figure}[htbp]
    \centering
    \includegraphics[width=0.9\textwidth]{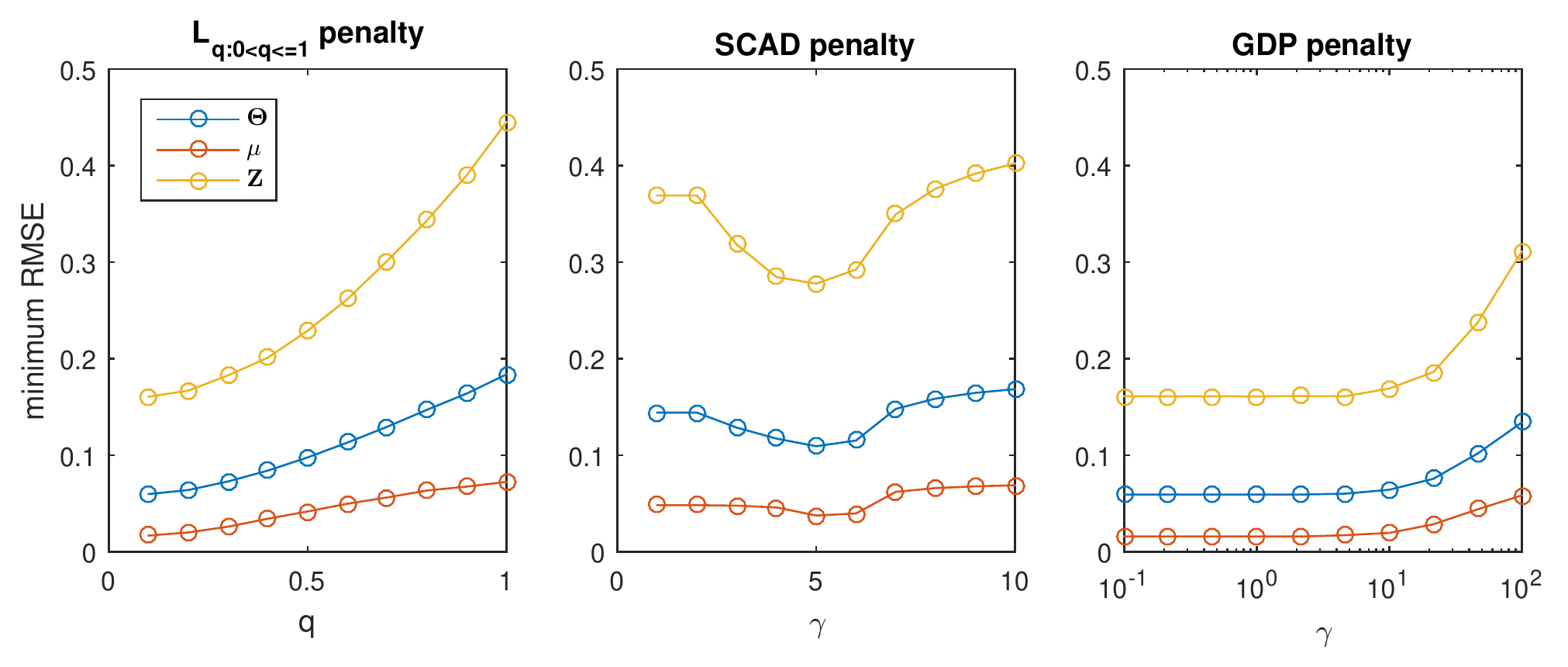}
    \caption{The minimum $\text{RMSE}(\mathbf{\Theta})$ achieved and the corresponding $\text{RMSE}(\bm{\mu})$ and $\text{RMSE}(\mathbf{Z})$ for different values of hyper-parameter for $L_q$ penalty (left), for SCAD penalty (center) and for GDP penalty (right). The legends indicate the RMSEs in estimating $\mathbf{\Theta}$, $\bm{\mu}$ and $\mathbf{Z}$ respectively. The $x$-axis of the left and center subplots has a linear scale, while right subplot has a log scale.}
    \label{chapter4_fig:3}
\end{figure}

For concave penalties, different values of the hyper-parameters, $q$ in $L_q$, $\gamma$ in SCAD and GDP, are selected according to their thresholding properties. For each value of the hyper-parameter, values of tuning parameter $\lambda$ are selected in the same manner as described above. The minimum $\text{RMSE}(\mathbf{\Theta})$ achieved and the corresponding $\text{RMSE}(\bm{\mu})$ and $\text{RMSE}(\mathbf{Z})$ for different values of hyper-parameter of the GSCA models with different penalty functions are shown in Fig.~\ref{chapter4_fig:3}. Here all GSCA models with concave penalties can achieve much lower RMSEs in estimating $\mathbf{\Theta}$, $\bm{\mu}$ and $\mathbf{Z}$ compared to the convex nuclear norm penalty ($L_{q:q=1}$ in the plot). Among the three concave penalties used, $L_{q}$ and GDP have better performance.

\begin{figure}[htbp]
    \centering
    \includegraphics[width=0.5 \textwidth]{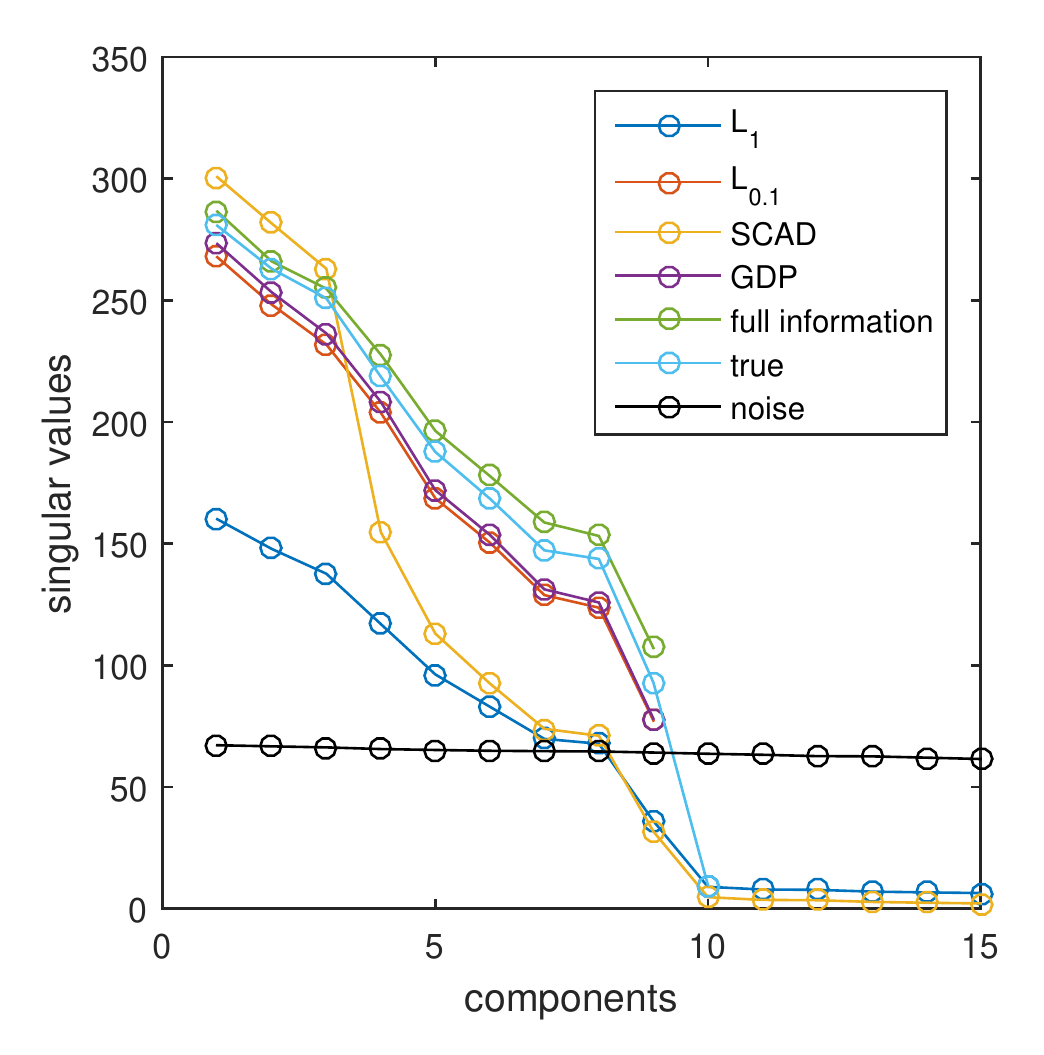}
    \caption{Approximation of the singular values using different penalties in the simulation experiment. Labels ``$L_1$'', ``$L_{0.1}$'', ``SCAD'', ``GDP'', ``full information'' indicate the singular values of estimated $\hat{\mathbf{Z}}$ from the corresponding models; ``$\text{true}$'' indicates the singular values of the simulated $\mathbf{Z}$; ``$\text{noise}$'' indicates the singular values of the noise term $\mathbf{E}$, which has full rank.}
    \label{chapter4_fig:4}
\end{figure}

If we get access to the full information, the underlying quantitative data $\mathbf{X}_1^{\ast}$ rather than the binary observation $\mathbf{X}_1$, the SCA model on $\mathbf{X}_1^{\ast}$ and $\mathbf{X}_2$ is simply a PCA model on $[\mathbf{X}_1^{\ast} ~ \mathbf{X}_2]$. From this model, we can get an estimation of $\mathbf{\Theta}$, $\bm{\mu}$ and $\mathbf{Z}$. We compared the results derived from the SCA model on the full information, the GSCA models with nuclear norm, $L_{q:q=0.1}$, SCAD ($\gamma=5$) and GDP ($\gamma=1$) penalties. All the models are selected to achieve the minimum $\text{RMSE}(\mathbf{\Theta})$. The RMSEs of estimating $\mathbf{\Theta}$, $\mathbf{\Theta}_1$, $\mathbf{\Theta}_2$, $\bm{\mu}$ and $\mathbf{Z}$ and the rank of estimated $\hat{\mathbf{Z}}$ from different models are shown in Table \ref{chapter4_tab:1}. Here we can see that the GSCA models with $L_{q:q=0.1}$ and GDP ($\gamma=1$) penalties have better performance in almost all criteria compared to the nuclear norm and SCAD penalties, and even comparable with the SCA model on full information. The singular values of the true $\mathbf{Z}$, estimated $\hat{\mathbf{Z}}$ from the above models and the noise terms $\mathbf{E} = [\mathbf{E}_1~\mathbf{E}_2]$ are shown in Fig.~\ref{chapter4_fig:4}. Only the first 15 singular values are shown to have higher resolution of the details. Since the $10^{\text{th}}$ singular value of the simulated data $\mathbf{Z}$ is smaller than the noise level, the best achievable rank estimation is 9. Both the $L_{q:q=0.1}$ and GDP ($\gamma=1$) penalties successfully find the correct rank 9, and they have a very good approximation of the first 9 singular values of $\mathbf{Z}$. On the other hand, the nuclear norm penalty shrinks all the singular values too much. Furthermore, the SCAD penalty overestimates the first three singular values and therefore shrinks all the other singular values too much. These results are easily understandable if taking their thresholding properties in Fig.~\ref{chapter3_fig:1} into account. Both the $L_{q}$ and the GDP penalties have very good performance in this simulation experiment.

\begin{table}[htbp]
\centering
\caption{The RMSEs of estimating $\mathbf{\Theta}$, $\bm{\mu}$ and $\mathbf{Z}$ and the rank of estimated $\hat{\mathbf{Z}}$ from different models.}
\label{chapter4_tab:1}
\begin{tabular}{|l|l|l|l|l|l|l|}
 \hline
    & $\text{RMSE}(\mathbf{\Theta})$ & $\text{RMSE}(\mathbf{\Theta}_1)$ & $\text{RMSE}(\mathbf{\Theta}_2)$ & $\text{RMSE}(\bm{\mu})$ & $\text{RMSE}(\mathbf{Z})$ & $\text{rank}(\hat{\mathbf{Z}})$\\
 \hline
  $L_{q:q=1}$   & 0.1840 & 0.2288 & 0.0537 & 0.0724 & 0.4456 & 52 \\
  $L_{q:q=0.1}$ & 0.0598 & 0.0682 & 0.0353 & 0.0168 & 0.1606 & 9\\
  SCAD($\gamma=5$) & 0.1093 & 0.1334 & 0.0395 & 0.0376 & 0.2777 & 24 \\
  GDP($\gamma=1$) & 0.0593 & 0.0675 & 0.0354 & 0.0160 & 0.1610 & 9 \\
  full information & 0.0222 & 0.0675 & 0.0354 & 0.0030 & 0.0674 & 9 \\
  \hline
\end{tabular}
\end{table}

\subsubsection{Comparing the GSCA model with GDP penalty and the iClusterPlus model}
\begin{figure}[htbp]
    \centering
    \includegraphics[width=0.9\textwidth]{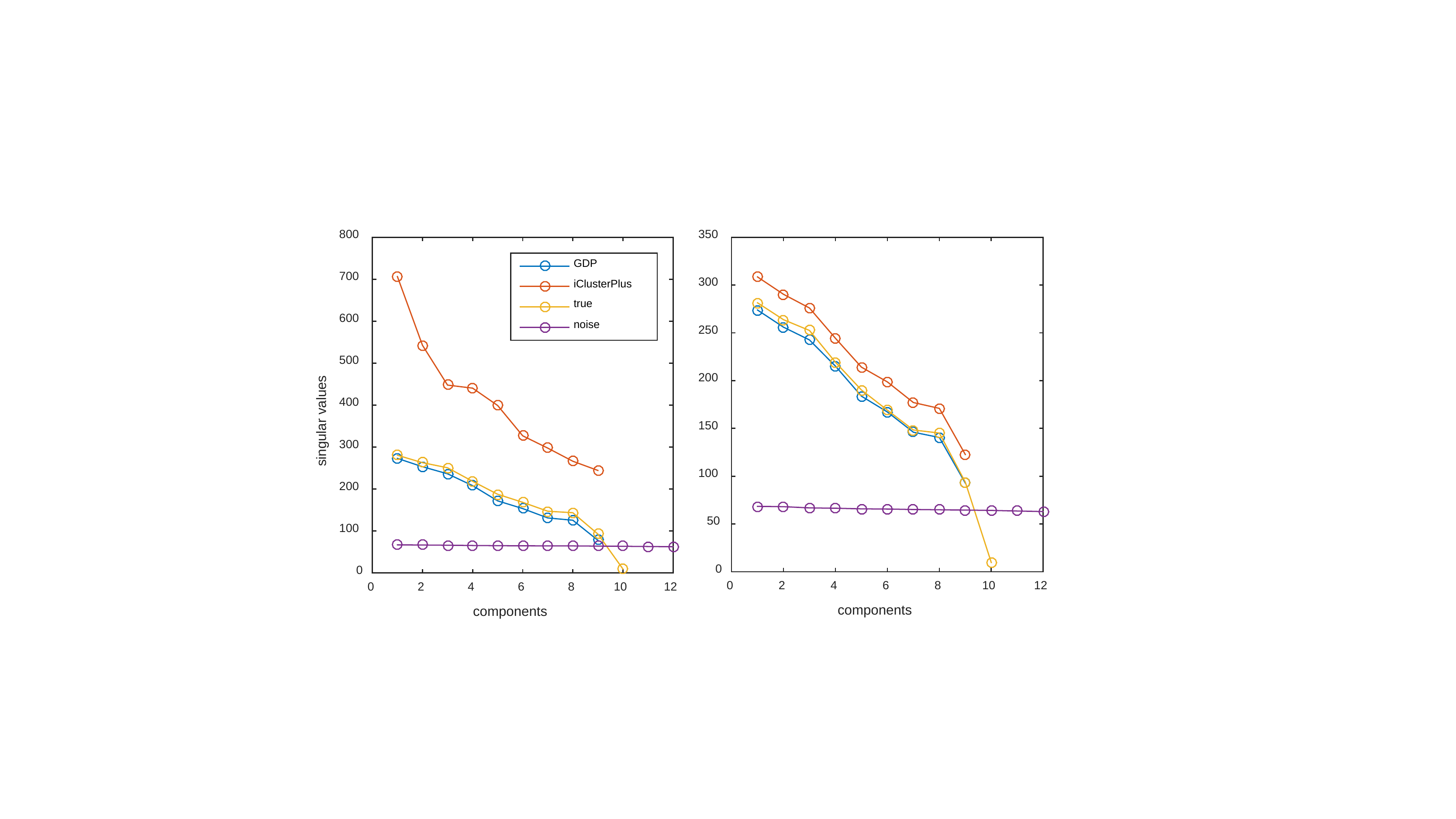}
    \caption{The singular values of estimated $\hat{\mathbf{Z}}$ using the iClusterPlus model and the GSCA model with GDP penalty on the simulation with imbalanced binary data (left) and with balanced binary data (right).}
    \label{chapter4_fig:5}
\end{figure}

We compared our GSCA model with GDP penalty to the iClusterPlus model on the simulated data sets. The parameters for the GSCA model with GDP penalty is the same as described above. The running time is 60.61s when $\epsilon_f=10^{-8}$, and 9.98s when $\epsilon_f=10^{-5}$. For the iClusterPlus model, 9 latent variables are specified. The tuning parameter of the lasso type constraint on the data specific coefficient matrices are set to 0. The default convergence criterion is used, that is the maximum of the absolute changes of the estimated parameters in two subsequent iterations is less than $10^{-4}$. The running time of the iClusterPlus model is close to 3 hours. The constructed iClusterPlus model provides the estimation of column offset $\hat{\bm{\mu}}$, the common latent variables $\hat{\mathbf{A}}$, and data set specific coefficient matrices $\hat{\mathbf{B}}_1$ and $\hat{\mathbf{B}}_2$. The estimated $\hat{\mathbf{Z}}$ and $\hat{\mathbf{\Theta}}$ are computed in the same way as defined in the model section. The RMSEs in estimating $\mathbf{\Theta}$, $\bm{\mu}$ and $\mathbf{Z}$ for iClusterPlus are 2.571, 2.473 and 3.060 respectively. Compared to the results from the GSCA models in Table \ref{chapter4_tab:1}, iClusterPlus is unable to provide good results on the simulated data sets. Supplemental Fig.~S4.2 compares the estimated $\hat{\mu}_1$ from the GSCA model with GDP penalty and iClusterPlus model. As shown in supplemental Fig.~S4.2(right), the iClusterPlus model is unable to estimate the offset $\bm{\mu}$ correctly. Many elements of estimated $\hat{\bm{\mu}_1}$ are exactly 0, which corresponds to an estimated marginal probability of 0.5. In addition, as shown in Fig.~\ref{chapter4_fig:5}(left), the singular values of the estimated $\hat{\mathbf{Z}}$ from the iClusterPlus model are clearly overestimated. These undesired results from the iClusterPlus model are due mainly to the imbalancedness of the simulated binary data set. If the offset term $\bm{\mu}_1$ in the simulation is set to 0, which corresponds to balanced binary data simulation, and fix all the other parameters in the same way as in the above simulation, the results of iClusterPlus and the GSCA with GDP penalty are more comparable. In that case the RMSEs of estimating $\mathbf{\Theta}$, $\mathbf{Z}$ in the GSCA model with GDP penalty are 0.071 and 0.091 respectively, while the RMSEs of the iClusterPlus model are 0.107 and 0.142 respectively. As shown in Fig.~\ref{chapter4_fig:5}(right), the singular values of estimated $\hat{\mathbf{Z}}$ from the iClusterPlus model are much more accurate compared to the imbalanced case. However, iClusterPlus still overestimates the singular values compared to the GSCA model with GDP penalty. This phenomenon is related to the fact that exact low rank constraint is also used in the iClusterPlus model. These results suggest that compared to iClusterPlus, the GSCA model with GDP penalty is more robust to the imbalanced binary data and has better performance in recovering the underlying structure in the simulation experiment.

\subsubsection{The performance of the GSCA model for the simulation with different SNRs}
\begin{figure}[htbp]
    \centering
    \includegraphics[width=0.9\textwidth]{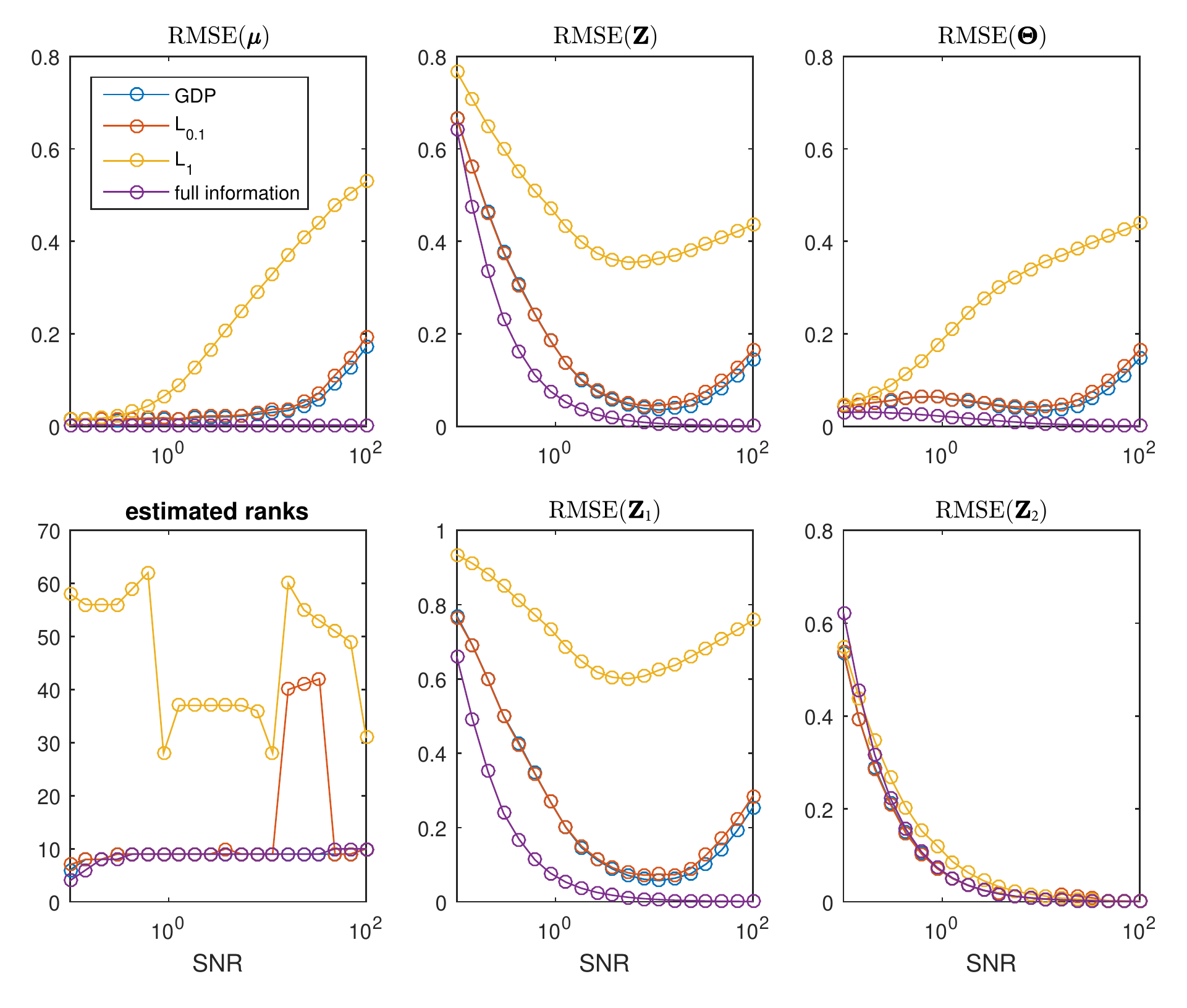}
    \caption{Minimum $\text{RMSE}(\mathbf{\Theta})$ (top right), and the corresponding $\text{RMSE}(\bm{\mu})$ (top left), $\text{RMSE}(\mathbf{Z})$ (top center), rank estimation of $\hat{\mathbf{Z}}$ (bottom left), $\text{RMSE}(\mathbf{Z}_1)$ (bottom center) and $\text{RMSE}(\mathbf{Z}_2)$ (bottom right) of the GSCA models with nuclear norm penalty (legend ``$L_{1}$''), GDP penalty (legend GDP), $L_{0.1}$ penalty (legend ``$L_{0.1}$'') and SCA model on full information (legend ``full information'') for different SNR levels.}
\label{chapter4_fig:6}
\end{figure}

We will explore the performance of the GSCA model for the simulated binary and quantitative data sets with varying noise levels in the following experiment. Equal SNR levels are used in the simulation for $\mathbf{X}_1$ and $\mathbf{X}_2$. 20 log spaced SNR values are equally selected from the interval $[0.1, 100]$. Then we simulated coupled binary data $\mathbf{X}_1$ and quantitative $\mathbf{X}_2$ using the different SNRs in the same way as described above. During this process, except for the parameters $c_1$ and $c_2$, which are used to adjust the SNRs, all other parameters used in the simulation were kept the same. The GSCA models with GDP penalty ($\gamma=1$), $L_{q}$ penalty ($q=0.1$), nuclear norm penalty, and the SCA model on the full information (defined above) are used in these simulation experiments. For these three models, the model selection process was done in the same way as described in above experiment. The models with the minimum $\text{RMSE}(\mathbf{\Theta})$ are selected. As shown in Fig.~\ref{chapter4_fig:6}, the GSCA models with concave GDP and $L_{q}$ penalties always have better performance than the convex nuclear norm penalty, and they are comparable to the situation where the full information is available. With the increase of SNR, the $\text{RMSE}(\mathbf{Z})$ derived from the GSCA model, which is used to evaluate the performance of the model in recovering the underlying low dimensional structure, first decreases to a minimum and then increases. As shown in bottom center and right, this pattern is mainly caused by how $\text{RMSE}(\mathbf{Z}_1)$ changes with respect to SNRs. Although this result counteracts the intuition that larger SNR means higher quality of data, it is in line with our previous results in Chapter \ref{chapter:3}.

\subsubsection{Assessing the model selection procedure}
\begin{figure}[htbp]
    \centering
    \includegraphics[width=0.9\textwidth]{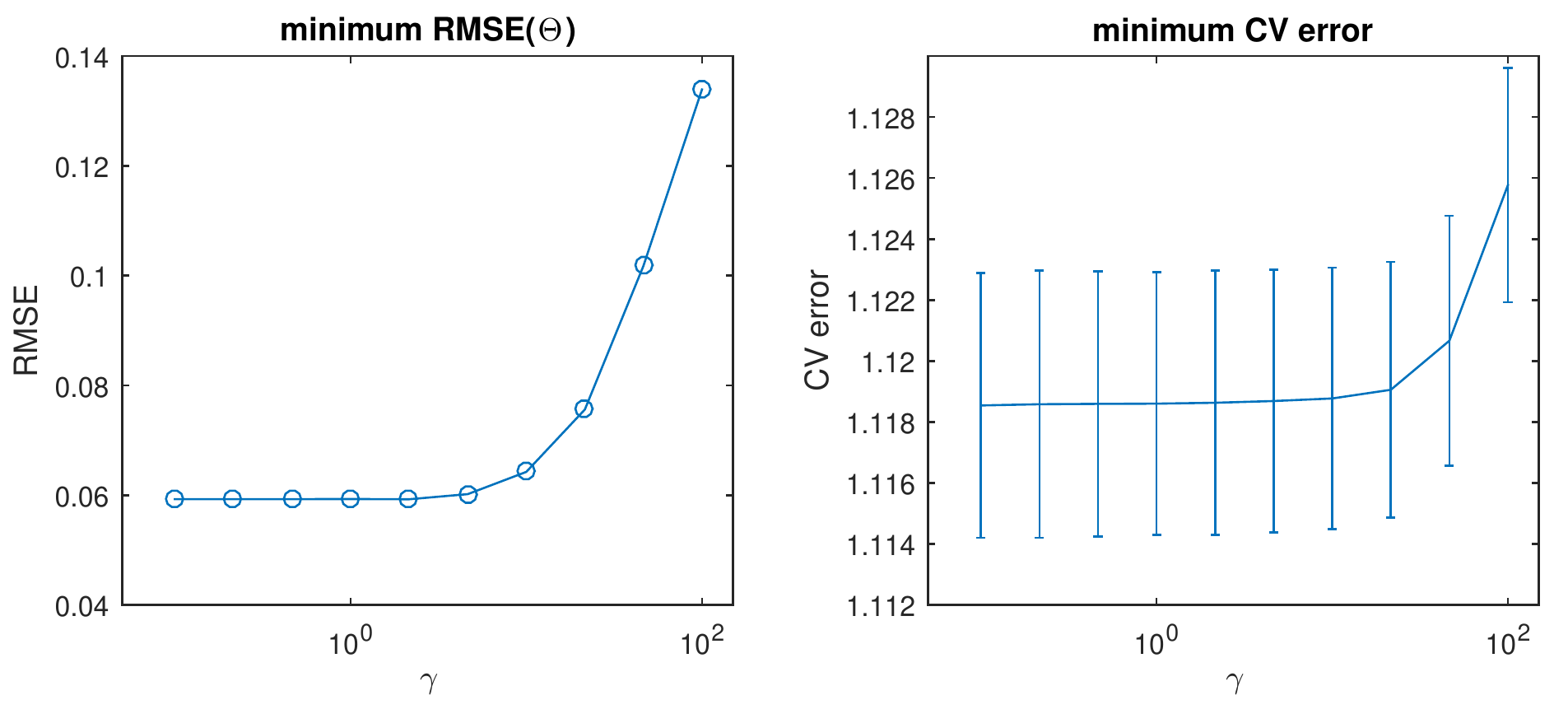}
    \caption{Minimum $\text{RMSE}(\mathbf{\Theta})$ (left) and minimum CV error (right) for different values of $\gamma$ from the GSCA model with GDP penalty. One standard error bars are added to the CV error plot.}
\label{chapter4_fig:7}
\end{figure}

The cross validation procedure and the cross validation error have been defined in the model selection section. The GSCA model with GDP penalty is used as an example to assess the model selection procedure. $\epsilon_f=10^{-5}$ is used as the stopping criteria for all the following experiments to save time. The values of $\lambda$ and $\gamma$ are selected in the same way as was described in Section \ref{section:4.4}. Fig.~\ref{chapter4_fig:7} shows the minimum $\text{RMSE}(\mathbf{\Theta})$ and minimum CV error achieved for different values of the hyper-parameter $\gamma$. The minimum CV error changes in a similar way as the minimum $\text{RMSE}(\mathbf{\Theta})$ with respect to the values of $\gamma$. However, taking into account the uncertainty of estimated CV errors, the difference of the minimum CV errors for different $\gamma$ is very small. Thus, we recommend to fix $\gamma$ to be 1, rather than using cross validation to select it. Furthermore, setting $\gamma = 1$ as the default value for the GDP penalty has a probabilistic interpretation, see in \cite{armagan2013generalized}.

\begin{figure}[htbp]
    \centering
    \includegraphics[width=0.9\textwidth]{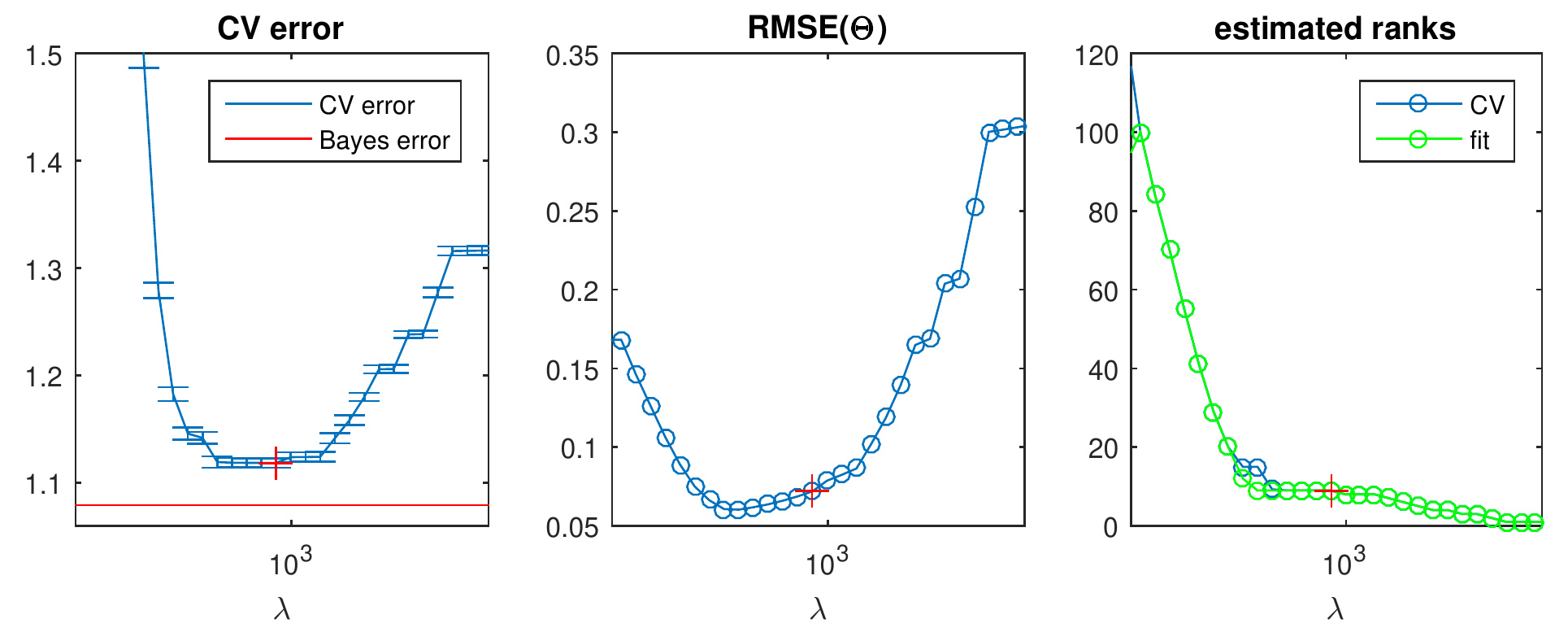}
    \caption{CV error, RMSE and estimated rank for different values of the tuning parameter $\lambda$. One standard error bars are added to the CV error plot. ``Bayes error'' indicates the mean log negative likelihood using simulated $\mathbf{\Theta}$ and $\sigma^2$ to fit the simulated data sets $\mathbf{X}_1$ and $\mathbf{X}_2$. The red cross marker indicates the point where the minimum CV error is achieved. ``CV'' and ``fit'' (right plot) indicate the mean $\text{rank}(\hat{\mathbf{Z}})$ derived from the models constructed in 7-fold cross validation and the $\text{rank}(\hat{\mathbf{Z}})$ derived from a model constructed on full data set without missing elements (the outputs of the constructed model during cross validation are set as the initialization).}
    \label{chapter4_fig:8}
\end{figure}

Whenever the GSCA model is used for exploratory data analysis, there is no need to select $\lambda$ explicitly. It is sufficient to find a proper value to achieve a two or three component GSCA model, in order to visualize the estimated score and loading matrices. If the goal is confirmatory data analysis, it is possible to select the tuning parameter $\lambda$ explicitly by the proposed cross validation procedure. Fig.~\ref{chapter4_fig:8} shows how the tuning parameter $\lambda$ affects the CV errors, $\text{RMSE}(\mathbf{\Theta})$ and the estimated ranks. The minimum CV error obtained is close to the Bayes error, which is the scaled negative log likelihood in cases where the true parameters $\mathbf{\Theta}$ and $\sigma^2$ are known. Even though, inconsistence exists between CV error plot (Fig.~\ref{chapter4_fig:8}, left) and the $\text{RMSE}(\mathbf{\Theta})$ plot (Fig.~\ref{chapter4_fig:8}, center), the selected model corresponding to minimum CV error can achieve very low $\text{RMSE}(\mathbf{\Theta})$ and correct rank estimation (Fig.~\ref{chapter4_fig:8}, right). Therefore, we suggest to use the proposed CV procedure to select the value of $\lambda$ at which the minimum CV error is obtained. Finally, we fit a model on full data set without missing elements using the selected value of $\lambda$ and the outputs of the selected model with minimum CV error as the initialization.

\section{Empirical illustration}  \label{section:4.5}
\subsection{Real data set}
The Genomic Determinants of Sensitivity in Cancer 1000 (GDSC1000) \cite{iorio2016landscape} contains 926 tumor cell lines with comprehensive measurements of point mutation, CNA, methylation and gene expression. We selected the binary CNA and quantitative gene expression measurements on the same cell lines (each cell line is a sample) as an example to demonstrate the GSCA model. To simplify the interpretation of the derived model, only the cell lines of three cancer types are included: BRCA (breast invasive carcinoma, 48 cell lines), LUAD (lung adenocarcinoma, 62 cell lines) and SKCM (skin cutaneous melanoma, 50 cell lines). The CNA data set has 410 binary variables. Each variable is a copy number region, in which ``1'' indicates the presence and ``0'' indicates the absence of an aberration. Note that, the CNA data is very imbalanced: only $6.66\%$ the elements are ``1''. The empirical marginal probabilities of binary CNA variables are shown in supplemental Fig.~S4.1. The quantitative gene expression data set contains 17,420 variables, of which 1000 gene expression variables with the largest variance are selected. After that, the gene expression data is column centered and scaled by the standard deviation of the each variable to make it more consistent with the assumption of the GSCA model.

\subsection{Exploratory data analysis of the coupled CNA and gene expression data sets}
We applied the GSCA model (with GDP penalty and $\gamma$=1) to the GDSC data set of 160 tumor cell lines that have been profiled for both binary CNA ($160 \times 410$) and quantitative gene expression ($160 \times 1000$). The results of model selection (supplemental Fig.~S4.3) validate the existence of a low dimensional common structure between CNA and gene expression data sets. For exploratory purposes, we will construct a three component model instead.

\begin{figure}[htbp]
    \centering
    \includegraphics[width=0.9\textwidth]{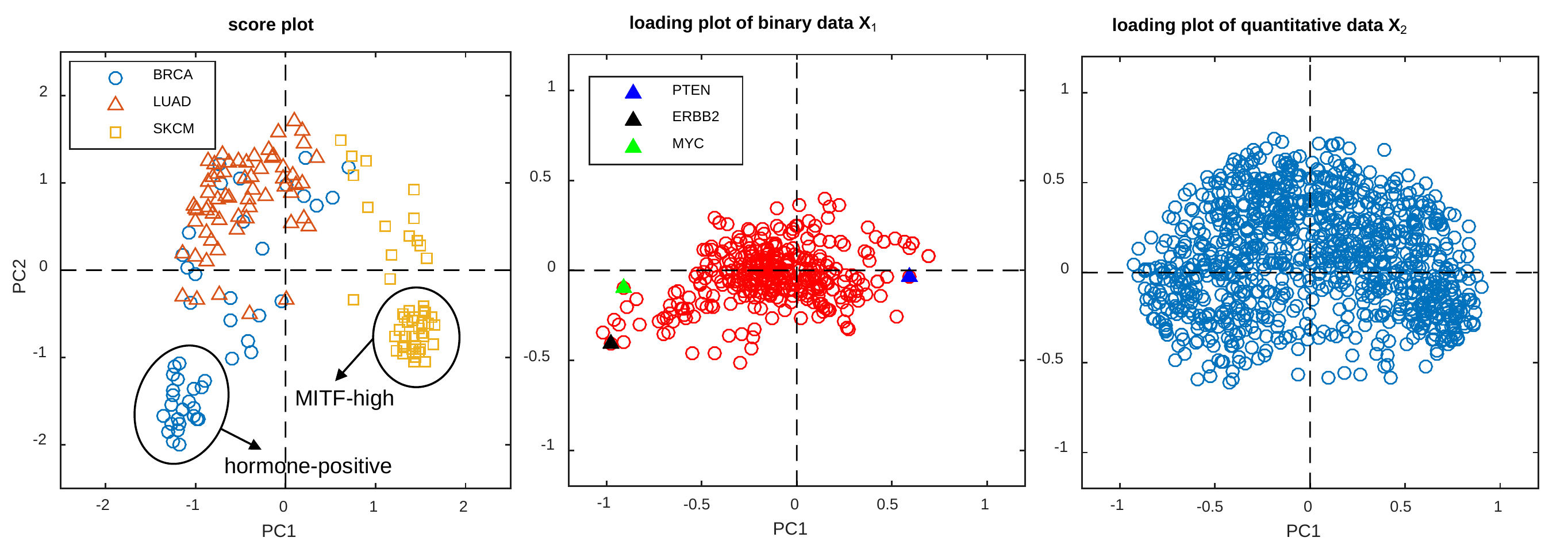}
    \caption{Score plot (left), loading plot for binary CNA data $\mathbf{X}_1$ (center) and loading plot for gene expression data $\mathbf{X}_2$ (right) derived from the constructed GSCA model.}
    \label{chapter4_fig:9}
\end{figure}

We first considered the score plot resulting from this GSCA model. The first two PCs show a clear clustering by cancer type (Fig.~\ref{chapter4_fig:9}, left), and in some cases even subclusters (i.e. hormone-positive breast cancer, MITF-high melanoma). These results suggest that the GSCA model captures the relevant biology in these data. Interestingly, when we performed PCA on the gene expression data, we obtained score plots that were virtually identical to those resulting from the GSCA model (supplemental Fig.~S4.4, left; modified RV coefficient of the scores matrices derived from these two models: 0.9998), suggesting that this biological relevance is almost entirely derived from the gene expression data.

We then wondered whether the GSCA model could leverage the gene expression data to help us gain insight into the CNA data. To test this, we first established how much insight could be gained from the CNA data in isolation. Supplemental Fig.~S4.5 shows the scores and loadings of the first two components from a three component logistic PCA model \cite{de2006principal} applied to the CNA data. While these do seem to contain structure in the loading plot, we believe that they mostly explain technical characteristics of the data. For example, deletions and amplifications are almost perfectly separated from each other by the PC1=0 line in the loading plot (supplemental Fig.~S4.6). Additionally, the scores on PC1 are strongly associated to the number of copy number aberrations (i.e., to the number of ones) in a given sample (supplemental Fig.~S4.7). Finally, the clusters towards the left of the loading plot suggested two groups of correlated features, but these could trivially be explained by genomic position, that is, these features correspond with regions on the same chromosome arm, which are often completely deleted or amplified (supplemental Fig.~S4.8). Following these observations, we believe that a study of the CNA data in isolation provides little biological insight.

On the other hand, using the GCSA model’s CNA loadings (Fig.~\ref{chapter4_fig:9}, center), we could more easily relate the features to the biology. Let us focus on the features with extreme values on PC1 and for which the corresponding chromosomal region contains a known driver gene. For example, the position of MYC amplifications in the loading plot indicates that MYC amplifications occur mostly in lung adenocarcinoma and breast cancer samples (Fig.~\ref{chapter4_fig:9}, center; supplemental Fig.~S4.9). Similarly, ERBB2 amplifications occur mainly in breast cancer samples (Fig.~\ref{chapter4_fig:9}, center; supplemental Fig.~S4.9). Finally, PTEN deletions were enriched in melanomas, though the limited size of the loading also indicates that they are not exclusive to melanomas (Fig.~\ref{chapter4_fig:9}, center; supplemental Fig.~S4.9). Importantly, these three findings are in line with known biology \cite{akbani2015genomic,cancer2014comprehensive,cancer2012comprehensive} and hence exemplify how GSCA could be used to interpret the CNA data. Altogether, using the GSCA model, we were able to 1) capture the biological relevance in the gene expression data, and 2) leverage that biological relevance from the gene expression to gain a better understanding of the CNA data.

\section{Discussion}  \label{section:4.6}
In this chapter, we generalized the standard SCA model to explore the dependence between coupled binary and quantitative data sets. However, the GSCA model with exact low rank constraint overfits the data, as some estimated parameters tend to diverge to positive infinity or negative infinity. Therefore, concave penalties are introduced in the low rank approximation framework to achieve low rank approximation and to mitigate the overfitting issues of the GSCA model. An efficient algorithm framework with analytical form updates for all the parameters is developed to optimize the GSCA model with any concave penalties. All concave penalties used in our experiments have better performance with respect to generalization error and estimated low rank of the constructed GSCA model compared to the nuclear norm penalty. Both $L_{q}$ and GDP penalties with proper model selection can recover the simulated low rank structures almost exactly only from indirect binary observation $\mathbf{X}_1$ and noisy quantitative observation $\mathbf{X}_2$. Furthermore, we have shown that the GSCA model outperforms the iClusterPlus model with respect to speed and accuracy of the estimation of the model parameters.

The superior performance of the GSCA models with the concave penalties compared to the models with an exact low rank constraint or a nuclear norm penalty is related to their different thresholding properties. The exact low rank constraint thresholds the singular values in a hard manner and, therefore, only the largest $R$ singular values are kept. On the other hand, the nuclear norm penalty works in a soft manner, in which all the singular values are shrunk by the same amount of $\lambda$. The thresholding properties of the concave penalties discussed in this chapter lie in between these two approaches. As $\mathbf{Z}=\mathbf{A}\mathbf{B}^{\text{T}}$ and $\mathbf{A}^{\text{T}}\mathbf{A}= \mathbf{I}_R$, the scale of the loadings is related to the scale of the singular values of $\mathbf{Z}$. Thus, we can shrink the singular values of $\mathbf{Z}$ to control the scale of estimated loading matrices in an indirect way. The exact low rank constraint kept the $R$ largest singular values but without control of the scale of the estimated singular values, leading to overfitting. On the other hand, nuclear norm penalty shrinks all the singular values by the same amount of $\lambda$, leading to biased estimation of the singular values. A concave penalty, like $L_q$ or GDP, achieves a balance in thresholding the singular values. Among the concave penalties we used in the experiment, the SCAD penalty does not work well in the simulation study. The reason is that the SCAD penalty does not shrink the large singular values, which therefore tend to be overfitted, while the smaller singular values are shrunk too much.

Compared to the iClusterPlus method, only the option of binary and quantitative data sets are included in our GSCA model, and at the moment no sparsity can be imposed for the integrative analysis of binary and quantitative data sets. However, the GSCA model with GDP penalty is optimized by a more efficient algorithm, it is much more robust to the imbalanced nature of the biological binary data and it provides a much better performance for the simulation experiments in this chapter. Furthermore, the exploratory analysis of the GDSC coupled CNA and gene expression data sets provided important information on the binary CNA data that was not obtained by a separate analysis.

\section*{Acknowledgements}
Y.S. gratefully acknowledges the financial support from China Scholarship Council (NO.201504910809).

\clearpage
\section{Supplementary information}
\subsection{GSCA model with exact low rank constraint}
The exact low rank constraint on $\mathbf{Z}$ can be expressed as the multiplication of two low rank matrices $\mathbf{A}$ and $\mathbf{B}$. The optimization problem related to the GSCA model with exact low rank constraint can be expressed as
\begin{equation*}
\begin{aligned}
    \min_{\bm{\mu},\mathbf{Z},\sigma^2} \quad & f_1(\mathbf{\Theta}_1) + f_2(\mathbf{\Theta}_2,\sigma^2) \\
    \text{subject to} \quad \mathbf{\Theta} &= \mathbf{1}\bm{\mu}^{\text{T}} + \mathbf{Z} \\
     \mathbf{\Theta} &= [\mathbf{\Theta}_1 ~ \mathbf{\Theta}_2] \\
    \text{rank}(\mathbf{Z}) &= R
\end{aligned}
\end{equation*}

The developed algorithm in the paper can be slightly modified to fit this model. Similar to the paper, the above optimization problem can majorized to the following problem.
\begin{equation*}
\begin{aligned}
          \min_{\bm{\mu},\mathbf{Z},\sigma^2} \quad & \frac{L}{2} ||\mathbf{\Theta} - \mathbf{H}^k||_F^2+ c\\
     \text{subject to} \quad   \mathbf{\Theta} &= \mathbf{1}\bm{\mu}^{\text{T}} + \mathbf{Z} \\
         \mathbf{H}^k &= \mathbf{\Theta}^k - \frac{1}{L} (\mathbf{W} \odot \nabla f(\mathbf{\Theta}^k))\\
            \mathbf{1}^{\text{T}}\mathbf{Z} &= 0 \\
             \text{rank}(\mathbf{Z}) &= R.
\end{aligned}
\end{equation*}

The analytical solution of $\bm{\mu}$ is also the column mean of $\mathbf{H}^k$. After deflating out the offset term $\bm{\mu}$, the majorized problem becomes $\min_{\mathbf{Z}} \frac{L}{2} ||\mathbf{Z} - \mathbf{J} \mathbf{H}^k||_F^2$ subject to $\text{rank}(\mathbf{Z}) = R$ and $\mathbf{1}^{\text{T}}\mathbf{Z} = 0$. The global optimal solution is the $R$ truncated SVD of $\mathbf{J} \mathbf{H}^k$. Other steps in the algorithm to fit the GSCA model with exact low rank constraint are exactly the same as in the paper to fit the GSCA model with concave penalties.

\subsection{Supplemental figures}
\begin{figure}[htbp]
    \centering
    \includegraphics[ width= 0.5\textwidth]{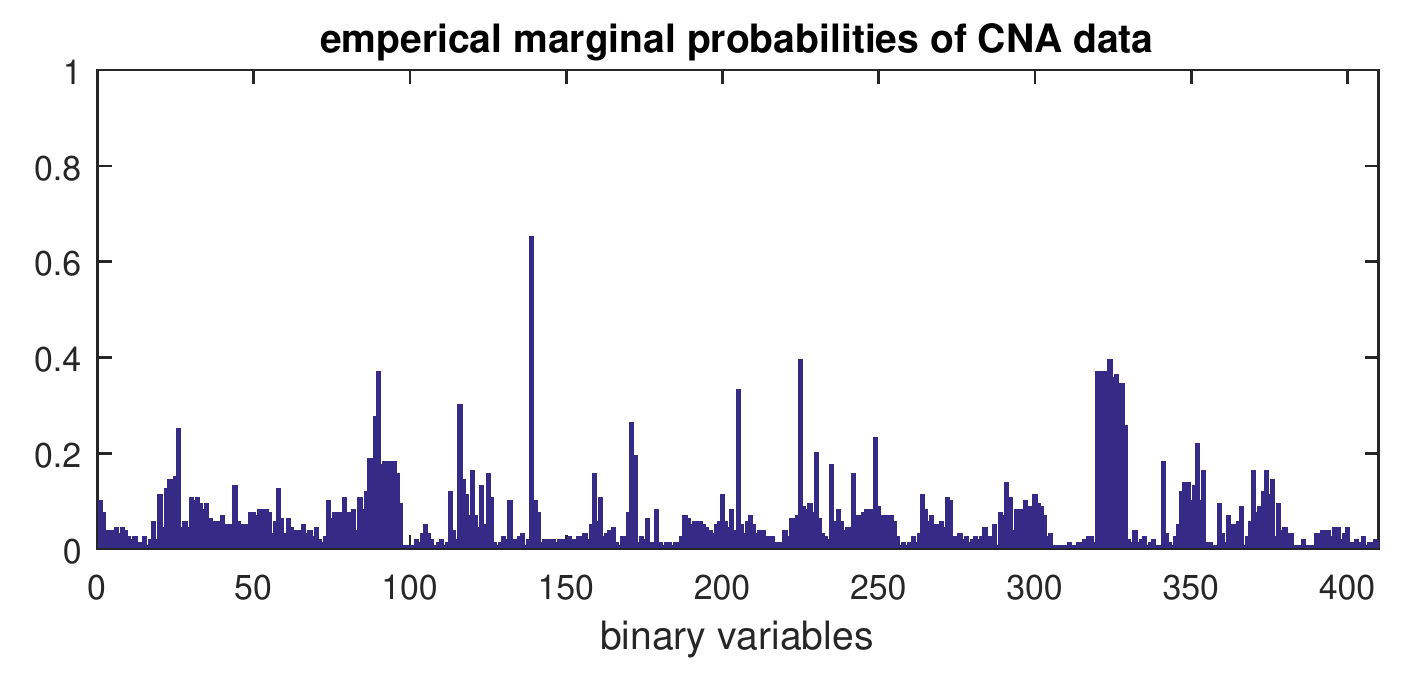}
    \caption*{Figure S4.1: Empirical marginal probabilities of the binary CNA data set.}
	\label{chapter4_fig:S1}
\end{figure}

\begin{figure}[htbp]
    \centering
    \includegraphics[width=0.9\textwidth]{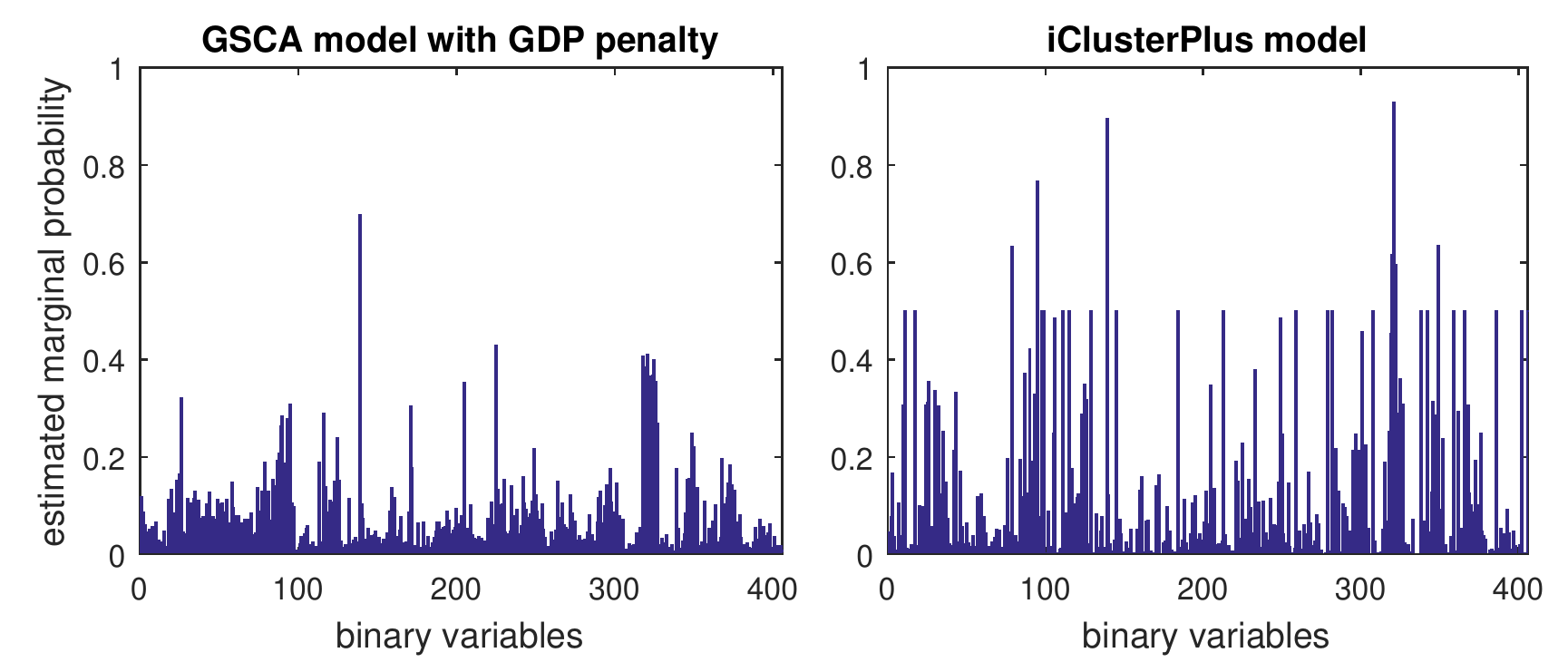}
    \caption*{Figure S4.2: Estimated marginal probabilities (the logit transform of the estimated $\hat{\bm{\mu}}_1$) from the GSCA model with GDP penalty (left) and iClusterPlus model (right).}
	\label{chapter4_fig:S2}
\end{figure}

\begin{figure}[htbp]\label{Fig:S6}
    \centering
    \includegraphics[width=0.9\textwidth]{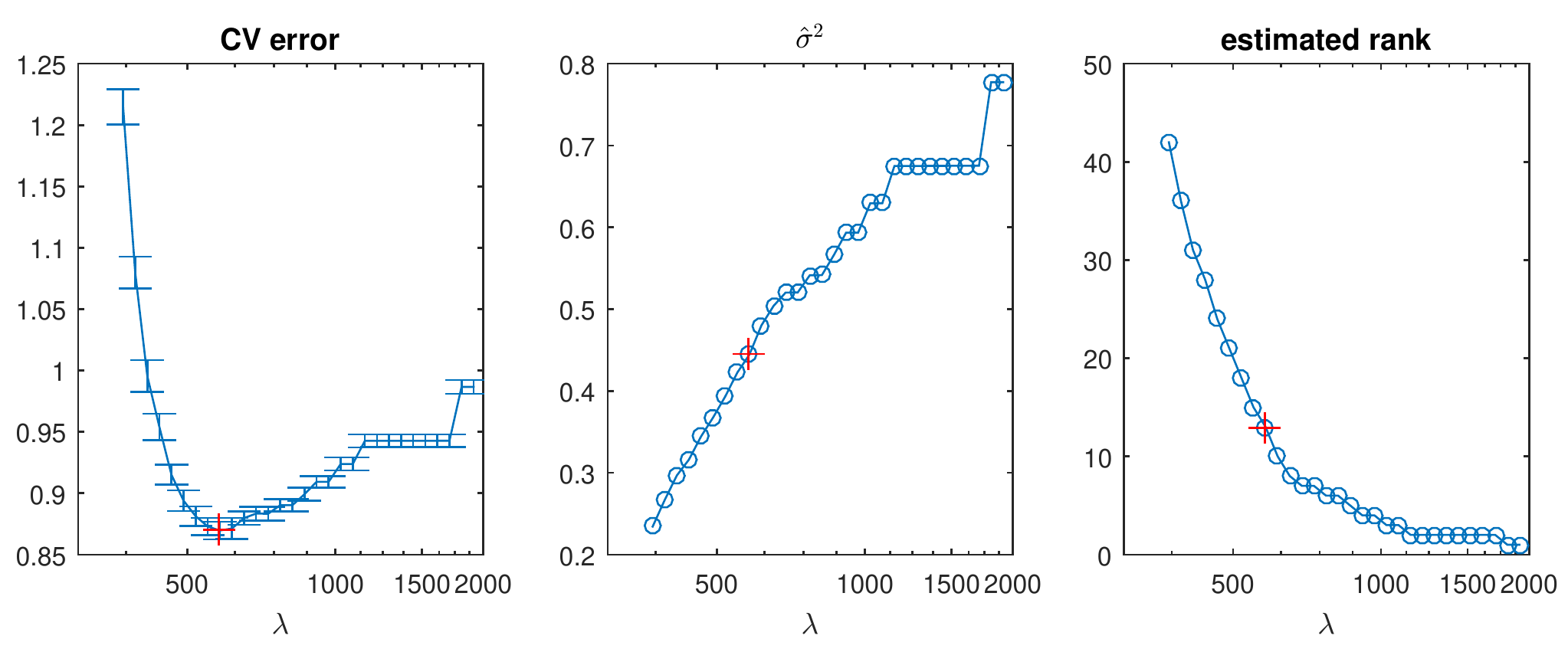}
    \caption*{Figure S4.3: Model selection of the GSCA model with GDP penalty on the GDSC data sets: CV error (left); estimated $\hat{\sigma}^2$ (center); $\text{rank}(\hat{\mathbf{Z}})$ (right).}
	\label{chapter4_fig:S3}
\end{figure}

\begin{figure}[htbp]
    \centering
    \includegraphics[width= 0.9\textwidth]{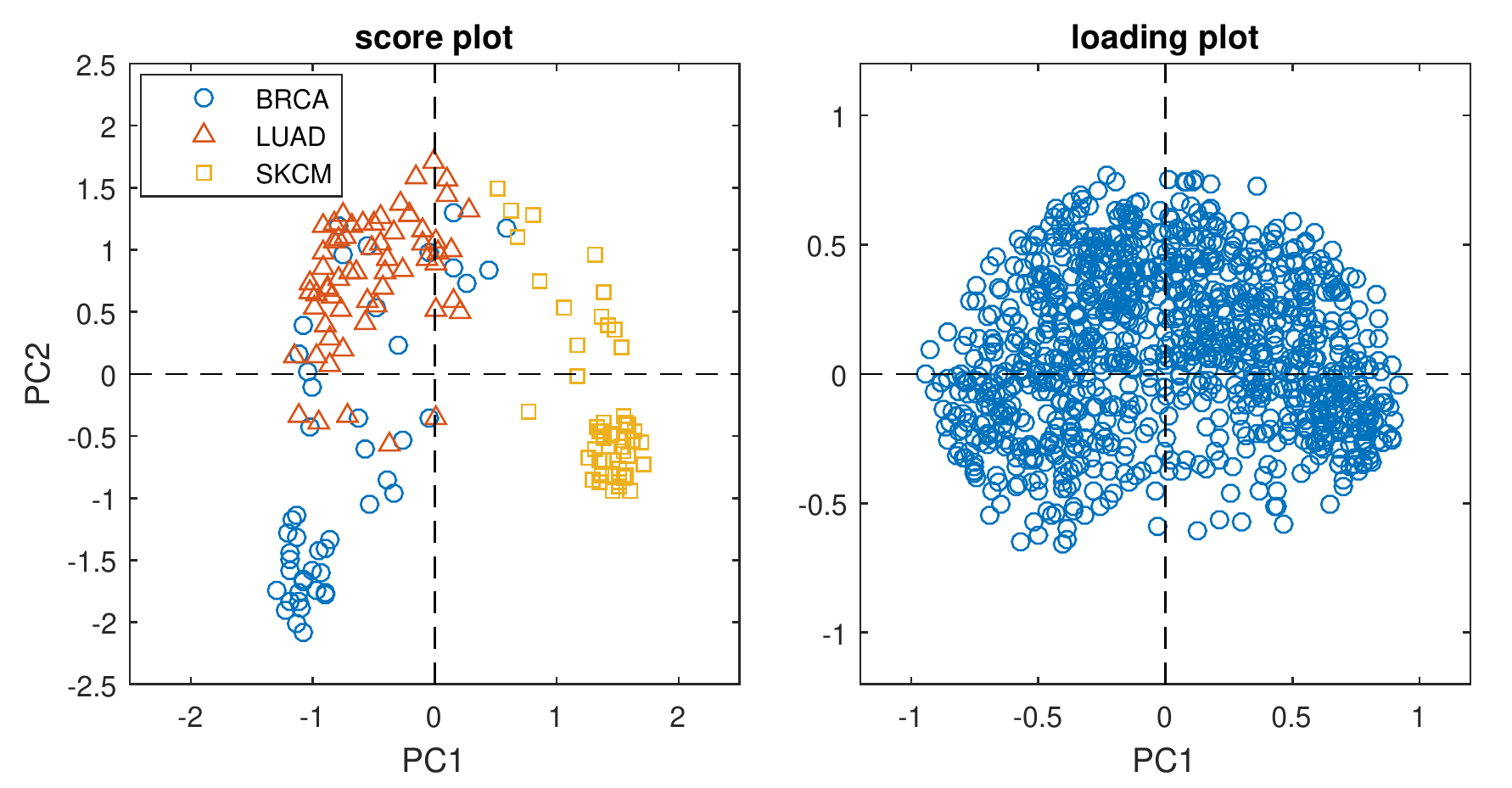}
    \caption*{Figure S4.4: Score plot (left) and loading plot (right) derived from a PCA model on the gene expression data $\mathbf{X}_2$. $\mathbf{X}_2$ are centered and scaled in the same as in the GSCA model. SVD algorithm is used to solve the PCA model.}
	\label{chapter4_fig:S4}
\end{figure}

\begin{figure}[htbp]
    \centering
    \includegraphics[width= 0.9\textwidth]{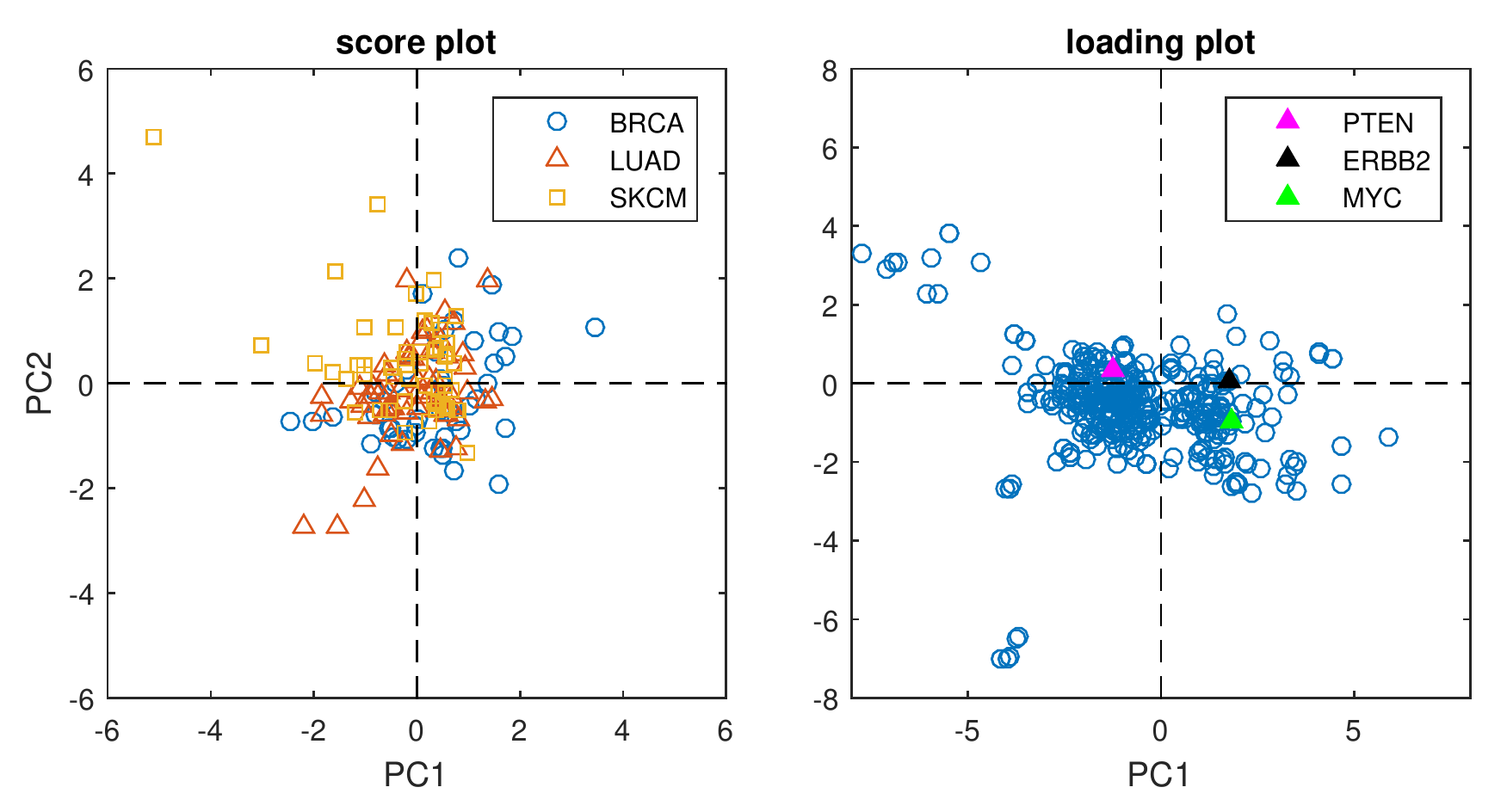}
    \caption*{Figure S4.5: Score plot (left) and loading plot (right) are derived from a three components logistic PCA model on the CNA data $\mathbf{X}_1$.}
	\label{chapter4_fig:S5}
\end{figure}

\begin{figure}[htbp]
    \centering
    \includegraphics[width=0.5 \textwidth]{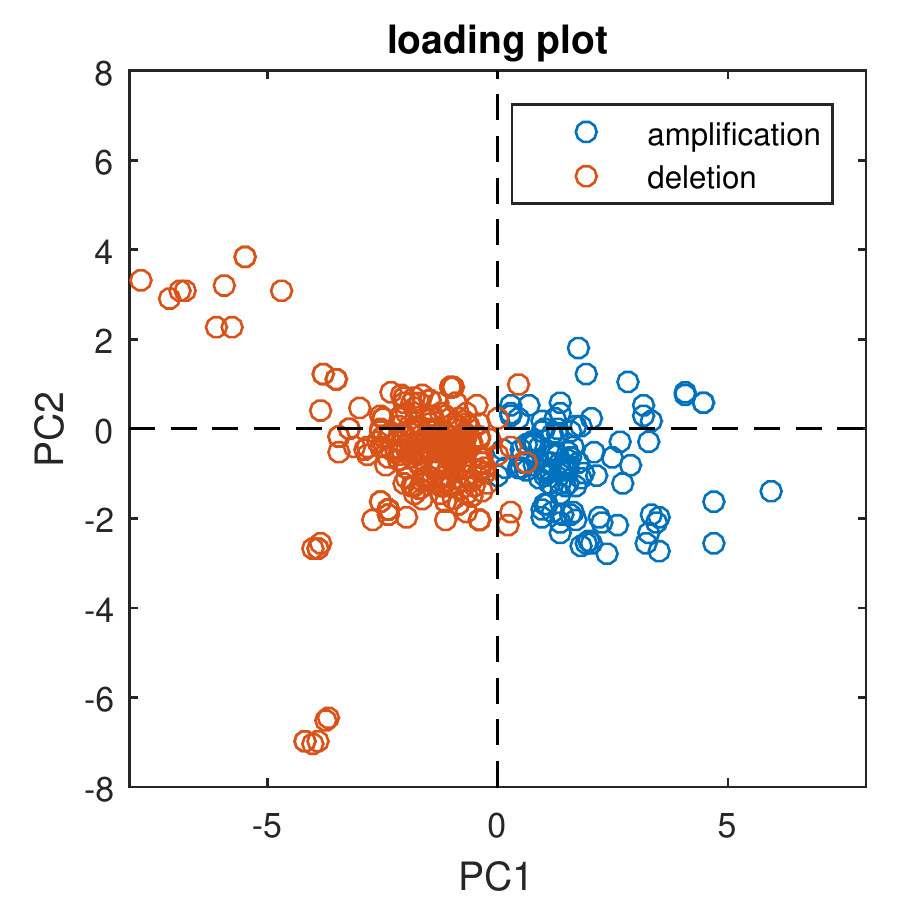}
    \caption*{Figure S4.6: Loading plot derived from the three components logistic PCA model on the CNA data $\mathbf{X}_1$. The legend indictates the amplification or deletion of CNA feature.}
	\label{chapter4_fig:S6}
\end{figure}

\begin{figure}[htbp]
    \centering
    \includegraphics[width=0.5 \textwidth]{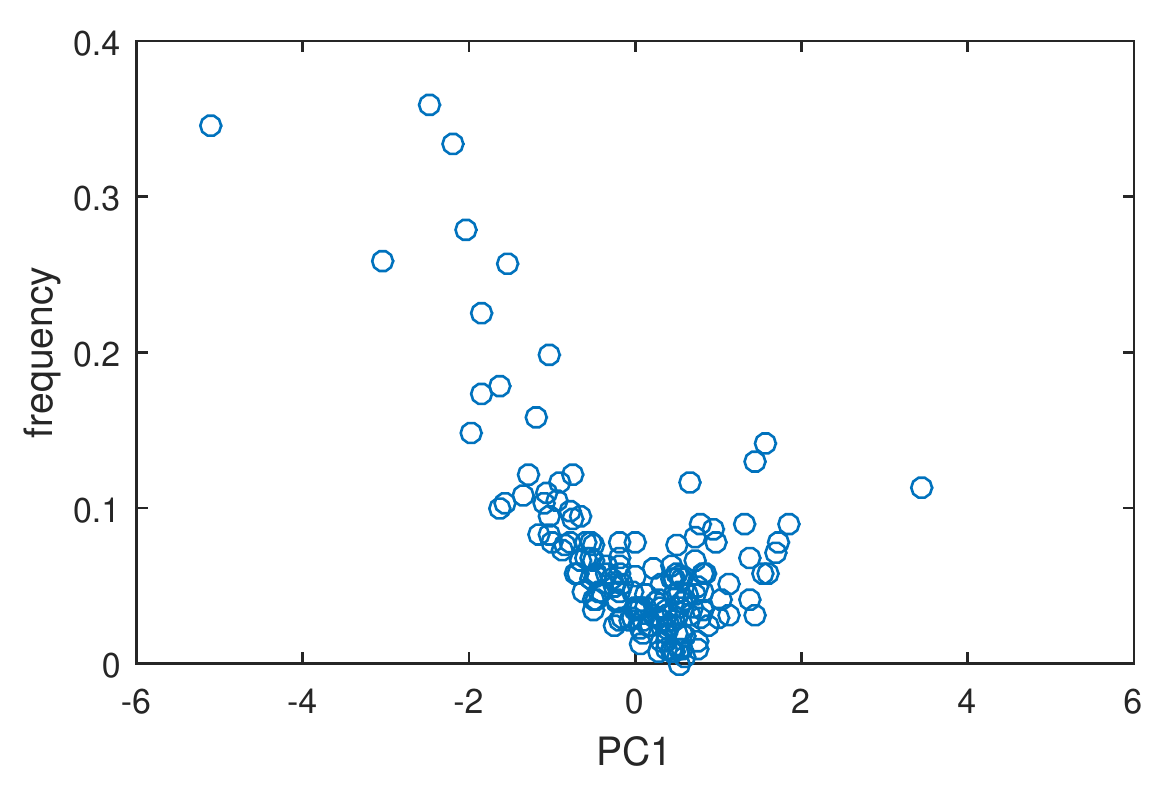}
    \caption*{Figure S4.7: The relationship between PC1 scores and the frequency of aberrations of given samples derived from the three components logistic PCA model on the CNA data $\mathbf{X}_1$. }
	\label{chapter4_fig:S7}
\end{figure}

\begin{figure}[htbp]
    \centering
    \includegraphics[width= 0.5 \textwidth]{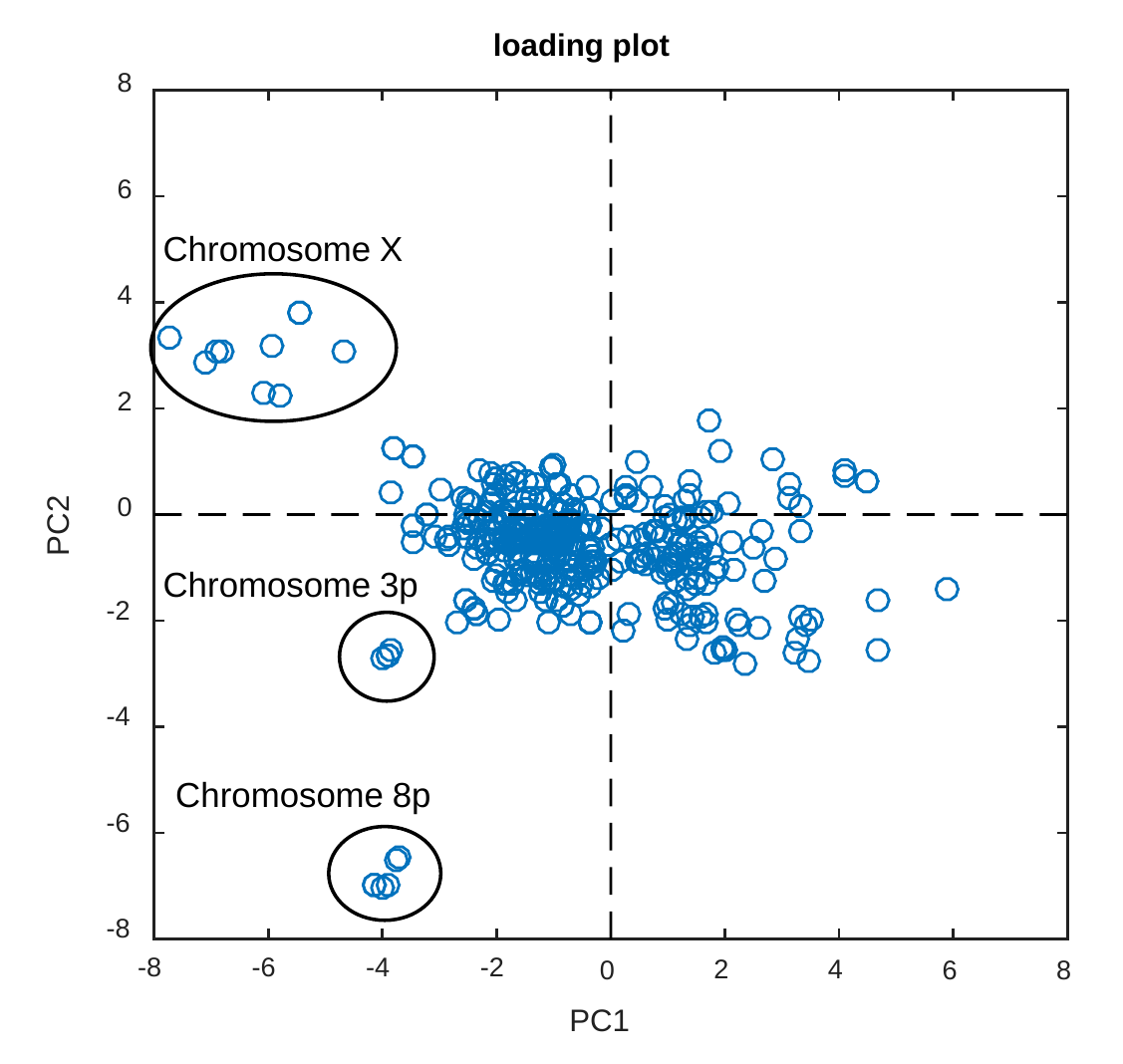}
    \caption*{Figure S4.8: Loading plot derived from the three components logistic PCA model on the CNA data $\mathbf{X}_1$. The annotation indicates those features are in the same chromosome region.}
	\label{chapter4_fig:S8}
\end{figure}

\begin{figure}[htbp]
    \centering
    \includegraphics[width= 0.9\textwidth]{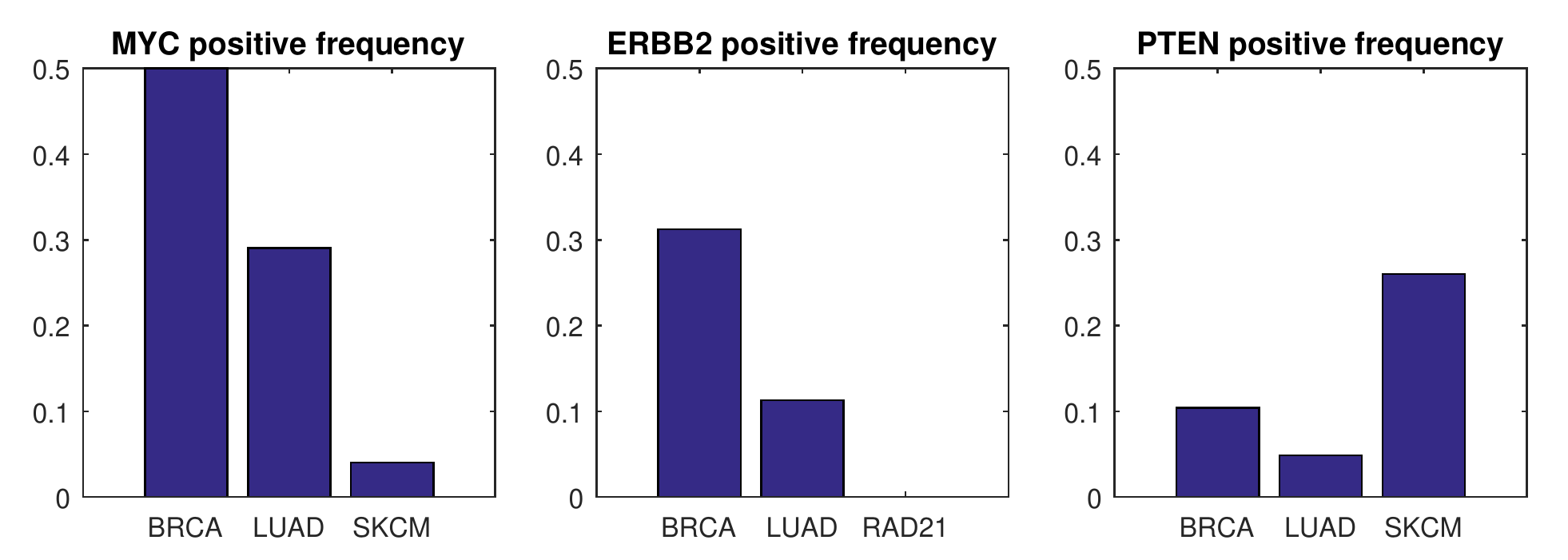}
    \caption*{Figure S4.9: Positive frequencies of MYC, ERBB2 and PTEN features in three different cancer types.}
	\label{chapter4_fig:S9}
\end{figure}

\chapter{Fusing multiple data sets with two types of heterogeneity} \label{chapter:5}
Multiple sets of measurements on the same objects obtained from different
platforms may reflect partially complementary information of the studied system. The
integrative analysis of such data sets not only provides us with the opportunity of
a deeper understanding of the studied system, but also introduces some new
statistical challenges. First, the separation of information that is common
across all or some of the data sets, and the information that is specific to
each data set is problematic. Furthermore, these data sets are often a mix of
quantitative and qualitative (binary or categorical) data types, while commonly used data
fusion methods require all data sets to be quantitative. These two types of heterogeneity existed in multiple data sets should be taken into account in the data fusion. In this chapter, we propose an exponential family simultaneous component analysis (ESCA) model
to tackle the potential mixed data types problem of multiple data sets. In
addition, a structured sparse pattern of the loading matrix is induced through a
nearly unbiased group concave penalty to disentangle the global, local
common and distinct information of the multiple data sets. A Majorization-Minimization based algorithm is derived to fit the proposed model. Analytic solutions are derived for updating all the parameters of the model in each iteration, and the algorithm will decrease the objective function in each iteration monotonically. For model selection, a missing value based cross validation procedure is implemented. The advantages of the proposed method in comparison with other approaches are assessed using comprehensive simulations as well as the analysis of real data from a chronic lymphocytic leukaemia (CLL) study.
\footnote{This chapter is based on Song, Y., Westerhuis, J.A. and Smilde, A.K., 2019. Separating common (global and local) and distinct variation in multiple mixed types data sets. arXiv preprint arXiv:1902.06241}

\section{Background} \label{section:5.1}
Multiple data sets measured on the same samples are becoming increasingly common in different research areas, from biology, food science to psychology. One typical example from biological research is the GDSC1000 study, in which 926 cell lines are fully characterized with respect to point mutation, copy number alternation (CNA), methylation, gene expression and drug responses \cite{iorio2016landscape}. However, these comprehensive measurements from the same cell lines not only provide the opportunity for a deeper understanding of the studied biological system, but also introduce statistical challenges.

The first challenge is how to separate the information that is common across all or some of the data sets, and the information which is specific to each data set (often called distinct). These different sources of information have to be disentangled from every data set to have a holistic understanding of the studied system. The second challenge is that measurements from different platforms can be of different data type, such as binary, quantitative or counts. These different data types have different mathematical properties, which should be taken into account in the data analysis. For example, the measurement of a binary variable only has two possible exclusive outcomes, often classified as ``1'' and ``0''. Examples of binary data in biology include point mutation, and the binarized CNA and methylation data sets \cite{iorio2016landscape}. Taking binary measurements ``1'', ``0'' as the quantitative values 1, 0, and casting them into the classical data fusion methods that assume data sets to be quantitative, clearly neglects their binary nature.

In this chapter, we focus on the component or latent variable based data fusion approaches, although other approaches exist such as undirected graphical model based methods which are able to explore the association between data sets of different data types \cite{aben2018itop}, or between variables of different data types \cite{lee2015learning,cheng2017high}. Two commonly used latent variable based data fusion methods are simultaneous component analysis (SCA) \cite{van2009structured} and iCluster \cite{shen2009integrative}, which both focus on using low dimensional structures to approximate the common variation across all the data sets. Both of these approaches have already been generalized to qualitative data sets \cite{mo2013pattern,song2018generalized}. In addition, the concept of common and distinct variation in data fusion has been framed in \cite{van2016separating,smilde2017common}, and several methods \cite{lock2013joint,lofstedt2013global,schouteden2014performing,maage2012preference,smilde2017common} have been proposed. One typical example is JIVE \cite{lock2013joint}. The JIVE model directly specifies the components for the global common variation (variation across all the data sets) and the distinct variation (variation specific to each data set) in the model, and estimates them simultaneously. However, JIVE model is incapable to tackle the qualitative data sets and the local common variation (variation across some of the data sets) is ignored. Direct generalization of JIVE to account for the local common variation is infeasible as with the increase of the number of data sets, the possible combinations of local common variation blows up exponentially. Other methods \cite{schouteden2014performing,lofstedt2013global} encounter similar problems with respect to the estimation of local common variation. In addition, the model selection procedure in these methods is still an unsolved issue \cite{maage2018performance}. A promising solution was provided in \cite{klami2015group,gaynanova2017structural}, in which a group regularization procedure was applied to provide structured sparsity on the loading matrix where the loadings of all variables of a given data set are forced to 0 to disentangle the common (global and local) and distinct variation indirectly. Details will be shown in the following model section. In the SLIDE model \cite{gaynanova2017structural}, first a series of structured sparsity patterns on the loading matrix of a SCA model are learned using a group lasso penalty. Then, these learned structured sparsity patterns are imposed as hard constraints on the loading matrix of a SCA model, and the appropriate model is selected by Bi-cross-validation \cite{perry2009cross}. The Bayesian counterpart of the SLIDE model is the group factor analysis model \cite{klami2015group}, and the generalization of the group factor analysis to mixed data types is the MOFA model \cite{argelaguet2018multi}. These two Bayesian models use an automatic relevance determination procedure to induce the structured sparsity.

The first contribution in this chapter is the generalization of the SCA model to the exponential family SCA (ESCA) model by exploiting the exponential family distribution to account for potentially different data types, such as binary, quantitative or count data. The generalization is done in a similar way as the extension of principal component analysis (PCA) to exponential family PCA \cite{collins2002generalization}. The second contribution is the use of a nearly unbiased group concave penalty to induce a structured sparse pattern on the loading matrix of the ESCA model to disentangle the common (global and local) and distinct variation of multiple data sets of mixed data types. In the SLIDE model \cite{gaynanova2017structural}, the structured sparse pattern is induced by the group lasso penalty, which shrinks in the group level (the groups) as a lasso penalty, and in the individual level (the individual elements inside a group) as a ridge regression penalty. However, a lasso type penalty leads to biased parameter estimation, as the same degree of shrinkage is applied to all the parameters. This will shrink the nonzero parameters too much and as a result makes the prediction or cross validation error based model selection procedures inconsistent \cite{meinshausen2006high,leng2006note}. On the other hand, concave penalties, such as generalized double Pareto (GDP) shrinkage \cite{armagan2013generalized} or bridge ($L_{q: 0<q \leq 1}$) penalty \cite{fu1998penalized}, are capable to achieve nearly unbiased estimation of the parameters while producing sparse solutions. Therefore, we replaced the group lasso penalty by a group concave penalty on the loadings. The group concave penalty shrinks the group level as a concave penalty, and it shrinks on the individual level as ridge regression penalty. The third contribution lies in the derived model fitting algorithm and the model selection procedure. A Majorization-Minimization based algorithm is derived to fit the proposed penalized ESCA (P-ESCA) model. Analytical form solutions for updating all the parameters of the model in each iteration are derived, and the algorithm will decrease the objective function in each iteration monotonically. Furthermore, the missing value problem is tackled in the developed algorithm, and this option is used in the cross validation procedure for the model selection. The proposed model is similar to the MOFA model, but differences exist in the way how the model is derived, how the structured sparsity is achieved, and how the model is selected. These differences are detailed out in the supplementary material.

Both the performance of the proposed P-ESCA model, and the effectiveness of the model selection procedure are validated by extensive simulations under different situations. The performance of the P-ESCA method is compared with SLIDE and MOFAs. Finally, P-ESCA is exemplified by the explorative analysis of the chronic lymphocytic leukaemia (CLL) data sets \cite{dietrich2018drug,argelaguet2018multi}.

\section{P-ESCA model} \label{section:5.2}
In this section, we will introduce the generalization of the ESCA model. Then we will show how to use the group concave penalty to induce the structured sparse pattern on the loading matrix of an ESCA model to disentangle the common (global and local) variation and distinct variation of multiple data sets.

\subsection{Exponential family SCA}
The quantitative measurements from $L$ different platforms on the same $I$ objects result into $L$ quantitative data sets, $\left\{\mathbf{X}_l \right\}_{l=1}^{L}$, and the $l^{\text{th}}$ data set $\mathbf{X}_l$($I \times J_l$) has $J_l$ variables. In the classical SCA model, we decompose the $L$ data sets as $\mathbf{X}_l = \mathbf{1}\bm{\mu}_l^{\text{T}} + \mathbf{AB}_l^{\text{T}} + \mathbf{E}_l$, in which $\mathbf{1}$($I\times 1$) is a column vector with ones; $\bm{\mu}_l$($J_l \times 1$) is the column offset term; $\mathbf{A}$($I\times R$) is the common score matrix; $\mathbf{B}_l$($J_l\times R$) and $\mathbf{E}_l$($I\times J_l$) are the loading matrix and residual term respectively for $\mathbf{X}_l$ and $R$ is the number of components. In addition, constraints $\mathbf{A}^{\text{T}}\mathbf{A} = \mathbf{I}$ and $\mathbf{1}^{\text{T}}\mathbf{A} = \mathbf{0}$, in which $\mathbf{I}$ is the identity matrix, are imposed to make the model identifiable. The SCA model tries to discover the common column subspace, which is spanned by the columns of the score matrix $\mathbf{A}$, in $L$ data sets to represent their common information. The column offset terms $\left\{ \bm{\mu}_l \right\}_{l=1}^{L}$ can be removed by column centering of the corresponding data sets $\left\{\mathbf{X}_l\right\}_{l=1}^{L}$. The parameters in the SCA model can be estimated by minimizing the sum of squares $\sum_{l}^{L} w_l ||\mathbf{X}_l - \mathbf{1}\bm{\mu}_l^{\text{T}} - \mathbf{AB}_l^{\text{T}}||_F^2$, in which $w_l$ is the relative weight of the $l^{\text{th}}$ data set $\mathbf{X}_l$.

The least squares loss criterion in the classical SCA model is only appropriate for quantitative data sets. When some or all data sets are of another data type, such as binary data, classical SCA model is not appropriate anymore. Motivated by the previous research on exponential family PCA model \cite{schein2003generalized}, we use the exponential family distribution to account for the different data types of multiple data sets, such as Bernoulli distribution for binary data, Poisson distribution for count data and Gaussian distribution for quantitative data.

Assume $x \in \mathbf{R}$ follows the exponential dispersion family distribution \cite{agresti2013categorical}, and $\theta$ and $\alpha$ are the natural parameter and the dispersion parameter respectively. The probability density or mass function can be specified as $p(x|\theta,\alpha) = \exp \left[(x\theta - b(\theta))/\alpha \right] h(x,\alpha)$, in which $b(\theta)$ is the log-partition function, and $h(x,\alpha)$ is a function which does not depend on the natural parameter $\theta$. Supplemental Table S5.1 lists the log-partition function $b(\theta)$ and its first and second order derivatives $b^{'}(\theta)$, $b^{''}(\theta)$ for Gaussian, Bernoulli and Poisson distributions. The relationship $\text{E}(x|\theta) = b^{'}(\theta)$ always hold in the exponential family distribution. Supplemental Fig.~S5.1 visualizes this relationship for the Gaussian, Bernoulli and Poisson distributions. If the $l^{\text{th}}$ data set $\mathbf{X}_l$ is quantitative, according to the probabilistic interpretation of the PCA model \cite{tipping1999probabilistic}, we assume there is a natural parameter matrix $\mathbf{\Theta}_l$($I\times J_l$) underlying $\mathbf{X}_l$, and the low dimensional structure exists in $\mathbf{\Theta}_l$, $\mathbf{X}_l = \mathbf{\Theta}_l + \mathbf{E}_l$ and $\mathbf{\Theta}_l = \mathbf{1}\bm{\mu}_l^{\text{T}} + \mathbf{A}_l\mathbf{B}_l^{\text{T}}$, and elements in the error term $\mathbf{E}_l$ follows a normal distribution $\epsilon_{ij}^{l} \sim \text{N}(0,\sigma^2)$. The conditional mean of the observed $\mathbf{X}_l$ given the low dimensional structure assumption is $\text{E}(\mathbf{X}_l| \mathbf{\Theta}_l) = b^{'}(\mathbf{\Theta}) = \mathbf{\Theta}_l$, in which $b^{'}()$ is the first order derivative of the log-partition function for the Gaussian distribution (supplemental Table S5.1). In exponential family PCA, the same idea has been generalized to other members of exponential family distributions by assuming $\text{E}(\mathbf{X}_l| \mathbf{\Theta}_l) = b^{'}(\mathbf{\Theta}_l)$ and $\mathbf{\Theta}_l = \mathbf{1}\bm{\mu}_l^{\text{T}} + \mathbf{A}_l\mathbf{B}_l^{\text{T}}$, in which the function form of $b^{'}()$ depends on the used probability distribution (supplemental Table S5.1).

In the exponential family PCA model, the elements in $\mathbf{X}_l$ are conditionally independent given the low dimensional structure assumption as $\mathbf{\Theta}_l = \mathbf{1}\bm{\mu}_l^{\text{T}} + \mathbf{A}_l\mathbf{B}_l^{\text{T}}$. Take $x_{ij}^{l}$ and $\theta_{ij}^{l}$ as the $ij^{\text{th}}$ element of $\mathbf{X}_l$ and $\mathbf{\Theta}_l$ respectively. The conditional log-likelihood of observing $\mathbf{X}_l$ is $\log(p(\mathbf{X}_l|\mathbf{\Theta}_l,\alpha_l)) =  \sum_{i}^{I}\sum_{j}^{J_l} \log(p(x_{ij}^{l}|\theta_{ij}^{l},\alpha_l)) = \sum_{i}^{I}\sum_{j}^{J_l} \frac{1}{\alpha_l}  (x_{ij}^{l}\theta_{ij}^{l} - b_l(\theta_{ij}^{l})) + c = \frac{1}{\alpha_l} \left[<\mathbf{X}_{l},\mathbf{\Theta}_l> - <\mathbf{1}\mathbf{1}^{\text{T}}, b_l(\mathbf{\Theta}_l)>\right] + c$, in which $<,>$ indicates the inner product, for matrices, $<\mathbf{X}_l,\mathbf{\Theta}_l> = \text{trace}(\mathbf{X}_l^{\text{T}}\mathbf{\Theta}_l)$; $c$, a constant that does not depend on the unknown parameter $\mathbf{\Theta}_l$; $b_l()$ and $\alpha_l$ are the element-wise log-partition function and the known dispersion parameter respectively for the $l^{\text{th}}$ data set $\mathbf{X}_l$. In the ESCA model, we assume that the natural parameter matrices $\left\{ \mathbf{\Theta}_l \right\}_{l=1}^{L}$ lie in the same column subspace, which is spanned by the common score matrix $\mathbf{A}$. To make the model identifiable, constraints $\mathbf{A}^{\text{T}}\mathbf{A} = \mathbf{I}$ and $\mathbf{1}^{\text{T}}\mathbf{A} = \mathbf{0}$ are imposed. The optimization problem associated with this ESCA model can be expressed as follows,
\begin{equation}\label{chapter5_eq:1}
\begin{aligned}
    \min_{ \left\{\bm{\mu}_l\right\}_{l}^{L}, \mathbf{A}, \left\{\mathbf{B}_l\right\}_{l}^{L}} \quad & \sum_{l=1}^{L} -\log(p(\mathbf{X}_l|\mathbf{\Theta}_l, \alpha_l))\\
	& =\sum_{l=1}^{L} \frac{1}{\alpha_l} \left[ <\mathbf{1}\mathbf{1}^{\text{T}}, b_l(\mathbf{\Theta}_l)> - <\mathbf{X}_l, \mathbf{\Theta}_l> \right] + c\\
    \text{subject to} \quad \mathbf{\Theta}_l &= \mathbf{1}\bm{\mu}_l^{\text{T}} + \mathbf{AB}_l^{\text{T}}, \quad l = 1,\ldots,L \\
     \mathbf{1}^{\text{T}}\mathbf{A} &= \mathbf{0}\\
	 \mathbf{A}^{\text{T}}\mathbf{A} &= \mathbf{I}.
\end{aligned}
\end{equation}

\subsection{Separating common and distinct variation via structured sparsity}
The drawback of the SCA or ESCA models is that only the global common components, which account for the common variation across all the data sets, is allowed. However, the real situation in multiple data sets integration can be far more complex as local common variation across some of the data sets and distinct variation in each data set are expected as well. Directly specifying the components in the ESCA model for common (global and local) and distinct variation in the same way as JIVE model \cite{lock2013joint} is infeasible, as the number of possible combinations of local common variation will blow up exponentially with an increasing number of data sets. A promising solution is using structured sparsity on the loading matrix to disentangle the common (global and local) and distinct variation indirectly \cite{klami2015group,gaynanova2017structural}. Structured sparsity of the data set specific loading matrices in component based data fusion methods has been explored by \cite{van2011flexible,acar2015data}. The idea of using structured sparsity to disentangle the common (global and local) and distinct variation in multiple quantitative data sets is made explicit in \cite{klami2015group,gaynanova2017structural}. To illustrate the idea, we use an example with three quantitative data sets. Suppose we construct a SCA model on three column centered quantitative data sets $\left\{\mathbf{X}_l \right\}_{l=1}^{3}$, the common score matrix is $\mathbf{A}$, the corresponding loading matrices are $\left\{\mathbf{B}_l \right\}_{l=1}^{3}$, and $\mathbf{X}_l = \mathbf{A}\mathbf{B}_{l}^{\text{T}} + \mathbf{E}_l$, in which $\mathbf{E}_l$ is the residual term for $l^{\text{th}}$ data set. If the structured sparsity pattern in $\left\{\mathbf{B}_l \right\}_{l=1}^{3}$ is expressed as follows,
\begin{equation*}
\begin{aligned}
   \left(
                 \begin{array}{c}
                   \mathbf{B}_1 \\
                   \mathbf{B}_2 \\
                   \mathbf{B}_3 \\
                 \end{array}
               \right)
               = \left(
                   \begin{array}{cccccccc}
                     \mathbf{b}_{1,1} & \mathbf{b}_{1,2} & \mathbf{b}_{1,3} & \mathbf{0}       & \mathbf{b}_{1,5} & \mathbf{0}       & \mathbf{0}      \\
                     \mathbf{b}_{2,1} & \mathbf{b}_{2,2} & \mathbf{0}       & \mathbf{b}_{2,4} & \mathbf{0}       & \mathbf{b}_{2,6} & \mathbf{0}       \\
                     \mathbf{b}_{3,1} & \mathbf{0}       & \mathbf{b}_{3,3} & \mathbf{b}_{3,4} & \mathbf{0}       & \mathbf{0}       & \mathbf{b}_{3,7} \\
                   \end{array}
                 \right),
\end{aligned}
\end{equation*}
in which $\mathbf{b}_{l,r} \in \mathbf{R}^{J_l}$ indicates the $r^{\text{th}}$ column of the $l^{\text{th}}$ loading matrix $\mathbf{B}_l$, then we have the following relationships,
\begin{equation*}
\begin{aligned}
   \mathbf{X}_1 & = \mathbf{a}_1\mathbf{b}_{1,1}^{\text{T}} &+& \mathbf{a}_2\mathbf{b}_{1,2}^{\text{T}} &+& \mathbf{a}_3\mathbf{b}_{1,3}^{\text{T}} &+& \mathbf{0}                     &+& \mathbf{a}_5\mathbf{b}_{1,5}^{\text{T}} &+& \mathbf{0}                     &+& \mathbf{0}  &+& \mathbf{E}_1                   \\
   \mathbf{X}_2 & = \mathbf{a}_1\mathbf{b}_{2,1}^{\text{T}} &+& \mathbf{a}_2\mathbf{b}_{2,2}^{\text{T}} &+& \mathbf{0}                     &+& \mathbf{a}_4\mathbf{b}_{2,4}^{\text{T}} &+& \mathbf{0}                     &+& \mathbf{a}_6\mathbf{b}_{2,6}^{\text{T}} &+& \mathbf{0}  &+& \mathbf{E}_2                   \\
   \mathbf{X}_3 & = \mathbf{a}_1\mathbf{b}_{3,1}^{\text{T}} &+& \mathbf{0}                     &+& \mathbf{a}_3\mathbf{b}_{3,3}^{\text{T}} &+& \mathbf{a}_4\mathbf{b}_{3,4}^{\text{T}} &+& \mathbf{0}                     &+& \mathbf{0}                     &+& \mathbf{a}_7\mathbf{b}_{3,7}^{\text{T}} &+& \mathbf{E}_3 .
\end{aligned}
\end{equation*}
Here $\mathbf{a}_r$ indicates the $r^{\text{th}}$ column of the common score matrix $\mathbf{A}$. The first component represents the global common variation across three data sets; the $2^{\text{nd}}$, $3^{\text{nd}}$ and $4^{\text{nd}}$ components represent the local common variation across two data sets and the $5^{\text{nd}}$, $6^{\text{nd}}$ and $7^{\text{nd}}$ components represent the distinct variation specific to each single data set. In this way, the structured sparsity pattern in the loading matrices $\left\{\mathbf{B}_l \right\}_{l=1}^{3}$ can be used to separate the common (global and local) and distinct variation of multiple quantitative data sets.

\subsection{Group concave penalty}
In \cite{van2011flexible,acar2015data,gaynanova2017structural}, the structured sparsity is induced by a group lasso penalty on the columns of $\{ \mathbf{B}_l \}_{1}^{L}$. The used group lasso penalty is $\lambda \sum_{l}\sum_{r} ||\mathbf{b}_{l,r}||_{2}$, in which $\lambda$ is the tuning parameter, $\mathbf{b}_{l,r}$ indicates the $r^{\text{th}}$ column of the $l^{\text{th}}$ loading matrix $\mathbf{B}_l$, and $||\quad||_{2}$ indicates the $L_2$ norm of a vector. This group lasso penalty shrinks $||\mathbf{b}_{l,r}||_{2}$ as a lasso penalty and the elements inside $\mathbf{b}_{l,r}$ as a ridge penalty. However, lasso type penalty leads to biased parameter estimation as the same degree of shrinkage is applied to all the parameters, which will shrink the nonzero parameters too much and makes the prediction or cross validation error based model selection procedures inconsistent \cite{meinshausen2006high,leng2006note}. This leads in general to the selection of too complex models. The SLIDE model \cite{gaynanova2017structural} solves the model selection problem in a two stages manner. First, varying degrees of regularization are imposed to induce a series of structured sparse loading patterns. Then these structured sparse patterns are taken as hard constraints on a new SCA model, in which a Bi-cross validation procedure \cite{perry2009cross} is used for the final selection. This two stages approach is similar to the often used re-estimation trick in lasso regression. However, such a two-step strategy cannot easily be generalized to the ESCA model. For example, if a binary data set is used and the structured sparse pattern is imposed as a hard constraint on the loading matrices in a ESCA model, the estimated loadings of the binary data set can easily go to infinity \cite{groenen2016multinomial,song2018generalized}.

\begin{figure}[htbp]
    \centering
    \includegraphics[width=0.9\textwidth]{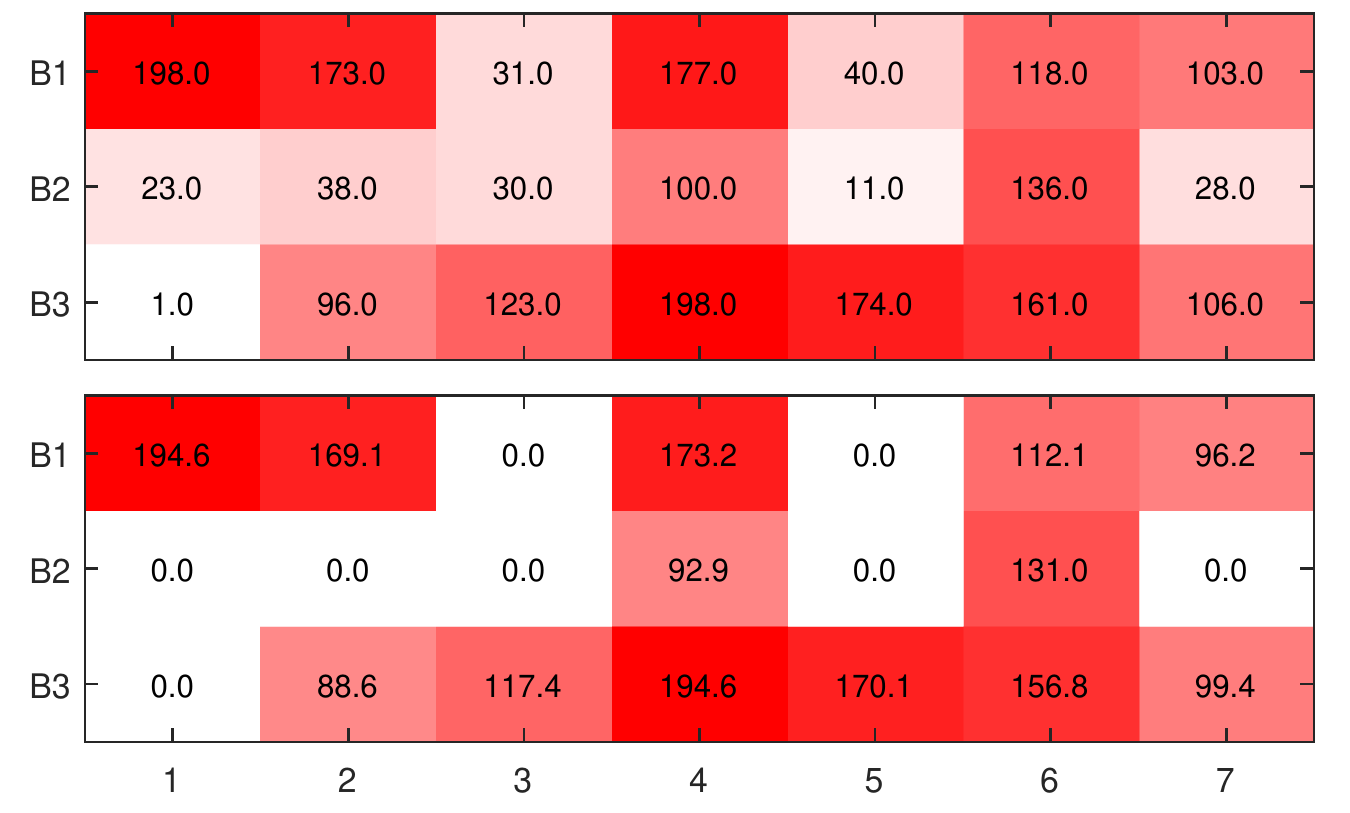}
    \caption{How the group GDP ($\gamma = 1$) penalty induces structured sparse pattern on $\left\{\mathbf{B} \right\}_{l=1}^3$. Values inside the plot indicate the $\text{L}_2$ norm of the corresponding loading vector $\mathbf{b}_{l,r}$. Top: loading matrix before thresholding; bottom: loading matrix after thresholding.}
	\label{chapter5_fig:1}
\end{figure}

The above issue introduced by the biased estimation of lasso type penalties can be alleviated by using concave penalties \cite{fu1998penalized,armagan2013generalized}, which can achieve sparse solutions and nearly unbiased parameter estimation simultaneously. Therefore, in this chapter, we applied group concave penalties, generalized double Pareto (GDP) shrinkage \cite{armagan2013generalized} and bridge ($L_{q: 0<q \leq 1}$) penalty \cite{fu1998penalized} are included as special cases, on the loading matrices of the ESCA model to induce structured sparse pattern. Take $\sigma_{lr} = ||\mathbf{b}_{l,r}||_2$, in which $\mathbf{b}_{l,r}$ is the $r^{\text{th}}$ column of $\mathbf{B}_l$, and $g()$ is a general concave penalty function in Table \ref{chapter5_tab:1}. The penalty on $\mathbf{B}_l$ can be expressed as $\lambda_l \sum_{r} g(\sigma_{lr})$, in which $\lambda_l$ is the tuning parameter. The group lasso penalty is a special case of the group $L_{q}$ (bridge) penalty by setting $q=1$. The thresholding properties of the group $L_{\text{q}}$ penalty, group GDP penalty and group lasso can be found in supplemental Fig.~S5.2. In order to account for the situation that the data sets have an unequal number of variables, we add the weights in the same way as in the standard group lasso regression problem, i.e. $\lambda_l \sqrt{J_l} \sum_{r} g(\sigma_{lr})$. The group concave penalty on $\{ \mathbf{B}_l \}_{1}^{L}$ can be expressed as $\sum_{l} \Big[\lambda_l \sqrt{J_l} \sum_{r} g(\sigma_{lr}) \Big]$. Based on successful results in previous work \cite{song2018generalized} we will focus on the GDP penalty, which is differentiable everywhere in its domain and its performance is insensitive to the selection of the hyper-parameter $\gamma$. Fig.~\ref{chapter5_fig:1} gives an example to show how the group GDP ($\gamma = 1$) penalty induces structured sparsity pattern on the loading matrices $\left\{\mathbf{B} \right\}_{l=1}^3$.

\begin{table}[htbp]
\centering
\caption{Three commonly used group penalty functions. Take $\sigma$ as the $L_2$ norm of a group of elements. $q$ and $\gamma$ are the tuning parameters. The supergradient is the counter concept of the subgradient for a concave function. When the concave function is differentiable everywhere, the supergradient is the gradient.}
\label{chapter5_tab:1}
\begin{tabular}{lll}
  \toprule
penalty & formula & supergradient \\
  \midrule
 group lasso & $ \sigma $ & $1$ \\

group $L_{q: 0<q \leq 1}$ & $ \sigma^q $ & $\left\{ \begin{array}{ll} +\infty &\textrm{$\sigma=0$}\\
                                 q \sigma^{q-1} &\textrm{$\sigma>0$}\\ \end{array} \right.$ \\
group GDP & $ \log(1+\frac{\sigma}{\gamma}) $ & $\frac{1}{\gamma + \sigma}$ \\
  \bottomrule
\end{tabular}
\end{table}

\subsection{Identifiability}
The constraint $\mathbf{1}^{\text{T}}\mathbf{A} = \mathbf{0}$ makes the column offset terms $\left\{ \bm{\mu} \right\}_{l=1}^{L}$ identifiable. The columns of the score matrix $\mathbf{A}$ span the joint subspace $\bigcup_{l=1}^L \text{col}(\mathbf{\Theta}_l)$, in which $\text{col}()$ indicates the column subspace. The structured sparse pattern on the loading matrices and the multiplication of the score and loading matrices provide a way to separate the joint subspace $\bigcup_{l=1}^L \text{col}(\mathbf{\Theta}_l)$ into subspaces $ \text{col(GC)}$, $\text{col(LC)}$, $\text{col(D)}$ corresponding to the global common, local common and distinct variation, $ \text{col(GC)} \bigcup \text{col(LC)} \bigcup \text{col(D)} = \bigcup_{l=1}^L \text{col}(\mathbf{\Theta}_l)$. If the orthogonality constraint $\mathbf{A}^{\text{T}}\mathbf{A} = \mathbf{I}$ is imposed, the separated subspaces $ \text{col(GC)}$, $\text{col(LC)}$, $\text{col(D)}$, corresponding to the global common, local common and distinct variation, are orthogonal to each other, and unique as $\text{col(GC)} \bigcap \text{col(LC)} \bigcap \text{col(D)} = \emptyset$. However, there is still a rotation freedom for the components within the subspace corresponding to the global common or local common or distinct variation.

\subsection{Regularized likelihood criterion}
The regularized likelihood criterion of fitting the proposed P-ESCA model can be derived as follows. To tackle the missing value problem, $L$ weight matrices are introduced. For the $l^{\text{th}}$ data set $\mathbf{X}_l$, we introduce a same size weight matrix $\mathbf{W}_l$, in which $w_{ij}^{l} = 0$ if the corresponding element in $\mathbf{X}_l$ is missing, while $w_{ij}^{l} = 1$  \textit{vise versa}. This option is the basis for different missing value based cross validation approaches. The corresponding optimization problem can be expressed as follows,
\begin{equation}\label{chapter5_eq:2}
\begin{aligned}
    \min_{ \left\{\bm{\mu}_l\right\}_{l}^{L}, \mathbf{A}, \left\{\mathbf{B}_l\right\}_{l}^{L}} \quad & \sum_{l=1}^{L} \Big[ -\log(p(\mathbf{X}_l|\mathbf{\Theta}_l, \alpha_l)) + \lambda_l \sqrt{J_l} \sum_{r} g(\sigma_{lr}) \Big] \\
	&= \sum_{l=1}^{L} \Big[ \frac{1}{\alpha_l}( <\mathbf{W}_l, b_l(\mathbf{\Theta}_l)> - <\mathbf{W}_l \odot \mathbf{X}_l,\mathbf{\Theta}_l>) + \lambda_l \sqrt{J_l} \sum_{r} g(\sigma_{lr}) \Big] + c\\
    \text{subject to} \quad \mathbf{\Theta}_l &= \mathbf{1}\bm{\mu}_l^{\text{T}} + \mathbf{AB}_l^{\text{T}}, \quad l = 1,\ldots,L \\
     \mathbf{1}^{\text{T}}\mathbf{A} &= \mathbf{0}\\
	 \mathbf{A}^{\text{T}}\mathbf{A} &= \mathbf{I} \\
	 \sigma_{lr} &= ||\mathbf{b}_{l,r}||_2, l = 1...L; r = 1,\ldots, R,
\end{aligned}
\end{equation}
in which $\odot$ indicates the element-wise matrix multiplication.

\section{Algorithm} \label{section:5.3}
The original problem in equation \ref{chapter5_eq:2} is difficult to optimize directly because of the non-convex orthogonality constraint $\mathbf{A}^{\text{T}}\mathbf{A} = \mathbf{I}$ and the group concave penalty $g()$. However, by using the Majorization-Minimization (MM) principle, the original difficult problem can be majorized to a simpler problem, for which analytical form solutions can be derived for all the parameters. According to the MM principle, the derived algorithm will monotonously decrease the loss function in each iteration. Further details of the MM principle can be found in \cite{de1994block,hunter2004tutorial}.

\subsection{The majorization of the regularized likelihood criterion}
Take $f_l(\mathbf{\Theta}_l) = \frac{1}{\alpha_l} \left[<\mathbf{W}_l, b_l(\mathbf{\Theta}_l)> - <\mathbf{W}_l \odot \mathbf{X}_l, \mathbf{\Theta}_l>\right]$ as the loss function for fitting the $l^{\text{th}}$ data set $\mathbf{X}_l$, and $g_l(\mathbf{B}_l) = \sum_{r} g(\sigma_{lr})$ as the group concave penalty for the $l^{\text{th}}$ loading matrix $\mathbf{B}_l$. We can majorize $f_l(\mathbf{\Theta}_l)$ and $g_l(\mathbf{B}_l)$ respectively as follows.

\subsubsection*{The majorization of $f_l(\mathbf{\Theta}_l)$}
Given $\tilde{f_l}(\theta_{ij}^{l}) = b_l(\theta_{ij}^{l}) - x_{ij}^{l}\theta_{ij}^{l}$, we have $f_l(\mathbf{\Theta}_l) = \frac{1}{\alpha_l} \sum_{i}\sum_{j} w_{ij}^{l} \tilde{f_l}(\theta_{ij}^{l})$. The first and second gradients of $\tilde{f_l}(\theta_{ij}^{l})$ with respect to $\theta_{ij}^{l}$ are $\nabla \tilde{f_l}(\theta_{ij}^{l}) = b_l^{'}(\theta_{ij}^{l}) - x_{ij}^{l}$ and $\nabla^2 \tilde{f_l}(\theta_{ij}^{l}) = b_l^{''}(\theta_{ij}^{l})$. Assume that $\nabla^2 \tilde{f_{1}}(\theta_{ij}^{l})$ is upper bounded by a constant $\rho_{l}$, which will be detailed below. If $\theta^{l}$ represents the general representation of $\theta_{ij}^{l}$, then according to the Taylor's theorem and the assumption that $\nabla^2 \tilde{f_{l}}(\theta^{l}) \leq \rho_{l}$ for all $\theta^{l} \in \text{domain}(\tilde{f_l})$, we have the following inequality,
\begin{equation}\label{chapter5_eq:3}
\begin{split}
\tilde{f_l}(\theta^{l}) = &\tilde{f_l}((\theta^{l})^k) + <\nabla \tilde{f_l}((\theta^{l})^k), \theta^{l}-(\theta^{l})^k> + \\ &\frac{1}{2}(\theta^{l}-(\theta^{l})^k)^{\text{T}} \nabla^{2} \tilde{f_l}\left[(\theta^{l})^k + t(\theta^{l}-(\theta^{l})^k)\right](\theta^{l}-(\theta^{l})^k)\\
\leq& \tilde{f_l}((\theta^{l})^k) + <\nabla \tilde{f_l}((\theta^{l})^k), \theta^{l}-(\theta^{l})^k> + \frac{\rho_{l}}{2}(\theta^{l}-(\theta^{l})^k)^2 \\
=& \frac{\rho_{l}}{2}\left[\theta^{l}-(\theta^{l})^k + \frac{1}{\rho_l}\nabla \tilde{f_l}((\theta^{l})^k)\right]^2 + c.
\end{split}
\end{equation}
Here $(\theta^{l})^k$ is an approximation of $\theta^{l}$ at the $k^{\text{th}}$ iteration and $t\in[0,1]$ is an unknown constant. Combining the above inequality and the majorization step \cite{kiers1997weighted} of transforming a weighted least square problem to a least squares problem, we have the following inequality,
\begin{equation}\label{chapter5_eq:4}
\begin{aligned}
     f_l(\mathbf{\Theta}_l) &= \frac{1}{\alpha_l} \sum_{i}\sum_{j} w_{ij}^{l} \tilde{f_l}(\theta_{ij}^{l})\\
     &\leq \frac{\rho_l}{2\alpha_l} ||\mathbf{W}_l \odot (\mathbf{\Theta}_l - \mathbf{\Theta}_l^k + \frac{1}{\rho_l}(b_l^{'}(\mathbf{\Theta}_{l}^{k}) - \mathbf{X}_l)) ||_F^2 + c\\
     &\leq 	\frac{\rho_l}{2\alpha_l} ||\mathbf{\Theta}_l - \mathbf{H}_{l}^{k}||_F^2 + c \\
\mathbf{H}_{l}^{k} &= \mathbf{W}_l \odot(\mathbf{\Theta}_l^k - \frac{1}{\rho_l} (b_l^{'}(\mathbf{\Theta}_{l}^{k}) - \mathbf{X}_l)) + (\mathbf{1}\mathbf{1}^{\text{T}}-\mathbf{W}_l)\odot \mathbf{\Theta}_l^k \\
&= \mathbf{\Theta}_l^k - \frac{1}{\rho_l} \mathbf{W}_l \odot (b_l^{'}(\mathbf{\Theta}_{l}^{k}) - \mathbf{X}_l)), 	
\end{aligned}
\end{equation}
in which $\mathbf{\Theta}_l^k$ is the approximation of $\mathbf{\Theta}_l$ during the $k^{\text{th}}$ iteration. For the Bernoulli distribution, an elegant bound $b^{''}(\theta) \leq 0.25$ can be used \cite{de2006principal}; for the Gaussian likelihood, $b^{''}(\theta) = 1$; for the Poisson distribution, $b^{''}(\theta)$ is unbounded, however, we can always set $\rho_l = \text{max}(b^{''}(\mathbf{\Theta}_l^k))$.

\subsubsection*{The majorization of $g_l(\mathbf{B}_l)$}
Assume $\mathbf{B}_l^k$ is the $k^{\text{th}}$ approximation of $\mathbf{B}_l$, and $\sigma_{lr}^k = ||\mathbf{b}_{l,r}^k||_2$. According to the definition of a concave function \cite{boyd2004convex}, we always have the inequality $g(\sigma_{lr}) \leq g(\sigma_{lr}^k) + \omega_{lr}^k(\sigma_{lr} - \sigma_{lr}^k) = \omega_{lr}^k \sigma_{lr} + c$, in which $\omega_{lr}^k \in \partial g(\sigma_{lr}^k)$ and $\partial g(\sigma_{lr}^k)$ is the set of supergradients (the counterpart concept of the subgradient for a concave function) of the function $g()$ at $\sigma_{lr}^k$. When the supergradient is unique, then $\omega_{lr}^k = \partial g(\sigma_{lr}^k)$. Therefore, $g_l(\mathbf{B}_l) = \sum_{r} g(\sigma_{lr})$ can be majorized as follows,
\begin{equation}\label{chapter5_eq:5}
\begin{aligned}
g_l(\mathbf{B}_l) &= \sum_{r} g(\sigma_{lr})\\
              &\leq \sum_{r} \omega_{lr}^k \sigma_{lr} + c\\
       \omega_{lr}^k &\in \partial g(\sigma_{lr}^k).
\end{aligned}
\end{equation}

\subsubsection*{The majorization of the regularized likelihood criterion}
Combining the above two majorization steps, we have majorized the original complex problem in the equation \ref{chapter5_eq:2} to a simper problem in each iteration as follows,
\begin{equation}\label{chapter5_eq:6}
\begin{aligned}
    \min_{ \left\{\bm{\mu}_l\right\}_{l}^{L}, \mathbf{A}, \left\{\mathbf{B}_l\right\}_{l}^{L}} \quad & \sum_{l=1}^{L}\Big[ \frac{\rho_l}{2\alpha_l} ||\mathbf{\Theta}_l - \mathbf{H}_{l}^{k}||_F^2  + \lambda_l \sqrt{J_l} \sum_{r} \omega_{lr}^k \sigma_{lr} \Big] \\
    \text{subject to} \quad \mathbf{\Theta}_l &= \mathbf{1}\bm{\mu}_l^{\text{T}} + \mathbf{AB}_l^{\text{T}}, l = 1 \ldots L \\
     \mathbf{1}^{\text{T}}\mathbf{A} &= \mathbf{0}\\
	 \mathbf{A}^{\text{T}}\mathbf{A} &= \mathbf{I} \\
	 \sigma_{lr} &= ||\mathbf{b}_{l,r}||_2, l = 1 \ldots L, r = 1 \ldots R\\
\mathbf{H}_{l}^{k} &= \mathbf{\Theta}_l^k - \frac{1}{\rho_l} \mathbf{W}_l \odot (b_l^{'}(\mathbf{\Theta}_{l}^{k}) - \mathbf{X}_l), l = 1 \ldots L \\
\omega_{lr}^k &\in \partial g(\sigma_{lr}^k), l = 1 \ldots L, r = 1 \ldots R.
\end{aligned}
\end{equation}

\subsection{Block coordinate descent}
The majorized optimization problem in equation \ref{chapter5_eq:6} can be solved by the block coordinate descent approach, and the analytic solution can be derived for all the parameters.

\subsubsection*{Updating $\{ \bm{\mu}_l \}_{1}^{L}$}
When fixing all other parameters except $\bm{\mu}_l$, the analytic solution of $\bm{\mu}_l$ in equation \ref{chapter5_eq:6} is simply the column mean of $\mathbf{H}_l^k$, $\bm{\mu}_l = \frac{1}{I} (\mathbf{H}_l^k)^{\text{T}} \mathbf{1}$.

\subsubsection*{Updating $\mathbf{A}$}
When fixing all other parameters except $\mathbf{A}$, and deflating the offset term $\{ \bm{\mu}_l \}_{1}^{L}$, the loss function in equation \ref{chapter5_eq:6} becomes $\sum_{l=1}^{L} \frac{\rho_l}{2\alpha_l} ||\mathbf{AB}_l^{\text{T}} - \mathbf{JH}_{l}^{k}||_F^2 + c$, in which $\mathbf{J} = \mathbf{I} - \frac{1}{I} \mathbf{1} \mathbf{1}^{\text{T}}$ is the column centering matrix. If we take $d_l = \sqrt{\rho_l /\alpha_l}$, the above equation can also be written in this way $\sum_{l=1}^{L} \frac{1}{2} ||\mathbf{A}d_l\mathbf{B}_l^{\text{T}} - d_l\mathbf{JH}_{l}^{k}||_F^2$. To simplify the equations, we set $\widetilde{\mathbf{B}}_l = d_l \mathbf{B}_l$ and $\widetilde{\mathbf{JH}}_l^k = d_l \mathbf{JH}_{l}^{k}$. Then, we take $\widetilde{\mathbf{B}}$ as the row concatenation of $\left\{ \widetilde{\mathbf{B}}_l \right\}_{l=1}^{L}$, $\widetilde{\mathbf{B}}^{\text{T}} = [\widetilde{\mathbf{B}}_1^{\text{T}}  \ldots \widetilde{\mathbf{B}}_l^{\text{T}} \ldots \widetilde{\mathbf{B}}_L^{\text{T}}]$, and take $\widetilde{\mathbf{JH}}^k$ as the column concatenation of $\left\{\widetilde{\mathbf{JH}}_l^k \right\}_{l=1}^{L}$, $\widetilde{\mathbf{JH}}^k = [\widetilde{\mathbf{JH}}_1^k \ldots \widetilde{\mathbf{JH}}_l^k \ldots \widetilde{\mathbf{JH}}_L^k]$. After that, we have $\sum_{l=1}^{L} \frac{\rho_l}{2\alpha_l} ||\mathbf{AB}_l^{\text{T}} - \mathbf{JH}_{l}^{k}||_F^2 = \sum_{l=1}^{L} \frac{1}{2} ||\mathbf{A}\widetilde{\mathbf{B}}_l^{\text{T}} - \widetilde{\mathbf{JH}}_{l}^{k}||_F^2 = \frac{1}{2}||\mathbf{A}\widetilde{\mathbf{B}}^{\text{T}} - \widetilde{\mathbf{JH}}^{k}||_F^2$. Updating $\mathbf{A}$ equivalents to minimizing $\frac{1}{2}||\mathbf{A}\widetilde{\mathbf{B}}^{\text{T}} - \widetilde{\mathbf{JH}}^{k}||_F^2, \text{s.t.} \mathbf{A}^{\text{T}}\mathbf{A} = \mathbf{I}$. Assume the SVD decomposition of $\widetilde{\mathbf{JH}}^{k}\widetilde{\mathbf{B}}$ is $\widetilde{\mathbf{JH}}^{k}\widetilde{\mathbf{B}} = \mathbf{UDV}^{\text{T}}$, the analytic solution for $\mathbf{A}$ is $\mathbf{A} = \mathbf{UV}^{\text{T}}$. The derivation of the above solution is shown in the following paragraph.

To simplify the derivation, we take $\mathbf{B} = \widetilde{\mathbf{B}}$ and $\mathbf{C} = \widetilde{\mathbf{JH}}^{k}$. So the optimization problem is $\min_{\mathbf{A}} ||\mathbf{A}\mathbf{B}^{\text{T}} - \mathbf{C}||_F^2, \text{s.t.} \mathbf{A}^{\text{T}} \mathbf{A} = \mathbf{I}$. This equation can be expanded as $||\mathbf{A}\mathbf{B}^{\text{T}} - \mathbf{C}||_F^2 = \text{tr}(\mathbf{BA}^{\text{T}}\mathbf{A}\mathbf{B}^{\text{T}}) - 2\text{tr}(\mathbf{B}\mathbf{A}^{\text{T}}\mathbf{C}) + \text{tr}(\mathbf{C}^{\text{T}}\mathbf{C})$. Since $\mathbf{A}^{\text{T}}\mathbf{A} = \mathbf{I}$, the above optimization problem equivalents to maximizing a trace function problem, $\max_{\mathbf{A}} \text{tr}(\mathbf{B}\mathbf{A}^{\text{T}}\mathbf{C}), \text{s.t.} \mathbf{A}^{\text{T}} \mathbf{A} = \mathbf{I}$. Assume the SVD decomposition of $\mathbf{C}\mathbf{B}$ is $\mathbf{C}\mathbf{B} = \mathbf{UDV}^{\text{T}}$, we have $\text{tr}(\mathbf{B}\mathbf{A}^{\text{T}}\mathbf{C}) = \text{tr}(\mathbf{A}^{\text{T}}\mathbf{C}\mathbf{B}) = \text{tr}(\mathbf{A}^{\text{T}}\mathbf{UDV}^{\text{T}}) = \text{tr}(\mathbf{V}^{\text{T}}\mathbf{A}^{\text{T}}\mathbf{UD})$. According to the Kristof theorem \cite{ten1993least}, we have $\text{tr}(\mathbf{V}^{\text{T}}\mathbf{A}^{\text{T}}\mathbf{UD}) \leq \sum_{r}d_{rr}$, in which $d_{rr}$ is the $r^{\text{th}}$ diagonal element of $\mathbf{D}$, and this upper-bound can be achieved by setting $\mathbf{A} = \mathbf{UV}^{\text{T}}$.

\subsubsection*{Updating $\{ \mathbf{B}_l \}_{1}^{L}$}
Because $\mathbf{A}^{\text{T}}\mathbf{A} = \mathbf{I}$, it is easy to prove that $||\mathbf{AB}_l^{\text{T}} - \mathbf{JH}_{l}^{k}||_F^2 = ||\mathbf{A}^{\text{T}}\mathbf{AB}_l^{\text{T}} - \mathbf{A}^{\text{T}}\mathbf{JH}_{l}^{k}||_F^2 = ||\mathbf{B}_l - (\mathbf{JH}_{l}^{k})^{\text{T}} \mathbf{A}||_F^2$. Also, because of that the least squares problems are decomposable, we have $||\mathbf{B}_l - (\mathbf{JH}_{l}^{k})^{\text{T}} \mathbf{A}||_F^2 = \sum_{r} ||\mathbf{b}_{l,r} - (\mathbf{JH}_{l}^{k})^{\text{T}} \mathbf{a}_r||_2^2$, in which $\mathbf{a}_r$ is the $r^{\text{th}}$ column of $\mathbf{A}$. In this way, we have the following optimization problem,
\begin{equation}\label{chapter5_eq:7}
\begin{aligned}
    \min_{\mathbf{B}_l} \quad & \frac{\rho_l}{2\alpha_l} ||\mathbf{AB}_l^{\text{T}} - \mathbf{JH}_{l}^{k}||_F^2  + \lambda_l \sqrt{J_l} \sum_{r} \omega_{lr}^k \sigma_{lr} \\
	&= \frac{\rho_l}{2\alpha_l}||\mathbf{B}_l - (\mathbf{JH}_{l}^{k})^{\text{T}} \mathbf{A}||_F^2 + \lambda_l \sqrt{J_l} \sum_{r} \omega_{lr}^k \sigma_{lr}\\
    &= \sum_{r} \Big[ \frac{\rho_l}{2\alpha_l}(\mathbf{b}_{l,r} - (\mathbf{JH}_{l}^{k})^{\text{T}} \mathbf{a}_r)^2 + \lambda_l \sqrt{J_l}\omega_{lr}^k \sigma_{lr} \Big]\\
    \text{subject to} \quad \sigma_{lr} &= ||\mathbf{b}_{l,r}||_2, l = 1 \ldots L, r = 1 \ldots R,
\end{aligned}
\end{equation}
The above optimization problem is equivalent to finding the proximal operator of a $L_2$ (or Euclidean) norm, and the analytic solution exists \cite{parikh2014proximal}. Take $\tilde{\lambda}_{lr} = \lambda_l \sqrt{J_l} \omega_{lr}^k \alpha_l/\rho_l$, the analytical solution of $\mathbf{b}_{l,r}$ is $\mathbf{b}_{l,r} = \max(0, 1- \frac{\tilde{\lambda}_{lr}}{||(\mathbf{JH}_{l}^{k})^{\text{T}} \mathbf{a}_r||_2}) (\mathbf{JH}_{l}^{k})^{\text{T}} \mathbf{a}_r$. To update the parameter $\mathbf{B}_l$, we can simply apply this proximal operator to all the columns of $\mathbf{B}_l$.

\subsubsection*{Initialization and stopping criteria}
The initialization of the parameters $\left\{\bm{\mu}_l^0\right\}_{l=1}^{L}$, $\mathbf{A}^0$, $\left\{\mathbf{B}_l^0\right\}_{l=1}^{L}$ can be set to the results of a classical SCA model on $\left\{\mathbf{X}_l\right\}_{l=1}^{L}$ or to accept user imputed initializations. The relative change of the objective function is used as the stopping criteria. Pseudocode of the algorithm described above is shown in Algorithm \ref{alg:pESCA}, in which $f^k$ is the value of the objective function in $k^{\text{th}}$ iteration, $\epsilon_f$ is the tolerance of relative change of the objective function.

\begin{algorithm}[h!]
  \caption{An MM algorithm for fitting the P-ESCA model.}
  \label{alg:pESCA}
  \begin{algorithmic}[1]
    \Require
      $\left\{\mathbf{X}_l\right\}_{l=1}^{L}$, $\left\{\alpha_l\right\}_{l=1}^{L}$, $g()$, $\left\{\lambda_l\right\}_{l=1}^{L}$, $\gamma$;
    \Ensure
      $\hat{\bm{\mu}}$, $\hat{\mathbf{A}}$, $\hat{\mathbf{B}}$;
    \State Compute $\left\{\mathbf{W}_l\right\}_{l=1}^{L}$ for missing values in $\left\{\mathbf{X}_l\right\}_{l=1}^{L}$;
    \State Initialize $\left\{\bm{\mu}_l^0\right\}_{l=1}^{L}$, $\mathbf{A}^0$, $\left\{\mathbf{B}_l^0\right\}_{l=1}^{L}$;
    \State $\mathbf{\Theta}_l^0 = \mathbf{1}(\bm{\mu}_l^0)^{\text{T}} + \mathbf{A}^0(\mathbf{B}_l^0)^{\text{T}}, l = 1 \ldots L$;
    \State $k = 0$;
    \While{$(f^{k-1}-f^{k})/f^{k-1}>\epsilon_f$}
        \For{$l=1\ldots L$}
            \State \text{Estimate} $\rho_l$ \text{according to the data type of} $\mathbf{X}_l$;
            \State  $\mathbf{H}_{l}^{k} = \mathbf{\Theta}_l^k - \frac{1}{\rho_l} \mathbf{W}_l \odot (b^{'}(\mathbf{\Theta}_{l}^{k}) - \mathbf{X}_l))$;
            \State $\bm{\mu}_l^{k+1} = \frac{1}{I} (\mathbf{H}_l^k)^{\text{T}} \mathbf{1}$;
            \State $\widetilde{\mathbf{B}}_l^{k} = \sqrt{\frac{\rho_l}{\alpha_l}} \mathbf{B}_l^{k}$;
            \State $\widetilde{\mathbf{JH}}_l^k = \sqrt{\frac{\rho_l}{\alpha_l}} \mathbf{JH}_{l}^{k}$;
        \EndFor
        \State $(\bm{\mu}^{k+1})^{\text{T}} = [(\bm{\mu}_1^{k+1})^{\text{T}} \ldots (\bm{\mu}_l^{k+1})^{\text{T}} \ldots (\bm{\mu}_L^{k+1})^{\text{T}}]$;
        \State $(\widetilde{\mathbf{B}}^k)^{\text{T}} = [(\widetilde{\mathbf{B}}_1^k)^{\text{T}}  \ldots (\widetilde{\mathbf{B}}_l^k)^{\text{T}} \ldots (\widetilde{\mathbf{B}}_L^k)^{\text{T}}]$;
        \State $\widetilde{\mathbf{JH}}^k = [\widetilde{\mathbf{JH}}_1^k \ldots \widetilde{\mathbf{JH}}_l^k \ldots \widetilde{\mathbf{JH}}_L^k]$;
        \State $\mathbf{UDV}^{\text{T}} = \widetilde{\mathbf{JH}}^{k} \widetilde{\mathbf{B}}^k$;
        \State $\mathbf{A}^{k+1} = \mathbf{UV}^{\text{T}}$;
        \For{$l=1\ldots L$}
            \For{$r=1\ldots R$}
                \State $\sigma_{lr}^k = ||\mathbf{b}_{l,r}^k||_2$;
                \State $\omega_{lr}^k \in \partial g(\sigma_{lr}^k)$;
                \State $\tilde{\lambda}_{lr} = \lambda_l \sqrt{J_l} \omega_{lr}^k \alpha_l/\rho_l$;
                \State $\mathbf{b}_{l,r}^{k+1} = \max(0, 1- \frac{\tilde{\lambda}_{lr}}{||(\mathbf{JH}_{l}^{k})^{\text{T}} \mathbf{a}_r^{k+1}||_2}) (\mathbf{JH}_{l}^{k})^{\text{T}} \mathbf{a}_r^{k+1}$;
             \EndFor
             \State $\mathbf{B}_{l}^{k+1} = [\mathbf{b}_{l,1}^{k+1} \ldots \mathbf{b}_{l,r}^{k+1} \ldots \mathbf{b}_{l,R}^{k+1}]$;
        \EndFor
        \State $(\mathbf{B}^{k+1})^{\text{T}} = [(\mathbf{B}_{1}^{k+1})^{\text{T}} \ldots (\mathbf{B}_{l}^{k+1})^{\text{T}} \ldots (\mathbf{B}_{L}^{k+1})^{\text{T}}]$
        \State $k=k+1$;
    \EndWhile
    \State Compute variation explained ratios.
  \end{algorithmic}
\end{algorithm}

\subsection{Variation explained ratio of the P-ESCA model}
For the quantitative data set $\mathbf{X}_l$, the parameters are $\bm{\mu}_l$, $\mathbf{A}$ and $\mathbf{B}_l$. The total variation explained ratio of the model for $\mathbf{X}_l$ is defined as $\text{varExp}_{l} = 1 - ||\mathbf{W}_l \odot (\mathbf{X}_l - \mathbf{1} \bm{\mu}_l^{\text{T}} - \mathbf{AB}_l^{\text{T}})||_F^2/||\mathbf{W}_l \odot (\mathbf{X}_l - \mathbf{1}\bm{\mu}_l^{\text{T}})||_F^2$. And the variation explained ratio for the $r^{\text{th}}$ component on $\mathbf{X}_l$ is defined as $\text{varExp}_{lr} = 1 - ||\mathbf{W}_l \odot (\mathbf{X}_l - \mathbf{1}\bm{\mu}_l^{\text{T}} - \mathbf{a}_r \mathbf{b}_{l,r}^{\text{T}})||_F^2/||\mathbf{W}_l \odot (\mathbf{X}_l - \mathbf{1}\bm{\mu}_l^{\text{T}})||_F^2$. For the binary data set, we use a similar strategy as the MOFA model \cite{argelaguet2018multi}, where the $\mathbf{H}_{l}^k$ is taken as the pseudo $\mathbf{X}_l$ during the $k^{\text{th}}$ iteration, and $\mathbf{H}_{l}^k$ rather than $\mathbf{X}_l$ is used to compute the variation explained ratios. The multiple data sets can also be taken as a single full data set. In that case the $\{1/\sqrt{\alpha_l} \}_{1}^{L}$ values are taken as the weights for them, and then we can compute the variation explained ratios of each component for this full data set. The full single data set $\widetilde{\mathbf{X}}$ and the weight matrix $\widetilde{\mathbf{W}}$ are the column concatenation of $\{(1/\sqrt{\alpha_l}) \mathbf{X}_l \}_{1}^{L}$ and $\{ \mathbf{W}_l \}_{1}^{L}$, in which $\mathbf{X}_l$ is replaced by $\mathbf{H}^{k}_l$ if the $l^{\text{th}}$ data set is not quantitative. The offset term $\widetilde{\bm{\mu}}$ and the loading matrix $\widetilde{\mathbf{B}}$ are the row concatenation of $\{ (1/\sqrt{\alpha_l}) \bm{\mu}_l \}_{1}^{L}$ and $\{ (1/\sqrt{\alpha_l}) \mathbf{B}_l \}_{1}^{L}$ and the score matrix $\widetilde{\mathbf{A}} = \mathbf{A}$.

\section{Simulation process} \label{section:5.4}
To evaluate the proposed model and the model selection procedure, three data sets of different data types with underlying global, local common and distinct structures are simulated. The following simulations and experiments focus on the quantitative and binary data types. We will first show the simulation of the structure $\mathbf{A}\mathbf{B}^{\text{T}}$, in which $\mathbf{B}$ is the row concatenation of $\left\{\mathbf{B}_l\right\}_{l=1}^3$, $\mathbf{B}^{\text{T}} = [\mathbf{B}_1^{\text{T}} \quad \mathbf{B}_2^{\text{T}} \quad \mathbf{B}_3^{\text{T}}]$. The structure $\mathbf{A}\mathbf{B}^{\text{T}}$ can be expressed in the SVD type as $\mathbf{A}\mathbf{B}^{\text{T}} = \mathbf{U}\mathbf{D}\mathbf{V}^{\text{T}}$ ($\mathbf{A}=\mathbf{U}$, $\mathbf{B}=\mathbf{V}\mathbf{D}$), in which $\mathbf{U}^{\text{T}}\mathbf{U} = \mathbf{I}$, $\mathbf{D}$ is a diagonal matrix, and the structured sparse pattern exists in the matrix $\mathbf{V}$. First, all the elements in $\mathbf{U}$ and $\mathbf{V}$ are simulated from the standard normal distribution. To make sure that $\mathbf{1}^{\text{T}}\mathbf{U} = \mathbf{0}$, simulated $\mathbf{U}$ is first column centered, and then it is orthogonalized by the SVD algorithm to have $\mathbf{U}^{\text{T}}\mathbf{U} = \mathbf{I}$. Also, $\mathbf{V}$ is orthogonalized by the QR algorithm to obtain $\mathbf{V}^{\text{T}}\mathbf{V} = \mathbf{I}$. In this example 21 components are predefined, 7 groups of global, local common and distinctive nature, 3 components each. The structure of these components are set in $\mathbf{V}$ as indicated below,
\begin{equation*}
\begin{aligned}
   \mathbf{V}= \left(
                 \begin{array}{c}
                   \mathbf{V}_1 \\
                   \mathbf{V}_2 \\
                   \mathbf{V}_3 \\
                 \end{array}
               \right)
               = \left(
                   \begin{array}{cccccccc}
                     \mathbf{V}_{1,1:3} & \mathbf{V}_{1,4:6} & \mathbf{V}_{1,7:9} & \mathbf{0}       & \mathbf{V}_{1,13:15} & \mathbf{0}       & \mathbf{0}      \\
                     \mathbf{V}_{2,1:3} & \mathbf{V}_{2,4:6} & \mathbf{0}       & \mathbf{V}_{2,10:12} & \mathbf{0}       & \mathbf{V}_{2,16:18} & \mathbf{0}       \\
                     \mathbf{V}_{3,1:3} & \mathbf{0}       & \mathbf{V}_{3,7:9} & \mathbf{V}_{3,10:12} & \mathbf{0}       & \mathbf{0}       & \mathbf{V}_{3,19:21} \\
                   \end{array}
                 \right),
\end{aligned}
\end{equation*}
in which $\mathbf{V}_{1,1:3}$ indicates the loadings for the first three components for data set 1, etc. After that, 21 values are sampled from $\text{N}(1,0.5)$, and their absolute values are taken as the diagonal elements of $\mathbf{D}$. Furthermore, an extra diagonal matrix $\mathbf{C}$, which has the same size as matrix $\mathbf{D}$, is used to adjust the signal to noise ratios (SNRs) in simulating different global, local common and distinct structures. Then we have $\mathbf{A}\mathbf{B}^{\text{T}} = \mathbf{U}(\mathbf{C}\odot \mathbf{D})\mathbf{V}^{\text{T}}$. In order to define the SNR, we have to specify the noise term $\mathbf{E}_l$ for the $l^{\text{th}}$ data set $\mathbf{X}_l$. If $\mathbf{X}_l$ is quantitative, all the elements in $\mathbf{E}_l$ can be sampled from $N(0,\alpha_l)$. If $\mathbf{X}_l$ is binary, according to the latent variable interpretation of logistic PCA \cite{song2019logistic}, we assume there is a continuous latent matrix $\mathbf{X}_{l}^{\ast}$ underlying the binary observation $\mathbf{X}_l$, and the elements of the noise term $\mathbf{E}_l$ follow the standard logistic distribution. After the specification of the noise terms, we can adjust the diagonal elements in $\mathbf{C}$ to satisfy the predefined SNRs in simulating the global, local common and distinct structures. We restrict the diagonal elements of $\mathbf{C}$ for the same structure to share a single value to have a unique solution. For example, for the global structure $\text{C}123 = \mathbf{U}_{:,1:3}(\mathbf{C}_{1:3,1:3} \odot \mathbf{D}_{1:3,1:3}) \mathbf{V}_{:,1:3}^{\text{T}}$, the corresponding noise term is $\mathbf{E}_{123} = [\mathbf{E}_{1} \quad \mathbf{E}_{2} \quad \mathbf{E}_{3}]$, and the SNR of the global structure as defined as $\text{SNR} = \frac{||\text{C}123||_F^2}{||\mathbf{E}_{123}||_F^2}$. The SNRs for the simulation of the local common (C12, C13, C23) and distinct (D1, D2, D3) structures are defined in the same way.

If $\mathbf{X}_l$ is quantitative, we simply sample all the elements in $\bm{\mu}_l$ from the standard normal distribution. If $\mathbf{X}_l$ is binary, the column offset $\bm{\mu}_l$ represents the logit transformation of the marginal probabilities of binary variables. In our simulation, we will first sample $J_l$ marginal probabilities from a Beta distribution. The Beta distribution can be specified in the following way. For example, if we have 100 samples of a binary variable and we assume the marginal probability to be 0.1, this means we only observe $100 \times 0.1 = 10$ ``1''s. If we model them as Binomial observations with parameter $\pi$, and use a uniform prior distribution for $\pi$, then the posterior distribution of $\pi$ is $\pi \sim \text{Beta}(11, 91)$ \cite{gelman2013bayesian}. After generating $J_l$ marginal probabilities from this Beta distribution, the logit transformation of this vector of probabilities are set as $\bm{\mu}_l$. If $\mathbf{X}_l$ is quantitative, $\mathbf{X}_l$ is simulated as $\mathbf{X}_l = \mathbf{1}\bm{\mu}_l^{\text{T}} + \mathbf{A}\mathbf{B}_l^{\text{T}} + \mathbf{E}_l$, and all the elements of $\mathbf{E}_l$ are sampled from $N(0,\alpha_l)$. If $\mathbf{X}_l$ is binary, we have $\mathbf{\Theta}_l = \mathbf{1}\bm{\mu}_l^{\text{T}} + \mathbf{A}\mathbf{B}_l^{\text{T}}$, and all the elements of $\mathbf{X}_l$ are sampled from the Bernoulli distributions, whose probabilities are the corresponding elements in the inverse logit transformation of $\mathbf{\Theta}_l$. An equivalent way to simulate the binary $\mathbf{X}_l$ is to first generate $\mathbf{X}_l^{\ast} = \mathbf{1}\bm{\mu}_l^{\text{T}} + \mathbf{A}\mathbf{B}_l^{\text{T}} + \mathbf{E}_l$, in which all the elements in $\mathbf{E}_l$ are sampled from the standard logistic distribution. Then, all the elements in $\mathbf{X}_l$ are the binary observations of the corresponding elements of $\mathbf{X}_l^{\ast}$, $x_{ij}^{l}=1$ if $(x_{ij}^{\ast})^{l} > 0$, and $x_{ij}^{l}=0$ \textit{vise versa}. In the following sections, we will use Gaussian-Gaussian-Gaussian (G-G-G) to represent the simulation of three quantitative data sets; Bernoulli-Bernoulli-Bernoulli (B-B-B) for the simulation of three binary data sets; G-B-B for a quantitative data set and two binary data sets and G-G-B for two quantitative data sets and a binary data set.

\section{Evaluation matrices and model selection} \label{section:5.5}
To evaluate the accuracy of the model in estimating the simulated parameters, such as $\mathbf{\Theta}_l$ and $\bm{\mu}_l$, the relative mean squared error (RMSE) is used. If, for example, the simulated parameter is $\mathbf{\Theta}$, $\mathbf{\Theta} = [\mathbf{\Theta}_{1} \quad \mathbf{\Theta}_{2} \quad \mathbf{\Theta}_{3}]$, and its estimation is $\hat{\mathbf{\Theta}}$, the RMSE is defined as $\text{RMSE}(\mathbf{\Theta}) = \frac{||\mathbf{\Theta}-\hat{\mathbf{\Theta}}||_F^2}{||\mathbf{\Theta}||_F^2}$. All of the following evaluation matrices $\text{RMSE}(\mathbf{\Theta}_l)$, $\text{RMSE}(\mathbf{\Theta})$ and $\text{RMSE}(\bm{\mu})$ will be used in the experimental section. To evaluate the recovered subspaces with respect to the simulated global common, local common and distinct structures, the modified RV coefficient \cite{smilde2008matrix} is used. If the simulated global structure is $\text{C}123$, and its estimation is $\widehat{\text{C}123}$, the similarity between the subspaces of $\text{C}123$ and $\widehat{\text{C}123}$ is calculated by the modified RV coefficient.

For the real data sets, we can use the cross validation (CV) error as the proxy of the prediction error to estimate the performance of the model. The K-fold CV procedure used in Chapter \ref{chapter:4} can be quite slow for P-ESCA model. Therefore we set up the CV procedure as follows in a similar way as Chapter \ref{chapter:3}. From each data set $\mathbf{X}_l$, we will randomly select 10\% non-missing elements as $\mathbf{X}_l^{\text{test}}$, and these selected elements in $\mathbf{X}_l$ are set to missing values. The remaining elements form the training set $\mathbf{X}_l^{\text{train}}$. For binary data, the selection of the test set samples is performed in a stratified manner to tackle the situation of unbalanced binary data. Here the test set consist of 10\% ``1''s and ``0''s which are randomly selected from $\mathbf{X}_l$ as $\mathbf{X}_l^{\text{test}}$. A P-ESCA model is constructed on the training sets $\left\{\mathbf{X}_l^{\text{train}} \right\}_{l=1}^{L}$, to obtain an estimation of $\{\hat{\mathbf{\Theta}}_l \}_1^{L}$, in which $\hat{\mathbf{\Theta}}_l = \mathbf{1}\hat{\bm{\mu}_l}^{\text{T}} + \hat{\mathbf{A}}\hat{\mathbf{B}}_l^{\text{T}}$. Then the parameters $\{\hat{\mathbf{\Theta}}_l^{\text{test}} \}_1^{L}$ corresponding to $\{ \hat{\mathbf{X}^{\text{test}}}_l \}_1^{L}$ are indexed. The CV error for $\mathbf{X}_l$ is obtained as the negative log likelihood of using $\hat{\mathbf{\Theta}}_l^{\text{test}}$ to predict $\mathbf{X}_l^{\text{test}}$.

If the data sets $\left\{ \mathbf{X}_l \right\}_{l=1}^{L}$ are of the same data type, a single tuning parameter $\lambda$ is used to replace the $\left\{ \lambda_l \right\}_{l=1}^{L}$ during the model selection. First, $\left\{ \mathbf{X}_l \right\}_{l=1}^{L}$ are split into $\left\{ \mathbf{X}_l^{\text{train}} \right\}_{l=1}^{L}$ and $\left\{ \mathbf{X}_l^{\text{test}} \right\}_{l=1}^{L}$ in the same way as described above. Then $N$ $\lambda$ values are selected (with equal distance in log-space) and for each $\lambda$ value a P-ESCA model is constructed on the training sets $\left\{ \mathbf{X}_l^{\text{train}} \right\}_{l=1}^{L}$. A warm start strategy is used, in which the outputs of a previous model are used to initialize the next model with a slightly higher regularization strength. The warm start strategy has a special meaning in the current context. If some component loadings are shrunk to 0 in the previous model, they will also be 0 in the next models with higher $\lambda$ values. Thus, the search space of the next model will be constrained based on the learned structured sparse pattern in the previous model. In this way, with increasing $\lambda$, components are removed adaptively. We prefer to select the model with the minimum CV error on $\left\{\mathbf{X}_l^{\text{test}} \right\}_{l=1}^{L}$ and the corresponding value of $\lambda$ is $\lambda_{\text{opt}}$. After that we re-fit a P-ESCA model with $\lambda_{\text{opt}}$ on the full data sets $\left\{ \mathbf{X}_l \right\}_{l=1}^{L}$ and the outputs derived from the selected model with minimum CV error are used for initialization in order to preserve the learned structured sparse pattern.

If the data sets are of mixed data types, we prefer to use distinct tuning parameters for each data type. Suppose we have three data sets $\left\{\mathbf{X}_l \right\}_{l=1}^3$, of which $\mathbf{X}_1$ is quantitative and  $\left\{\mathbf{X}_l \right\}_{l=2}^3$ are binary. We specify two tuning parameters $\lambda_{g}$ and $\lambda_{b}$ for the loading matrices corresponding to the quantitative and binary data sets. A heuristic model selection approach, which has the same computational complexity as tuning a single parameter, can be used for the model selection. The splitting of $\left\{ \mathbf{X}_l \right\}_{l=1}^{L}$ into the training and test sets is the same as discussed above. Then again, $N$ values of $\lambda_{g}$ and $\lambda_{b}$ are selected with equal distance in log-space. For the first model, we fix $\lambda_{g}$ to be 0 or a very small value, and tune $\lambda_{b}$ in the same way as above. The model with the minimum CV error on the binary test sets $\left\{\mathbf{X}_l^{\text{test}}\right\}_{l=2}^3$ is selected, and the corresponding value of $\lambda_b$ is $\lambda_{\text{opt}}^b$. After that, $\lambda_{b}$ is fixed to $\lambda_{\text{opt}}^b$, and the outputs of the above selected model are set as the initialization for the models in the model selection of $\lambda_{g}$, which is done in the same way as described above. The model with the minimum CV error on the quantitative test set $\mathbf{X}_1^{\text{test}}$ is selected, and the corresponding value of $\lambda_g$ is $\lambda_{\text{opt}}^g$. After the model selection, we re-fit the P-ESCA model on the full data sets $\left\{\mathbf{X}_l \right\}_{l=1}^3$ with the $\lambda_{\text{opt}}^g$ and $\lambda_{\text{opt}}^b$ and again the outputs of the selected model in the model selection process are used for initialization.

\section{Experiments} \label{section:5.6}
\subsection{Evaluating the dispersion parameter estimation procedure}
The dispersion parameters of the Bernoulli and Poisson distributions can always set to $1$, while for the Binomial distribution with $n$ experiments, it can always be set to $n$. However, for a Gaussian distribution, the dispersion parameter $\alpha$ represents the variance of the noise term, and is assumed to be known. Suppose we have a data set $\mathbf{X}_l$, we prefer to use a PCA model to estimate the $\alpha_l$ before constructing a P-ESCA model. The rank of the PCA model is selected by a missing value based cross validation procedure similar as described above. Details of the $\alpha$ estimation procedure are shown in the supplementary material. After obtaining an estimation of $\hat{\alpha_l}$, it can be casted into the model or the data set can be scaled by $\sqrt{\hat{\alpha_l}}$, which is the estimated standard deviation. We simulated G-G-G, G-G-B and G-B-B data sets to test the $\alpha$ estimation procedure. The parameters in the simulation are set as $I = 100$, $J_1 = 5000$, $J_2 = 500$, $J_3 = 50$; the SNRs of the global, local common and distinct structures are all set to 1; the marginal probability is set to $0.1$ to simulate unbalanced binary data sets. The $\alpha$ estimation procedure was repeated 3 times and the average is taken as the estimation. As shown in supplemental Table S5.2, the mean estimated dispersion parameters in different situations are quite accurate, and the estimations derived from the 3 times repetitions are very stable.

\subsection{An example of CV error based model selection}
We use the simulated G-G-G data sets as an example to show how the model selection is performed when multiple data sets are of the same data type. The following parameters are used in the simulation, $I = 100$, $J_1 = 1000$, $J_2 = 500$, $J_3 = 100$; the SNRs of global, local common and distinct structures are all set to 1; all the dispersion parameters $\{\alpha_l\}_1^3$ are set to be 1. The signals, which are taken as the singular values of the simulated structures, and the noise terms, which are taken as the singular values of the corresponding residual terms, are characterized in supplemental Fig.~S5.3. The true variation explained ratios of each component in every data set is computed using the simulated parameters, and is visualized in supplemental Fig.~S5.4. For the model selection procedure, the maximum number of iterations is set to 500; the stopping criteria is set to $\epsilon_{f} = 10^{-6}$; 30 $\lambda$ values are selected from the interval $[1,500]$ equidistant in log-space; 50 components are used in the initialization. The values of $\{ \alpha_{l} \}_{1}^{L}$ in the P-ESCA model are set to the estimated values from the above $\alpha$ estimation procedure.

\begin{figure}[h]
    \center
    \includegraphics[width=0.9\textwidth]{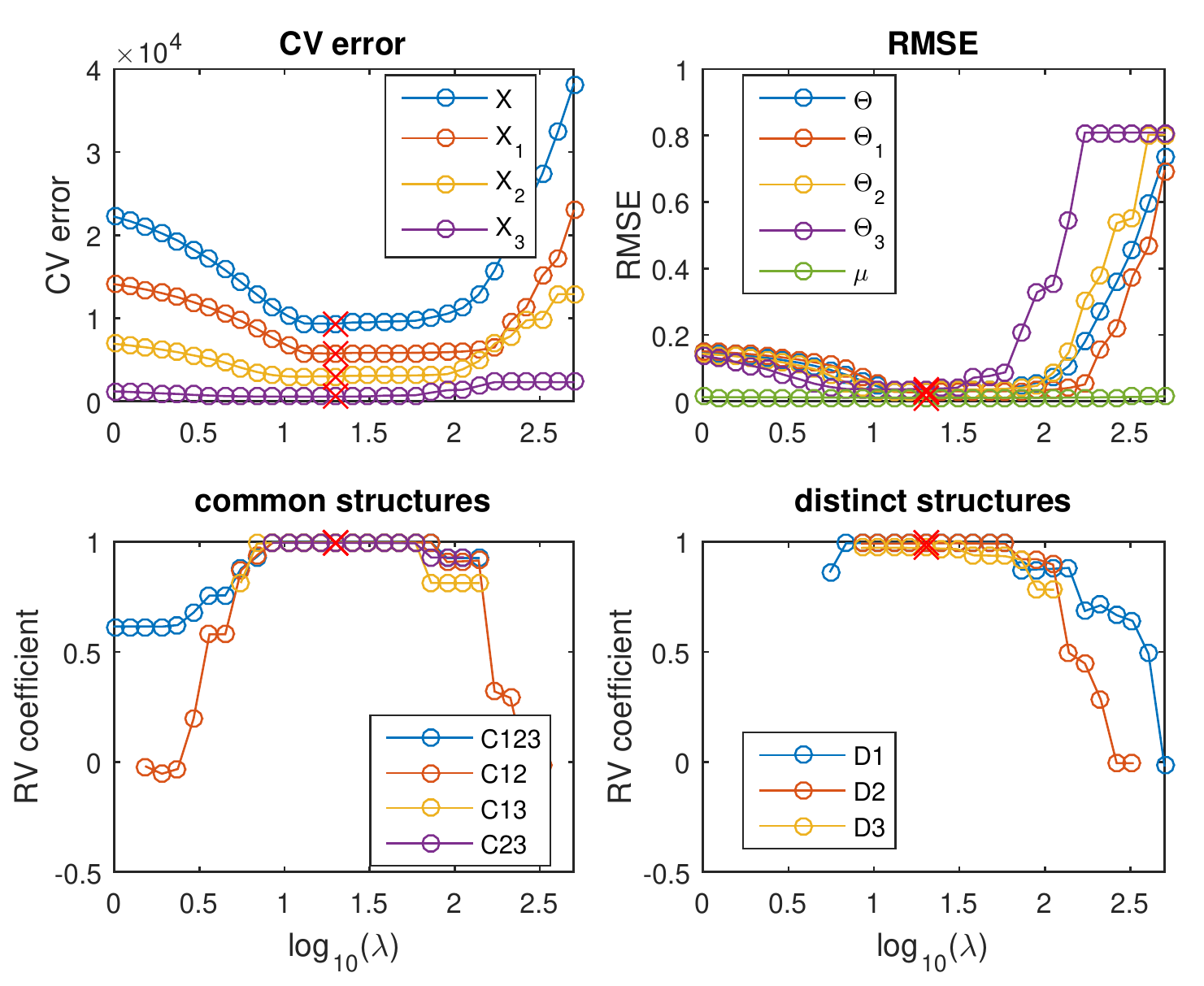}
    \caption{The CV errors (top left), RMSEs (top right), RV coefficients of the common structures (bottom left), and of distinct structures (bottom right) for varying $\lambda$ values for the P-ESCA model with a group GDP ($\gamma = 1$) penalty. The red cross marker indicates the model with minimum CV error.}
	\label{chapter5_fig:2}
\end{figure}

Fig.~\ref{chapter5_fig:2} shows how the CV errors, RMSEs and the RV coefficients change with respect to $\lambda$ when a P-ESCA model with a group GDP ($\gamma=1$) penalty is used. The top figures in Fig.~\ref{chapter5_fig:2} show that the CV errors change in a similar way as the RMSEs. The model with minimum CV error has low RMSEs in estimating the simulated parameters (Fig.~\ref{chapter5_fig:2} top right) and correctly identifies the dimensions of the subspaces for the global, local common and distinct structures (Fig.~\ref{chapter5_fig:2} bottom). However, when the group lasso penalty is used this was not the case. Supplemental Fig.~S5.5 shows that when a group lasso penalty is used, the models with minimal CV error do not coincide with the correct dimensions of the subspaces. In the model with minimum CV error, almost all the components are assigned to the global structure. This result relates to the fact that the lasso type penalty over-shrinks the non-zero parameters, and then the CV error based model selection procedure tends to select a too complex model to compensate to the biased parameter estimation. On the other hand, as the GDP penalty achieves nearly unbiased parameter estimation, the CV error based model selection procedure correctly identifies the correct model.

After the model selection, a high precision P-ESCA model ($\epsilon_{f} = 10^{-8}$) with a group GDP penalty is re-fitted on the full data sets with the value of $\lambda$ corresponding to the minimum CV error and the selected structured sparse pattern. For this selected model, the RMSEs in estimating $\mathbf{\Theta}$, $\mathbf{\Theta}_1$, $\mathbf{\Theta}_2$, $\mathbf{\Theta}_2$ and $\bm{\mu}$ are 0.0259, 0.0239, 0.0285, 0.0335 and 0.0096 respectively. The RV coefficients in estimating the global common structure $\text{C123}$ is 0.9985; local common structures $\text{C12}$, $\text{C13}$ and $\text{C23}$, 0.9977, 0.9969, 0.9953; the distinct structures $\text{D1}$, $\text{D2}$ and $\text{D3}$, 0.9961, 0.9937, 0.9779. The variation explained ratios of each component on the three data sets computed using the estimated parameters, visualized in Fig.~\ref{chapter5_fig:3}, are very similar to the true ones in supplemental Fig.~S5.4. These values can be very useful in exploring the constructed model.

\begin{figure}[h]
    \centering
    \includegraphics[width=0.9\textwidth]{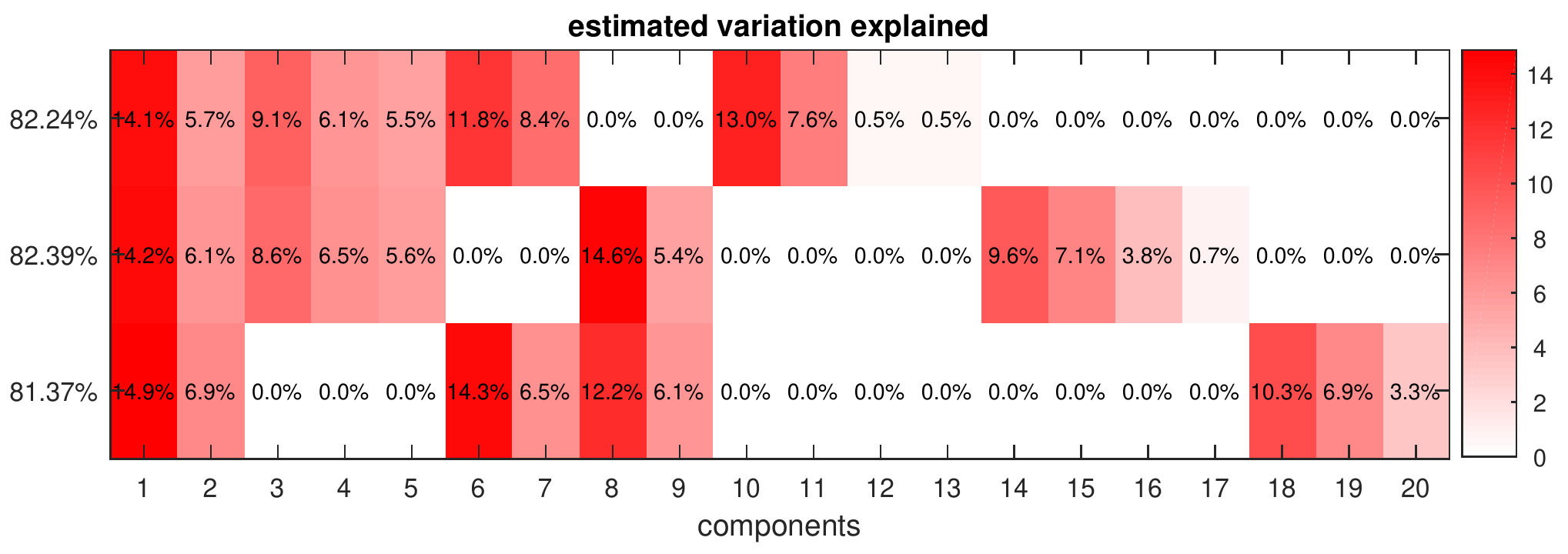}
    \caption{Variation explained ratios computed using the estimated parameters from the selected P-ESCA model with a group GDP penalty. From the top to the bottom, we have data sets $\mathbf{X}_1$, $\mathbf{X}_2$ and $\mathbf{X}_3$; from the left to the right, we have 20 components corresponding to the global, local common and distinct structures. The total variation explained ratios for each data set are shown on the left side of the plot, while he variation explained ratio for each component is shown inside the plot.}
	\label{chapter5_fig:3}
\end{figure}

\subsection{Full characterization of the P-ESCA model when applied to multiple quantitative data sets}
When applied to multiple quantitative data sets, our model is similar as the SLIDE model, except that we use different penalties and a different model selection procedure. The details of the differences between the two approaches are summarized in the supplementary material. Since the concave GDP penalty is capable to achieve a nearly unbiased estimation of the parameters, the P-ESCA model with a group GDP penalty is expected to achieve similar performance to the two stages procedure used in the SLIDE model. Therefore, we simulated seven realistic cases by adjusting the SNRs of the simulated structures to compare the performance of these two models and their model selection procedures. The SNRs of the simulated structures corresponding to these seven cases are listed in supplemental Table S5.3. Case 1: only the local common structures exist and they have unequal SNRs; case 2: the JIVE case, only the global common and distinct structures exist, and they are all of low SNRs; case3: all the simulated structures are of low SNRs; case 4: global common structure dominate the simulation; case 5: local common structures dominate the simulation; case 6: distinct structures dominate the simulation; case 7: none of the global, local common and distinct structures exist.

The following parameters are used in the G-G-G data simulations, $I = 100$, $J_1 = 1000$, $J_2 = 500$, $J_3 = 100$, all of the $\{ \alpha_l \}_1^3$ are set to 1. In order to have exactly 3 components for all the simulated structures, we reject the simulations of which the singular values of the three components of any specific structure are not 2 times larger than the singular value of the corresponding residual term. The P-ESCA model with a group GDP ($\gamma=1$) penalty is selected and re-fitted on the full data sets in the same way as above. For the SLIDE model, the simulated data sets $\{ \mathbf{X}_l \}_1^{3} $ are column centered and block-scaled by the Frobenius norm of each data set. Then the SLIDE model is selected and fitted using the default parameters. The deflated column offset term is taken as the estimated $\hat{\bm{\mu}}$. The derived loading matrices $\{ \mathbf{B}_l \}_1^3$ are re-scaled by the corresponding Frobenius norm of each data set. G-G-G data sets are simulated for all the 7 cases, and for each case, the simulation experiment (data simulation, model selection, fitting the final model) is repeated 10 times for both the P-ESCA model and the SLIDE model. The mean RV coefficients in evaluating the estimated global, local common and distinct structures and the corresponding mean estimated ranks are shown in Table \ref{chapter5_tab:2}, and the mean RMSEs in estimating the simulated parameters are shown in supplemental Table S5.4. In all 7 cases, these two methods have very accurate estimation of the subspaces corresponding to the global, local common and distinct structures, and of the simulated parameters $\mathbf{\Theta}$, which is the column concatenation of $\{ \mathbf{\Theta}_l \}_1^{3} $, $\{ \mathbf{\Theta}_l \}_1^{3} $, and $\bm{\mu}$, which is row concatenation of $\{ \bm{\mu}_l \}_1^L$. For some of the cases there is a slight advantage for the P-ESCA model.

\begin{table}[h]
\small
\centering
\caption{Mean RV coefficients and the mean rank estimates in evaluating the recovered subspaces derived from 10 experiments using the P-ESCA model and the SLIDE model for seven G-G-G simulated cases. The results are shown as in mean RV coefficient(mean rank estimation) form. The row names $1^p$ and $1^s$ indicate P-ESCA and SLIDE models applied to simulation case 1. Same rule applies to other row names.}
\label{chapter5_tab:2}
\begin{tabular}{llllllll}
  \toprule
    & C123 & C12 & C13 & C23 & D1 & D2 & D3 \\
  \midrule
  $1^{p}$ & 0        (0)  & 0.998(3)   & 0.999(3)   & 0.999(3)   & 0        (0)  & 0        (0) & 0        (0)           \\
  $1^{s}$ & 0        (0)  & 0.998(3)   & 0.998(3)   & 0.998(3)   & 0        (0)  & 0        (0) & 0        (0)           \\
 \hline
 $2^{p}$  & 0.996(3)   & 0        (0)  & 0        (0)  & 0        (0)  & 0.997(3)   & 0.985(3)   & 0.973(3)   \\
 $2^{s}$  & 0.996(3)   & 0        (0)  & 0        (0)  & 0        (0)  & 0.997(3)   & 0.985(3)   & 0.973(3)   \\
 \hline
 $3^{p}$  & 0.998(3)   & 0.997(3)   & 0.997(3)   & 0.995(3)   & 0.996(3)   & 0.994(3)   & 0.976(3)   \\
 $3^{s}$  & 0.995(3)   & 0.996(3)   & 0.993(3)   & 0.991(3)   & 0.995(3)   & 0.994(3)   & 0.976(3)   \\
 \hline
 $4^{p}$  & 1        (3)  & 1        (3)  & 1        (3)  & 0.999(3)   & 0.997(3)   & 0.995(3)   & 0.977(3)   \\
 $4^{s}$  & 1        (3)  & 1        (3)  & 0.999(3)   & 0.999(3)   & 0.997(3)   & 0.994(3)   & 0.977(3)   \\
 \hline
 $5^{p}$  & 1        (3)  & 1        (3)  & 1        (3)  & 1        (3)  & 0.998(3)   & 0.995(3)   & 0.977(3)   \\
 $5^{s}$  & 0.999(3)   & 1        (3)  & 0.999(3)   & 0.999(3)   & 0.997(3)   & 0.995(3)   & 0.977(3)   \\
 \hline
 $6^{p}$ & 0.998(3)   & 1        (3)  & 0.994(3.1) & 0.999(3)   & 0.998(2.9) & 0.999(3)   & 0.998(3)   \\
 $6^{s}$ & 0.996(3)   & 0.999(3)   & 0.998(3)   & 0.998(3)   & 0.999(3)   & 0.999(3)   & 0.997(3)   \\
 \hline
 $7^{p}$  & 0        (0) & 0        (0)  & 0        (0)  & 0        (0)  & 0        (0)  & 0        (0)  & 0        (0)       \\
 $7^{s}$  & 0        (0) & 0        (0)  & 0        (0)  & 0        (0)  & 0        (0)  & 0        (0)  & 0        (1.4) \\
  \bottomrule
\end{tabular}
\end{table}

\subsection{Full characterization of the P-ESCA model when applied to multiple binary data sets}
The performance of the proposed P-ESCA model is fully characterized with respect to multiple binary data sets. Here we make a comparison to the MOFA model, which is the Bayesian counterpart of P-ESCA. In the P-ESCA model, the structured sparse pattern is induced through a group concave penalty, and the model selection is done through missing value based cross validation, while in the MOFA model, the structured sparse pattern is induced through the automatic relevance determination approach and the model is selected through maximizing the marginal likelihood. In addition, MOFA model also shrinks a component to be 0 when its variation explained ratios for all the data sets are less than a threshold, the default value of which is 0. The details of the differences are summarized in the supplementary material. For the model selection of the P-ESCA model, the range of $\lambda$ values is $[1,100]$, and the other parameters are the same as before. To give an impression of the model selection process, we also characterized how the CV errors, RMSEs and the RV coefficients change with respect to $\lambda$ in the P-ESCA model with a group GDP penalty on the simulated B-B-B data sets in supplemental Fig.~S5.6. For the MOFA model, the default parameters are used, but as exact sparsity cannot be achieved by the automatic relevance determination procedure used in the MOFA model, we take a component for a single data set to be 0 when the variation explained ratio of this component on this data set is less than $0.1\%$.

In the seven B-B-B simulations cases, we set $I=200$, and the marginal probability to be $0.1$ to simulate very unbalanced binary data sets. Other parameters are the same as in the G-G-G simulation cases. The mean RV coefficients in evaluating the estimated global, local common and distinct structures and the corresponding mean estimated ranks are shown in Table \ref{chapter5_tab:3}, and the mean RMSEs in estimating the simulated parameters are shown in supplemental Table S5.5. Compared to the results derived from the P-ESCA model on the G-G-G data sets (Table \ref{chapter5_tab:2}), the recovered subspaces related to the global, local common and distinct structures from P-ESCA model on B-B-B data sets are less accurate with respect to RV coefficient and rank estimation, especially when the SNR of a specific structure is much lower than others (in case 4, 5, 6). However, given the fact that all the three data sets only have binary observations, the recovered subspaces are accurate enough. Furthermore, it is interesting to find that such low RMSEs in estimating $\{ \mathbf{\Theta}_l \}_1^{3} $, $\bm{\mu}$ (supplemental Table S5.5) can be achieved solely from a model on multiple binary data sets. Although these results are a little bit counter intuitive, it is coordinate with the previous research \cite{davenport20141,song2018generalized}. According to our previous research \cite{song2018generalized}, this result mainly relates to the fact that the GDP penalty can achieve nearly unbiased parameter estimation. On the other hand, the RMSEs in estimating the simulated parameters from the MOFA model (supplemental Table S5.5) are much larger. Especially for the estimation of the simulated column offset term, all the elements in the estimated $\hat{\bm{\mu}}$ from the MOFA model are very close to 0, and are far away from the simulated $\bm{\mu}$. However, the recovered subspaces from the MOFA model are comparable to the results derived from the P-ESCA model (Table \ref{chapter5_tab:3}).

\begin{table}[htbp]
\small
\centering
\caption{Mean RV coefficients and mean rank estimations of recovered subspaces derived from 10 repeated simulation experiments using the P-ESCA model and the MOFA model for seven B-B-B cases. For case 7, a one component MOFA model is selected, however, results cannot be extracted when the offset term is included. The row names $1^p$ and $1^m$ indicate P-ESCA and MOFA models applied to simulation case 1. Same rule applies to other row names.}
\label{chapter5_tab:3}
\begin{tabular}{llllllll}
  \toprule
  & C123 & C12 & C13 & C23 & D1 & D2 & D3 \\
  \midrule
 $1^{p}$  & 0(0)        & 0.993(2.9) & 0.994(2.9) & 0.991(2.5) & 0(0.2)     & 0(0.5)       & 0(0) \\
 $1^{m}$  & 0(0)        & 0.834(2.1) & 0.984(3.2) & 0.989(3.2) & 0(1.4)     & 0(1  )       & 0(0)  \\
 \hline
 $2^{p}$ & 0.993(2.6)  & 0(0.4)     & 0(0)       & 0(0  )     & 0.990(3  ) & 0.982(3)     & 0.914(3) \\
 $2^{m}$ & 0.959(2.7)  & 0(0  )     & 0(0.1)     & 0(0.2)     & 0.964(3.2) & 0.975(3.1)   & 0.885(2.6) \\
 \hline
 $3^{p}$ & 0.956(1.9)  & 0.959(4) & 0.972(1.8) & 0.939(1.9) & 0.967(4.3) & 0.945(4.1)   & 0.878(2.6) \\
 $3^{m}$ & 0.940(2.6)  & 0.925(2.3) & 0.977(3.2)   & 0.956(3.1) & 0.936(3.6) & 0.934(3.7)   & 0.848(2.3) \\
 \hline
 $4^{p}$ & 0.992(2.3)  & 0.988(3.3) & 0.981(2.4) & 0.980(2.2) & 0.831(3.6) & 0.838(3.2 )   & 0.151(0.2) \\
 $4^{m}$ & 0.986(2.9)  & 0.955(2.9) & 0.990(3  ) & 0.985(2.9) & 0.960(2.6) & 0.929(2.3)   & 0.220(0.3)      \\
 \hline
 $5^{p}$ & 0.980(2.1)  & 0.990(3.8) & 0.991(2.5)     & 0.986(2.6)   & 0.916(3.4) & 0.808(2.3)   & 0.074(0.1)\\
 $5^{m}$ & 0.915(3.1)  & 0.956(2.8) & 0.991(2.9)     & 0.984(3  )   & 0.878(2.6)   & 0.917(2  )   & 0.193(0.3)      \\
 \hline
 $6^{p}$ & 0.192(0.2)  & 0.981(4.7) & 0.984(2.3) & 0.979(2.6) & 0.991(4.6) & 0.988(3   )   & 0.963(2.8) \\
 $6^{m}$ & 0.525(1.1)  & 0.949(2.1) & 0.980(3.7) & 0.979(3.4)  & 0.978(4.3) & 0.977(4.2)   & 0.953(3.1) \\
 \hline
 $7^{p}$ & 0(0)        & 0(0)        & 0(0)        & 0(0)        & 0(0)        & 0(0)        & 0(0) \\
 $7^{m}$ & NA          & NA           & NA           & NA         & NA         &NA         & NA \\
  \bottomrule
\end{tabular}
\end{table}

\subsection{Full characterization of the P-ESCA model when applied to multiple data sets of mixed data types}
The proposed P-ESCA model is also fully characterized on the simulated multiple data sets of mixed quantitative and binary data types. Both G-B-B and G-G-B data sets are simulated for all the seven simulation cases. We set $I=200$, all of $\{ \alpha_l \}_1^3$ to be 1, the marginal probability in simulating unbalanced data sets to be $0.1$. Other parameters are the same as above. The range of $\lambda$ values for loadings related to the quantitative data sets is $[1,500]$, and for loadings related to binary data sets is $[1,100]$. The mean RV coefficients of the estimated global, local common and distinct structures and the corresponding mean ranks estimation from the P-ESCA and the MOFA model in the seven G-B-B simulation cases are shown in Table \ref{chapter5_tab:4}, for the G-G-B simulation the results are shown in Table \ref{chapter5_tab:5}. The mean RMSEs in estimating the simulated parameters are shown in supplemental Table S5.6, for the G-B-B simulations are in supplemental Table S5.7. Similar to the previous results of B-B-B simulations, the P-ESCA model can achieve quite accurate estimates of the subspaces related to the global, local common and distinct structures (Table \ref{chapter5_tab:4}, Table \ref{chapter5_tab:5}) when the SNRs of different structures are relatively equal. However, when the SNR of a specific structure is very low compared to others (in case 4, 5, 6), the P-ESCA model has difficulty for its recovery. However, compared to the MOFA model, P-ESCA can achieve better results with respect to the recovered subspaces (Table \ref{chapter5_tab:4}, Table \ref{chapter5_tab:5}) and estimation of the simulated parameters (supplemental Table S5.6, Table S5.7) in G-B-B and G-G-B simulations.

\begin{table}[htbp]
\small
\centering
\caption{Mean RV coefficients and mean rank estimations of the recovered subspaces derived from simulation experiments using the P-ESCA model and the MOFA model for seven G-B-B cases. The row names $1^p$ and $1^m$ indicate P-ESCA and MOFA models applied to simulation case 1. Same rule applies to other row names.}
\label{chapter5_tab:4}
\begin{tabular}{llllllll}
  \toprule
 & C123 & C12 & C13 & C23 & D1 & D2 & D3 \\
  \midrule
 $1^{p}$  & 0(0)        & 0.997(2.8)   & 0.987(2.3)   & 0.993(3) & 0(0.9)     & 0(0)       & 0(0) \\
 $1^{m}$  & 0(0)        & 0.826(2.5)   & 0.978(3)   & 0.973(3.7) & 0(1.6)     & 0(0.5)     & 0(0) \\
 \hline
 $2^{p}$  & 0.978(2.3)  & 0(0.5)       & 0(0)       & 0(0)       & 0.993(3.2) & 0.981(3) & 0.918(2.9) \\
 $2^{m}$  & 0.984(2.7)  & 0(0.1)       & 0(0)       & 0(0.2)     & 0.533(4.2) & 0.975(3) & 0.895(2.7) \\
 \hline
 $3^{p}$  & 0.975(2)  & 0.972(3.9)   & 0.945(1.5)   & 0.974(2.2) & 0.932(4.6)   & 0.968(3.8)   & 0.892(2.6) \\
 $3^{m}$  & 0.914(3)  & 0.879(2.7)   & 0.962(2.7)   & 0.971(3.1) & 0.475(4.6)   & 0.970(2.9)   & 0.860(2.5) \\
 \hline
 $4^{p}$  & 0.998(2.7)  & 0.995(2.9)     & 0.917(1.6)     & 0.991(2.4)   & 0.547(4.8)   & 0.909(3.3)   & 0(0) \\
 $4^{m}$  & 0.856(3.7)  & 0.547(2)     & 0.788(3.7)     & 0.990(3)   & 0.378(4.9)   & 0.935(2.8)   & 0.398(0.6)   \\
 \hline
 $5^{p}$  & 0.982(2.1)  & 0.995(3.4)     & 0.994(2.2)     & 0.994(2.8)   & 0.698(4.3)   & 0.929(3.1) & 0.164(0.2)\\
 $5^{m}$  & 0.677(3.3)  & 0.691(2)     & 0.916(3.3)     & 0.991(3.1)   & 0.316(5.2)   & 0.835(2.8) & 0.475(0.8)   \\
 \hline
 $6^{p}$  & 0(0)      & 0.980(5.1)       & 0.971(1.8)     & 0.989(2.5)   & 0.989(5.1)   & 0.992(3.5)   & 0.966(2.9) \\
 $6^{m}$  & 0.624(1.4)  & 0.750(1.9)     & 0.899(3.9)     & 0.985(3.4)   & 0.837(6.2)   & 0.978(4.3)   & 0.954(2.9) \\
 \hline
 $7^{p}$  & 0(0)        & 0(0)         & 0(0)         & 0(0)       & 0(0)       & 0(0)       & 0(0) \\
 $7^{m}$  & NA          & NA           & NA           & NA         & NA         &NA         & NA    \\
  \bottomrule
\end{tabular}
\end{table}

\begin{table}[htbp]
\small
\centering
\caption{Mean RV coefficients and mean rank estimations of the recovered subspaces derived from 10 simulation experiments using the P-ESCA model and the MOFA model for seven G-G-B cases. The row names $1^p$ and $1^m$ indicate P-ESCA and MOFA models applied to simulation case 1. Same rule applies to other row names.}
\label{chapter5_tab:5}
\begin{tabular}{llllllll}
  \toprule
     & C123 & C12 & C13 & C23 & D1 & D2 & D3 \\
  \midrule
 $1^{p}$ & 0(0)        & 0.998(3) & 0.997(2.4) & 0.999(2.9) & 0(0.6)     & 0(0.1)       & 0(0) \\
 $1^{m}$ & 0(0)        & 0.528(3.8) & 0.998(2.8) & 0.998(2.9) & 0(0.4)     & 0(0.3)       & 0(0)  \\
 \hline
 $2^{p}$ & 0.971(2.4)  & 0(0.6)     & 0(0)       & 0(0)     & 0.998(3) & 0.995(3)     & 0.920(2.9) \\
 $2^{m}$ & 0.987(2.8)  & 0(1.2)     & 0(0)       & 0(0)       & 0.997(3)   & 0.994(3)     & 0.899(2.8) \\
 \hline
 $3^{p}$ & 0.984(2.2)  & 0.977(3.8) & 0.979(2.2) & 0.981(2.2) & 0.970(3.8) & 0.968(3.8)   & 0.922(3) \\
 $3^{m}$ & 0.977(2.7)  & 0.524(4.3) & 0.993(2.8)   & 0.989(2.9) & 0.989(3.2) & 0.980(3.1)   & 0.888(2.3) \\
 \hline
 $4^{p}$ & 0.996(3)    & 0.965(3  ) & 0.997(2.6) & 0.996(2.4) & 0.941(3.4) & 0.899(3.6)   & 0.844(2.4) \\
 $4^{m}$ & 0.955(4)    & 0.673(3.1) & 0.903(2.9) & 0.997(2.9) & 0.920(3) & 0.983(3.1)   & 0.703(1.2)      \\
 \hline
 $5^{p}$ & 0.996(2.4)  & 0.998(3.6) & 1(2.6)     & 0.999(2.6)   & 0.978(3.4) & 0.952(3.4)   & 0.808(1.8)\\
 $5^{m}$ & 0.761(3.9)  & 0.715(3.2) & 0.999(3)   & 0.998(2.9) & 0.995(3.1)   & 0.982(3.2)   & 0.494(0.7)      \\
 \hline
 $6^{p}$ & 0.348(0.5)  & 0.984(5.5) & 0.992(2.4) & 0.996(2.5) & 0.997(3.6) & 0.997(3.5)   & 0.970(3) \\
 $6^{m}$ & 0.894(2)  & 0.949(5) & 0.925(3.2) & 0.972(3)  & 0.933(3) & 0.973(3.1)   & 0.960(2.9) \\
 \hline
 $7^{p}$ & 0(0)        & 0(0)        & 0(0)        & 0(0)        & 0(0)        & 0(0)        & 0(0) \\
 $7^{m}$ & NA          & NA           & NA           & NA         & NA         &NA         & NA \\
  \bottomrule
\end{tabular}
\end{table}

\section{Real data analysis}
\begin{figure}[h]
    \centering
    \includegraphics[width=0.9\textwidth]{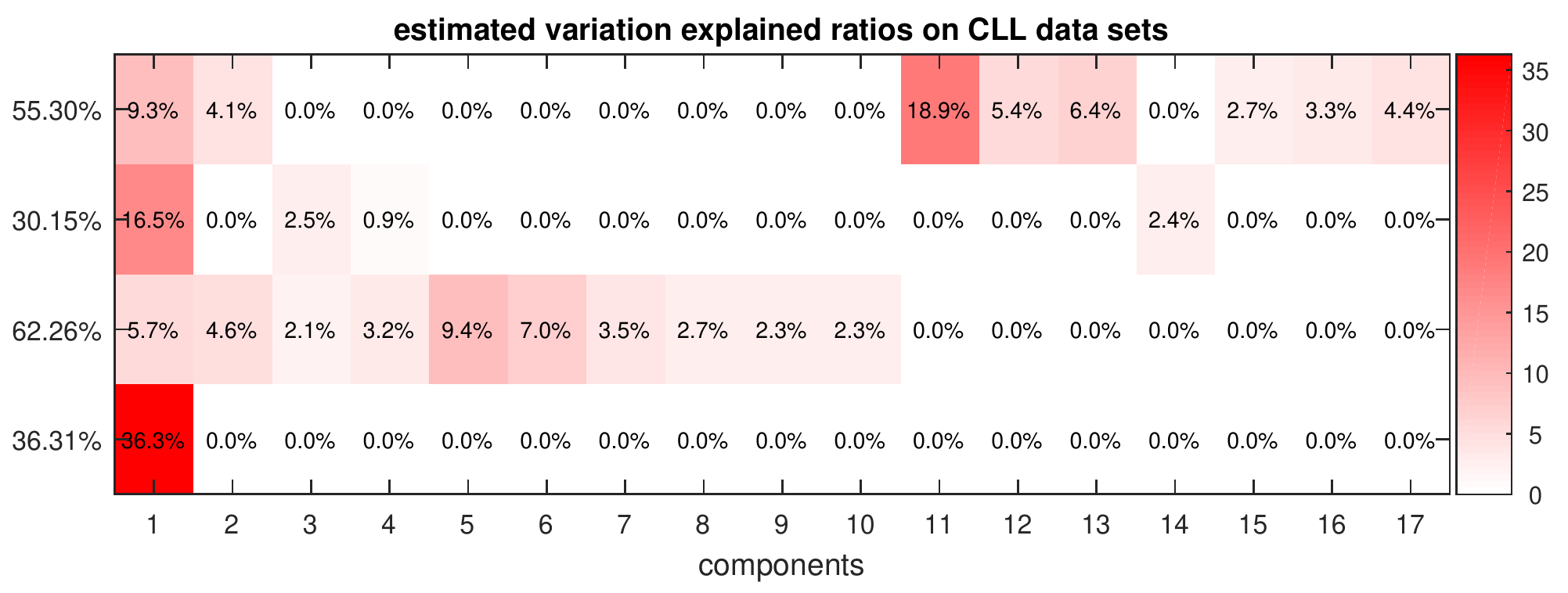}
    \caption{Variation explained ratios computed using the estimated parameters from the selected P-ESCA model on CLL data sets. From the top to the bottom, the data sets are drug response, methylation, transcriptome and mutation data.}
	\label{chapter5_fig:4}
\end{figure}

We applied the P-ESCA model on the chronic lymphocytic leukaemia (CLL) data set \cite{dietrich2018drug,argelaguet2018multi}, which was used in the chapter of the MOFA model, to give an example of the real data analysis. For the 200 samples in the CLL data set, not all of them are fully characterized for all the measurements. Drug response data has 184 samples and 310 variables; DNA methylation data, 196 samples and 4248 variables; transcriptome data, 136 samples and 5000 variables; mutation data, 200 samples and 69 binary variables. The missing pattern of the CLL data sets is visualized in supplemental Fig.~S5.7. Except for the missing values related to the samples that were not measured by a specific platform, there are also some selected variables missing in the mutation data (supplemental Fig.~S5.7). All the quantitative data sets are first column centered and scaled by the sample standard deviation of each variable. After that, the dispersion parameters of the quantitative data sets are estimated by the $\alpha$ estimation procedure. Rank estimation of each single data set was performed three times and results are shown in supplemental Table S5.8. The P-ESCA model with a GDP ($\gamma=1$) is selected and re-fitted on the CLL data sets in the same way as described above. The initial number of components is set to 50. The selected model has 41 components, and if we take each loading vector related to a single data set in a component as a group, there are 51 non-zero loading groups. The model selection results are shown in supplemental Fig.~S5.8. Since the variation explained ratios of 41 components are difficult to visualize, we only show the components (Fig.~\ref{chapter5_fig:4}), whose variation explained ratio are larger than 2\% for at least one data set. The above procedure (processing, model selection, fitting the final model) is repeated 5 times to test its stability. The Pearson coefficient matrix for the 5 estimations of the $\hat{\bm{\mu}}$ and the RV coefficient matrices for the 5 estimations of the $\hat{\mathbf{A}}$, $\hat{\mathbf{B}}$ and $\hat{\mathbf{\Theta}}$ are shown in supplemental Fig.~S5.9.

In \cite{argelaguet2018multi}, a 10 components MOFA model is selected on the CLL data sets. The variation explained plots of the 10 components MOFA model, reproduced from \cite{argelaguet2018multi}, is shown in supplemental Fig.~S5.10. There is some overlap between the two models (Fig.~\ref{chapter5_fig:4}, supplemental Fig.~S5.10). Both models have one strong common component in which all data sets participate, and a common component in which two (P-ESCA) or three (MOFA) data sets participate. Furthermore the drug response and the transcriptomic (mRNA) data have extra distinct components. The variation explained is somewhat higher for the P-ESCA model which also uses extra components. The amount of variation explained is the highest for the drug response and mRNA data sets. The main difference between the models is the fact that P-ESCA only finds a single component relevant for the binary mutation data while MOFA finds two. The comparison of the two models with respect to the estimated $\hat{\bm{\mu}}$ is infeasible because the column offset term is not included in this 10 components MOFA model. In general the P-ESCA result is more complex than the results in \cite{argelaguet2018multi} in terms of number of selected components and variation explained. However, this is mainly because, during the model selection of \cite{argelaguet2018multi}, the minimum variation explained threshold is set to 2\%. If we set the threshold to the default value 0\%, and set the initial number of components to be 50, and other parameters are kept the same, a 50 components MOFA model is selected.

\section{Discussion}
In this chapter, we generalized an exponential family SCA (ESCA) model for the data integration of multiple data sets of mixed data types. Then, we introduced the nearly unbiased group concave penalty to induce structured sparsity pattern on the loading matrices of the ESCA model to separate the global, local common and distinct variation. An efficient MM algorithm with analytical form updates for all the parameters was derived to fit the proposed group concave penalty penalized ESCA (P-ESCA) model. In addition, a missing value based cross validation procedure is developed for the model selection. In many different realistic simulations (different SNR levels, and combinations of quantitative and or binary data sets of different), the P-ESCA model and the model selection procedure work well with respect to recovering the subspaces related to the global, local common and distinct structures, and the estimation of the simulated parameters.

The performance of the P-ESCA model and the cross validation based model selection procedure relate to the fact that the used group concave penalty can achieve nearly unbiased estimation of the parameters while generating sparse solutions. The nearly unbiased parameter estimation makes the P-ESCA model have high accuracy in the estimation of the simulated parameters, and the cross validation error based model selection procedure is consistent. Another key point of the model selection procedure is that the randomly sampled $10\%$ non-missing elements are usually a typical set of elements from the population. This makes the CV error a good proxy of the prediction error of the model. The rank estimation in different repetitions of the model selection procedure is robust and only differ slightly with respect to the very weak components.

When applied to multiple quantitative data sets, the proposed P-ESCA model can achieve slightly better performance than the SLIDE model in recovering the subspaces of the simulated structures and in estimating the simulated parameters. Also, since missing value problems (missing values in a single data set, or missing complete samples in one or some of the data sets) are very common in practice, the option of tackling missing values is a big advantage. In the P-ESCA model and its model selection procedure, the effect of missing values is masked by using the weight matrices, making full use of the available data sets. When applied to the multiple binary data sets or the mixed quantitative and binary data sets, the proposed P-ESCA model has better performance than the MOFA model in recovering the subspaces of the simulated structures and in estimating the simulated parameters. Furthermore, the exact orthogonality constraint can be achieved in the P-ESCA model, which is crucial for the uniqueness of the recovered subspaces related to the global, local common and distinct variation.

\section*{Acknowledgements}
Y.S. gratefully acknowledges the financial support from China Scholarship Council (NO.201504910809).

\clearpage
\section{Supplementary information}

\subsection{Dispersion parameter estimation using PCA}
The notations of this subsection is the same as the Chapter \ref{chapter:5}. Before constructing an ESCA or P-ESCA model, the dispersion parameter $\alpha$ of a quantitative data set $\mathbf{X}$, which is the variance of the residual term, is assumed to be known. Assume the column centered quantitative data set is $\mathbf{X}$($I \times J$), and the PCA model of $\mathbf{X}$ can be expressed as $\mathbf{X} = \mathbf{AB}^{\text{T}} + \mathbf{E}$. $\mathbf{A}$($I \times R$) and $\mathbf{B}$($J \times R$) are the score and loading matrix respectively; $\mathbf{E}$($I \times J$) is the residual term and elements in $\mathbf{E}$, $\epsilon_{ij} \sim N(0,\alpha)$; $R$ is the true low rank of $\mathbf{X}$. In order to tackle the potential missing value problem, we also introduce the weight matrix $\mathbf{W}$ in the same way as above. The maximum likelihood estimation of $\alpha$ can be expressed as $\hat{\alpha}_{\text{mle}} = \frac{1}{||\mathbf{W}||_0} ||\mathbf{W} \odot (\mathbf{X} - \mathbf{AB}^{\text{T}})||_{F}^2$, in which $||\mathbf{W}||_0$ is the number of non-missing elements in $\mathbf{W}$. Since this is a biased estimation of $\alpha$, we can adjust the estimation according to the degree of freedom as $\hat{\alpha} = \frac{1}{||\mathbf{W}||_0 - (I+J)R} ||\mathbf{W} \odot (\mathbf{X} - \mathbf{AB}^{\text{T}})||_{F}^2$. The parameters $R$, $\mathbf{A}$ and $\mathbf{B}$ are estimated as follows.

We select the rank $R$ using a similar model selection strategy as in the main text. We first split $\mathbf{X}$ into $\mathbf{X}^{\text{test}}$ and $\mathbf{X}^{\text{train}}$ in the same way as in the main text. Then, a series of PCA models with different number of components are constructed on $\mathbf{X}^{\text{train}}$, and the CV error is defined as the least square error in fitting $\mathbf{X}^{\text{test}}$. After that $\hat{R}$ is set to the number of components of the model with the minimum CV error. Then a rank $\hat{R}$ PCA model is constructed on the full data $\mathbf{X}$, and we get an estimate of $\hat{\mathbf{A}}$ and $\hat{\mathbf{B}}$. Then $\hat{\alpha}$ is set to $\hat{\alpha} = \frac{1}{||\mathbf{W}||_0 - (I+J)\hat{R}} ||\mathbf{W} \odot (\mathbf{X} - \hat{\mathbf{A}}\hat{\mathbf{B}}^{\text{T}})||_{F}^2$. The EM type algorithm used to fit the PCA model with the option of missing values is implemented in Matlab in the same way as in \cite{kiers1997weighted}.

\subsection{The difference between the SLIDE model and the P-ESCA model when applied to multiple quantitative data sets}
The notations of this subsection is the same as the Chapter \ref{chapter:5}.

\begin{itemize}
  \item Different processing steps. The SLIDE model does column centering and block scaling using the Frobenius norm of the corresponding data set to preprocess the data. Then the relative weights of the data sets in the SCA model are set to 1. On the other hand, we estimate the dispersion parameter (variation of the noise term) of each data set and the inverse of the estimated dispersion parameter is equivalent to the relative weight of the data sets in the SCA model.
  \item Different penalty terms. The SLIDE model uses the group lasso penalty to induce the structured sparsity. Because of the block scaling processing step, there is no weight $\left\{ \sqrt{J_l} \right\}_{l=1}^L$ on the group lasso penalty to accommodate for the potential unequal number of variables in different data sets. On the other hand, the weighted group concave penalty is used in the P-ESCA model.
  \item Option for missing values. The option of tackling the missing value problem is not included in the SLIDE model.
  \item Different model selection procedures. The SLIDE model uses a two stages approach to do model selection, while our model selection approach is as described as in the main text.
\end{itemize}

\subsection{The difference between the MOFA model and the P-ESCA model}
The notations of this subsection is the same as the Chapter \ref{chapter:5}.

\begin{itemize}
  \item Different origins. Although these two methods are similar with respect to what they can do, they have different origins. The MOFA model is developed in the Bayesian probabilistic matrix factorization framework in the same line as the group factor analysis model and the factor analysis model, while the P-ESCA model is derived in the deterministic matrix factorization framework in the same line as the SLIDE model, the SCA model and the PCA model.
  \item Different ways in inducing structured sparsity. In the P-ESCA model, the structured sparse pattern is induced through a group concave penalty, while in the MOFA model, it is induced through the automatic relevance determination approach. The group concave penalty can shrink a group of elements to be exactly 0, while the automatic relevance determination cannot achieve exact sparsity. In addition, MOFA model also shrinks a component to be 0 when its variation explained ratios for all the data sets are less than a threshold, whose default value is 0.
  \item Different model selection procedures. The P-ESCA model is selected by a missing value based CV approach; while the selection of a MOFA model relies on maximizing the marginal likelihood. In theory, maximizing the marginal likelihood has no difficulty in tuning multiple parameters, while the CV based model selection procedure is infeasible for such task.
  \item Orthogonality constraint. The orthogonality constraint $\mathbf{A}^{\text{T}}\mathbf{A} = \mathbf{I}$ can only be achieved in the P-ESCA model. Whether this property is meaningful or not depends on the specific research question. However, the constraint is crucial for the proof of the uniqueness of the recovered subspaces corresponding to the global, local common and distinct variation.
\end{itemize}

\subsection{Supplemental tables}
\begin{table}[htbp]
\centering
\caption*{Table S5.1: A list of log-partition functions and their first and second order derivatives for the Gaussian, Bernoulli and Poisson
distributions. $\theta$ indicates the natural parameter.}
\label{chapter5_tab:S1}
\begin{tabular}{cccc}
  \toprule
Distribution & $b(\theta)$ & $b^{'}(\theta)$ & $b^{''}(\theta)$ \\
  \midrule
Gaussian     & $\frac{\theta^2}{2}$    & $\theta$ & 1 \\
Bernoulli    & $\log(1+\exp(\theta))$ & $\frac{\exp(\theta)}{1+\exp(\theta)}$ & $\frac{\exp(\theta)}{(1+\exp(\theta))^2}$ \\
Poisson      & $\exp(\theta)$ & $\exp(\theta)$ & $\exp(\theta)$ \\
  \bottomrule
\end{tabular}
\end{table}

\begin{table}[htbp]
\centering
\caption*{Table S5.2: Results of the $\alpha$ estimation procedure. $1^{g}$ indicates that a Gaussian distribution is used and $\alpha_l = 1$; $b$ indicates the Bernoulli distribution.  The estimated dispersion parameter $\hat{\alpha_l}$ and the corresponding times are shown as $\text{mean} \pm \text{std} (\text{seconds})$. When the estimated ranks are the same in each of the three times CV procedure is repeated, the corresponding standard deviation is 0.}
\label{chapter5_tab:S2}
\begin{tabular}{llllll}
 \hline
   $\alpha_1$ & $\alpha_2$ & $\alpha_3$ & $\hat{\alpha_1}$($\text{time}$) & $\hat{\alpha_2}$($\text{time}$) & $\hat{\alpha_3}$($\text{time}$)) \\
 \hline
  $1^{g}$   & $1^{g}$  & $1^{g}$ & 0.9920  $\pm$ 0 (9.01)  & 1.0029  $\pm$ 0 (2.46) & 1.0183 $\pm$ 0 (17.06) \\
  $100^{g}$ & $25^{g}$ & $1^{g}$ & 99.7148 $\pm$ 0.7469 (10.66) & 24.9525 $\pm$ 0 (2.96) & 1.1609 $\pm$ 0.2437 (76.33) \\
  \hline
  $1^{g}$   & $1^{g}$  & $b$     & 0.9892  $\pm$ 0 (10.92)  & 0.9793  $\pm$ 0 (2.80) &  \\
  $100^{g}$ & $25^{g}$ & $b$     & 99.8457 $\pm$ 0 (10.79)  & 24.7688 $\pm$ 0 (3.43) &  \\
  \hline
  $1^{g}$   & $b$ & $b$     & 0.9896   $\pm$ 0 (10.90) &  &  \\
  $100^{g}$ & $b$ & $b$     & 100.1774 $\pm$ 0 (10.75) &  &  \\

  \hline
\end{tabular}
\end{table}

\begin{table}[htbp]
\centering
\caption*{Table S5.3: Seven simulation cases used to evaluate the proposed P-ESCA model. For each simulation case, the corresponding SNRs in simulating the global structure $\text{C123}$, local common structures, $\text{C12}$, $\text{C13}$, $\text{C23}$, and distinct structures $\text{D1}$, $\text{D2}$, $\text{D3}$, are give. If the SNR of a specific structure is 0, it means this structure does not exist in the simulation.}
\label{chapter5_tab:S3}
\begin{tabular}{llllllll}
  \toprule
case & $\text{C123}$ & $\text{C12}$ & $\text{C13}$ & $\text{C23}$ & $\text{D1}$ & $\text{D2}$ & $\text{D3}$ \\
  \midrule
 1   & 0  & 1  & 2  & 3   & 0  & 0  & 0   \\
 2   & 1  & 0  & 0  & 0   & 1  & 1  & 1   \\
 3   & 1  & 1  & 1  & 1   & 1  & 1  & 1   \\
 4   & 10 & 5  & 5  & 5   & 1  & 1  & 1   \\
 5   & 5  & 10 & 10 & 10  & 1  & 1  & 1   \\
 6   & 1  & 5  & 5  & 5   & 10 & 10 & 10   \\
 7   & 0  &  0 & 0  & 0   & 0  & 0  & 0   \\
  \bottomrule
\end{tabular}
\end{table}

\begin{table}[htbp]
\centering
\caption*{Table S5.4: Mean RMSEs in estimating the simulated parameters $\mathbf{\Theta}$, $\left\{ \mathbf{\Theta} \right\}_{l=1}^3$ and $\bm{\mu}$, derived from repeating the experiments 10 time using the P-ESCA model and the SLIDE model for seven G-G-G simulation cases. The row names $1^p$ and $1^s$ indicate P-ESCA and SLIDE models applied to simulation case 1. Same rule applies to other row names.}
\label{chapter5_tab:S4}
\begin{tabular}{llllll}
  \toprule
 & $\text{RMSE}(\mathbf{\Theta})$ & $\text{RMSE}(\mathbf{\Theta}_1)$ & $\text{RMSE}(\mathbf{\Theta}_2)$ & $\text{RMSE}(\mathbf{\Theta}_3)$ & $\text{RMSE}(\bm{\mu})$ \\
  \midrule
 $1^{p}$  &0.0167    &0.0181    &0.0152    &0.0135    &0.0102  \\
 $1^{s}$  &0.0178    &0.0194    &0.0161    &0.0145    &0.0102  \\
 \hline
 $2^{p}$  &0.0251    &0.0241    &0.0255    &0.0334    &0.0100 \\
 $2^{s}$ &0.0269    &0.0259    &0.0273    &0.0349    &0.0100 \\
 \hline
 $3^{p}$  &0.0274    &0.0266    &0.0278    &0.0333    &0.0097 \\
 $3^{s}$ &0.0298    &0.0290    &0.0301    &0.0366    &0.0097 \\
 \hline
 $4^{p}$  &0.0064    &0.0062    &0.0065    &0.0076    &0.0099 \\
 $4^{s}$ &0.0068    &0.0066    &0.0069    &0.0081    &0.0099 \\
 \hline
 $5^{p}$  &0.0052    &0.0051    &0.0052    &0.0063    &0.0099  \\
 $5^{s}$ &0.0055    &0.0054    &0.0056    &0.0067    &0.0099 \\
 \hline
 $6^{p}$  &0.0064    &0.0062    &0.0065    &0.0078    &0.0100 \\
 $6^{s}$ &0.0068    &0.0066    &0.0068    &0.0082    &0.0100 \\
 \hline
 $7^{p}$  &0.0099    &0.0097    &0.0104    &0.0098    &0.0099 \\
 $7^{s}$ &0.0132    &0.0097    &0.0104    &0.0658    &0.0099 \\
  \bottomrule
\end{tabular}
\end{table}

\begin{table}[htbp]
\centering
\caption*{Table S5.5: Mean RMSEs in estimating the simulated parameters $\mathbf{\Theta}$, $\left\{ \mathbf{\Theta} \right\}_{l=1}^3$ and $\bm{\mu}$ derived from repeating the experiments 10 times using the P-ESCA model and the MOFA model for seven B-B-B simulation cases. The row names $1^p$ and $1^m$ indicate P-ESCA and MOFA models applied to simulation case 1. Same rule applies to other row names.}
\label{chapter5_tab:S5}
\begin{tabular}{llllll}
  \toprule
  & $\text{RMSE}(\mathbf{\Theta})$ & $\text{RMSE}(\mathbf{\Theta}_1)$ & $\text{RMSE}(\mathbf{\Theta}_2)$ & $\text{RMSE}(\mathbf{\Theta}_3)$ & $\text{RMSE}(\bm{\mu})$ \\
  \midrule
 $1^{p}$ &0.0530    &0.0450    &0.0498    &0.1218    &0.0265  \\
 $1^{m}$   &0.4762    &0.5004    &0.4518    &0.4130    &0.9999  \\
 \hline
 $2^{p}$ &0.0528    &0.0488    &0.0511    &0.1009    &0.0223 \\
 $2^{m}$   &0.5951    &0.5911    &0.5936    &0.6432    &1.0000 \\
  \hline
 $3^{p}$   &0.0830    &0.0651    &0.0775    &0.2922    &0.0331 \\
 $3^{m}$   &0.5037    &0.4965    &0.5077    &0.5558    &0.9999 \\
 \hline
 $4^{p}$ &0.1080    &0.0673    &0.1240    &0.4298    &0.0731 \\
 $4^{m}$   &0.3297    &0.3233    &0.3322    &0.3805    &0.9999 \\
 \hline
 $5^{p}$ &0.1225    &0.0750    &0.1506    &0.4546    &0.0860  \\
 $5^{m}$   &0.3267    &0.3196    &0.3302    &0.3802    &0.9998 \\
 \hline
 $6^{p}$ &0.1066    &0.0662    &0.1275    &0.4123    &0.0752 \\
 $6^{m}$   &0.3324    &0.3259    &0.3364    &0.3788    &0.9999 \\
 \hline
 $7^{p}$ &0.0130    &0.0129    &0.0129    &0.0133    &0.0130 \\
 $7^{m}$   &NA    &NA    &NA    &NA    &NA \\
  \bottomrule
\end{tabular}
\end{table}

\begin{table}[htbp]
\centering
\caption*{Table S5.6: Mean RMSEs in estimating the simulated parameters $\mathbf{\Theta}$, $\left\{ \mathbf{\Theta} \right\}_{l=1}^3$ and $\bm{\mu}$ derived from repeating the experiments 10 times using the P-ESCA model and the MOFA model for seven G-B-B simulation cases. The row names $1^p$ and $1^m$ indicate P-ESCA and MOFA models applied to simulation case 1. Same rule applies to other row names.}
\label{chapter5_tab:S6}
\begin{tabular}{llllll}
  \toprule
 & $\text{RMSE}(\mathbf{\Theta})$ & $\text{RMSE}(\mathbf{\Theta}_1)$ & $\text{RMSE}(\mathbf{\Theta}_2)$ & $\text{RMSE}(\mathbf{\Theta}_3)$ & $\text{RMSE}(\bm{\mu})$ \\
  \midrule
 $1^{p}$  &0.0376    &0.0078    &0.0463    &0.0855    &0.0210  \\
 $1^{m}$  &0.1674    &0.0023    &0.3422    &0.4241    &1.0000  \\
 \hline
 $2^{p}$  &0.0415    &0.0105    &0.0544    &0.0985    &0.0167 \\
 $2^{m}$  &0.1663    &0.0020    &0.3259    &0.3894    &1.0000 \\
 \hline
 $3^{p}$  &0.0552    &0.0110    &0.0708    &0.1874    &0.0231 \\
 $3^{m}$  &0.2008    &0.0029    &0.3346    &0.3847    &1.0000 \\
 \hline
 $4^{p}$  &0.0712    &0.0021    &0.0986    &0.4200    &0.0750 \\
 $4^{m}$  &0.1674    &0.0023    &0.3422    &0.4241    &1.0000 \\
 \hline
 $5^{p}$  &0.0775    &0.0018    &0.1107    &0.4023    &0.0806  \\
 $5^{m}$  &0.1663    &0.0020    &0.3259    &0.3894    &1.0000  \\
 \hline
 $6^{p}$  &0.0731    &0.0027    &0.0878    &0.3086    &0.0626 \\
 $6^{m}$  &0.2008    &0.0029    &0.3346    &0.3847    &1.0000 \\
 \hline
 $7^{p}$  &0.0107    &0.0050    &0.0129    &0.0116    &0.0107 \\
 $7^{m}$  & NA       & NA       & NA       & NA       & NA \\
  \bottomrule
\end{tabular}
\end{table}

\begin{table}[htbp]
\centering
\caption*{Table S5.7: Mean RMSEs in estimating the simulated parameters $\mathbf{\Theta}$, $\left\{ \mathbf{\Theta} \right\}_{l=1}^3$ and $\bm{\mu}$ derived from repeating the experiments 10 times using the P-ESCA model and the MOFA model for seven G-G-B simulation cases. The row names $1^p$ and $1^m$ indicate P-ESCA and MOFA models applied to simulation case 1. Same rule applies to other row names.}
\label{chapter5_tab:S7}
\begin{tabular}{llllll}
  \toprule
& $\text{RMSE}(\mathbf{\Theta})$ & $\text{RMSE}(\mathbf{\Theta}_1)$ & $\text{RMSE}(\mathbf{\Theta}_2)$ & $\text{RMSE}(\mathbf{\Theta}_3)$ & $\text{RMSE}(\bm{\mu})$ \\
  \midrule
 $1^{p}$ &0.0143    &0.0089    &0.0069    &0.0555    &0.0092  \\
 $1^{m}$   &0.0831    &0.0091    &0.0071    &0.5825    &1.0000  \\
 \hline
 $2^{p}$  &0.0266    &0.0126    &0.0136    &0.0955    &0.0091 \\
 $2^{m}$  &0.1314    &0.0129    &0.0139    &0.7284    &1.0000 \\
  \hline
 $3^{p}$  &0.0268    &0.0137    &0.0143    &0.1158    &0.0095 \\
 $3^{m}$    &0.0973    &0.0139    &0.0145    &0.6701    &1.0000 \\
 \hline
 $4^{p}$ &0.0149    &0.0030    &0.0032    &0.1505    &0.0233 \\
 $4^{m}$   &0.0381    &0.0031    &0.0032    &0.4400    &1.0000 \\
 \hline
 $5^{p}$ &0.0174    &0.0025    &0.0025    &0.1897    &0.0314  \\
 $5^{m}$   &0.0359    &0.0025    &0.0026    &0.4197    &1.0000  \\
 \hline
 $6^{p}$ &0.0249    &0.0032    &0.0033    &0.1719    &0.0281 \\
 $6^{m}$   &0.0527    &0.0033    &0.0034    &0.3869    &1.0000 \\
 \hline
 $7^{p}$ &0.0068    &0.0050    &0.0048    &0.0123    &0.0068 \\
 $7^{m}$   &NA    &NA    &NA    &NA    &NA \\
  \bottomrule
\end{tabular}
\end{table}

\begin{table}[htbp]
\centering
\caption*{Table S5.8: Rank estimations of the CLL data sets. Drug: drug response data; meth: DNA methylation data; mRNA: transcriptome data; mut: mutation data.}
\label{chapter5_tab:S8}
\begin{tabular}{llllll}
  \toprule
  data set & data type    & size              & $k=1$ & $k=2$ & $k=3$ \\
  \midrule
  drug     & quantitative & $184 \times 310$  & 17 & 17 & 18   \\
  meth     & quantitative & $196 \times 4248$ & 8  & 9  & 9   \\
  mRNA     & quantitative & $136 \times 5000$ & 16 & 18 & 17   \\
  mut      & binary       & $200 \times 69$   & 1  & 1  & 0   \\
  \bottomrule
\end{tabular}
\end{table}

\subsection{Supplemental figures}
\begin{figure}[htbp]
    \centering
    \includegraphics[width=0.9\textwidth]{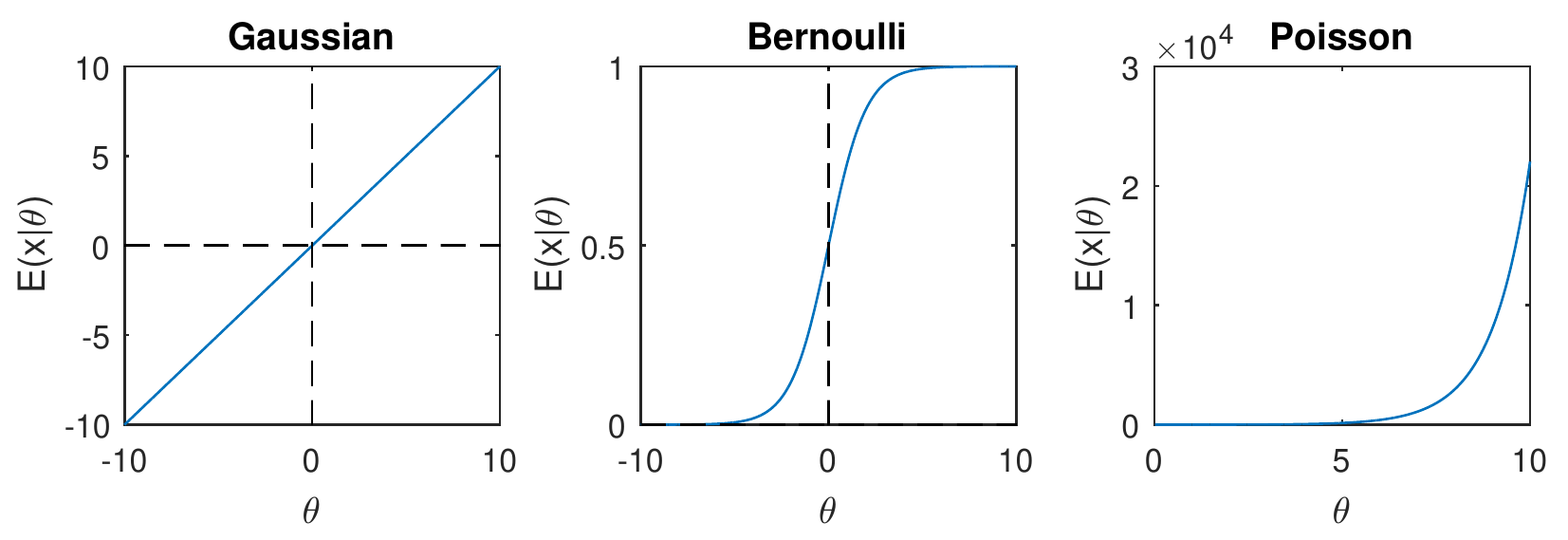}
    \caption*{Figure S5.1: The conditional mean of $x$, $\text{E}(x|\theta)$, for varying $\theta$ values for Gaussian, Bernoulli, Poisson distributions}
	\label{chapter5_fig:S1}
\end{figure}

\begin{figure}[h]
    \centering
    \includegraphics[width=0.9\textwidth]{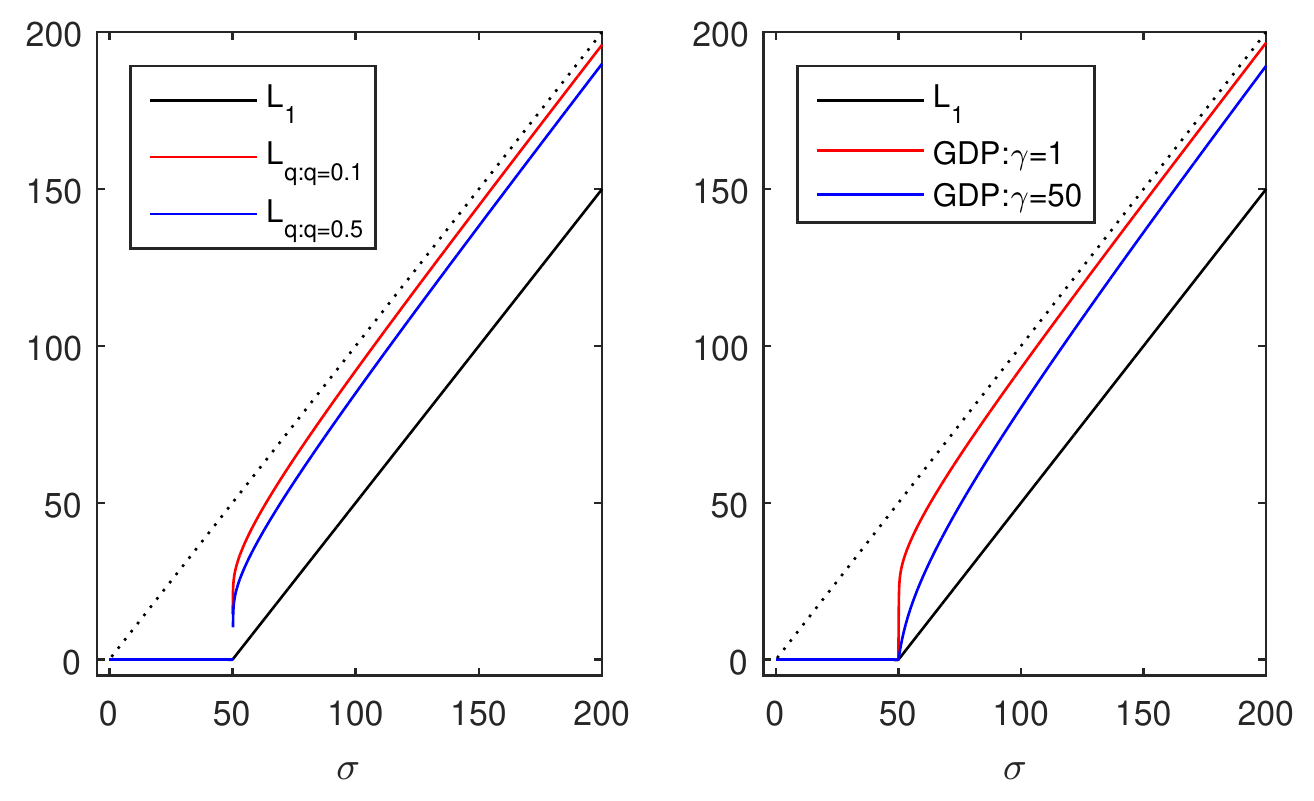}
    \caption*{Figure S5.2: The thresholding properties of the group lasso ($\text{L}_1$), the group $\text{L}_{q}$ and the group GDP penalty. $\sigma$ is taken as the $L_2$ norm of a group of elements. $q$ and $\gamma$ are the hyper-parameters of the corresponding penalties. $x$ axis indicates the value of $\sigma$ before thresholding; $y$ axis indicates the value after threhsolding. $\text{L}_{q: 0<q < 1}$ penalty is non-differentiable at 0.}
	\label{chapter5_fig:S2}
\end{figure}

\begin{figure}[htbp]
    \centering
    \includegraphics[width=0.9\textwidth]{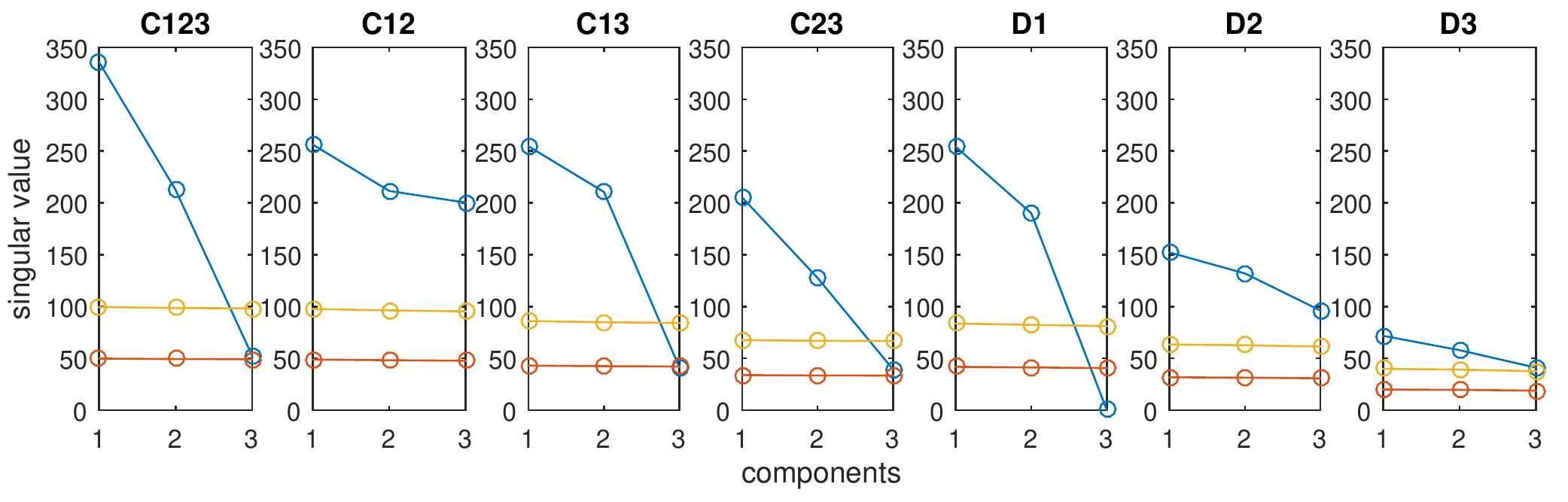}
    \caption*{Figure S5.3: The singular values of the simulated common, local common and distinct structures and of the corresponding residual terms. Blue dots: singular values of the simulated structure; red dots: singular values of the residual term; yellow dots: 2 times of the singular values of the residual term.}
	\label{chapter5_fig:S3}
\end{figure}

\begin{figure}[htbp]
    \centering
    \includegraphics[width=0.9\textwidth]{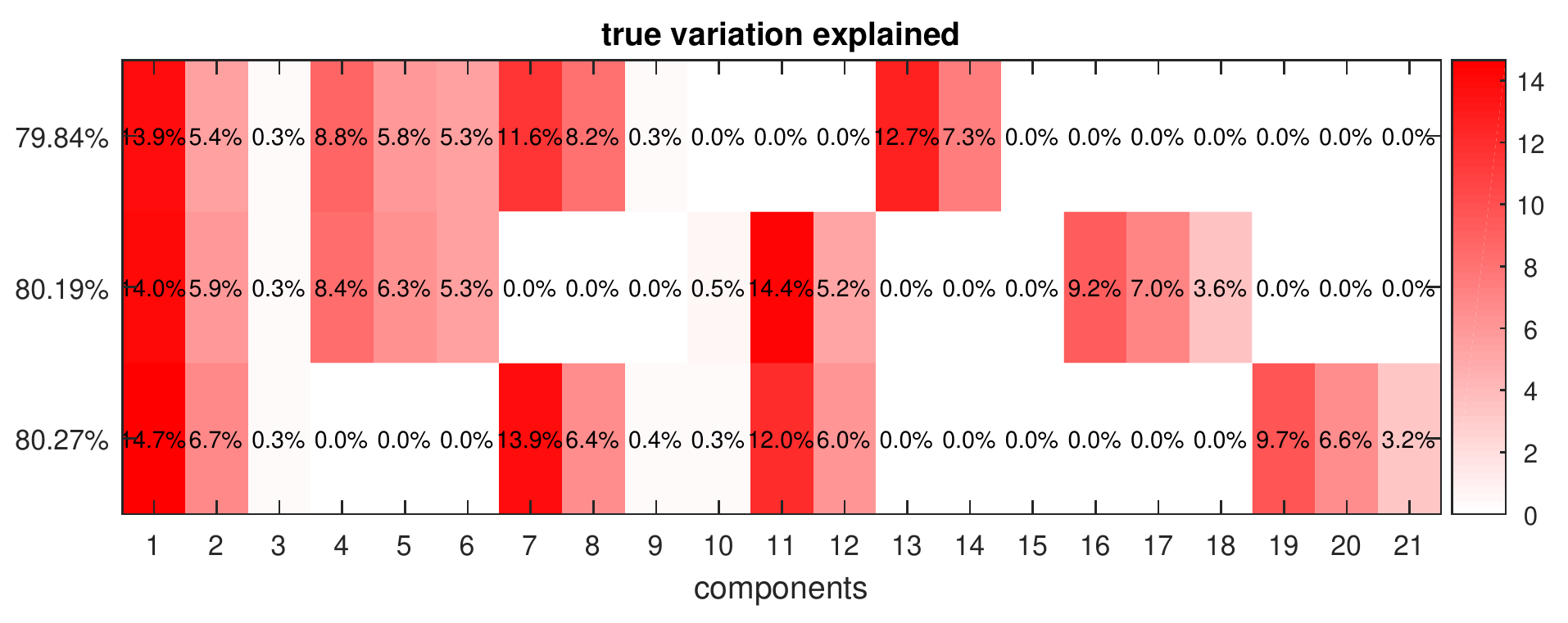}
    \caption*{Figure S5.4: The variation explained ratios computed using the simulated parameters. From the top to the bottom, we have data sets $\mathbf{X}_1$, $\mathbf{X}_2$ and $\mathbf{X}_3$; from the left to the right, we have 21 components corresponding to the global, local common and distinct structures. The total variation explained ratios for each data set are shown in the left of the plot, while the variation explained ratio of each component for each data set is shown inside the plot.}
	\label{chapter5_fig:S4}
\end{figure}

\begin{figure}[htbp]
    \centering
    \includegraphics[width=0.9\textwidth]{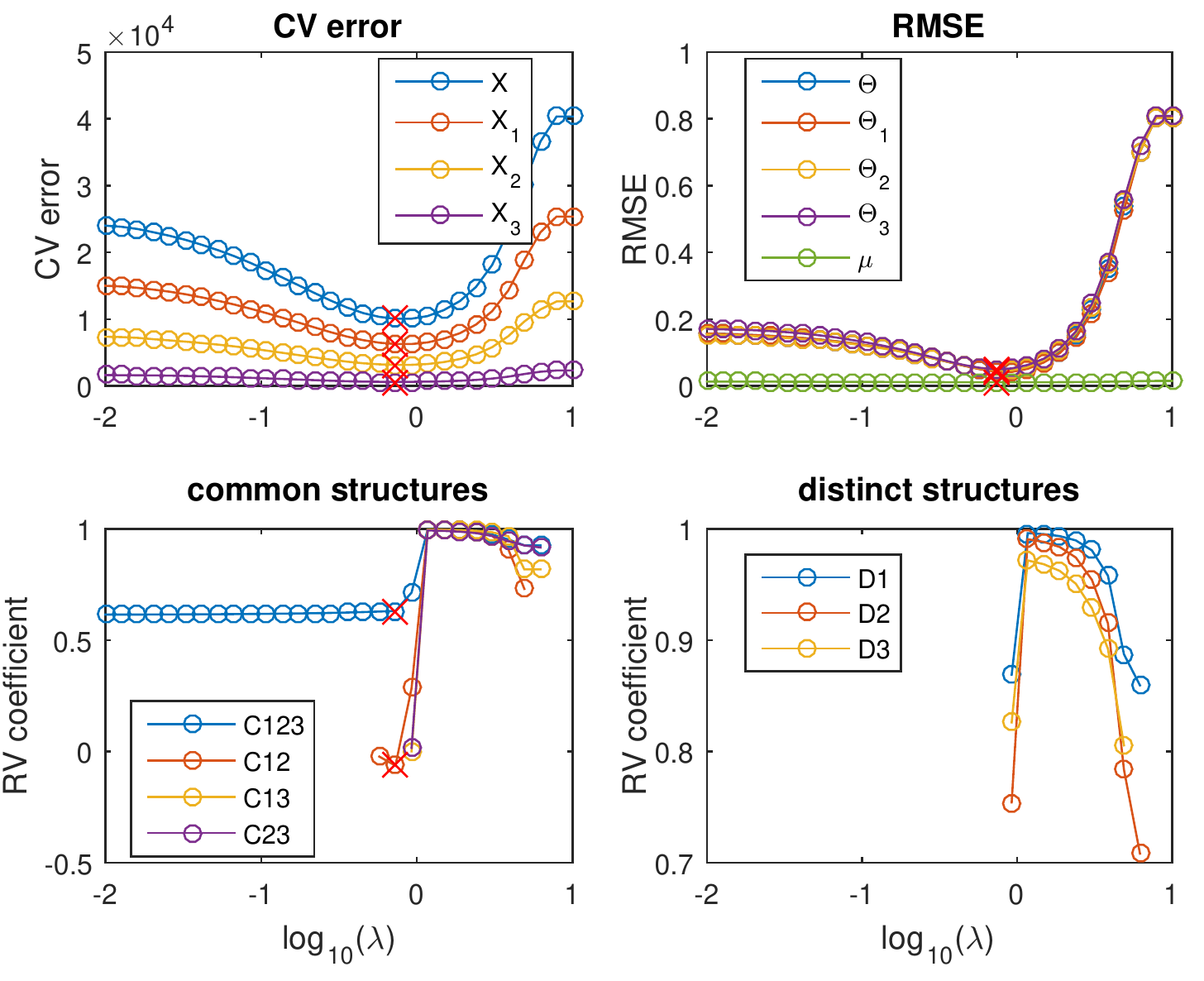}
    \caption*{Figure S5.5: CV errors (top left), RMSEs (top right) and the RV coefficients in estimating the common structures (bottom left), and distinct structures (bottom right) as a function of the regularization strength $\lambda$ when the P-ESCA model with a group lasso penalty is used. The red cross marker indicates the point corresponding to the minimum CV error.}
	\label{chapter5_fig:S5}
\end{figure}

\begin{figure}[htbp]
    \centering
    \includegraphics[width=0.9\textwidth]{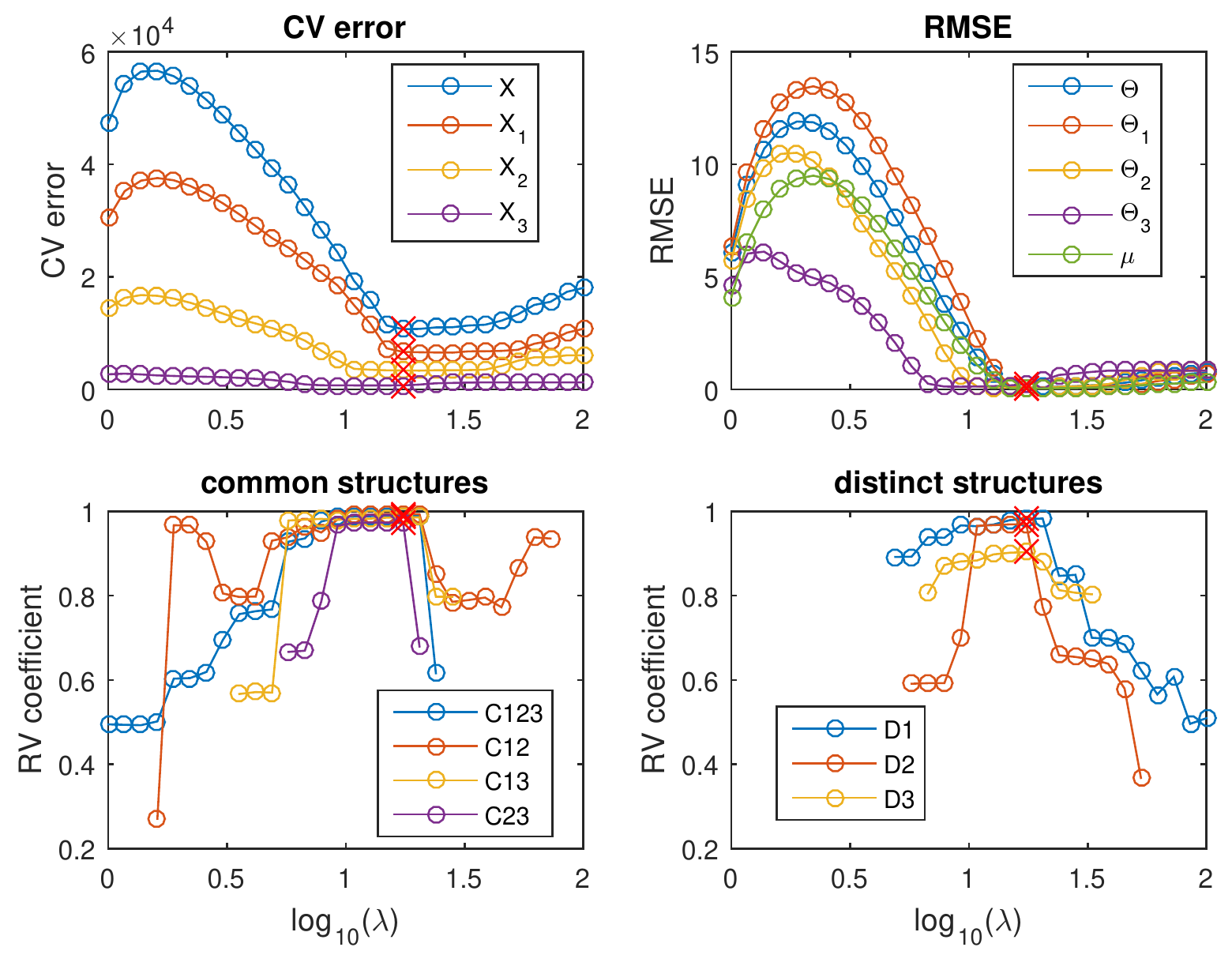}
    \caption*{Figure S5.6: CV errors (top left), RMSEs (top right) and the RV coefficients in estimating the common structures (bottom left), and distinct structures (bottom right) as a function of the regularization strength $\lambda$ for the P-ESCA model with a group GDP penalty on the simulated B-B-B data sets. The red cross marker indicates the point corresponding to the minimum CV error. The SNRs of global, local common and distinct structures in the B-B-B simulation are set to be 1. The reason for the increased CV errors at the early stage (top left) is that these models have not convergenced in 500 iterations.}
	\label{chapter5_fig:S6}
\end{figure}

\begin{figure}[htbp]
    \centering
    \includegraphics[width=0.9\textwidth]{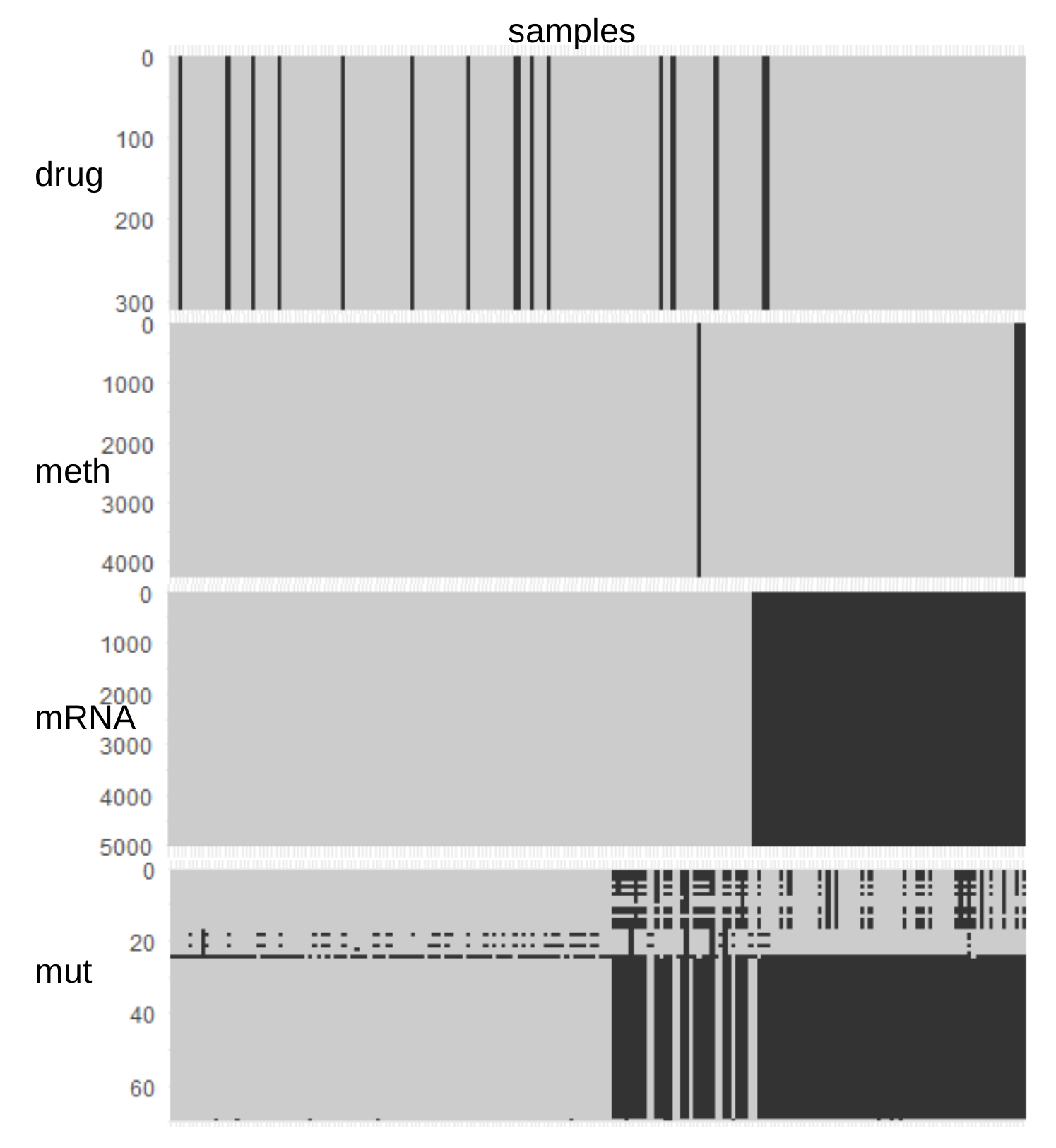}
    \caption*{Figure S5.7: Missing pattern of the CLL data sets. Black color indicates the data is missing, while gray color, the data is present. Drug: drug response data; meth: DNA methylation data; mRNA: transcriptome data; mut: mutation data.}
	\label{chapter5_fig:S7}
\end{figure}

\begin{figure}[htbp]
    \centering
    \includegraphics[width=0.9\textwidth]{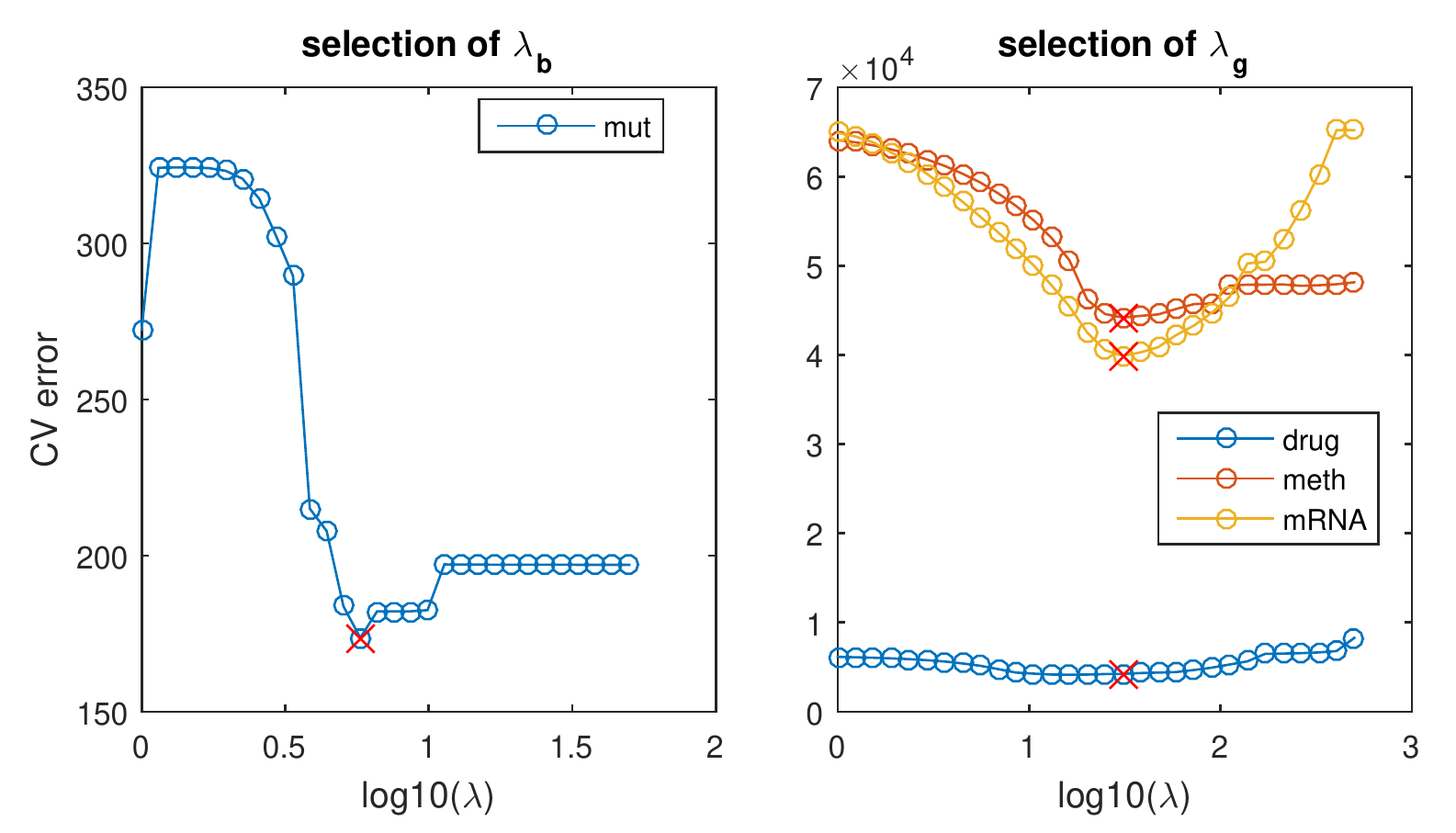}
    \caption*{Figure S5.8: Model selection of the P-ESCA model with a group GDP penalty on the CLL data sets. Drug: drug response data; meth: DNA methylation data; mRNA: transcriptome data; mut: mutation data. The red cross marker indicates the selected model.}
	\label{chapter5_fig:S8}
\end{figure}

\begin{figure}[htbp]
    \centering
    \includegraphics[width=0.9\textwidth]{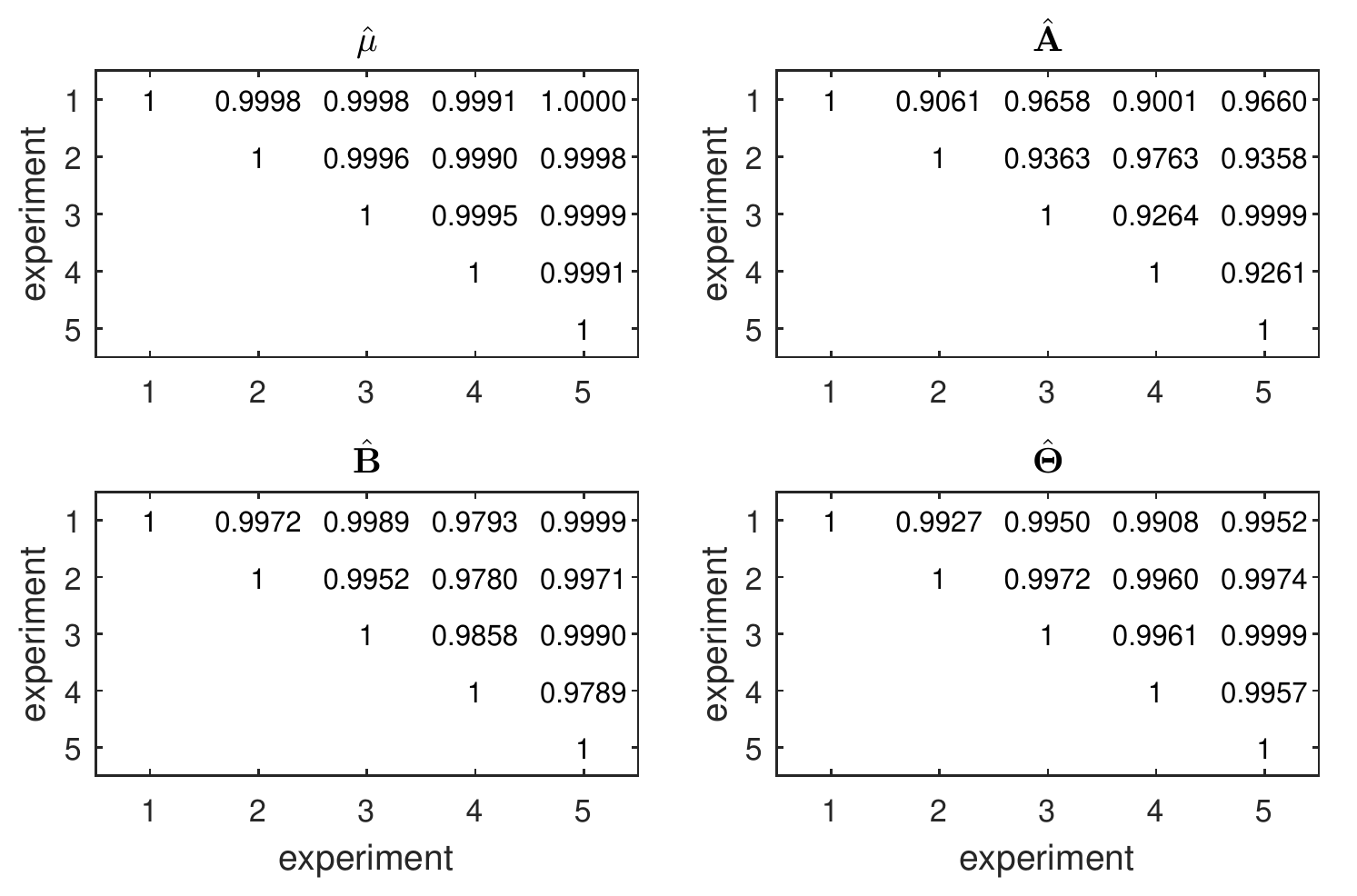}
    \caption*{Figure S5.9: The Pearson coefficient matrix for the 5 estimations of the $\hat{\bm{\mu}}$ and the RV coefficient matrices for the 5 estimations of the $\hat{\mathbf{A}}$, $\hat{\mathbf{B}}$ and $\hat{\mathbf{\Theta}}$ derived from the P-ESCA model.}
	\label{chapter5_fig:S9}
\end{figure}

\begin{figure}[htbp]
    \centering
    \includegraphics[width=0.9\textwidth]{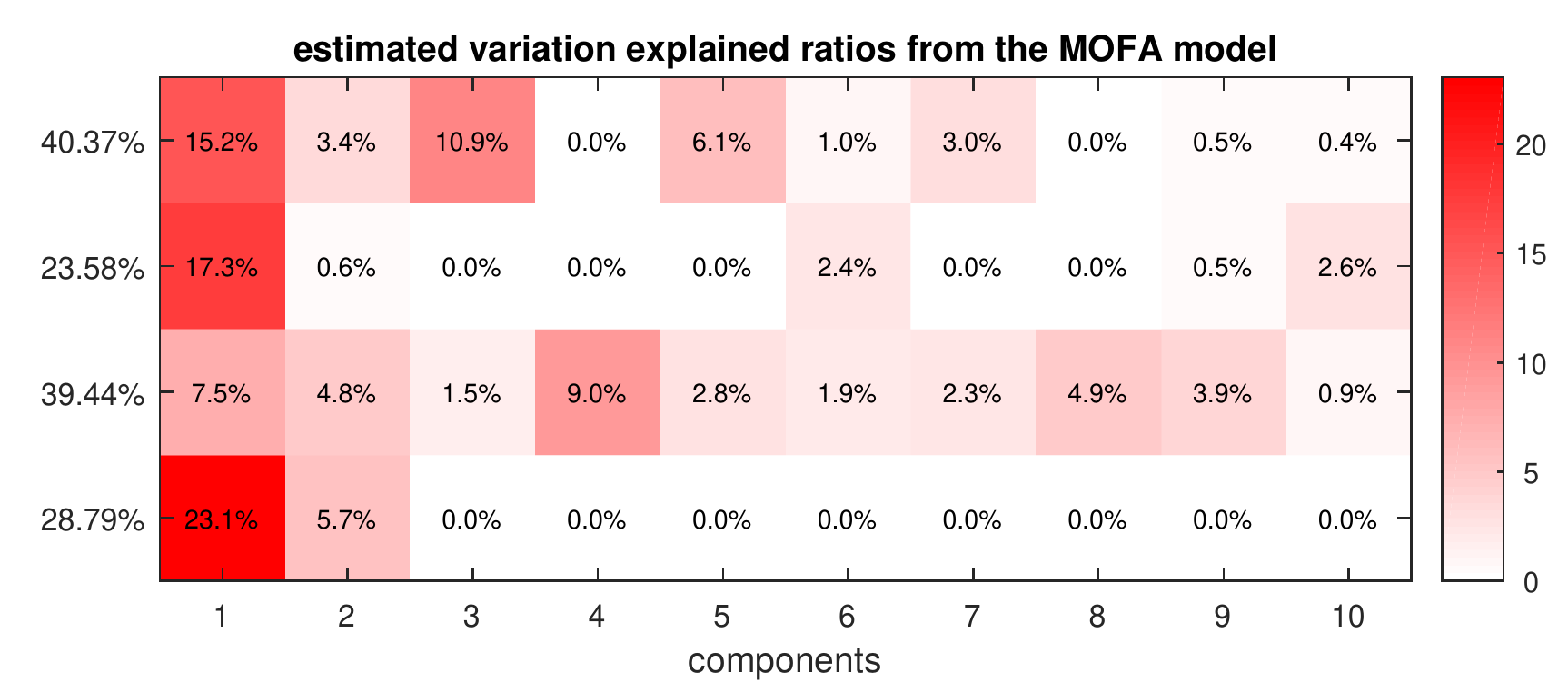}
    \caption*{Figure S5.10: Variation explained plots reproduced from the MOFA paper. From the top to the bottom, the data sets are drug response, methylation, mRNA and mutation}
	\label{chapter5_fig:S10}
\end{figure}

\chapter{Outlook} \label{chapter:6}
In the proposed penalized exponential family SCA (P-ESCA) model (Chapter \ref{chapter:5}), a group concave penalty is used to induce group-wise sparse pattern on the loading matrix to disentangle the global, local common and distinct components. The P-ESCA model and the associated MM algorithm can be further generalized to include other types of penalties to induce more interesting sparse patterns on the loading matrix, which may be useful for the data analyst. In addition, the currently developed model selection procedure has difficulties in tuning multiple tuning parameters. It will be worthwhile to explore other types of model selection approaches to address this issue. Also, in the current thesis, the parametric exponential family distribution is used to tackle the heterogeneous measurement scales. There also exist other possible non-parametric and semi-parametric approaches. It is worthwhile to generalize them for the data fusion of multiple data sets with the two types of heterogeneity. Furthermore, it is also interesting to generalize the P-ESCA model for prediction tasks or to taking into account the experimental design underlining the used multiple data sets.

\section{Including other types of sparse patterns}
The developed P-ESCA model and the associated MM algorithm have a lot of potential for further generalization. The options for inducing element-wise sparsity on the loading matrix or the composition of both group-wise and element-wise sparsity or other types of penalties can be easily included. Some examples will be shown in the following subsections. These P-ESCA model extensions can be selected using the developed missing value based CV procedure (Chapter \ref{chapter:5}). A possible alternative model selection approach is the Bayesian optimization \cite{frazier2018tutorial} framework, which is appropriate for the tuning of multiple (usually less than 20) continuous tuning parameters. Since this framework has been successfully applied in the automatic tuning of various machine learning algorithms \cite{snoek2012practical}, it will be worthwhile to explore its usage for the P-ESCA models with multiple tuning parameters.

The following notations are the same as the P-ESCA model in Chapter \ref{chapter:5}. Suppose the $r^{\text{th}}$ column of the $l^{\text{th}}$ loading matrix $\mathbf{B}_l$ is $\mathbf{b}_{l,r}$. The group concave penalty on the $l^{\text{th}}$ loading matrix $\mathbf{B}_l$ is imposed on the $L_2$ norm of $\mathbf{b}_{l,r}$ as $\lambda_l \sum_{r} g(||\mathbf{b}_{l,r}||_2)$, in which $g()$ is a concave function. This group concave penalty (or concave $L_2$ norm penalty) will shrink on the group level ($L_2$ norm of $\mathbf{b}_{l,r}$) as a concave penalty, and in the element level (elements inside $\mathbf{b}_{l,r}$) as a ridge regression type penalty. Sometimes, we may need other types of penalties, such as the element-wise sparsity on the elements of $\mathbf{B}_l$ or the composition of both element-wise and group-wise sparsity patterns on $\mathbf{B}_l$. All these options can be easily included into the P-ESCA model by a slightly modification of the developed MM algorithm.

\subsection{P-ESCA model with an element-wise concave penalty}
When a single data set is used, the P-ESCA model with an element-wise concave penalty is an approach for the sparse exponential family PCA model. When multiple data sets are used, the model is an approach for the sparse exponential family SCA model.

\subsubsection{Element-wise concave penalty}
An element-wise concave penalty can be imposed on the elements of $\mathbf{B}_l$ to induce the element-wise sparsity on $\mathbf{B}_l$. Suppose the $jr^{\text{th}}$ element of $\mathbf{B}_l$ is $b_{jr}^{l}$, and its absolute value is $\sigma_{ljr}$, $\sigma_{ljr} = |b_{jr}^{l}|$. The concave penalty on $\mathbf{B}_l$ can be expressed as $\lambda_l \sum_{j}^{J_l}\sum_{r}^{R} g(\sigma_{ljr})$, in which $g()$ is a concave function in Table \ref{chapter5_tab:1}. The optimization problem associated with this P-ESCA model with an element-wise concave penalty can be expressed as follows,
\begin{equation}\label{chapter6_eq:1}
\begin{aligned}
    \min_{ \left\{\bm{\mu}_l\right\}_{l}^{L}, \mathbf{A}, \left\{\mathbf{B}_l\right\}_{l}^{L}} \quad & \sum_{l=1}^{L} \Big[ -\log(p(\mathbf{X}_l|\mathbf{\Theta}_l, \alpha_l)) + \lambda_l \sum_{j}^{J_l}\sum_{r}^{R} g(\sigma_{ljr}) \Big] \\
    \text{subject to} \quad \mathbf{\Theta}_l &= \mathbf{1}\bm{\mu}_l^{\text{T}} + \mathbf{AB}_l^{\text{T}}, \quad l = 1,\ldots,L \\
     \mathbf{1}^{\text{T}}\mathbf{A} &= \mathbf{0}\\
	 \mathbf{A}^{\text{T}}\mathbf{A} &= \mathbf{I} \\
	 \sigma_{ljr} &= |b_{jr}^{l}|, l = 1...L; j=1,\dots,J_l; r = 1,\ldots, R.
\end{aligned}
\end{equation}

\subsubsection{Algorithm}
The algorithm developed in Section \ref{section:5.3} can be modified only with respect to the majorization step of the penalty function and the updating $\mathbf{B}_l$ step to fit the P-ESCA model with an element-wise penalty in equation \ref{chapter6_eq:1}. Similar to the equation \ref{chapter5_eq:5}, the element-wise penalty $g(\sigma_{ljr})$ is a concave function with respect to $\sigma_{ljr}$ and can be majorized as $g(\sigma_{ljr}) \leq \omega_{ljr}^k \sigma_{ljr} + c$, in which $\omega_{ljr}^k = \partial g(\sigma_{ljr}^k)$ and $\sigma_{ljr}^k$ is the absolute value of the $jr^{\text{th}}$ element of $\mathbf{B}_l^k$ (the $k^{\text{th}}$ approximation of $\mathbf{B}_l$ during the $k^{\text{th}}$ iteration). After majorizing the original problem in equation \ref{chapter6_eq:1}, updating the offset terms $\{\mu_l \}_1^L$ and score matrix $\mathbf{A}$ in exactly the same way as in Section \ref{section:5.3}, the optimization problem associated with the updating of $\mathbf{B}_l$ is
\begin{equation*}
\begin{aligned}
    \min_{\mathbf{B}_l} \quad & \frac{\rho_l}{2\alpha_l} ||\mathbf{AB}_l^{\text{T}} - \mathbf{JH}_{l}^{k}||_F^2  + \lambda_l \sum_{j}^{J_l}\sum_{r}^{R} \omega_{ljr}^k \sigma_{ljr} \\
	&= \frac{\rho_l}{2\alpha_l}||\mathbf{B}_l - (\mathbf{JH}_{l}^{k})^{\text{T}} \mathbf{A}||_F^2 + \lambda_l \sum_{j}^{J_l}\sum_{r}^{R} \omega_{ljr}^k \sigma_{ljr}\\
    &= \sum_{j}^{J_l} \sum_{r}^R \Big[ \frac{\rho_l}{2\alpha_l}(b_{jr}^l - ((\mathbf{JH}_{l}^{k})^{\text{T}} \mathbf{A})_{jr})^2 + \lambda_l \omega_{ljr}^k \sigma_{ljr} \Big]\\
    \sigma_{ljr} &= |b_{jr}^{l}|, l = 1...L; j=1,\dots,J_l; r = 1,\ldots, R,
\end{aligned}
\end{equation*}
in which $((\mathbf{JH}_{l}^{k})^{\text{T}} \mathbf{A})_{jr}$ indicates the $jr^{\text{th}}$ element of the matrix $(\mathbf{JH}_{l}^{k})^{\text{T}} \mathbf{A}$. The above optimization problem is simple the proximal operator of the $L_1$ norm, and the analytical solution exists \cite{parikh2014proximal}. Take $\tilde{\lambda}_{ljr} = \lambda_l \omega_{ljr}^k \alpha_l/\rho_l$ and $v_{ljr} = ((\mathbf{JH}_{l}^{k})^{\text{T}} \mathbf{A})_{jr}$, the analytical solution of $b_{jr}^{l}$ is $b_{jr}^{l} = \text{sign}(v_{ljr}) \max(0, |v_{ljr}| -\tilde{\lambda}_{ljr})$. To update the parameter $\mathbf{B}_l$, we can simply apply this proximal operator to all the elements of $\mathbf{B}_l$. The other parts of the algorithm are the same as in Section \ref{section:5.3}.

\subsection{P-ESCA model with a concave L1 norm penalty}
\subsubsection{Concave L1 norm penalty}
Another way to induce group sparsity on $\mathbf{B}_l$ is through the concave $L_1$ norm penalty \cite{huang2012selective}. Suppose the $r^{\text{th}}$ column of the $l^{\text{th}}$ loading matrix $\mathbf{B}_l$ is $\mathbf{b}_{l,r}$, and its $L_1$ norm is $\sigma_{lr}$, $\sigma_{lr} = ||\mathbf{b}_{l,r}||_1 = \sum_{j}^{J_l} |b_{jr}^l|$. The concave $L_1$ norm penalty on $\mathbf{B}_l$ can be expressed as $\lambda_l J_l \sum_{r}^{R} g(\sigma_{lr})$, in which weight $J_l$ is used to accommodate the potential different number of variables in different data set, and $g()$ is a concave function in Table \ref{chapter5_tab:1}. This concave $L_1$ norm penalty will shrink on the group level ($L_1$ norm of $\mathbf{b}_{l,r}$) as a concave penalty, and in the element level (elements inside $\mathbf{b}_{l,r}$) as a lasso type penalty. The optimization problem associated with this P-ESCA model with a concave $L_1$ norm penalty can be expressed as follows,
\begin{equation}\label{chapter6_eq:2}
\begin{aligned}
    \min_{ \left\{\bm{\mu}_l\right\}_{l}^{L}, \mathbf{A}, \left\{\mathbf{B}_l\right\}_{l}^{L}} \quad & \sum_{l=1}^{L} \Big[ -\log(p(\mathbf{X}_l|\mathbf{\Theta}_l, \alpha_l)) + \lambda_l J_l \sum_{r}^{R} g(\sigma_{lr}) \Big] \\
    \text{subject to} \quad \mathbf{\Theta}_l &= \mathbf{1}\bm{\mu}_l^{\text{T}} + \mathbf{AB}_l^{\text{T}}, \quad l = 1,\ldots,L \\
     \mathbf{1}^{\text{T}}\mathbf{A} &= \mathbf{0}\\
	 \mathbf{A}^{\text{T}}\mathbf{A} &= \mathbf{I} \\
	 \sigma_{lr} &= ||\mathbf{b}_{l,r}||_1, l = 1...L; r = 1,\ldots, R.
\end{aligned}
\end{equation}

\subsubsection{Algorithm}
The algorithm developed in Section \ref{section:5.3} can be modified only with respect to the majorization step of the penalty function and the updating $\mathbf{B}_l$ step to fit the P-ESCA model with the concave $L_1$ norm penalty in equation \ref{chapter6_eq:2}. Similar to the equation \ref{chapter5_eq:5}, the penalty function $g(\sigma_{lr})$ is concave with respect to $\sigma_{lr}$ and can be majorized as $g(\sigma_{lr}) \leq \omega_{lr}^k \sigma_{lr} + c$, in which $\omega_{lr}^k = \partial g(\sigma_{lr}^k)$ and $\sigma_{lr}^k$ is the $L_1$ norm of the $r^{\text{th}}$ column of $\mathbf{B}_l^k$ (the $k^{\text{th}}$ approximation of $\mathbf{B}_l$ during the $k^{\text{th}}$ iteration). After majorizing the original problem in equation \ref{chapter6_eq:2}, updating the offset terms $\{\mu_l \}_1^L$ and score matrix $\mathbf{A}$ in exactly the same way as in Section \ref{section:5.3}, the optimization problem associated with the updating of $\mathbf{B}_l$ is
\begin{equation*}
\begin{aligned}
    \min_{\mathbf{B}_l} \quad & \frac{\rho_l}{2\alpha_l} ||\mathbf{AB}_l^{\text{T}} - \mathbf{JH}_{l}^{k}||_F^2  + \lambda_l J_l \sum_{r}^{R} \omega_{lr}^k \sigma_{lr} \\
	&= \frac{\rho_l}{2\alpha_l}||\mathbf{B}_l - (\mathbf{JH}_{l}^{k})^{\text{T}} \mathbf{A}||_F^2 + \lambda_l J_l \sum_{r}^{R} \omega_{lr}^k (\sum_{j}^{J_l} |b_{jr}^l|)\\
    &= \sum_{j}^{J_l} \sum_{r}^R \Big[ \frac{\rho_l}{2\alpha_l}(b_{jr}^l - ((\mathbf{JH}_{l}^{k})^{\text{T}} \mathbf{A})_{jr})^2 + \lambda_l J_l \omega_{lr}^k |b_{jr}^l| \Big].
\end{aligned}
\end{equation*}
Take $\tilde{\lambda}_{ljr} = \lambda_l J_l \omega_{lr}^k \alpha_l/\rho_l$ and $v_{ljr} = ((\mathbf{JH}_{l}^{k})^{\text{T}} \mathbf{A})_{jr}$, the analytical solution of $b_{jr}^{l}$ is $b_{jr}^{l} = \text{sign}(v_{ljr}) \max(0, |v_{ljr}| -\tilde{\lambda}_{ljr})$. To update the parameter $\mathbf{B}_l$, we can simply apply this proximal operator to all the elements of $\mathbf{B}_l$. The other parts of the algorithm are the same as in Section \ref{section:5.3}.

\subsection{P-ESCA model with a composite concave penalty}
\subsubsection{Composite concave penalty}
There also exists a composite concave penalty to induce both group and element-wise sparsity \cite{huang2012selective}. The composite concave penalty on $\mathbf{B}_l$ can be expressed as $\lambda_l J_l \sum_{r}^{R} g_{out}(\sum_{j}^{J_l} g_{inner}(|b_{jr}^l|))$, in which weight $J_l$ is used to accommodate the potential different number of variables in different data set, $g_{out}()$ and $g_{inner}()$ are two concave functions for the group level and element level respectively. We will use the same concave function $g()$ from Table \ref{chapter5_tab:1} for both $g_{out}()$ and $g_{inner}()$. This composite concave penalty will shrink both on the group level ($\sum_{j}^{J_l} g_{inner}(|b_{jr}^l|)$) and in the element level (elements inside $\mathbf{b}_{l,r}$) as a concave penalty. The optimization problem associated with this P-ESCA model with a composite concave penalty can be expressed as follows,
\begin{equation}\label{chapter6_eq:3}
\begin{aligned}
    \min_{ \left\{\bm{\mu}_l\right\}_{l}^{L}, \mathbf{A}, \left\{\mathbf{B}_l\right\}_{l}^{L}} \quad & \sum_{l=1}^{L} \Big[ -\log(p(\mathbf{X}_l|\mathbf{\Theta}_l, \alpha_l)) + \lambda_l J_l \sum_{r}^{R} g_{out}(\sum_{j}^{J_l} g_{inner}(|b_{jr}^l|)) \Big] \\
    \text{subject to} \quad \mathbf{\Theta}_l &= \mathbf{1}\bm{\mu}_l^{\text{T}} + \mathbf{AB}_l^{\text{T}}, \quad l = 1,\ldots,L \\
     \mathbf{1}^{\text{T}}\mathbf{A} &= \mathbf{0}\\
	 \mathbf{A}^{\text{T}}\mathbf{A} &= \mathbf{I}. \\
\end{aligned}
\end{equation}

\subsubsection{Algorithm}
The algorithm developed in Section \ref{section:5.3} can be modified only with respect to the majorization step of the penalty function and the updating $\mathbf{B}_l$ step to fit the P-ESCA model with a composite concave penalty in equation \ref{chapter6_eq:3}. Here we take $\sigma_{lr} = \sum_{j}^{J_l} g_{inner}(|b_{jr}^l|)$ and $\sigma_{ljr} = |b_{jr}^l|$. Since both $g_{out}(\sigma_{lr})$ and $g_{inner}(\sigma_{ljr})$ are concave function and they are monotonically non-decreasing, their composition is also a concave function with respect to $\sigma_{ljr}$. Therefore, we can majorize the composite function $g_{out}(\sum_{j}^{J_l} g_{inner}(\sigma_{ljr}))$ in a similar way  as the equation \ref{chapter5_eq:5}, $g_{out}(\sum_{j}^{J_l} g_{inner}(\sigma_{ljr})) \leq \sum_{j}^{J_l} \omega_{ljr}^k \sigma_{ljr} + c$, in which $\omega_{ljr}^k = \partial g_{out}(\sigma_{lr}^k) \partial g_{inner}(\sigma_{ljr}^k)$ and $\sigma_{ljr}^k$ is the absolute value of the $jr^{\text{th}}$ element of $\mathbf{B}_l^k$ (the $k^{\text{th}}$ approximation of $\mathbf{B}_l$ during the $k^{\text{th}}$ iteration), $\sigma_{lr}^k = \sum_{j}^{J_l} g_{inner}(\sigma_{ljr}^k)$. After majorizing the original problem in equation \ref{chapter6_eq:3}, updating the offset terms $\{\mu_l \}_1^L$ and score matrix $\mathbf{A}$ in exactly the same way as in Section \ref{section:5.3}, the optimization problem associated with the updating of $\mathbf{B}_l$ is
\begin{equation*}
\begin{aligned}
    \min_{\mathbf{B}_l} \quad & \frac{\rho_l}{2\alpha_l} ||\mathbf{AB}_l^{\text{T}} - \mathbf{JH}_{l}^{k}||_F^2  + \lambda_l J_l \sum_{r}^{R} (\sum_{j}^{J_l} \omega_{ljr}^k \sigma_{ljr}) \\
	&= \frac{\rho_l}{2\alpha_l}||\mathbf{B}_l - (\mathbf{JH}_{l}^{k})^{\text{T}} \mathbf{A}||_F^2 + \lambda_l J_l \sum_{r}^{R} (\sum_{j}^{J_l} \omega_{ljr}^k \sigma_{ljr})\\
    &= \sum_{j}^{J_l} \sum_{r}^R \Big[ \frac{\rho_l}{2\alpha_l}(b_{jr}^l - ((\mathbf{JH}_{l}^{k})^{\text{T}} \mathbf{A})_{jr})^2 + \lambda_l J_l \omega_{ljr}^k \sigma_{ljr} \Big],
\end{aligned}
\end{equation*}
Take $\tilde{\lambda}_{ljr} = \lambda_l J_l \omega_{ljr}^k \alpha_l/\rho_l$ and $v_{ljr} = ((\mathbf{JH}_{l}^{k})^{\text{T}} \mathbf{A})_{jr}$, the analytical solution of $b_{jr}^{l}$ is $b_{jr}^{l} = \text{sign}(v_{ljr}) \max(0, |v_{ljr}| -\tilde{\lambda}_{ljr})$. To update the parameter $\mathbf{B}_l$, we can simply apply this proximal operator to all the elements of $\mathbf{B}_l$. The other parts of the algorithm are the same as in Section \ref{section:5.3}.

\subsection{P-ESCA model with other types of penalties}
All the algorithms for the above P-ESCA models with different penalties are based on the fact that the updating of $\mathbf{B}_l$ in equation \ref{chapter5_eq:7} can be re-expressed as a problem of finding the proximal operator for the $L_2$ norm or the $L_1$ norm penalty. Therefore, P-ESCA model can also be extended to include other types of penalty whose proximal operator has a simple or analytical solution. For example, there is no difficulty in including concave penalties on the rows of the loading matrix $\mathbf{B}_l$ to induce row-wise sparsity, which could be useful for the feature selection. Furthermore, we can also add cardinality constraints (pseudo $L_{0}$ norm) on the number of nonzero elements, the number of nonzero rows, or the number of nonzero columns of the loading matrix $\mathbf{B}_l$ to induce the desired sparsity pattern. These various $L_0$ norm penalties are non-convex, however, there are heuristic solutions for the corresponding proximal operator \cite{boyd2011distributed}. These various $L_0$ penalties can be useful if all our data sets are quantitative. However, when discrete data sets are used, the derived model with the $L_0$ norm type of penalty will have problems in constraining the scale of estimated parameters. The standard logistic PCA model, in which the exact low rank constraint can be regarded as applying $L_0$ norm penalty on the singular values, is a good example to illustrate this point.

\section{Other directions of tackling heterogeneous measurement scales}
In the current thesis, the heterogeneous measurement scales are accounted for by assuming a parametric exponential family distribution in a similar way as the generalized linear models. There also existed other possible directions \cite{kiers1989three, anderson2018xpca, de2009gifi} to tackle the problems induced by the heterogeneous measurement scales. One promising alternative is the semi-parametric XPCA method \cite{anderson2018xpca}. In the probabilistic interpretation of a PCA model on a matrix $\mathbf{X}$($I\times J$), we assume we have $I$ samples from a $J$ dimensional multivariate normal distribution and therefore normal marginal distribution for each column. On the contrary, XPCA model is based on a semi-parametric $J$ dimensional multivariate distribution, which is the combination of nonparametric marginals of all the $J$ quantitative or discrete columns and a Gaussian copula. The assumptions of parametric marginal distributions (normal distribution for quantitative data, Bernoulli distribution for binary data) for the columns of the observed data set $\mathbf{X}$ are relaxed in the XPCA model. Therefore, when the exponential family distribution is not a good approximation of the observed data, for example, the empirical distribution of a quantitative variable is far from symmetric, XPCA model has a clear advantage. Another interesting alternative is the non-parametric representation matrices approach \cite{kiers1989three}, in which each variable (continuous or discrete) is represented by a representation matrix and the resulting representation matrices can be used in a three way model for symmetric data. The advantage of the representation matrices approach is that no probabilistic assumption is made for the model.

\section{Using data fusion for supervised learning}
All the methods developed in this thesis are unsupervised learning approaches. It is worthwhile to extend these methods in the supervised learning framework for prediction tasks. A simple approach, same as the extension of PCA to principal component regression model for prediction tasks, is as follows. These various unsupervised data fusion methods are taken as feature extraction approaches for multiple data sets. The derived low dimensional score matrix can be regarded as the extracted features and can be used as inputs for any other supervised learning methods. However, the extracted low dimensional features are not necessarily optimal for the prediction tasks. Therefore, when label information is available, it is better to make full use of it to make the extracted features more informative to the prediction tasks. The P-ESCA model can be extended from this perspective in a similar way as extending the PCA model to the partial least squares regression model \cite{geladi1986partial}. The extracted low dimensional structures from the P-ESCA model should not only represent the multiple data sets well but also have high covariance with the label information.

\section{Incorporating the information of experimental design}
Sometimes, the multiple sets of measurements on the same objects result from carefully designed experimental studies rather than observational studies. Such an experimental design always contains several factors, such as different treatments or different time points or their combinations, which are of interest with respect to the research question. Therefore, these experimental factors are underlying the multiple data sets on the same objects. To study the effects of these experimental factors or to remove their effects on the explorative data analysis, the used data fusion approaches should take the experimental design structure into account. The proposed P-ESCA can be extended from this direction by including extra low dimensional structures to account for these experimental factors in a similar way as the ANOVA-simultaneous component analysis (ASCA) model \cite{smilde2005anova}.

\summary
Multiple high dimensional measurements from different platforms on the same biological system are becoming increasingly common in biological research. These different sources of measurements not only provide us with the opportunity of a deeper understanding of the studied system, but they also introduce some new statistical challenges. All these challenges are related to the heterogeneity of the data sets. The first type of heterogeneity is the type of \emph{data}, such as metabolomics, proteomics and RNAseq data in genomics. These different omics data reflect the properties of the studied biological system from different perspectives. The second type of heterogeneity is the type of \emph{scale}, which indicates the measurements are obtained at different scales, such as binary, ordinal, interval and ratio-scaled variables. Within this thesis, various data fusion approaches are developed to tackle either one or two types of heterogeneity that exist in multiple data sets.

In Chapter \ref{chapter:2}, we reviewed and compared various parametric and nonparametric extensions of principal component analysis (PCA) specifically geared for binary data. The special mathematical characteristics of binary data are taken into account from different perspectives in these different extensions of PCA. We explored their performance with respect to finding the correct number of components, overfitting, retrieving the correct low dimensional structure, variable importance, etc, using both realistic simulations of binary data as well as mutation, copy number aberrations (CNA) and methylation data of the GDSC1000 project. Our results indicate that if a low dimensional structure exists in the data, most of the methods can find it. We recommend to use the parametric logistic PCA model (projection based approach) if the probabilistic generating process can be assumed underlying the data, and to use the nonparametric Gifi model if such an assumption is not valid and the data is considered as given.

In Chapter \ref{chapter:3}, we developed a robust logistic PCA model via non-convex singular value thresholding. The promising logistic PCA model for binary data has an overfitting issue because of the used exact low rank constraint. We proposed to fit a logistic PCA model via non-convex singular value thresholding to alleviate the overfitting issue. An efficient majorization-minimization (MM) algorithm is implemented to fit the model and a missing value based cross validation (CV) procedure is introduced for the model selection. Furthermore, we re-expressed the logistic PCA model based on the latent variable interpretation of the generalized linear models (GLMs) on binary data. The latent variable interpretation of the logistic PCA model not only makes the assumption of low rank structure easier to understand, but also provides us a way to define signal to noise ratio (SNR) in the simulation of multivariate binary data. Our experiments on realistic simulations of imbalanced binary data and low SNR show that the CV error based model selection procedure is successful in selecting the proposed model. And the selected model demonstrates superior performance in recovering the underlying low rank structure compared to models with exact low rank constraint and convex nuclear norm penalty.

In the Chapter \ref{chapter:4}, we developed a generalized simultaneous component analysis (GSCA) model for the data fusion of binary and quantitative data sets. Simultaneous component analysis (SCA) model is one of the standard tools for exploring the underlying dependence structure present in multiple quantitative data sets measured on the same objects. However, it does not have any provisions when a part of the data are binary. To this end, we propose the GSCA model, which takes into account the distinct mathematical properties of binary and quantitative measurements in the maximum likelihood framework. In the same way as in the SCA model, a common low dimensional subspace is assumed to represent the shared information between these two distinct types of measurements. However, the GSCA model can easily be overfitted when a rank larger than one is used, which can lead to the problem that some of the estimated parameters can become very large. To achieve a low rank solution and combat overfitting, we propose to use non-convex singular value thresholding. An efficient majorization algorithm is developed to fit this model with different concave penalties. Realistic simulations (low SNR and highly imbalanced binary data) are used to evaluate the performance of the proposed model in recovering the underlying structure. Also, a missing value based CV procedure is implemented for the model selection. We illustrate the usefulness of the GSCA model for exploratory data analysis of quantitative gene expression and binary CNA measurements obtained from the GDSC1000 data sets.

In Chapter \ref{chapter:5}, we proposed a penalized exponential family SCA (P-ESCA) model for the data fusion of multiple data sets with two types of heterogeneity. Multiple sets of measurements on the same objects obtained from different platforms may reflect partially complementary information of the studied system. However, the heterogeneity of such data sets introduces some new statistical challenges for their data fusion. First, the separation of information that is common across all or some of the data sets, and the information that is specific to each data set is problematic. Furthermore, these data sets are often a mix of quantitative and discrete (binary or categorical) data types, while commonly used data fusion methods require all data sets to be quantitative. Therefore, we proposed an exponential family simultaneous component analysis (ESCA) model to tackle the potential mixed data types problem of multiple data sets. In addition, a structured sparse pattern of the loading matrix is induced through a nearly unbiased group concave penalty to disentangle the global, local common and distinct information of the multiple data sets. An efficient MM algorithm is derived to fit the proposed model. Analytic solutions are derived for updating all the parameters of the model in each iteration, and the algorithm will decrease the objective function in each iteration monotonically. For model selection, a missing value based CV procedure is implemented. The advantages of the proposed method in comparison with other approaches are assessed using comprehensive simulations as well as the analysis of real data from a chronic lymphocytic leukaemia (CLL) study.

In Chapter \ref{chapter:6}, we considered various extensions of the developed P-ESCA model with respect to new penalties (element-wise, group-wise and their composition) and new model selection approach. Also, we remarked the potential of the semi-parametric XPCA model and non-parametric representation matrices approach in tackling the data sets of heterogeneous measurement scales. Furthermore, it is also interesting to generalize the P-ESCA model for prediction tasks or to tacking into account the experimental design underlining the used multiple data sets.

\samenvatting
In biologisch onderzoek wordt het steeds gebruikelijker meerdere hoog-di\-men\-sionale metingen op verschillende platformen aan hetzelfde biologische systeem uit te voeren. Deze van verschillende bronnen afkomstige metingen bieden de mogelijkheid tot een beter begrip van het bestudeerde systeem, maar brengen ook nieuwe statistische uitdagingen met zich mee. Al deze uitdagingen houden verband met de heterogeniteit van de dataverzamelingen. De eerste vorm van heterogeniteit ligt in het type van de gegevens. Zo zijn er verschillende typen omics-data, metabolomics, proteomics en RNAseq data in genomics, die ieder een eigen perspectief op de eigenschappen van het biologische systeem bieden. De tweede vorm van heterogeniteit ligt in de meetschaal van de data. De data kunnen op verschillende schalen gemeten worden, zoals binair, ordinale schaal, intervalschaal en ratioschaal. In dit proefschrift worden verschillende datafusiemethoden ontwikkeld waarmee één of beide soorten dataheterogeniteit aangepakt kunnen worden.

In hoofdstuk \ref{chapter:2} wordt een aantal, zowel parametrische als niet-para\-me\-tri\-sche, uitbreidingen van principale componenten analyse (PCA) vergeleken die specifiek betrekking hebben op binaire data. In deze uitbreidingen van PCA wordt op verschillende manieren rekening gehouden met de speciale wiskundige karakteristieken van binaire gegevens. We onderzochten de prestaties van deze uitbreidingen van PCA met betrekking tot het vinden van het juiste aantal componenten, overfitting, het vinden van de juiste laag-dimensionale structuur, het belang van variabelen, enz. door gebruik van zowel realistische simulaties van binaire data als van mutatiedata, `copy number aberrations' (CNA) en methylatiedata van het GDSC1000 project. Onze resultaten laten zien dat als er een laag-dimensionale structuur in de data aanwezig is, de meeste methoden deze kunnen vinden. Wij adviseren om het parametrisch logistisch PCA-model (op projectie gebaseerde benadering) te gebruiken als verondersteld wordt dat een stochastisch proces aan de data ten grondslag ligt. Als dit niet het geval is en de data als vast gegeven kan worden beschouwd, raden we aan het niet-parametrische Gifi-model te gebruiken.

In hoofdstuk \ref{chapter:3} hebben we een robuust logistisch PCA-model ontwikkeld met behulp van een niet-convexe drempelwaardenfunctie voor de singuliere waarden. Het veelbelovende logistische PCA-model voor binaire data heeft een probleem met over-fitting vanwege de gebruikte randvoorwaarde van een  exacte lage rang. We stellen voor om het over-fitten te verminderen door het logistische PCA-model te fitten met gebruik van een niet-convexe drempelwaardenfunctie voor singuliere waarden. Een efficiënt majorisatie-minimalisatie (MM) algoritme is geïmplementeerd om het model te fitten en een op missende waarden gebaseerde kruisvalidatie (KV) procedure is geïntroduceerd voor modelselectie. Bovendien hebben we het logistische PCA-model uitgedrukt op basis van de latente variabelen interpretatie van gegeneraliseerde lineaire modellen (GLMs). Niet alleen maakt de aanname van een structuur met lage rang het model beter te begrijpen, maar biedt ook een manier om de signaal-ruis verhouding in de simulatie van multivariate binaire data te definiëren. Onze experimenten met realistische simulaties van ongebalanceerde binaire data met een lage signaal-ruis verhouding laten zien dat modelselectie gebaseerd op KV-fouten goed in staat is het voorgestelde model te selecteren. Dit geselecteerde model is uitstekend in staat de onderliggende lage-rang structuur terug te vinden en werkt beter dan modellen met een exacte lage-rang randvoorwaarde die een convexe spoornormboete gebruiken.

In hoofdstuk \ref{chapter:4} ontwikkelden we een gegeneraliseerd simultaan componenten analyse (GSCA) model voor de fusie van binaire en kwantitatieve dataverzamelingen. Simultane componenten analyse (SCA) is \'e\'en van de standaard hulpmiddelen om de onderliggende afhankelijkheidsstructuur te onderzoeken die aanwezig is in meerdere kwantitatieve data sets die aan hetzelfde object gemeten zijn. Echter, SCA is niet geschikt als een deel van de data binair is. Daarom stellen we een GSCA-model voor dat rekening houdt met de specifieke mathematische eigenschappen van binaire data en kwantitatieve metingen binnen het kader van grootste aannemelijkheid. Op dezelfde manier als voor het SCA-model veronderstellen we het bestaan van een laag-dimensionale deelruimte waarin de gedeelde informatie van de twee typen van metingen wordt gerepresenteerd. Het GSCA-model is evenwel geneigd tot overfitten wanneer een rang groter dan 1 wordt gebruikt. Hierdoor kunnen sommige parameters bijzonder groot geschat worden. Om een oplossing met lage rang zonder overfitting te vinden, stellen we voor een niet-convexe drempelwaardefunctie te gebruiken voor de selectie van singuliere waarden. We ontwikkelden een effici\"ent majorisatie algoritme om dit model te fitten voor verschillende concave boetefuncties. Realistische simulaties (lage signaal-ruis verhouding en sterk ongebalanceerde binaire data) werden gebruikt om te beoordelen hoe goed het model de onderliggende structuur kan reproduceren. Ook is een op missende waarden gebaseerde kruisvalidatie geimplementeerd om modellen te selecteren. De bruikbaarheid van het GSCA-model als exploratieve tool wordt gedemonstreerd aan de hand van kwantitatieve genexpressiedata en binaire CNA-metingen uit de GDSC1000 dataverzameling.

In hoofdstuk \ref{chapter:5} stellen we een exponentieel SCA-model met boeteoptie (P-ESCA) voor om meerdere dataverzamelingen met twee typen heterogeniteit samen te voegen. Meerdere metingen aan hetzelfde object maar uitgevoerd op verschillende platforms kunnen complementaire informatie over het bestudeerde systeem opleveren. De heterogeniteit van deze data biedt interessante nieuwe statistische uitdagingen als de dataverzamelingen gefuseerd worden. Ten eerste is de scheiding van informatie die gemeenschappelijk is voor alle (of enkele) van de dataverzamelingen van de informatie die specifiek is voor iedere dataverzameling afzonderlijk, lastig. Bovendien zijn deze dataverzamelingen vaak een mix van kwantitatieve en discrete (binair of categorisch) data typen, terwijl gebruikelijke datafusiemethoden vereisen dat alle dataverzamelingen kwantitatief zijn. Met het door ons voorgestelde exponenti\"ele simultane componenten analyse (ESCA) model kunnen we dergelijke gemengde dataverzamelingen wel analyseren. Om de globale, lokaal gemeenschappelijke en verzameling-specifieke informatie in de verschillende dataverzamelingen te ontwarren, hebben we op de componentenladingsmatrix, via een groep van concave boetefuncties bijna zonder systematische fout, een gestructureerd, bijna leeg patroon opgelegd. Om het voorgestelde model te fitten hebben we een algoritme gebaseerd op majorisatie-minimalisatie gemaakt. Dit algoritme gebruikt analytische oplossingen om de modelparameters na iedere iteratie te actualiseren; de doelfunctie wordt door het algoritme iedere iteratie monotoon verminderd. Voor de modelselectie gebruiken we een op missende waarden gebaseerde kruisvalidatie. De voordelen van de voorgestelde methode in vergelijking met andere methoden werden beoordeeld met uitgebreide simulaties en de analyse van echte data uit een chronische lymfatische leukemie (CCL) studie.

In hoofdstuk \ref{chapter:6} beschouwen we verschillende uitbreidingen van het ontwikkelde P-ESCA-model met nieuwe boetesystemen (per element, per groep en element-groep samenstelling) en een nieuwe benadering voor modelselectie. We kijken ook naar de mogelijkheden die het semi-parametrische XPCA-model en de niet-parametrische matrixmethode bieden om data verzamelingen met heterogene meetschalen aan te pakken.

\acknowledgments
My four years of Ph.D. life in Amsterdam is difficult but productive. In the beginning, I didn't have much confidence in myself for the transition from an experimentalist to a data analyst (statistician). However, during this process, I learned a lot on statistics, and have developed several statistical methods for the data fusion of heterogeneous data sets. These cannot be done without the help and support of a lot of people. Here I would like to express my gratitude to you.

First, I would like to say thanks to Age Smilde and Johan Westerhuis for your four years' supervision. I appreciate that you accepted me to do my Ph.D. research in the BDA group. And thanks so much for all the discussions we have during the last four years and your insightful comments on my work. These helped me a lot in learning statistics and doing research.

I would also like to say thanks to Huub Hoefsloot and Age Smilde for the matrix algebra course. Matrix algebra is the foundation of statistics and numerical computation. A solid understanding of this subject helps me a lot on the algorithm development for my research. Also, I would like to acknowledge Gooitzen Zwanenburg for your help in solving my problems on the computer, software, Internet, the Dutch translation of the thesis summary, and many other things. Furthermore, I want to thank Dicle, Chloie, Sandra, and Maryam. I am glad to be in the same office as you. Thanks for the discussion we have about life, research and many other topics. Also, thanks for all the members in the BDA group for the coffee break conversations, the presentations, and discussions during the group meetings.

Part of my Ph.D. research has collaborated with Lodewyk Wessels, Nanne Aben from the Netherlands Cancer Institute (NKI) and Patrick J.F. Groenen from the Erasmus University. Thanks to Lodewyk Wessels and Nanne Aben for your real biological data sets, your informative comments on our paper and all the biological interpretations. Also, thanks to Patrick for your enlightening comments on the algorithm section of our paper and for your work on the GDP penalty, which is used throughout my thesis.

I would also like to acknowledge the China Scholarship Council for financial support. The scholarship makes it possible for me to study in Amsterdam.

Lastly, I would like to thank my family and my girlfriend for your love and encouragement.

\vskip 4em
\hfill Yipeng Song
\vskip 1em
\hfill May 21, 2019

\bibliographystyle{ieeetr}
\bibliography{reference}

\begin{thebibliography}{100}

\bibitem{zimmermann2017integration}
M.~Zimmermann, M.~Kogadeeva, M.~Gengenbacher, G.~McEwen, H.-J. Mollenkopf,
  N.~Zamboni, S.~H.~E. Kaufmann, and U.~Sauer, ``Integration of metabolomics
  and transcriptomics reveals a complex diet of mycobacterium tuberculosis
  during early macrophage infection,'' {\em mSystems}, vol.~2, no.~4,
  pp.~e00057--17, 2017.

\bibitem{steinmetz1999methodology}
V.~Steinmetz, F.~Sevila, and V.~Bellon-Maurel, ``A methodology for sensor
  fusion design: Application to fruit quality assessment,'' {\em Journal of
  Agricultural Engineering Research}, vol.~74, no.~1, pp.~21--31, 1999.

\bibitem{doeswijk2011increase}
T.~Doeswijk, A.~Smilde, J.~Hageman, J.~Westerhuis, and F.~Van~Eeuwijk, ``On the
  increase of predictive performance with high-level data fusion,'' {\em
  Analytica Chimica Acta}, vol.~705, no.~1-2, pp.~41--47, 2011.

\bibitem{Stevens1946}
S.~Stevens, ``On the theory of scales of measurement,'' {\em Science},
  vol.~103, pp.~677--680, June 1946.

\bibitem{Suppes1962}
P.~Suppes and J.~Zinnes, ``Basic measurement theory,'' Psychology Series~45,
  Stanford University, Institute for Mathematical Studies in the Social
  Sciences, March 1962.

\bibitem{Krantz1971}
D.~Krantz, R.~Luce, P.~Suppes, and A.~Tversky, {\em Foundations of Measurement
  (Volume I)}.
\newblock Dover, 1971.

\bibitem{Narens1981}
L.~Narens, ``On the scales of measurement,'' {\em Journal of Mathematical
  Psychology}, vol.~24, no.~3, pp.~249--275, 1981.

\bibitem{Narens1986}
L.~Narens and R.~D. Luce, ``Measurement - the theory of numerical
  assignments,'' {\em Psychological Bulletin}, vol.~99, pp.~166--180, Mar.
  1986.

\bibitem{Luce1987}
R.~D. Luce and L.~Narens, ``Measurement scales on the continuum,'' {\em
  Science}, vol.~236, no.~4808, pp.~1527--1532, 1987.

\bibitem{Hand1996}
D.~J. Hand, ``Statistics and the theory of measurement,'' {\em Journal of the
  Royal Statistical Society Series A-statistics in Society}, vol.~159,
  pp.~445--473, 1996.

\bibitem{Adams1965}
E.~Adams, R.~Fagot, and R.~Robinson, ``A theory of appropriate statistics,''
  {\em Psychometrika}, vol.~30, no.~2, pp.~99--127, 1965.

\bibitem{Michell1986}
J.~Michell, ``Measurement scales and statistics - a clash of paradigms,'' {\em
  Psychological Bulletin}, vol.~100, no.~3, pp.~398--407, 1986.

\bibitem{agresti2013categorical}
A.~Agresti, {\em Categorical data analysis}.
\newblock John Wiley \& Sons, 2013.

\bibitem{smilde2017common}
A.~K. Smilde, I.~M{\aa}ge, T.~Naes, T.~Hankemeier, M.~A. Lips, H.~A. Kiers,
  E.~Acar, and R.~Bro, ``Common and distinct components in data fusion,'' {\em
  Journal of Chemometrics}, vol.~31, no.~7, p.~e2900, 2017.

\bibitem{gaynanova2017structural}
I.~Gaynanova and G.~Li, ``Structural learning and integrative decomposition of
  multi-view data,'' {\em arXiv preprint arXiv:1707.06573}, 2017.

\bibitem{fan2001variable}
J.~Fan and R.~Li, ``Variable selection via nonconcave penalized likelihood and
  its oracle properties,'' {\em Journal of the American Statistical
  Association}, vol.~96, no.~456, pp.~1348--1360, 2001.

\bibitem{jolliffe2002principal}
I.~Jolliffe, {\em {Principal component analysis}}.
\newblock Wiley Online Library, 2002.

\bibitem{gavish2017optimal}
M.~Gavish and D.~L. Donoho, ``Optimal shrinkage of singular values,'' {\em IEEE
  Transactions on Information Theory}, vol.~63, no.~4, pp.~2137--2152, 2017.

\bibitem{friedman2008sparse}
J.~Friedman, T.~Hastie, and R.~Tibshirani, ``Sparse inverse covariance
  estimation with the graphical lasso,'' {\em Biostatistics}, vol.~9, no.~3,
  pp.~432--441, 2008.

\bibitem{tibshirani2005sparsity}
R.~Tibshirani, M.~Saunders, S.~Rosset, J.~Zhu, and K.~Knight, ``Sparsity and
  smoothness via the fused lasso,'' {\em Journal of the Royal Statistical
  Society: Series B (Statistical Methodology)}, vol.~67, no.~1, pp.~91--108,
  2005.

\bibitem{witten2009penalized}
D.~M. Witten, R.~Tibshirani, and T.~Hastie, ``A penalized matrix decomposition,
  with applications to sparse principal components and canonical correlation
  analysis,'' {\em Biostatistics}, vol.~10, no.~3, pp.~515--534, 2009.

\bibitem{huang2012selective}
J.~Huang, P.~Breheny, and S.~Ma, ``A selective review of group selection in
  high-dimensional models,'' {\em Statistical Science: {A} Review Journal of
  The Institute of Mathematical Statistics}, vol.~27, no.~4, 2012.

\bibitem{tibshirani1996regression}
R.~Tibshirani, ``Regression shrinkage and selection via the lasso,'' {\em
  Journal of the Royal Statistical Society: Series B (Methodological)},
  vol.~58, no.~1, pp.~267--288, 1996.

\bibitem{efron2004least}
B.~Efron, T.~Hastie, I.~Johnstone, R.~Tibshirani, {\em et~al.}, ``Least angle
  regression,'' {\em The Annals of Statistics}, vol.~32, no.~2, pp.~407--499,
  2004.

\bibitem{parikh2014proximal}
N.~Parikh, S.~Boyd, {\em et~al.}, ``Proximal algorithms,'' {\em Foundations and
  Trends{\textregistered} in Optimization}, vol.~1, no.~3, pp.~127--239, 2014.

\bibitem{meinshausen2010stability}
N.~Meinshausen and P.~B{\"u}hlmann, ``Stability selection,'' {\em Journal of
  the Royal Statistical Society: Series B (Statistical Methodology)}, vol.~72,
  no.~4, pp.~417--473, 2010.

\bibitem{armagan2013generalized}
A.~Armagan, D.~B. Dunson, and J.~Lee, ``Generalized double {P}areto
  shrinkage,'' {\em Statistica Sinica}, vol.~23, no.~1, p.~119, 2013.

\bibitem{maage2019performance}
I.~M{\aa}ge, A.~K. Smilde, and F.~M. van~der Kloet, ``Performance of methods
  that separate common and distinct variation in multiple data blocks,'' {\em
  Journal of Chemometrics}, vol.~33, no.~1, p.~e3085, 2019.

\bibitem{mclendon2008comprehensive}
C.~G. A. T.~R. Network {\em et~al.}, ``{Comprehensive genomic characterization
  defines human glioblastoma genes and core pathways},'' {\em Nature},
  vol.~455, no.~7216, pp.~1061--1068, 2008.

\bibitem{iorio2016landscape}
F.~Iorio, T.~A. Knijnenburg, D.~J. Vis, G.~R. Bignell, M.~P. Menden,
  M.~Schubert, N.~Aben, E.~Gon{\c{c}}alves, S.~Barthorpe, H.~Lightfoot, {\em
  et~al.}, ``A landscape of pharmacogenomic interactions in cancer,'' {\em
  Cell}, vol.~166, no.~3, pp.~740--754, 2016.

\bibitem{wu2014detecting}
H.-T. Wu, I.~Hajirasouliha, and B.~J. Raphael, ``{Detecting independent and
  recurrent copy number aberrations using interval graphs},'' {\em
  Bioinformatics}, vol.~30, pp.~i195--i203, June 2014.

\bibitem{young1980quantifying}
F.~W. Young, J.~de~Leeuw, and Y.~Takane, ``Quantifying qualitative data,'' {\em
  Lantermann and H. Feger (Eds.): Similarity and Choice. Papers in Honour of
  Clyde Coombs. Berne: Hans Huber}, 1980.

\bibitem{collins2001generalization}
M.~Collins, S.~Dasgupta, and R.~E. Schapire, ``{A generalization of principal
  component analysis to the exponential family},'' in {\em Advances in Neural
  Information Processing Systems}, MIT Press, 2001.

\bibitem{schein2003generalized}
A.~I. Schein, L.~K. Saul, and L.~H. Ungar, ``A generalized linear model for
  principal component analysis of binary data.,'' in {\em AISTATS}, vol.~3,
  p.~10, 2003.

\bibitem{landgraf2015generalized}
A.~J. Landgraf, {\em {Generalized principal component analysis:
  {D}imensionality reduction through the projection of natural parameters}}.
\newblock PhD thesis, The Ohio State University, 2015.

\bibitem{de2009gifi}
J.~de~Leeuw and P.~Mair, ``{Gifi methods for optimal scaling in R: the package
  homals},'' {\em {Journal of Statistical Software}}, vol.~{31}, pp.~{1--21},
  {AUG} {2009}.

\bibitem{mori2016nonlinear}
Y.~Mori, M.~Kuroda, and N.~Makino, {\em {Nonlinear principal component analysis
  and its applications}}.
\newblock Springer, 2016.

\bibitem{kiers1989three}
H.~A. Kiers, {\em Three-way methods for the analysis of qualitative and
  quantitative two-way data}.
\newblock DSWO press Leiden, 1989.

\bibitem{pearson1901lines}
K.~Pearson, ``{On lines and planes of closest fit to systems of points in
  space},'' {\em Philosophical Magazine Series 6}, vol.~2, no.~11,
  pp.~559--572, 1901.

\bibitem{zou2006sparse}
H.~Zou, T.~Hastie, and R.~Tibshirani, ``{Sparse principal component
  analysis},'' {\em Journal of Computational and Graphical Statistics},
  vol.~15, no.~2, pp.~265--286, 2006.

\bibitem{ten1993least}
J.~M. ten Berge, {\em Least squares optimization in multivariate analysis}.
\newblock DSWO Press, Leiden University Leiden, 1993.

\bibitem{tipping1999probabilistic}
M.~E. Tipping and C.~M. Bishop, ``Probabilistic principal component analysis,''
  {\em Journal of the Royal Statistical Society: Series B (Statistical
  Methodology)}, vol.~61, no.~3, pp.~611--622, 1999.

\bibitem{de2006principal}
J.~De~Leeuw, ``Principal component analysis of binary data by iterated singular
  value decomposition,'' {\em Computational Statistics \& Data analysis},
  vol.~50, no.~1, pp.~21--39, 2006.

\bibitem{udell2016generalized}
M.~Udell, C.~Horn, R.~Zadeh, {\em et~al.}, ``{Generalized low rank models},''
  {\em Foundations and Trends{\textregistered} in Machine Learning}, vol.~9,
  no.~1, pp.~1--118, 2016.

\bibitem{gifi_B_90}
A.~Gifi, {\em {Nonlinear multivariate analysis}}.
\newblock New York, N.Y.: Wiley, 1990.
\newblock This is a publication under a collective pseudonym.

\bibitem{wei2013role}
Q.~Wei and R.~L. Dunbrack~Jr, ``{The role of balanced training and testing data
  sets for binary classifiers in bioinformatics},'' {\em PloS One}, vol.~8,
  no.~7, p.~e67863, 2013.

\bibitem{wold1978cross}
S.~Wold, ``Cross-validatory estimation of the number of components in factor
  and principal components models,'' {\em Technometrics}, vol.~20, no.~4,
  pp.~397--405, 1978.

\bibitem{Bro2008}
R.~Bro, K.~Kjeldahl, A.~K. Smilde, {\em et~al.}, ``Cross-validation of
  component models: {A} critical look at current methods,'' {\em Analytical and
  Bioanalytical Chemistry}, vol.~390, no.~5, pp.~1241--1251, 2008.

\bibitem{RProject}
{R Development Core Team}, {\em {R: {A} language and environment for
  statistical computing}}.
\newblock R Foundation for Statistical Computing, Vienna, Austria, 2008.
\newblock {ISBN} 3-900051-07-0.

\bibitem{stacklies2007pcamethods}
W.~Stacklies, H.~Redestig, M.~Scholz, {\em et~al.}, ``{pcaMethods--a
  bioconductor package providing PCA methods for incomplete data},'' {\em
  Bioinformatics}, vol.~23, no.~9, pp.~1164--1167, 2007.

\bibitem{groenen2016multinomial}
P.~J. Groenen and J.~Josse, ``Multinomial multiple correspondence analysis,''
  {\em arXiv preprint arXiv:1603.03174}, 2016.

\bibitem{davenport20141}
M.~A. Davenport, Y.~Plan, E.~Van Den~Berg, and M.~Wootters, ``1-{B}it matrix
  completion,'' {\em Information and Inference: {A} Journal of the IMA},
  vol.~3, no.~3, pp.~189--223, 2014.

\bibitem{shabalin2013reconstruction}
A.~A. Shabalin and A.~B. Nobel, ``Reconstruction of a low-rank matrix in the
  presence of gaussian noise,'' {\em Journal of Multivariate Analysis},
  vol.~118, pp.~67--76, 2013.

\bibitem{josse2016adaptive}
J.~Josse and S.~Sardy, ``Adaptive shrinkage of singular values,'' {\em
  Statistics and Computing}, vol.~26, no.~3, pp.~715--724, 2016.

\bibitem{candes2009exact}
E.~J. Cand{\`e}s and B.~Recht, ``Exact matrix completion via convex
  optimization,'' {\em Foundations of Computational Mathematics}, vol.~9,
  no.~6, p.~717, 2009.

\bibitem{mazumder2010spectral}
R.~Mazumder, T.~Hastie, and R.~Tibshirani, ``Spectral regularization algorithms
  for learning large incomplete matrices,'' {\em Journal of Machine Learning
  Research}, vol.~11, no.~Aug, pp.~2287--2322, 2010.

\bibitem{fu1998penalized}
W.~J. Fu, ``Penalized regressions: {T}he bridge versus the lasso,'' {\em
  Journal of Computational and Graphical Statistics}, vol.~7, no.~3,
  pp.~397--416, 1998.

\bibitem{de1994block}
J.~De~Leeuw, ``Block-relaxation algorithms in statistics,'' in {\em Information
  Systems and Data Analysis}, pp.~308--324, Springer, 1994.

\bibitem{hunter2004tutorial}
D.~R. Hunter and K.~Lange, ``A tutorial on {MM} algorithms,'' {\em The American
  Statistician}, vol.~58, no.~1, pp.~30--37, 2004.

\bibitem{kiers1997weighted}
H.~A.~L. Kiers, ``Weighted least squares fitting using ordinary least squares
  algorithms,'' {\em Psychometrika}, vol.~62, no.~2, pp.~251--266, 1997.

\bibitem{boyd2004convex}
S.~Boyd and L.~Vandenberghe, {\em Convex optimization}.
\newblock Cambridge University Press, 2004.

\bibitem{lu2015generalized}
C.~Lu, C.~Zhu, C.~Xu, S.~Yan, and Z.~Lin, ``Generalized singular value
  thresholding,'' in {\em Twenty-Ninth AAAI Conference on Artificial
  Intelligence}, 2015.

\bibitem{le2012asymptotics}
L.~Le~Cam and G.~L. Yang, {\em Asymptotics in statistics: {S}ome basic
  concepts}.
\newblock Springer Science and Business Media, 2012.

\bibitem{song2018generalized}
Y.~Song, J.~A. Westerhuis, N.~Aben, L.~F. Wessels, P.~J. Groenen, and A.~K.
  Smilde, ``Generalized simultaneous component analysis of binary and
  quantitative data,'' {\em arXiv preprint arXiv:1807.04982}, 2018.

\bibitem{van2009structured}
K.~Van~Deun, A.~K. Smilde, M.~J. van~der Werf, H.~A.~L. Kiers, and
  I.~Van~Mechelen, ``A structured overview of simultaneous component based data
  integration,'' {\em BMC Bioinformatics}, vol.~10, no.~1, p.~246, 2009.

\bibitem{van2009integrating}
R.~A. van~den Berg, I.~Van~Mechelen, T.~F. Wilderjans, K.~Van~Deun, H.~A.~L.
  Kiers, and A.~K. Smilde, ``Integrating functional genomics data using maximum
  likelihood based simultaneous component analysis,'' {\em BMC Bioinformatics},
  vol.~10, no.~1, p.~340, 2009.

\bibitem{mo2013pattern}
Q.~Mo, S.~Wang, V.~E. Seshan, A.~B. Olshen, N.~Schultz, C.~Sander, R.~S.
  Powers, M.~Ladanyi, and R.~Shen, ``Pattern discovery and cancer gene
  identification in integrated cancer genomic data,'' {\em Proceedings of the
  National Academy of Sciences}, vol.~110, no.~11, pp.~4245--4250, 2013.

\bibitem{collins2002generalization}
M.~Collins, S.~Dasgupta, and R.~E. Schapire, ``A generalization of principal
  components analysis to the exponential family,'' in {\em Advances in Neural
  Information Processing Systems}, pp.~617--624, 2002.

\bibitem{koltchinskii2011nuclear}
V.~Koltchinskii, K.~Lounici, A.~B. Tsybakov, {\em et~al.}, ``Nuclear-norm
  penalization and optimal rates for noisy low-rank matrix completion,'' {\em
  The Annals of Statistics}, vol.~39, no.~5, pp.~2302--2329, 2011.

\bibitem{wu2015fast}
D.~Wu, D.~Wang, M.~Q. Zhang, and J.~Gu, ``Fast dimension reduction and
  integrative clustering of multi-omics data using low-rank approximation:
  {A}pplication to cancer molecular classification,'' {\em BMC Genomics},
  vol.~16, no.~1, p.~1022, 2015.

\bibitem{bro2008cross}
R.~Bro, K.~Kjeldahl, A.~K. Smilde, and H.~A.~L. Kiers, ``Cross-validation of
  component models: {A} critical look at current methods,'' {\em Analytical and
  Bioanalytical Chemistry}, vol.~390, no.~5, pp.~1241--1251, 2008.

\bibitem{liu2007support}
Y.~Liu, H.~H. Zhang, C.~Park, and J.~Ahn, ``Support vector machines with
  adaptive {$L_q$} penalty,'' {\em Computational Statistics \& Data Analysis},
  vol.~51, no.~12, pp.~6380--6394, 2007.

\bibitem{song2017principal}
Y.~Song, J.~A. Westerhuis, N.~Aben, M.~Michaut, L.~F. Wessels, and A.~K.
  Smilde, ``Principal component analysis of binary genomics data,'' {\em
  Briefings in Bioinformatics}, 2017.

\bibitem{akbani2015genomic}
R.~Akbani, K.~C. Akdemir, B.~A. Aksoy, M.~Albert, A.~Ally, S.~B. Amin,
  H.~Arachchi, A.~Arora, J.~T. Auman, B.~Ayala, {\em et~al.}, ``Genomic
  classification of cutaneous melanoma,'' {\em Cell}, vol.~161, no.~7,
  pp.~1681--1696, 2015.

\bibitem{cancer2014comprehensive}
C.~G. A.~R. Network {\em et~al.}, ``Comprehensive molecular profiling of lung
  adenocarcinoma,'' {\em Nature}, vol.~511, no.~7511, p.~543, 2014.

\bibitem{cancer2012comprehensive}
C.~G.~A. Network {\em et~al.}, ``Comprehensive molecular portraits of human
  breast tumours,'' {\em Nature}, vol.~490, no.~7418, p.~61, 2012.

\bibitem{aben2018itop}
N.~Aben, J.~A. Westerhuis, Y.~Song, H.~A.~L. Kiers, M.~Michaut, A.~K. Smilde,
  and L.~F.~A. Wessels, ``{iTOP}: {I}nferring the topology of omics data,''
  {\em Bioinformatics}, vol.~34, pp.~i988--i996, 09 2018.

\bibitem{lee2015learning}
J.~D. Lee and T.~J. Hastie, ``Learning the structure of mixed graphical
  models,'' {\em Journal of Computational and Graphical Statistics}, vol.~24,
  no.~1, pp.~230--253, 2015.

\bibitem{cheng2017high}
J.~Cheng, T.~Li, E.~Levina, and J.~Zhu, ``High-dimensional mixed graphical
  models,'' {\em Journal of Computational and Graphical Statistics}, vol.~26,
  no.~2, pp.~367--378, 2017.

\bibitem{shen2009integrative}
R.~Shen, A.~B. Olshen, and M.~Ladanyi, ``Integrative clustering of multiple
  genomic data types using a joint latent variable model with application to
  breast and lung cancer subtype analysis,'' {\em Bioinformatics}, vol.~25,
  no.~22, pp.~2906--2912, 2009.

\bibitem{van2016separating}
F.~M. van~der Kloet, P.~Sebasti{\'a}n-Le{\'o}n, A.~Conesa, A.~K. Smilde, and
  J.~A. Westerhuis, ``Separating common from distinctive variation,'' {\em BMC
  bioinformatics}, vol.~17, no.~5, p.~S195, 2016.

\bibitem{lock2013joint}
E.~F. Lock, K.~A. Hoadley, J.~S. Marron, and A.~B. Nobel, ``Joint and
  individual variation explained ({JIVE}) for integrated analysis of multiple
  data types,'' {\em The Annals of Applied Statistics}, vol.~7, no.~1, p.~523,
  2013.

\bibitem{lofstedt2013global}
T.~L{\"o}fstedt, D.~Hoffman, and J.~Trygg, ``Global, local and unique
  decompositions in {OnPLS} for multiblock data analysis,'' {\em Analytica
  Chimica Acta}, vol.~791, pp.~13--24, 2013.

\bibitem{schouteden2014performing}
M.~Schouteden, K.~Van~Deun, T.~F. Wilderjans, and I.~Van~Mechelen, ``Performing
  {DISCO-SCA} to search for distinctive and common information in linked
  data,'' {\em Behavior Research Methods}, vol.~46, no.~2, pp.~576--587, 2014.

\bibitem{maage2012preference}
I.~M{\aa}ge, E.~Menichelli, and T.~N{\ae}s, ``Preference mapping by {PO-PLS}:
  Separating common and unique information in several data blocks,'' {\em Food
  Quality and Preference}, vol.~24, no.~1, pp.~8--16, 2012.

\bibitem{maage2018performance}
I.~M{\aa}ge, A.~K. Smilde, and F.~M. van~der Kloet, ``Performance of methods
  that separate common and distinct variation in multiple data blocks,'' {\em
  Journal of Chemometrics}, p.~e3085, 2018.

\bibitem{klami2015group}
A.~Klami, S.~Virtanen, E.~Lepp{\"a}aho, and S.~Kaski, ``Group factor
  analysis,'' {\em IEEE Transactions on Neural Networks and Learning Systems},
  vol.~26, no.~9, pp.~2136--2147, 2015.

\bibitem{perry2009cross}
P.~O. Perry, ``Cross-validation for unsupervised learning,'' {\em arXiv
  preprint arXiv:0909.3052}, 2009.

\bibitem{argelaguet2018multi}
R.~Argelaguet, B.~Velten, D.~Arnol, S.~Dietrich, T.~Zenz, J.~C. Marioni,
  F.~Buettner, W.~Huber, and O.~Stegle, ``Multi-{O}mics {F}actor {A}nalysis—a
  framework for unsupervised integration of multi-omics data sets,'' {\em
  Molecular Systems Biology}, vol.~14, no.~6, p.~e8124, 2018.

\bibitem{meinshausen2006high}
N.~Meinshausen, P.~B{\"u}hlmann, {\em et~al.}, ``High-dimensional graphs and
  variable selection with the lasso,'' {\em The Annals of Statistics}, vol.~34,
  no.~3, pp.~1436--1462, 2006.

\bibitem{leng2006note}
C.~Leng, Y.~Lin, and G.~Wahba, ``A note on the lasso and related procedures in
  model selection,'' {\em Statistica Sinica}, pp.~1273--1284, 2006.

\bibitem{dietrich2018drug}
S.~Dietrich, M.~Ole{\'s}, J.~Lu, L.~Sellner, S.~Anders, B.~Velten, B.~Wu,
  J.~H{\"u}llein, M.~da~Silva~Liberio, T.~Walther, {\em et~al.},
  ``Drug-perturbation-based stratification of blood cancer,'' {\em The Journal
  of Clinical Investigation}, vol.~128, no.~1, pp.~427--445, 2018.

\bibitem{van2011flexible}
K.~Van~Deun, T.~F. Wilderjans, R.~A. Van~den Berg, A.~Antoniadis, and
  I.~Van~Mechelen, ``A flexible framework for sparse simultaneous component
  based data integration,'' {\em BMC Bioinformatics}, vol.~12, no.~1, p.~448,
  2011.

\bibitem{acar2015data}
E.~Acar, R.~Bro, and A.~K. Smilde, ``Data fusion in metabolomics using coupled
  matrix and tensor factorizations,'' {\em Proceedings of the IEEE}, vol.~103,
  no.~9, pp.~1602--1620, 2015.

\bibitem{song2019logistic}
Y.~Song, J.~A. Westerhuis, and A.~K. Smilde, ``Logistic principal component
  analysis via non-convex singular value thresholding,'' {\em arXiv preprint
  arXiv:1902.09486}, 2019.

\bibitem{gelman2013bayesian}
A.~Gelman, H.~S. Stern, J.~B. Carlin, D.~B. Dunson, A.~Vehtari, and D.~B.
  Rubin, {\em Bayesian data analysis}.
\newblock Chapman and Hall/CRC, 2013.

\bibitem{smilde2008matrix}
A.~K. Smilde, H.~A. Kiers, S.~Bijlsma, C.~Rubingh, and M.~Van~Erk, ``Matrix
  correlations for high-dimensional data: the modified {RV}-coefficient,'' {\em
  Bioinformatics}, vol.~25, no.~3, pp.~401--405, 2008.

\bibitem{frazier2018tutorial}
P.~I. Frazier, ``A tutorial on bayesian optimization,'' {\em arXiv preprint
  arXiv:1807.02811}, 2018.

\bibitem{snoek2012practical}
J.~Snoek, H.~Larochelle, and R.~P. Adams, ``Practical bayesian optimization of
  machine learning algorithms,'' in {\em Advances in Neural Information
  Processing Systems}, pp.~2951--2959, 2012.

\bibitem{boyd2011distributed}
S.~Boyd, N.~Parikh, E.~Chu, B.~Peleato, J.~Eckstein, {\em et~al.},
  ``Distributed optimization and statistical learning via the alternating
  direction method of multipliers,'' {\em Foundations and
  Trends{\textregistered} in Machine learning}, vol.~3, no.~1, pp.~1--122,
  2011.

\bibitem{anderson2018xpca}
C.~Anderson-Bergman, T.~G. Kolda, and K.~Kincher-Winoto, ``{XPCA}: {E}xtending
  {PCA} for a combination of discrete and continuous variables,'' {\em arXiv
  preprint arXiv:1808.07510}, 2018.

\bibitem{geladi1986partial}
P.~Geladi and B.~R. Kowalski, ``Partial least-squares regression: a tutorial,''
  {\em Analytica Chimica Acta}, vol.~185, pp.~1--17, 1986.

\bibitem{smilde2005anova}
A.~K. Smilde, J.~J. Jansen, H.~C. Hoefsloot, R.-J.~A. Lamers, J.~Van Der~Greef,
  and M.~E. Timmerman, ``{ANOVA}-simultaneous component analysis ({ASCA}): a
  new tool for analyzing designed metabolomics data,'' {\em Bioinformatics},
  vol.~21, no.~13, pp.~3043--3048, 2005.

\end{thebibliography}

\end{document}